\documentclass[11pt]{article}
\usepackage[T1]{fontenc}
\usepackage[utf8]{inputenc}
\usepackage{lmodern}
\usepackage[margin=1in]{geometry}
\usepackage{microtype}
\usepackage{amsmath,amsthm,amssymb}
\usepackage{mathtools}
\usepackage{booktabs}
\usepackage{array}
\usepackage{longtable}
\usepackage{graphicx}
\usepackage{url}
\makeatletter\providecommand{\path}{\begingroup\Url}\makeatother
\usepackage{enumitem}
\usepackage{pgfplots}
\pgfplotsset{compat=1.18}
\usetikzlibrary{positioning,arrows.meta}
\usepackage[numbers,sort&compress]{natbib}
\usepackage[colorlinks=true,linkcolor=blue!50!black,citecolor=blue!50!black,urlcolor=blue!50!black]{hyperref}

\theoremstyle{plain}
\newtheorem{theorem}{Theorem}[section]
\newtheorem{proposition}[theorem]{Proposition}
\newtheorem{lemma}[theorem]{Lemma}
\newtheorem{corollary}[theorem]{Corollary}
\newtheorem{conjecture}[theorem]{Conjecture}
\theoremstyle{definition}

\newcommand{\entry}[1]{\paragraph{\normalfont\bfseries #1}}

\newcommand{\PP}{\mathbb{P}}

\newcommand{\bstar}{\beta^{*}}
\newcommand{\ip}[2]{\left\langle #1, #2 \right\rangle}
\newcommand{\norm}[1]{\left\lVert #1 \right\rVert}
\newcommand{\Ltwo}{L^{2}}
\newcommand{\Fop}{\mathcal{F}}

\title{Coercive at the equilibrium, monotone nowhere: exact convergence certificates for learning in first-price auctions}
\author{Benjamin Lubin\\Boston University\\\texttt{blubin@bu.edu}}
\date{November 3, 2026}
\hypersetup{pdftitle={Coercive at the equilibrium, monotone nowhere: exact convergence certificates for learning in first-price auctions},pdfauthor={Benjamin Lubin}}

\newcommand{\LeanFileCount}{27}
\newcommand{\LeanLineCount}{6{,}315}

\newcommand{\LeanTheoremCount}{338}

\newcommand{\MathlibVersion}{v4.34.1}

\begin{document}
\maketitle

\begin{abstract}
The Bayes--Nash equilibria of an auction are the solutions of a
variational inequality, and for standard auctions its operator fails every
monotonicity notion. The existence, sensitivity and convergence theory of
monotone operators, which learning in auctions and mechanism design for
learnability would otherwise rest on, is therefore unavailable. We show
what replaces monotonicity for
the first-price auction: a certificate that is strong but local. For uniform
values the single-population operator has an exact closed form and an exact
Minty identity at the equilibrium; for two bidders its Gateaux derivative
there is $\Ltwo$-coercive with best constant exactly one in every
piecewise-linear bid space, uniformly in the number of pieces $m$, while the
Minty inequality holds only on a ball whose radius decays like $m^{-5/2}$.
The decay is exact: $m^{5/2}$ times the radius converges to
$1/(8\cdot 3^{1/4})$, and the worst perturbations converge to an alternating
geometric profile with ratio $-(2-\sqrt3)$, peaked $\Theta(\log m)$ knots
below the top; both follow from a sharp inequality for a leading-order
problem.
Refinement shrinks the ball, not the constant. Bidders who learn
independently, rather than as one population, see the same constant. For a
general prior and bidder count there are two certificates: the
function-space certificate at the true equilibrium, and the learners'
certificate at the equilibrium of the finite-dimensional bid space a
learning algorithm must run in. The first is an explicit weight $w_F$,
positive for every log-concave prior; it equals the second exactly when the
equilibrium bid is linear, that is, for the value distributions $F(v) = v^a$.
Otherwise the learners' certificate
differs from it by knot masses, because the discretized equilibrium's
slopes oscillate with a law
proved for two and three bidders and for four when a computer-verified
bottom layer covers the prior. The limit of the learners' certificate is
then proved in closed form. It is the minimum over values of the weight
$w_F$ corrected by what that oscillation contributes. This function of the
prior alone matches the computed constants and is the characterization: in
the limit the learners' certificate is positive uniformly in $m$ exactly
when that minimum is positive. The limit is approached at a known rate, so a
positive limit gives a positive certificate at every large $m$, and for one
prior at every $m$ by interval arithmetic. For more bidders the same limit
is conjectured.
The exact identities and the leading-order inequalities are machine-checked
in Lean.

\end{abstract}

\noindent\textbf{Keywords.} variational inequalities, learning in auctions, first-price auction, convergence certificates, mechanism design

\section{Introduction}
\label{sec:intro}

Whether bidders who learn their bids converge to a Bayes--Nash equilibrium
is a question about an operator. Each bidder follows the gradient of her
expected utility; the equilibria are the solutions of a variational
inequality (VI) over type-contingent strategies \citep{ui2016bayesian}; and
the convergence theory of such dynamics rests on monotonicity of the VI's
operator. \citet{bichler2025convergence}
showed that for standard auctions the operator fails every monotonicity
notion, even for two bidders with uniform values, and that for the
first-price auction there is no Minty solution at all, no bid function
$\bstar$ with $\ip{\Fop(\beta)}{\beta - \bstar} \ge 0$ for every admissible
$\beta$, which is the condition that makes an equilibrium a global
attractor of the dynamics; yet the same paper
and its journal version \citep{bichler2025convergenceJ} prove convergence of
the two-piece first-price flow by a Lyapunov function and observe convergence
numerically far beyond. Something replaces monotonicity. This paper says
what, computes it exactly where it can be computed, and characterizes how
far it reaches.

\paragraph{The object.} The operator is the single-population field of the
symmetric first-price auction on piecewise-linear bid functions with $m$
pieces, the finite-dimensional space any learning algorithm runs in, with
results uniform in the number of pieces. The results concern
its Gateaux derivative at the equilibrium and the Minty pairing
$\ip{\Fop(\bstar+\eta)}{\eta}$ in a neighbourhood of it: the field at a
perturbed bid, paired with the perturbation, whose positivity for every
admissible $\eta$ is the Minty inequality and says that the field points
back toward $\bstar$ from everywhere. Whoever fixes a
mechanism and relies on the bidders reaching its equilibrium, a designer
most of all, needs the convergence as a hypothesis that holds rather than
an observation; a \emph{certificate} is that hypothesis: a verifiable
sufficient condition, at a given mechanism and equilibrium, for the
learners' dynamics to converge to that equilibrium
(Section~\ref{sec:certificates}). Which certificate holds, and with what
constant, is then a property of the mechanism that a designer can carry as
data; \citet{lubin2026designing} investigates mechanism design under such a
constraint.

\paragraph{The answer in one sentence.} Monotonicity is replaced by a
certificate that is strong but local: at the equilibrium the operator's
derivative is coercive, that is, the quadratic form
$\ip{D\Fop(\bstar)\eta}{\eta}$ dominates a fixed multiple of $\norm\eta^2$,
the multiple being its constant, or modulus, with best constant exactly one for two
bidders with uniform values in every piecewise-linear space, while the
neighbourhood on
which the Minty inequality holds shrinks as the space is refined, at an
exact rate and in an exact direction. Refinement shrinks the neighbourhood,
not the modulus. For a general prior the certificate at the true
equilibrium is an explicit weight, and the certificate the learners see, at
the discretized equilibrium, differs from it by an explicit oscillation
whose closed-form limit decides positivity.

\paragraph{Main results.} In the order of the three arcs of the paper.
\begin{enumerate}[leftmargin=*,itemsep=2pt]
\item \emph{An exact field and an exact Minty identity} for uniform values
  (Lemma~\ref{lem:field}, Proposition~\ref{prop:minty}): the operator is a
  closed-form rational expression in the bid and its slope, and everything
  in the uniform case is read off one identity with no remainder.
\item \emph{Modulus exactly one} (Theorem~\ref{thm:modulus}): for two
  bidders the Gateaux derivative at the equilibrium is $\Ltwo$-coercive with
  best constant $1$ on every piecewise-linear space with at least two
  pieces, uniformly in $m$; for $n \ge 3$ the constant decays like
  $m^{-(n-2)}$.
\item \emph{What fails is a radius} (Theorems~\ref{thm:sharp},
  \ref{thm:radius}, \ref{thm:direction}): the Minty inequality holds on an
  $\Ltwo$ ball of radius $\asymp m^{-5/2}$, with $m^{5/2}$ times the radius
  converging to $1/(8\cdot3^{1/4})$ and the minimizers converging to an
  alternating geometric profile with ratio $-(2-\sqrt3)$, whose peak sits
  $\Theta(\log m)$ grid points (knots) below the top; the constant is the
  sharp constant
  of a leading-order inequality proved by a potential argument.
\item \emph{Independent learners see the same modulus}
  (Corollary~\ref{cor:twoplayer}, Proposition~\ref{prop:nplayer}): the
  two-bidder directional form has best constant $1$, and with $n$ bidders
  the weight is the single-population one; the term that keeps the form
  positive is the cap on the win probability.
\item \emph{The function-space certificate for a general prior is a
  weight} (Proposition~\ref{prop:general}, Theorem~\ref{thm:power}): the
  derivative is coercive uniformly in $m$ iff $\operatorname{ess\,inf}
  w_F/f > 0$; the equilibrium is linear exactly for the power priors, for
  which the discretized certificate equals the function-space one; every
  log-concave prior has a positive weight, and priors whose equilibrium bid
  is steeper than $n-1$ a negative one.
\item \emph{The discretized certificate has a closed-form limit}
  (Theorems~\ref{thm:oscillation2}, \ref{thm:limit2}, \ref{thm:four},
  \ref{thm:layer}; Theorems~\ref{thm:interior}, \ref{thm:airy},
  Corollaries~\ref{cor:eventual}, \ref{cor:pilot}): the discretized
  equilibrium's slopes oscillate with a law, proved for two and three
  bidders and for four when a computer-verified bottom layer covers the
  prior; the certificate's limit is $\min_{[0,1]}E_F^{\mathrm{osc}}$, the
  weight plus an explicit oscillation correction, and this is the
  characterization: the piecewise-linear certificate is positive uniformly
  in $m$ in the limit exactly when $\min E_F^{\mathrm{osc}} > 0$, a condition
  on $F$ alone. The approach to the limit has a known rate with a positive
  leading term, so a positive limit gives a positive certificate at every
  large $m$, and for one prior at every $m$ by interval arithmetic. For
  five or more bidders the same limit is conjectured.
\end{enumerate}
The exact identities behind these results, the leading-order
inequalities of item 3 and the Bloch inequality of item 6 are
machine-checked in Lean~4 (Appendix~\ref{app:lean}).

\paragraph{How to read the paper.} Section~\ref{sec:related} places the
work and Section~\ref{sec:setup} fixes the operator, the spaces and the
certificates. The results are organized in three arcs, each with its
framing, its key results, and a sketch of how the framing reaches them:
Arc~I, the uniform prior, where everything is exact
(Section~\ref{sec:exact}); Arc~II, the learners, where the one-population
operator is shown to be what independent learners see
(Section~\ref{sec:twoplayer}); Arc~III, general priors and bidder counts,
where the two certificates part and the discretization's oscillation is
the mechanism (Section~\ref{sec:general}). Section~\ref{sec:conclusion}
lists what is open. The appendices contain, for each arc, a map and the
chain of its results, with what each establishes and how it is proved
(Appendix~\ref{app:arcs}); the proofs (Appendix~\ref{app:proofs}); what
is machine-checked in Lean, and what each check takes as given
(Appendix~\ref{app:lean}); the computational records behind every number
(Appendix~\ref{app:numerics}); the three roles the computer played, in
the theory, in the proofs and in the checking, and what each establishes
(Appendix~\ref{app:computer}); and a primer on the concepts and
techniques used, for readers unfamiliar with any of them
(Appendix~\ref{app:primer}).

\section{Related work}
\label{sec:related}

\paragraph{Variational inequalities for equilibria.} Nash equilibria of
concave games are VI solutions \citep{rosen1965existence,facchinei2003finite}, and Bayes--Nash
equilibria of smooth concave Bayesian games are VI solutions in $\Ltwo$ over
type-contingent strategies \citep{ui2016bayesian}, with uniqueness under a
statewise negative-definite payoff Jacobian. For auctions, \citet{bichler2025convergence}
pose the VI on uniformly increasing bid functions, prove that the first- and
second-price operators are not monotone in any sense, show that second price
admits a Minty solution and first price does not, and, in the journal version \citep{bichler2025convergenceJ}, give the one
convergence certificate for first-price learning, a Lyapunov function for the
symmetric flow in a two-piece bid space by hand and sum-of-squares
certificates for three and four pieces numerically. Their reading of the
negative result is that its complexity persists in the piecewise-linear
class, where the Minty violators live, and that a convergence proof needs a
Lyapunov function rather than a monotonicity condition;
Theorem~\ref{thm:modulus} and the radius theorems of Arc~I say in what sense
that is right: in every piecewise-linear space the modulus at the
equilibrium is intact, and it is only the Minty neighbourhood that shrinks
with the number of pieces, which is why the function-space negative result
does not contradict the convergence of the flow. Sensitivity of VI solutions
in a parameter is classical under strong regularity
\citep{robinson1980strongly}, and differentiation through a VI layer under
local strong monotonicity is the device of \citet{li2020endtoend}; neither
had been available for an auction operator.

\paragraph{Certificates.} In optimization and control a \emph{certificate}
is a verifiable object that proves a property of a system without solving
the system: a dual feasible point certifies a bound on an optimal value
\citep{boyd2004convex}; a Lyapunov function certifies stability of an
equilibrium, a barrier function certifies that a set is never left
\citep{prajna2004safety}, and sum-of-squares programming turns the search
for such functions into a semidefinite program whose solution is the
certificate \citep{parrilo2003semidefinite,papachristodoulou2002construction}.
The word carries two commitments: the object is checkable independently of
how it was found, and it is sufficient for the property, not necessary. In
the VI and learning-in-games literature the same role is played by
hypotheses rather than by named objects: strong or strict monotonicity of
the operator, variational stability of the equilibrium, or a Lyapunov
function for the mean dynamics \citep{facchinei2003finite,
mertikopoulos2019learning}. We use the word for this whole family, and
Section~\ref{sec:certificates} fixes the ones we use and their strengths.
The certificate is indexed by the mechanism, so which certificate is
available is itself a property of the design; \citet{lubin2026designing}
uses that property as a constraint of the designer's program, in the way a
transient-stability condition enters stability-constrained optimal power
flow \citep{gan2000stability}, and finds that for asymmetric bidders only
the Hurwitz property survives.

\paragraph{Learning in games and the Bichler program.} Convergence of
no-regret and gradient dynamics to variationally stable equilibria under
noise is due to \citet{mertikopoulos2019learning}; the limit sets of
stochastic approximation are the chain-transitive sets of the mean ODE
\citep{benaim2005stochastic}. For auctions, the discretized
distributional-strategy game of \citet{bichler2025computing} with its
discretization bound, and the uniqueness of Bayesian coarse-correlated
equilibria of \citet{ahunbay2025uniqueness} on symmetric increasing supports,
are what make a learned outcome well defined. The broader program of
computing equilibria by learning and designing mechanisms by learning is
surveyed in \citet{bichler2025differentiable}; there the two problems are
separate pipelines, and the design side rests on strategy-proofness. The
competitive newsvendor game of \citet{koeck2026deep} is the one Bayesian
game with a strict-monotonicity proof in that program. The program's own
answer to what characterizes learnability beyond monotonicity is
\emph{potentialness} \citep{bichler2025potentialness}, the share of the
potential component in the Hodge decomposition of a finite normal-form
game's deviation flow: an empirical predictor of convergence in random
matrix games and in small discretized auctions, with no convergence theorem
and no reach to the operator studied here. The certificates below are the
per-operator, per-equilibrium, proved counterpart. When the followers' game
is strongly monotone at every admissible leader action, finite-time
tracking rates for a leader facing contracting stochastic followers follow
with no further certificate \citep{frankel2025finite}; the certificates
here are for the auction, where that hypothesis fails. The literature on
designing the induced game for learnability
\citep{mathevet2010supermodular,healy2012designing,parkes2004learnable}
asks the design question these certificates make precise.

\section{Setting}
\label{sec:setup}

\paragraph{The auction and the operator.} $n$ bidders have independent
private values with a common distribution $F$ on
$[0,1]$, uniform unless stated otherwise, and bid in a first-price auction
with uniform tie-breaking. A bid function $b : [0,1] \to [0,\bar b]$,
$\bar b \ge 1$, is
nondecreasing, absolutely continuous, $b(0) = 0$, with $b' \ge \delta > 0$
almost everywhere for some $\delta$. In the symmetric (single-population)
formulation all bidders use $b$; a deviator of value $v$ bidding $z$ wins
with probability $W(z)^{n-1}$, $W(z) = \PP(b(V') \le z) = \min(F(b^{-1}(z)),
1)$, and has payoff density $\varphi(v,z) = (v-z)W(z)^{n-1}$. The ex-ante
utility of a deviator using $b'$ is $U(b';b) = \int_0^1 \varphi(v, b'(v))
\,dF(v)$, and the operator is $\Fop(b) = -\varphi_z(\cdot, b(\cdot))$,
identified through the $\Ltwo(dF)$ pairing: $\ip{\Fop(b)}{\eta} = -\int
\varphi_z(v,b(v))\eta(v)\,dF(v)$. A symmetric Bayes--Nash equilibrium is a
$\bstar$ with $\ip{\Fop(\bstar)}{b - \bstar} \ge 0$ for every admissible $b$,
the VI of \citet{ui2016bayesian} in the form of \citet{bichler2025convergence}.
This is the Stampacchia form of the VI; its Minty form asks for
$\ip{\Fop(\beta)}{\beta - \bstar} \ge 0$ for every admissible $\beta$, which
for a continuous operator implies it, and which monotonicity would make
equivalent to it (Appendix~\ref{primer:vi}); the certificates of
Section~\ref{sec:certificates} are about how much of the Minty form
survives near $\bstar$ when monotonicity fails. For uniform values
$\bstar(v) = (n-1)v/n$. The independent-learners formulation gives each
bidder her own bid function and evaluates her win probability against the
others'; Section~\ref{sec:twoplayer}. Every result states its bidder
count: $n = 2$, a named $n$, or general $n$.

\paragraph{The spaces.} The $m$-piece space is the set of bid functions that
are piecewise linear on the grid $v_j = j/m$ with $b(0) = 0$. The grid
points $v_j$ are the \emph{knots}, the places where the slope may change,
and a function in the space is determined by its values there, its knot
values, so the space has dimension $m$ (Appendix~\ref{primer:galerkin});
it contains $\bstar$ for uniform values. Learning runs
in these spaces \citep{bichler2025convergence}. The discretization belongs
to the learner, not to the bidders: an algorithm can represent a bid
function only through finitely many parameters, the $m$-piece space is that
representation, and the results below hold for every $m$, with constants
uniform in $m$ where the statement says so, so that nothing about bidders'
behaviour is assumed beyond the dynamics. Every statement about
``uniformly in $m$'' is a statement over the family of them. The dynamics
we certify are the projected gradient flow and its stochastic
discretizations in the $m$-piece space, with the three notions of limit
(last iterate, empirical distribution, mean-ODE attractor) of the learner
model in \citet{lubin2026designing}, whose convergence statements assume
the certificates below as their hypotheses.

\subsection{Certificates}
\label{sec:certificates}

A \emph{certificate} is a verifiable sufficient condition, at a given
mechanism and equilibrium, for the projected gradient flow and its
stochastic discretizations introduced above to converge to that
equilibrium: the hypothesis a convergence theorem assumes, named so that
it can be stated as a property of the mechanism. The certificates used
here, from strongest to weakest, are the following.

\paragraph{Monotonicity.} $\Fop$ is \emph{monotone} if
\[
  \ip{\Fop(\beta) - \Fop(\beta')}{\beta - \beta'} \ge 0
\]
for all admissible $\beta, \beta'$, \emph{strictly} so if the inequality is
strict for $\beta \ne \beta'$, and \emph{strongly} so with modulus $\mu$ if
the right side is $\mu\norm{\beta - \beta'}^2$. Monotonicity says the field
never points back toward where a bidder came from: every first-order
dynamic then makes progress toward the solution from anywhere, which is
why it is the hypothesis of the classical VI convergence theory
(Appendix~\ref{primer:vi}), and under strong monotonicity the solution is
unique and depends Lipschitz-continuously on the mechanism. It is global,
and it is what standard auctions lack.

\paragraph{The two local Jacobian conditions.} At a solution $\bstar$ let
$D\Fop(\bstar)$ be the Jacobian, the linear part of the field near the
equilibrium (Appendix~\ref{primer:gateaux}), restricted to the critical
cone, the directions in which the solution can move without leaving the
feasible set. \textbf{(C-sym)}: the symmetric part of $D\Fop(\bstar)$ is
positive definite. This is monotonicity in a neighbourhood, hence
\emph{variational stability}: the field points away from the equilibrium
in every feasible direction, and every first-order dynamic introduced
above, deterministic or stochastic, converges locally. \textbf{(C-Hur)}:
the projected Jacobian is Hurwitz, every eigenvalue in the open left
half-plane. This gives exponential stability of the mean dynamics, the ODE
the stochastic learners follow on average, but not variational stability,
so it certifies the mean field only.

\paragraph{Lyapunov and finite-game conditions.} A \emph{Lyapunov
certificate} is a function with class-$\mathcal{K}$ bounds, zero at the
equilibrium and growing with the distance from it, that decreases along
the projected flow; it sits between the two Jacobian conditions according
to the region on which it decreases. A \emph{best-response contraction}
\citep{healy2012designing} is a diagonal-dominance condition, incomparable
with monotonicity. In the discretized game a pure equilibrium is
\emph{strict} if each best response to it is unique, which is equivalent
to variational stability there \citep[Proposition~5.2]{mertikopoulos2019learning}
and necessary for a unique pure Bayesian coarse correlated equilibrium,
the outcome no-regret learners converge to in the Bayesian game
\citep[Corollary~3.2]{ahunbay2025uniqueness}.

Theorem~\ref{thm:modulus} and Corollary~\ref{cor:twoplayer} establish
C-sym at the uniform-prior equilibrium with constant $1$; Arc~III is about
when C-sym holds uniformly in the grid for other priors.

What standard auctions satisfy is the obstruction this paper works around.
For two bidders with uniform values, the symmetric one-deviator operators of
the second- and first-price auctions on $K_\delta$ are neither monotone,
pseudo-monotone nor quasi-monotone for $\delta < 9/100$ and $\delta \le 1/10$
respectively; second price has a Minty solution (the truthful strategy, for
every $\delta \le 1$), first price has none for $\delta \le 1/5$
\citep{bichler2025convergence,bichler2025convergenceJ}.

\section{Arc I: the uniform prior, where everything is exact}
\label{sec:exact}

With uniform values the single-population operator is a rational function
of the bid and its slope, and the Minty pairing at a perturbed bid is one
explicit integral (Section~\ref{sec:field}). The arc answers two questions
about the certificate. Does it hold at the equilibrium, and with what
constant? Yes, with constant exactly one in every piecewise-linear space
with at least two pieces, so the operator is locally strongly monotone on
each of them (Section~\ref{sec:modulus}). On what neighbourhood does it
hold? On a ball around the equilibrium whose radius we determine exactly:
it decays like $m^{-5/2}$, with a limiting value of $m^{5/2}$ times the
radius that we compute in closed form, and the direction in which the
inequality is first violated is an explicit profile
(Section~\ref{sec:radius}). Together they are
the arc's message: the certificate is uniformly strong and uniformly
local. Throughout this arc values are uniform on
$[0,1]$, and bid functions are those of Section~\ref{sec:setup}, whose
slope bound $b' \ge \delta > 0$ makes $b^{-1}$ a $(1/\delta)$-Lipschitz
function. The map of the arc and its chain of results are
Appendix~\ref{app:arc1}; the proofs are in Appendix~\ref{app:proofs-arc1}.

\subsection{The field and the identity}
\label{sec:field}

Opponents use $b$; a deviator of value $v$ bids $z$. Since values are
uniform, $W(z) = \min(b^{-1}(z), 1)$, the win probability is $W(z)^{n-1}$
and the payoff density is $\varphi(v,z) = (v-z)W(z)^{n-1}$; the field is
$g(v;b) := \varphi_z(v,b(v))$ and $\Fop(b) = -g(\cdot;b)$ in the $\Ltwo(dv)$
pairing (Appendix~\ref{primer:gateaux}). Differentiating the payoff
density along the diagonal $z = b(v)$ gives the field in closed form.

\begin{lemma}[pointwise field]
\label{lem:field}
For almost every $v$,
\[
  g(v;b) := \varphi_z(v,b(v)) = -v^{n-1} + (n-1)\,v^{n-2}\,\frac{v - b(v)}{b'(v)}.
\]
\end{lemma}

The cap is inert along $z = b(v)$ and matters only when a bid is compared
with a \emph{different} opponent's bid function (Arc~II). For
piecewise-linear $b$ the field is a smooth function of the knot values
with no one-sided terms, and at $\bstar(v) = (n-1)v/n$ it vanishes
identically: this is the symmetric equilibrium, and it lies in every
piecewise-linear space on a uniform grid because it is linear.

Now perturb the equilibrium. For $b = \bstar + \eta$ the lemma gives the
field at the perturbed bid in closed form, since $v - b = v/n - \eta$ and
$b' = (n-1)/n + \eta'$, and pairing it with the perturbation turns the
Minty pairing into one explicit integral. The field's vanishing at $\bstar$
is what makes the pairing start at second order in $\eta$.

\begin{proposition}[the exact Minty identity]
\label{prop:minty}
Let $b = \bstar + \eta$ with $\eta$ absolutely continuous, $\eta(0) = 0$,
$\operatorname{ess\,inf}\eta' > -(n-1)/n$ and $(n-1)/n + \eta(1) \le \bar b$. Then
\[
  \ip{\Fop(\bstar+\eta)}{\eta}
  = \int_0^1 v^{n-2}\,\frac{\bigl(v\eta' + (n-1)\eta\bigr)\,\eta}{(n-1)/n + \eta'}\,dv,
  \qquad
  \ip{\Fop(\bstar+\eta)}{\eta} = \int_0^1 \frac{\eta^2 + v\eta\eta'}{1/2 + \eta'}\,dv
  \ \text{ for } n = 2.
\]
\end{proposition}

The identity is exact for every admissible $\eta$; it is the Minty pairing
with reference point $\bstar$, and everything in this arc is read off it.

\subsection{The modulus: the certificate at the equilibrium}
\label{sec:modulus}

For $\eta \in W^{1,\infty}(0,1)$ with $\eta(0) = 0$ and $\norm{\eta'}_\infty <
(n-1)/n$, the map $t \mapsto \ip{\Fop(\bstar + t\eta)}{t\eta}$ is analytic on
$|t| < 1$ by Proposition~\ref{prop:minty}, and its coefficient of $t^2$ is
$\ip{D\Fop(\bstar)\eta}{\eta}$, the Gateaux derivative of $\Fop$ at $\bstar$
applied to $\eta$ and paired with $\eta$.

\begin{theorem}[exact quadratic form]
\label{thm:modulus}
For such $\eta$,
\[
  \ip{D\Fop(\bstar)\eta}{\eta} = Q_n(\eta)
  := \frac n2 \int_0^1 v^{n-2}\eta^2\,dv + \frac{n}{2(n-1)}\,\eta(1)^2,
\]
and the right side extends $Q_n$ to $\{\eta \in H^1(0,1) : \eta(0) = 0\}$. In
particular
\[
  Q_2(\eta) = \int_0^1 \eta^2\,dv + \eta(1)^2 \;\ge\; \norm{\eta}_{\Ltwo}^2,
  \quad\text{with equality iff } \eta(1) = 0.
\]
Hence for $n = 2$ the Gateaux derivative of the single-population operator at
$\bstar$ is $\Ltwo(dv)$-coercive with best constant exactly $1$ on every
subspace of $\{\eta \in H^1 : \eta(0)=0\}$ containing a nonzero $\eta$ with
$\eta(1) = 0$, in particular on every piecewise-linear space on a uniform grid
with $m \ge 2$ pieces, uniformly in $m$; on each such finite-dimensional space
$\Fop$ is locally strongly monotone at $\bstar$. For $n \ge 3$ the best
constant on the $m$-piece space is at most $(n/2)(2/m)^{n-2}$.
\end{theorem}

The boundary term $\eta(1)^2$ is the top value's own-bid curvature and is
what keeps the constant from degrading with $m$; for $m = 1$ the only
direction is $\eta = cv$, with ratio $4$, so ``$m \ge 2$'' is needed. The
identity holds in function space, and $Q_2$ is coercive on all of $\{\eta
\in H^1 : \eta(0) = 0\}$; local strong monotonicity of $\Fop$ on an
infinite-dimensional space would require control of the remainder of
Proposition~\ref{prop:minty} uniformly in $\eta$, which the next subsection
shows is not available. On each finite-dimensional space the derivative is
coercive, so the Minty inequality holds on some ball around the
equilibrium there; the next subsection asks how large that ball is.

\subsection{The Minty ball: the domain of the certificate}
\label{sec:radius}

\paragraph{The size of the ball.}
Take $n = 2$ and a piecewise-linear $\eta$ with knots $v_j = j/m$ and slopes
$\eta_j'$ on piece $j$, so $w_j := 1/(1/2 + \eta_j')$ is constant there.
Integrating the $v\eta\eta'$ term of Proposition~\ref{prop:minty} by parts on
each piece (Appendix~\ref{primer:knots}),
\begin{equation}
\label{eq:knots}
  \ip{\Fop(\bstar+\eta)}{\eta}
  = \frac12\Bigl[\int_0^1 w\eta^2\,dv - \sum_{j=1}^{m-1} v_j\,(w_{j+1} - w_j)\,\eta(v_j)^2
  + w_m\,\eta(1)^2\Bigr].
\end{equation}
The only negative contributions are knot terms where the slope drops
($w_{j+1} > w_j$), weighted by the knot position and by $\eta(v_j)^2$. This is
the structure of the Minty violators of \citet[Lemma~4]{bichler2025convergenceJ},
which lie in the $m$-piece space for $m$ equal to their number of pieces.
The subsection describes the ball in three steps: its size up to a
constant, its exact radius, and the direction in which the inequality is
first violated at its boundary.

The simplest direction with a slope drop is the hat at the top knot,
$h_{m-1}$: the piecewise-linear function equal to $1$ at $v_{m-1}$ and $0$
at every other knot, a tent on the last two pieces with slopes $+m$ and
$-m$. Its multiples $a\,h_{m-1}$ form the top-hat ray, and along the ray
the pairing can be computed exactly: it has an exact zero at $a^* =
1/(6m(m-1))$, at distance $\sqrt{2/(3m)}/(6m(m-1))$ from the equilibrium,
positive below and negative above (Proposition~\ref{prop:ray}). The hat is
the first probe of the ball's boundary: the ray crosses it at a distance
of order $m^{-5/2}$, so that is an upper bound on the radius. The same
exponent is a lower bound.

\begin{proposition}[radius bounds]
\label{prop:radius}
Let $r_m$ be the largest $r$ such that $\ip{\Fop(\bstar+\eta)}{\eta} > 0$ for
every nonzero piecewise-linear $\eta$ on the $m$-grid with $\eta(0) = 0$ and
$\norm{\eta}_{\Ltwo} < r$. Then there are constants $0 < C_- \le C_+$,
independent of $m$, with $C_- m^{-5/2} \le r_m \le C_+ m^{-5/2}$.
\end{proposition}

The lower bound comes from crude estimates: on a small enough ball every
slope is small, the weights $w_j$ are close to $2$, and the knot terms are
dominated by the positive part, which gives $r_m \ge m^{-5/2}/288$
(Appendix~\ref{app:proofs-arc1}). Between this and the ray's upper bound
the size of the ball is settled up to a constant, and the two constants
differ by a factor of forty. The hat is not the nearest violation: an exact
rational knot vector at $m = 2$ lies closer to the equilibrium with a
negative pairing (Proposition~\ref{prop:notextremal}), and its alternating
tail is what a search over sampled directions does not produce. So the
boundary of the ball has to be found by a shape problem.

\paragraph{Its exact radius.} The exponent fixes the size of the ball up
to a constant; the rest of the subsection is the exact description of its
boundary, and the route is a shape problem: on the ball the pairing
expands to third order with a relative error of one over the grid, so the
constant is the maximum of a grid-free functional, attained by an
alternating geometric profile. Expanding the integrand of
Proposition~\ref{prop:minty} in $t$ along a direction $\eta$,
\begin{equation}
\label{eq:leading}
  \ip{\Fop(\bstar+t\eta)}{t\eta}
  = t^2 Q_2(\eta) - 4t^3\Bigl[\int_0^1 v\,\eta\,\eta'^2\,dv + \tfrac13\eta(1)^3\Bigr] + O(t^4),
\end{equation}
so for a direction supported on a bounded number of pieces near $v = v_c$,
in grid units $x_j$, the first violation is at $r_m \approx m^{-5/2}\,
\nu(x)^{3/2}/(4v_cJ(x))$ with $J(x) := \tfrac12\sum_j (x_{j+1}-x_j)^2(x_{j+1}+x_j)$
and $\nu(x) := \tfrac13\sum_j (x_j^2 + x_jx_{j+1} + x_{j+1}^2)$. The shape
problem is to maximize $J/\nu^{3/2}$ over finitely supported $x$; the
position enters through the weight $v_c$ and the boundary at $v = 1$, which
truncates the right tail. The two-sided profile $x_j = (-q)^{|j|}$ with $q
= 2 - \sqrt3$, the root in $(0,1)$ of $q^2 - 4q + 1 = 0$, is a stationary
point of $J/\nu^{3/2}$ with value $2\cdot3^{1/4}$
(Proposition~\ref{prop:stationary}), and it is the maximizer:

\begin{theorem}[the sharp inequality]
\label{thm:sharp}
$J(x) \le 2\cdot 3^{1/4}\,\nu(x)^{3/2}$ for every finitely supported $x$,
with equality for $x \ne 0$ iff $x_j = c\,(-q)^{|j - j_0|}$ for some $c > 0$
and $j_0$
($J$ is odd, so $-x$ never attains it). More strongly,
$J(x) \le 3\norm{x}_\infty\,\nu(x) - \norm{x}_\infty^3/\sqrt3$.
\end{theorem}

The proof normalizes by the sup norm rather than by $\nu$ and telescopes
a potential along the two chains from the peak; its Bellman inequality
reduces to four polynomial inequalities with explicit factorizations
(Appendix~\ref{primer:bellman}). With the sharp inequality in hand, the
limit follows: the exact knot identity, expanded with explicit
remainders, bounds the knot sum by the inequality applied to each tail of
the knot vector with free ends, and the geometric profile with its peak
pinned a few knots below the top gives the matching upper bound.

\begin{theorem}[the radius limit]
\label{thm:radius}
For $n = 2$, $m^{5/2} r_m \to 1/(8\cdot 3^{1/4}) = 0.09498$, which is $3^{5/4}/(4\sqrt2) =
0.698$ times the top-hat constant $\sqrt{2/3}/6$. Quantitatively,
$r_m \ge (1 - 18/m)\,m^{-5/2}/(8\cdot 3^{1/4})$ for $m \ge 3$, and with
$L := 1/(2\log(1/q)) = 0.3797$,
\[
  \frac{L\log m - 18}{m} \;\le\; \varepsilon_m := 8\cdot 3^{1/4}\,m^{5/2}r_m - 1 \;\le\;
  \frac{L\log m + 13}{m} \qquad \text{for all large } m:
\]
the logarithm is real and its coefficient is sharp. The quantitative form
with the ray bounds is Appendix~\ref{app:arc1}.
\end{theorem}

\begin{figure}[t]
\centering
\begin{tikzpicture}
  \begin{axis}[
    width=0.62\textwidth, height=0.42\textwidth,
    xmode=log, ymode=log, log basis x=2,
    xmin=1.7, xmax=300, ymin=5e-9, ymax=2e-1,
    xtick={2,4,8,16,32,64,128,256}, xticklabels={2,4,8,16,32,64,128,256},
    minor xtick={},
    xlabel={$m$ (pieces)}, ylabel={Minty radius $r_m$ in $L^2$},
    tick label style={font=\scriptsize}, label style={font=\small},
    legend style={font=\scriptsize, at={(0.97,0.97)}, anchor=north east, cells={anchor=west}, draw=none, fill=none},
    legend cell align=left,
  ]
    \addplot[color=black!75, thick, solid] coordinates {
      (2, 0.0481125)
      (2.18102, 0.0357733)
      (2.37841, 0.0269149)
      (2.59368, 0.0204422)
      (2.82843, 0.0156462)
      (3.08442, 0.0120519)
      (3.36359, 0.00933315)
      (3.66802, 0.00726052)
      (4, 0.00567012)
      (4.36203, 0.00444292)
      (4.75683, 0.00349145)
      (5.18736, 0.0027507)
      (5.65685, 0.00217194)
      (6.16884, 0.00171832)
      (6.72717, 0.0013618)
      (7.33603, 0.00108092)
      (8, 0.000859152)
      (8.72406, 0.000683722)
      (9.51366, 0.00054471)
      (10.3747, 0.000434391)
      (11.3137, 0.000346722)
      (12.3377, 0.000276967)
      (13.4543, 0.000221406)
      (14.6721, 0.000177106)
      (16, 0.000141753)
      (17.4481, 0.000113518)
      (19.0273, 9.09507e-05)
      (20.7494, 7.29019e-05)
      (22.6274, 5.84583e-05)
      (24.6754, 4.68933e-05)
      (26.9087, 3.76287e-05)
      (29.3441, 3.02036e-05)
      (32, 2.42503e-05)
      (34.8962, 1.94753e-05)
      (38.0546, 1.56441e-05)
      (41.4989, 1.25692e-05)
      (45.2548, 1.01005e-05)
      (49.3507, 8.11819e-06)
      (53.8174, 6.52593e-06)
      (58.6883, 5.24673e-06)
      (64, 4.21884e-06)
      (69.7925, 3.39273e-06)
      (76.1093, 2.72869e-06)
      (82.9977, 2.19484e-06)
      (90.5097, 1.76559e-06)
      (98.7015, 1.42042e-06)
      (107.635, 1.14281e-06)
      (117.377, 9.1953e-07)
      (128, 7.39919e-07)
      (139.585, 5.95427e-07)
      (152.219, 4.79178e-07)
      (165.995, 3.85644e-07)
      (181.019, 3.10382e-07)
      (197.403, 2.49818e-07)
      (215.269, 2.0108e-07)
      (234.753, 1.61856e-07)
      (256, 1.30288e-07)
    };
    \addlegendentry{top-hat ray, exact zero (upper bound)}
    \addplot[color=blue!70!black, thick, densely dotted] coordinates {
      (2, 0.0167902)
      (2.18102, 0.0135202)
      (2.37841, 0.0108871)
      (2.59368, 0.00876676)
      (2.82843, 0.00705939)
      (3.08442, 0.00568454)
      (3.36359, 0.00457745)
      (3.66802, 0.00368597)
      (4, 0.00296811)
      (4.36203, 0.00239005)
      (4.75683, 0.00192458)
      (5.18736, 0.00154976)
      (5.65685, 0.00124794)
      (6.16884, 0.00100489)
      (6.72717, 0.000809186)
      (7.33603, 0.000651593)
      (8, 0.000524692)
      (8.72406, 0.000422506)
      (9.51366, 0.000340221)
      (10.3747, 0.000273961)
      (11.3137, 0.000220606)
      (12.3377, 0.000177642)
      (13.4543, 0.000143045)
      (14.6721, 0.000115187)
      (16, 9.27534e-05)
      (17.4481, 7.46892e-05)
      (19.0273, 6.01431e-05)
      (20.7494, 4.843e-05)
      (22.6274, 3.8998e-05)
      (24.6754, 3.14029e-05)
      (26.9087, 2.52871e-05)
      (29.3441, 2.03623e-05)
      (32, 1.63966e-05)
      (34.8962, 1.32033e-05)
      (38.0546, 1.06319e-05)
      (41.4989, 8.56129e-06)
      (45.2548, 6.89394e-06)
      (49.3507, 5.55131e-06)
      (53.8174, 4.47016e-06)
      (58.6883, 3.59958e-06)
      (64, 2.89854e-06)
      (69.7925, 2.33404e-06)
      (76.1093, 1.87947e-06)
      (82.9977, 1.51344e-06)
      (90.5097, 1.21869e-06)
      (98.7015, 9.81342e-07)
      (107.635, 7.90221e-07)
      (117.377, 6.36322e-07)
      (128, 5.12395e-07)
      (139.585, 4.12603e-07)
      (152.219, 3.32247e-07)
      (165.995, 2.6754e-07)
      (181.019, 2.15436e-07)
      (197.403, 1.73478e-07)
      (215.269, 1.39693e-07)
      (234.753, 1.12487e-07)
      (256, 9.05795e-08)
    };
    \addlegendentry{limit $m^{-5/2}/(8\cdot 3^{1/4})$}
    \addplot[color=green!45!black, thick, dashed] coordinates {
      (2, 0.000613808)
      (2.18102, 0.000494266)
      (2.37841, 0.000398005)
      (2.59368, 0.000320492)
      (2.82843, 0.000258074)
      (3.08442, 0.000207813)
      (3.36359, 0.000167341)
      (3.66802, 0.00013475)
      (4, 0.000108507)
      (4.36203, 8.73747e-05)
      (4.75683, 7.0358e-05)
      (5.18736, 5.66555e-05)
      (5.65685, 4.56216e-05)
      (6.16884, 3.67365e-05)
      (6.72717, 2.95819e-05)
      (7.33603, 2.38207e-05)
      (8, 1.91815e-05)
      (8.72406, 1.54458e-05)
      (9.51366, 1.24377e-05)
      (10.3747, 1.00154e-05)
      (11.3137, 8.06483e-06)
      (12.3377, 6.49416e-06)
      (13.4543, 5.22939e-06)
      (14.6721, 4.21094e-06)
      (16, 3.39084e-06)
      (17.4481, 2.73046e-06)
      (19.0273, 2.19869e-06)
      (20.7494, 1.77048e-06)
      (22.6274, 1.42567e-06)
      (24.6754, 1.14802e-06)
      (26.9087, 9.24435e-07)
      (29.3441, 7.44397e-07)
      (32, 5.99422e-07)
      (34.8962, 4.82682e-07)
      (38.0546, 3.88677e-07)
      (41.4989, 3.1298e-07)
      (45.2548, 2.52026e-07)
      (49.3507, 2.02943e-07)
      (53.8174, 1.63419e-07)
      (58.6883, 1.31592e-07)
      (64, 1.05964e-07)
      (69.7925, 8.53268e-08)
      (76.1093, 6.8709e-08)
      (82.9977, 5.53276e-08)
      (90.5097, 4.45523e-08)
      (98.7015, 3.58755e-08)
      (107.635, 2.88886e-08)
      (117.377, 2.32624e-08)
      (128, 1.87319e-08)
      (139.585, 1.50838e-08)
      (152.219, 1.21462e-08)
      (165.995, 9.78063e-09)
      (181.019, 7.87581e-09)
      (197.403, 6.34196e-09)
      (215.269, 5.10683e-09)
      (234.753, 4.11225e-09)
      (256, 3.31137e-09)
    };
    \addlegendentry{lower bound $m^{-5/2}/288$}
    \addplot[only marks, mark=*, mark size=1.6pt, color=blue!70!black] coordinates {
      (2, 0.0425145)
      (3, 0.0102938)
      (4, 0.00449859)
      (6, 0.00149339)
      (8, 0.000698927)
      (12, 0.000229914)
      (16, 0.000106681)
      (24, 3.6959e-05)
      (32, 1.76054e-05)
      (48, 6.25036e-06)
      (64, 3.01217e-06)
      (96, 1.08149e-06)
      (128, 5.2406e-07)
      (192, 1.89e-07)
      (256, 9.17053e-08)
    };
    \addlegendentry{the radius $r_m$}
  \end{axis}
\end{tikzpicture}
\caption{The Minty radius $r_m$ of Proposition~\ref{prop:radius} at the
uniform-prior equilibrium, $n = 2$, versus the number of pieces $m$, on
log-log axes. The markers are the radii, computed by minimizing the norm
over the whole $m$-piece space subject to a zero pairing. The solid line is
the top-hat ray's exact zero of Proposition~\ref{prop:ray}, an upper bound;
the dotted line the limit $m^{-5/2}/(8\cdot 3^{1/4})$ of
Theorem~\ref{thm:radius}; the dashed line the lower bound $m^{-5/2}/288$ of
Appendix~\ref{app:proofs-arc1}. All three decay like $m^{-5/2}$, and the
radii approach the limit from above at the logarithmic rate of
Theorem~\ref{thm:radius}.}
\label{fig:radius}
\end{figure}
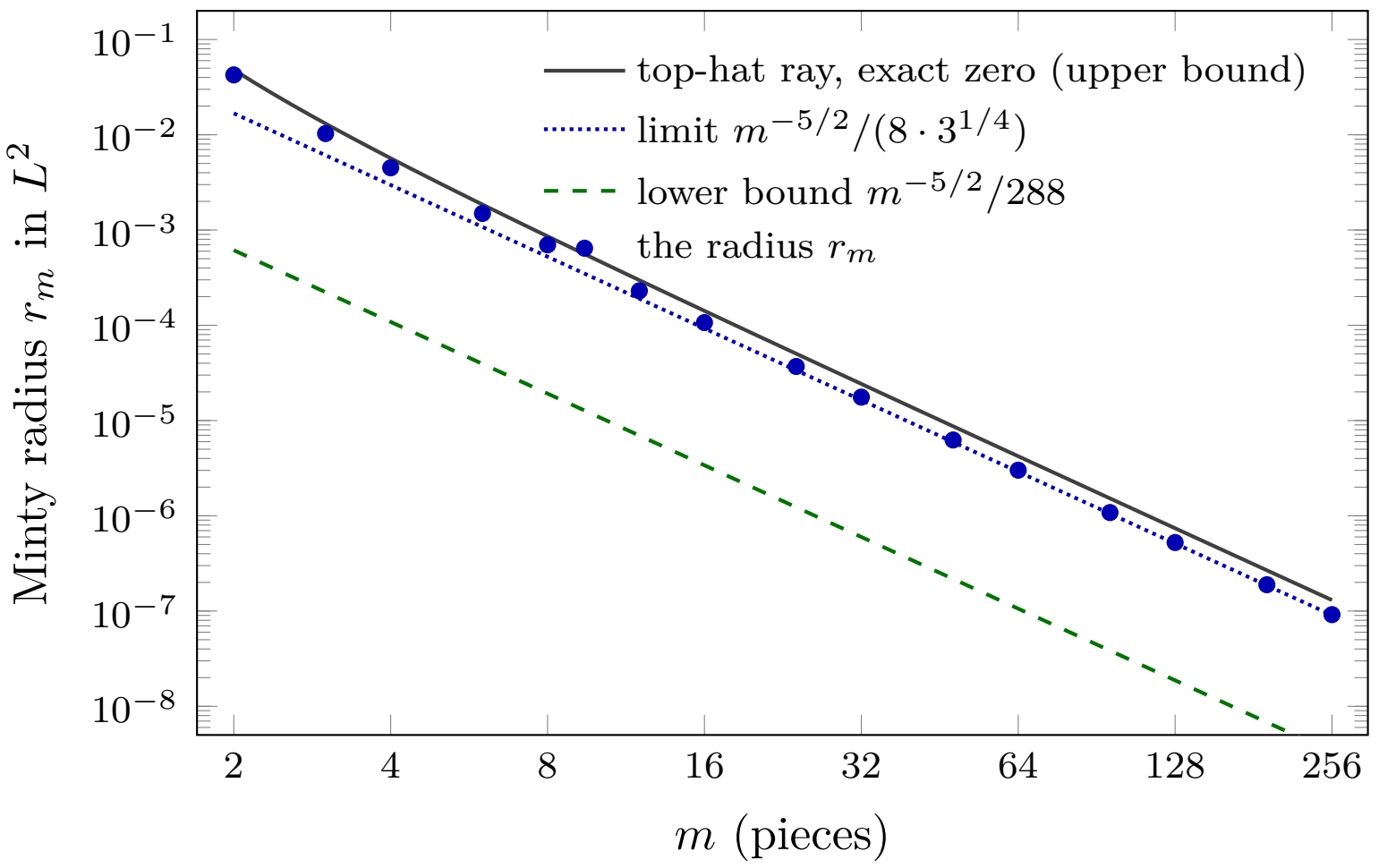

\paragraph{The radius against its bounds and its limit.}
Figure~\ref{fig:radius} illustrates the ball's size on one plot. The
markers are the radii $r_m$; the solid line above them is the top-hat
ray's exact zero, the first probe of the boundary, which the radii track
at a fixed ratio; the dotted line below them is the limit of
Theorem~\ref{thm:radius}, which the radii approach from above at its
logarithmic rate; the dashed line is the crude lower bound. The gap
between the solid line and the markers is the factor $0.698$ of
Theorem~\ref{thm:radius}: the hat overestimates the radius by a constant
factor, and the alternating profile closes it.

\paragraph{The direction of the first violation.} The minimizers that
attain the radius converge to the alternating geometric profile, with the
peak logarithmically many knots below the top; the proof is a stability
version of the sharp inequality, in which the Bellman slacks that the
telescoping discarded bound the minimizer's distance from the profile.

\begin{theorem}[the extremal direction]
\label{thm:direction}
For $n = 2$, let $\eta_m$ attain the radius, $y$ its knot vector divided by
its sup norm, $j_0$ the peak and $d_m = m - j_0$ its depth. Then for all
large $m$:
$y_{j_0} = +1$; $y_{j_0+i} \to (-q)^{|i|}$ for every fixed $i$; the end knot
$y_m \to 0$; and the peak sits $L\log m + O(1)$ knots below the top,
\[
  L\log m - 4 \;\le\; d_m \;\le\; L\log m + 29 \qquad \text{for all large } m,
\]
with the same $L$ as in Theorem~\ref{thm:radius}, because both come from
balancing the pinning loss $q^{2k}$ of a profile truncated $k$ knots below
the top against its position cost $k/m$. The quantitative form, with the
defect bounds, is Appendix~\ref{app:arc1}.
\end{theorem}

\subsection*{What the arc proves, and how}
The Minty pairing is one exact
integral, so its second-order term is the certificate and its full value,
integrated by parts piece by piece, is a sum whose only negative terms sit
at slope drops. On a small ball the slopes are small, so the pairing
reduces to a grid-free shape problem whose maximizer is an alternating
geometric profile; a potential argument proves that profile sharp, and
the stability version of the same argument shows the minimizers converge
to it. The certificate is therefore uniformly strong and uniformly local,
and the function-space failure of monotonicity established by
\citet{bichler2025convergenceJ} is the $m \to \infty$ limit of the radius
going to zero with the constant intact: refinement shrinks the
neighbourhood, not the modulus. The chain of results, with what each hands
to the next, is Appendix~\ref{app:arc1}.

\section{Arc II: the learners}
\label{sec:twoplayer}

Arcs~I and III certify the single-population operator, in which every
bidder's field is evaluated against the common bid function. Independent
learners do not see that operator: each bidder's field is evaluated
against the \emph{other} bidders' bid functions, and the win-probability
cap, the fact that a bid above every rival's top bid wins surely so that
the win probability never exceeds one, now matters. The question of this
arc is whether the learners see the certificate at all: whether the
second-order form of the learners' operator at the symmetric equilibrium
is as coercive as the single-population form $Q_n$ of
Theorem~\ref{thm:modulus}, and with the same weight. The answer is yes,
and exactly. For two bidders the form has best constant exactly $1$; for
$n$ bidders it is $nQ_n$ on symmetric directions, so its weight is the
single-population $(n/2)v^{n-2}$; and for a general prior its weight is
the one the single-population operator has for that prior, computed in
Section~\ref{sec:general} below. That is the arc's message: the
certificates of Arcs~I and III are the ones the learners see, and it is
the cap term, not the interior computation, that makes it so.

Throughout this arc the bidders use $b_i = \bstar + \eta_i$ with $\eta_i$
Lipschitz and $\eta_i(0) = 0$, and, where a grid is named, $\eta_i$
piecewise linear on a common grid. The map of the arc and its chain of
results are Appendix~\ref{app:arc2}; the proofs are in
Appendix~\ref{app:proofs-arc2}.

\subsection{The one-sided form: two bidders}
\label{sec:twoplayer-form}

Bidder 1's field against bidder 2's bid function is
\[
  \Fop_1(v) = W(b_1(v)) - (v - b_1(v))\,W'(b_1(v)), \qquad W = \min(b_2^{-1}, 1):
\]
when $b_1(v) \le b_2(1)$ this is $u - (v - b_1(v))/b_2'(u)$ with $u =
b_2^{-1}(b_1(v))$; when $b_1(v) > b_2(1)$ the bidder wins surely, $W = 1$,
$W' = 0$ and $\Fop_1 = 1$. The second regime has measure of order
$\varepsilon(\eta_1(1) - \eta_2(1))^+$ for perturbations of size $\varepsilon$
and differs by $O(1)$ from the first regime's formula, so it contributes at
second order and cannot be dropped. Because of it the two-bidder field is
only one-sidedly differentiable at $(\bstar,\bstar)$: the Jacobian of
Section~\ref{sec:certificates}, whose symmetric part is the quadratic form
$Q_n$ of Theorem~\ref{thm:modulus} for the single-population operator,
does not exist here, and its place is taken by the one-sided directional
form
\[
  B(\eta_1,\eta_2) := \lim_{\varepsilon\downarrow 0}\ \varepsilon^{-2}\,
  \Bigl(\ip{\Fop_1(\bstar+\varepsilon\eta_1,\bstar+\varepsilon\eta_2)}{\varepsilon\eta_1}
  + \ip{\Fop_2(\cdot)}{\varepsilon\eta_2}\Bigr),
\]
which is positively $2$-homogeneous but, because of the cap term below, not
a quadratic form. Computing the limit regime by regime on a common grid
gives it in closed form.

\begin{proposition}[directional second-order form of the two-bidder field]
\label{prop:twoplayer}
For $n = 2$ and piecewise-linear $\eta_1,\eta_2$ on a common grid with
$\eta_i(0) = 0$,
\[
  B(\eta_1,\eta_2) = 4\int_0^1(\eta_1^2 + \eta_2^2)\,dv - 6\int_0^1\eta_1\eta_2\,dv
  + 2\max\bigl(\eta_1(1),\eta_2(1)\bigr)^2.
\]
\end{proposition}

The matrix $\begin{pmatrix}4 & -3\\ -3 & 4\end{pmatrix}$ of the interior
form has eigenvalue $1$ on the symmetric direction $\eta_1 = \eta_2$ and
$7$ on the anti-symmetric one, and the cap term is nonnegative; the
constant of the form is therefore the symmetric eigenvalue.

\begin{corollary}
\label{cor:twoplayer}
For $n = 2$,
\[
  B(\eta_1,\eta_2) \;\ge\; \int_0^1(\eta_1^2 + \eta_2^2)\,dv + 2\max(\eta_1(1),\eta_2(1))^2
  \;\ge\; \norm{(\eta_1,\eta_2)}_{\Ltwo}^2,
\]
with equality iff $\eta_2 = \eta_1$ and $\eta_1(1) = 0$. On every common
grid with $m \ge 2$ pieces the infimum of $B$ over the unit sphere is
attained and equals $1$.
\end{corollary}

On symmetric directions $B = 2Q_2(\eta)$, with $Q_2$ as in
Theorem~\ref{thm:modulus}: the two-bidder form has best constant exactly
$1$, the single-population constant. On anti-symmetric directions the cap
term is what keeps the form positive: the interior formula alone would give
$14\int\eta^2 - 2\eta(1)^2$, negative for a hat at the top knot once $m \ge
3$, and an uncapped computation would conclude that independent learners
see an indefinite form. They do not. Two consequences: the
single-population certificate transfers to independent learners, and since
$B \ge \norm{\cdot}^2$ the $\Ltwo$ norm is a Lyapunov function for the
first-order dynamics at the equilibrium; and because the field is only
one-sidedly differentiable there, a finite-difference Jacobian at
$(\bstar,\bstar)$ is not a meaningful object, so neither of the two local
Jacobian conditions of Section~\ref{sec:certificates} applies to the
learners' operator as it stands, which is why the result is stated as a
form rather than a matrix.

\subsection{$n$ bidders}
\label{sec:twoplayer-n}

With $n$ bidders several cap regions coexist near the top, one for each
ordered pair with $\eta_i(1) > \eta_j(1)$, and they overlap. The
computation goes through because the capped field is linear in the number
of capped opponents, so overlapping regions add.

\begin{proposition}[directional second-order form of the $n$-bidder field]
\label{prop:nplayer}
Let $b_i = \bstar + \eta_i$, $i = 1, \dots, n$, with $\bstar = (n-1)v/n$,
$\eta_i$ piecewise linear on a common grid, $\eta_i(0) = 0$, $a_i :=
\eta_i(1)$, and let $B_n$ be the one-sided directional form of
$\sum_i \ip{\Fop_i}{\eta_i}$ in $\Ltwo(dv)$. Then
\[
  B_n(\eta) = \int_0^1 v^{n-2}\Bigl[\frac{n^2}{n-1}\sum_i \eta_i^2
  - \frac{n(n+1)}{(n-1)^2}\sum_{i<j}\eta_i\eta_j\Bigr]dv
  + \frac{n}{(n-1)^2}\sum_{i<j}\max(a_i,a_j)^2 .
\]
On symmetric directions $B_n = n\,Q_n(\eta)$, and in general
\[
  B_n(\eta) \;\ge\; \frac n2\int_0^1 v^{n-2}\sum_i\eta_i^2\,dv .
\]
\end{proposition}

The pointwise matrix of the integrand has eigenvalue $n/2$ on the symmetric
direction and $n(2n^2-n+1)/(2(n-1)^2)$ on the $n-1$ anti-symmetric ones, so
the integral alone dominates $\frac n2\int v^{n-2}\sum_i\eta_i^2$, with
equality iff all $\eta_i$ coincide, and the cap terms, one per pair, are
what keeps the form nonnegative on anti-symmetric directions. So the
single-population weight $(n/2)v^{n-2}$ of Theorem~\ref{thm:modulus} is
also what $n$ independent learners see.

\subsection{The learners' weight for a general prior}
\label{sec:twoplayer-F}

Arc~III, next, computes the single-population certificate for a general
prior; this subsection shows in advance that the learners see its weight,
so that Arc~III's results are theirs. The computation does not use the
uniform prior. For a prior $F$ with density $f$ and symmetric equilibrium
$\bstar$, Section~\ref{sec:general} pairs the single-population operator
in $L^2(f\,dv)$ and finds its second-order form $Q_F$, with a weight $w_F$
(Proposition~\ref{prop:general}).
The same expansion for the learners' operator, with the equilibrium ODE
$(n-1)F^{n-2}f(v-\bstar) = F^{n-1}\bstar{}'$ imposed, gives a form $B_n^F$
whose symmetric eigenvalue is exactly that weight, so $B_n^F = nQ_F$ on
symmetric directions, and whose anti-symmetric eigenvalue $\lambda_\perp$
exceeds it: with $s = v - \bstar$,
\[
  w_F = \frac{F^{n-1}(F - fs)}{2(n-1)s^2}, \qquad
  \lambda_\perp - w_F = \frac{nF^{n-1}(F + fs)}{2(n-1)^2 s^2} > 0.
\]
Independent learners see the weight $w_F$ for every prior, and $w_F < 0$
exactly where $\bstar{}' > n-1$, the criterion behind the negative weights
of Section~\ref{sec:general}. The form in full, with its cap term's
constant $c_F = f(1)/(2\bstar{}'(1))$, is Proposition~\ref{prop:generalform}
in Appendix~\ref{app:arc2}.

\subsection*{What the arc proves, and how}
The cap region has length of order the perturbation and carries an
order-one jump in the field, so it contributes at second order and the
form is one-sided; the interior part is an implicit-function computation
of the inverse bid plus one integration by parts, and the cap part is the
region's length times the jump, which for $n$ bidders is linear in the
number of capped opponents, so overlapping regions add. The matrix of the
interior form has its smaller eigenvalue on the symmetric directions, and
there the form is $n$ times the single-population form: the learners see
the certificate of Arc~I with its constant $1$ and its weight
$(n/2)v^{n-2}$, and for every prior the weight $w_F$ of Arc~III, while the
cap terms keep the form positive on the anti-symmetric directions that an
uncapped computation would call indefinite. The chain of results, with
what each hands to the next, is Appendix~\ref{app:arc2}.

\section{Arc III: general priors and bidder counts}
\label{sec:general}

The uniform prior is the case in which the operator is a rational
function of the knot values with the exact identity of
Proposition~\ref{prop:minty}, and its equilibrium is linear, so that the
function-space equilibrium lies in every $m$-piece space. This arc asks
two questions for a general prior and $n$ bidders. What is the
certificate at the function-space equilibrium? It has an explicit weight
$w_F$, a closed form in the prior and its density, and it holds uniformly
over the piecewise-linear spaces exactly when that weight is bounded below
by a positive multiple of the density (Section~\ref{sec:general-fs}). And
does the discretization that the learners run on see that certificate?
Not as it stands: when $\bstar$ is not linear the $m$-space equilibrium
$\bstar_m$ is a different point, the certificate at it differs from the
function-space one by knot masses, and the masses oscillate from knot to
knot with a law whose amplitude is explicit
(Section~\ref{sec:general-pl}). The limit of the $m$-space certificate is
then a closed form in $F$, the characterization theorem, proved for two
and three bidders and, with computer assistance, for four
(Sections~\ref{sec:general-23} and \ref{sec:general-4}); the limit is
approached from above at an explicit rate, so a positive limit means a
positive certificate from some $m$ on, and for one prior at every $m$
(Section~\ref{sec:general-finite}). Together they are the arc's message:
for a general prior the certificate has an explicit weight, and the
certificate the learners' discretization sees is that weight corrected by
the discretization's own oscillation, also in closed form. That the
learners see this arc's weight is Section~\ref{sec:twoplayer-F}.

Throughout this arc $F$ is a prior on $[0,1]$ with a density $f$ positive
on $(0,1)$, there are $n$ bidders, $\bstar$ is the symmetric equilibrium,
and $\bstar_m$ is the equilibrium of the $m$-piece space, the solution of
the field tested against the hat functions of the grid, a Galerkin
equilibrium (Appendix~\ref{primer:galerkin}). The theorems assume $f > 0$
on the closed interval, and the hypothesis is not technical: for a density
vanishing at the top the top slopes of $\bstar_m$ are of order $h$ and the
certificate diverges like $-m^2$; the leading-order account, with its one
unproved step, is Appendix~\ref{app:arc3}. The map of the arc and its
chain of results are Appendix~\ref{app:arc3}; the proofs are in
Appendix~\ref{app:proofs-arc3}.

\subsection{The function-space certificate}\label{sec:general-fs}

The ex-ante utility integrates against $dF$, so the operator is identified
through the $\Ltwo(dF)$ pairing, with the pointwise field of
Lemma~\ref{lem:field} for $W = F\circ b^{-1}$,
$g(v;b) = -F(v)^{n-1} + (n-1)F(v)^{n-2}f(v)\,(v - b(v))/b'(v)$,
and the symmetric equilibrium $\bstar(v) = v - \int_0^v F^{n-1}/F(v)^{n-1}$
solves $\bstar{}' = (n-1)(f/F)(v - \bstar)$. The exact identity of
Proposition~\ref{prop:minty} survives with the prior in the integrand.

\begin{proposition}[the exact identity and the quadratic form for a general prior]
\label{prop:general}
For $n$ bidders, let $b = \bstar + \eta$ with $\eta(0) = 0$ and
$\operatorname{ess\,inf}(\bstar{}' + \eta') > 0$. Then
\[
  \ip{\Fop(\bstar+\eta)}{\eta} = \int_0^1 \frac{F^{n-1}\eta\eta' + (n-1)F^{n-2}f\,\eta^2}{\bstar{}' + \eta'}\,f\,dv,
\]
and the $t^2$ coefficient of $\ip{\Fop(\bstar+t\eta)}{t\eta}$ is
\[
  Q_F(\eta) = \int_0^1 w_F(v)\,\eta(v)^2\,dv + c_F\,\eta(1)^2, \qquad
  w_F = \frac{F^{n-1}f}{v - \bstar} - \frac{1}{2(n-1)}\,\frac{d}{dv}\Bigl[\frac{F^n}{v - \bstar}\Bigr],
\]
\[
  c_F = \frac{1}{2(n-1)\,(1 - \bstar(1))} = \frac{f(1)}{2\bstar{}'(1)}
  \text{ when } f(1) > 0.
\]
Consequently $Q_F(\eta) \ge (\operatorname{ess\,inf} w_F/f)\,\norm\eta^2_{\Ltwo(f\,dv)}$
for every admissible $\eta$, and the Gateaux derivative at $\bstar$ is
coercive uniformly over all the piecewise-linear spaces iff
$\operatorname{ess\,inf} w_F/f > 0$.
\end{proposition}

On the $m$-piece space the best constant is at least
$\operatorname{ess\,inf} w_F/f$ and, for continuous $w_F/f$, tends to it as
$m \to \infty$, by hats with $\eta(1) = 0$ placed near the infimum.
Equality $Q_F = (\operatorname{ess\,inf} w_F/f)\norm\eta^2$ forces $\eta(1) =
0$ and $w_F/f$ constant on the support of $\eta$, which happens for the
uniform prior with $n = 2$ and for no prior with a non-constant weight.
For the uniform prior and $n = 2$, $w_F \equiv 1$: Theorem~\ref{thm:modulus}
is the statement that the weight is constant and equal to one. For $n$
bidders and the uniform prior, $w_F = \frac n2 v^{n-2}$, with infimum zero;
more generally, for any prior with $0 < f(0) < \infty$ and $n \ge 3$,
$w_F/f \sim C v^{n-2}$ near $0$, so with three or more bidders the
certificate is never uniform in $m$ for a prior with a positive finite
density at the bottom. The priors whose equilibrium is linear, so that
the function-space certificate is also the discretized one, are exactly
the power priors.

\begin{theorem}[power priors]
\label{thm:power}
For $n$ bidders, $\bstar$ is linear iff $F(v) = v^a$ for some $a > 0$, in which case $\bstar
= \kappa v/(\kappa+1)$ with $\kappa = a(n-1)$, $\bstar$ lies in every
$m$-piece space and is an equilibrium of it, and the symmetric part of the
Jacobian at it is the matrix of $Q_F$ in the hat basis, with
\[
  \frac{w_F}{f}(v) = C_{a,n}\,v^{\kappa - 1}, \qquad
  C_{a,n} = \frac{((n-2)a + 1)((n-1)a + 1)}{2(n-1)\,a}.
\]
Hence the piecewise-linear certificate holds uniformly in $m$ iff $\kappa \le
1$, with constant $C_{a,n}$ at $v = 1$ when $\kappa < 1$ and the constant
$(2n-3)/(n-1)$ when $\kappa = 1$; for $\kappa > 1$ the best constant on the
$m$-piece space decays like $m^{-(\kappa-1)}$.
\end{theorem}

Numerically $\bstar$ is the unique equilibrium of the $m$-piece space for
these priors (Appendix~\ref{app:arc3}).

\paragraph{The weight is not always positive.} Eliminating the derivative
through the equilibrium ODE, $w_F = F^{n-1}(F - f(v-\bstar))/(2(n-1)(v-\bstar)^2)$,
so $w_F < 0$ exactly where $\bstar{}' > n - 1$. If $F/f$ is nondecreasing
($F$ log-concave, which includes every decreasing or log-concave density)
then $\bstar{}' \le 1$ and $w_F > 0$ (Appendix~\ref{app:lean}): the failures
require an increasing reverse hazard rate $f/F$, not a steep density as
such. For $n = 2$ and $F = (v + cv^k)/(1+c)$, $w_F < 0$ on $[0.86, 1]$
already at $k = 6$, $c = 1$; for $n = 3$ the criterion $\bstar{}' > 2$ needs
a steeper density, met from $k = 15$ on. For these priors the
function-space certificate itself fails, and with it the certificate at
$\bstar_m$ for two and three bidders (Theorem~\ref{thm:limit2}, whose limit
is at most $\operatorname{ess\,inf} w_F/f$), for four bidders under the
hypotheses of Theorems~\ref{thm:four} and \ref{thm:layer}, and, if
Conjecture~\ref{conj:knot} holds, for more.

\subsection{The certificate at the $m$-space equilibrium}\label{sec:general-pl}

\paragraph{Knot masses.} When $\bstar$ is not linear it is not in the
$m$-piece space, and the certificate that governs the learners is the
symmetric part of the Jacobian at the $m$-space equilibrium $\bstar_m$, a
piecewise-linear approximation of $\bstar$ whose slopes $\sigma_k$ track
$\bstar{}'$ at the knots. Integrating the pairing by parts piece by piece
produces the point masses
$-\tfrac12 (v_k - \bstar_m(v_k))\,(n-1)F^{n-2}f(v_k)^2\,(1/\sigma_{k+1}^2 - 1/\sigma_k^2)\,\eta(v_k)^2$
at the interior knots (Appendix~\ref{primer:knots}; the identity is
checked in Lean, Appendix~\ref{app:lean}), negative when the slopes
decrease, that is, for a concave $\bstar$, and of the same order in $m$ as
the $\Ltwo$ mass of a hat. The two certificates therefore differ: for the
strictly concave prior $F_c(v) = (3v - v^2)/2$ and $n = 2$ the weight
satisfies $w_F/f \ge 1.5$, so the function-space derivative is coercive,
while the piecewise-linear certificate at $\bstar_m$ is negative from $m =
3$ on. The uniform prior is special in two ways at once: its weight is
constant, and its equilibrium is linear, so the two certificates coincide
and both equal one.

\paragraph{The slopes of $\bstar_m$ oscillate, with a law.} The
piecewise-linear equilibrium is not the interpolant of $\bstar$: its slopes
deviate from $\bstar{}'$ by $(-1)^{m-j} R(\mathrm{mid}_j)\,\bstar{}''(\mathrm{mid}_j)\,h$
on piece $j$, $h = 1/m$, with $R(1) = \tfrac16$ at leading order. The
equilibrium equations are Galerkin, $\int g(b_s)\,\mathrm{hat}_k\, f\,dv =
0$; the interpolation error tested against a full hat cancels at leading
order, but against the half hat at the top knot it leaves a residual that
only the slope of the top piece can balance. The checkerboard so created,
an error alternating in sign from knot to knot, propagates downward,
undamped by the central-difference stencil of the hat discretization, and
its amplitude obeys a first-order equation which the equilibrium ODE
integrates in closed form,
\[
  R(v) = \frac16\,\frac{Q(v)}{Q(1)}, \qquad
  Q = \frac{\bstar{}'^{\,4/3}}{\bstar{}''\,F^{(n-1)/3}\,f^{2/3}} .
\]
With the oscillation the knot masses alternate between $(1+\epsilon)$ and
$(1-\epsilon)$ times their smooth-slope value, $\epsilon(v) = 2R(v)$,
$\epsilon(1) = \tfrac13$. The law is a theorem once the error of the
Galerkin solution is granted its two-scale form, a slowly varying
checkerboard plus a slowly varying smooth part; the form below fixes the
decomposition up to $O(h^{2+\gamma})$, and it is observed with $\gamma =
1$ (Appendix~\ref{app:arc3}).

\begin{proposition}[the oscillation law, conditional]
\label{prop:oscillation}
For $n$ bidders, let $F \in C^4$ with $f > 0$ on $(0,1]$, fix $v_0 \in
(0,1)$, and let $\bstar_m$ be a solution of the Galerkin system whose knot
error $\delta_k = \bstar_m(v_k) - \bstar(v_k)$ on $[v_0, 1]$ has, for some
$\gamma \in (0,1]$, the form $\delta_k = (-1)^{m-k}\alpha_k + u_k$ with
$|\alpha_k| \le Ch^2$, $|\alpha_k - \alpha_{k-1}| \le Ch^3$, $|u_k| \le
Ch^{1+\gamma}$ and $|u_k - u_{k-1}| \le Ch^{2+\gamma}$. Then, uniformly on
$[v_0, 1]$,
\[
  \alpha_k = \frac{\bstar{}''(1)\,h^2}{12}\,\frac{\tilde Q(v_k)}{\tilde Q(1)}
  \,(1 + O(h^\gamma)), \qquad
  \tilde Q = \frac{\bstar{}'^{\,4/3}}{F^{(n-1)/3}f^{2/3}},
\]
so the slope oscillation is $\pm\bstar{}''(1)(h/6)\tilde Q(v)/\tilde Q(1)$,
$R(1) = \tfrac16$ and $R(v) = Q(v)/(6Q(1))$ as displayed, and the knot masses
alternate as $(1 \pm \epsilon(v))$ with $\epsilon = 2R$.
\end{proposition}

\subsection{Two and three bidders}\label{sec:general-23}

For two and three bidders the two-scale form of the knot error, a
checkerboard plus a smooth part, is a theorem, and with it the
oscillation law of Proposition~\ref{prop:oscillation}: the checkerboard's
amplitude follows the profile $\hat Q := \tilde Q/\tilde Q(1)$, which is
$1$ at the top.

\begin{theorem}[the oscillation law for two and three bidders]
\label{thm:oscillation2}
Let $n \in \{2, 3\}$ and $F \in C^5[0,1]$ with $f > 0$ on $[0,1]$. There are $h_0 > 0$
and $C$ such that for $h = 1/m \le h_0$ the Galerkin system has a solution
$\bstar_m$, unique in a ball of $h$-independent radius, in a weighted norm,
around the second Newton iterate from the interpolant, and its knot error
is
\[
  \delta_k = (-1)^{m-k}\,\frac{\bstar{}''(1)h^2}{12}\,\hat Q(v_k)\,(1 + \theta_k)
  + u_k + \rho_k, \qquad |\theta_k| \le \frac{C}{k} + C h^{(4-n)/3},
\]
with $|u_k| \le Ch^2\hat Q(v_k)$, $|u_k - u_{k-1}| \le C(1/k + hv_k^{-(n-1)/3})
h^2\hat Q(v_k)$ and $|\rho_k| \le Ch^{(4-n)/3}h^2\hat Q(v_k)$, uniformly in
$k$. Hence the oscillation law of Proposition~\ref{prop:oscillation} holds
on every $[v_0, 1]$ with error $O(h^{(4-n)/3})$, and the amplitude follows
$\hat Q$ down to the bottom knot with relative error $O(1/k)$. The norm of
the uniqueness ball and the sharper remainder for two bidders, $O(h)$, are
Appendix~\ref{app:arc3}.
\end{theorem}

The proof (Appendix~\ref{app:proofs-arc3}) is a Newton--Kantorovich
argument, which proves the existence of a root near an initial guess from the residual
there and a bound on the inverse Jacobian, in weighted norms
(Appendix~\ref{primer:newton}); its two obstacles are the bottom of the
support, where the entries of the Jacobian degenerate, and the fact that
the Jacobian's Lipschitz constant is $O(1)$ along alternating directions,
so that no contraction is available at the interpolant. The first is
handled by a global discrete Liouville--Green analysis, asymptotics for a
recurrence with slowly varying coefficients, of the Jacobian's two pinned
modes and its Green's function (Appendix~\ref{primer:recurrences}); the
second by taking two Newton steps with the pointwise structure of the
corrections tracked. The bookkeeping shows why the argument stops at four
bidders: the weighted inverse is then a logarithm and the second Newton
step is no longer small.

With two-periodic knot masses the knot part of the certificate is a form
with two-periodic coefficients, and a Bloch inequality
(Appendix~\ref{primer:bloch}) says that its extreme values per unit mass
over all profiles are attained by two-periodic ones: with $\zeta_F :=
\tau_F\,\epsilon$ and
\[
  \tau_F(v) = \frac{(v - \bstar)\,(n-1)F^{n-2} f\,\bstar{}''}{\bstar{}'^{\,3}}, \qquad
  g_\pm(\epsilon) = \frac{2 \pm \sqrt{1 + 3\epsilon^2}}{3},
\]
the knot part per unit mass ranges over $[3\tau_Fg_-, 3\tau_Fg_+]$, and in
both signs of $\tau_F$ the infimum is $2\tau_F - \sqrt{\tau_F^2 + 3\zeta_F^2}$,
so the limit function is
\[
  E_F^{\mathrm{osc}} = \frac{w_F}{f} + \tau_F - \sqrt{\tau_F^{\,2} + 3\zeta_F^{\,2}},
  \qquad
  \zeta_F = \frac{\bstar{}''(1)}{3}\,\frac{(v - \bstar)(n-1)F^{n-2}f}{\bstar{}'^{\,3}}\,
  \frac{\tilde Q(v)}{\tilde Q(1)},
\]
continuous on $[0,1]$ with $E_F^{\mathrm{osc}}(0) = w_F/f(0)$; the
smooth-slope value $(w_F - \kappa_F)/f$, $\kappa_F = 2(v - \bstar)(n-1)F^{n-2}
f^2(\bstar{}'')_-/\bstar{}'^{\,3}$, is an upper bound, and every quantity is
explicit in $F$, $f$, $f'$ and $v - \bstar$.

\begin{theorem}[the limit of the piecewise-linear certificate, two and three bidders]
\label{thm:limit2}
Let $n \in \{2, 3\}$ and $F \in C^5[0,1]$ with $f > 0$ on $[0,1]$. The best constant
$c_m$ of the piecewise-linear certificate at $\bstar_m$ satisfies
\[
  \lim_{m\to\infty} c_m = \min_{v \in [0,1]} E_F^{\mathrm{osc}}(v).
\]
In particular the certificate fails to hold uniformly in $m$ whenever
$\min E_F^{\mathrm{osc}} < 0$, which happens whenever the smooth-slope value
$\operatorname{ess\,inf}(w_F - \kappa_F)/f$ is negative.
\end{theorem}

\begin{corollary}[the certificate in the weighted geometry]
\label{cor:weighted}
Let $n \in \{2, 3\}$ and $F$ as in Theorem~\ref{thm:limit2}, and let
$\vartheta > 0$ be a continuous weight on $[0,1]$. The best constant
$c_m^\vartheta$ of the piecewise-linear certificate at $\bstar_m$ relative
to $\int\vartheta\eta^2f$ tends to $\min_{[0,1]}E_F^{\mathrm{osc}}/\vartheta$.
If $w_F > 0$ on $(0,1]$ and $\vartheta = w_F/f$,
\[
  \lim_{m\to\infty} c_m^{w_F/f} = \min_{v\in[0,1]}\Bigl[1 + \bigl(\tau_F - \sqrt{\tau_F^2 + 3\zeta_F^2}\bigr)\frac{f}{w_F}\Bigr],
\]
the bracket being read with its continuous extension, $1$, at $v = 0$.
\end{corollary}

The weight $w_F/f$ vanishes at the bottom when $n = 3$, a case the proof
covers (Appendix~\ref{app:arc3}). For the power priors $F = v^a$, outside
the hypotheses unless $a = 1$ but elementary, the $m$-space equilibrium is
$\bstar$ itself for every $n$ and $m$, and at it $c_m^{w_F/f} = 1$ exactly,
attained precisely on $\{\eta(1) = 0\}$. So in the geometry weighted by
$w_F$ the certificate is uniform in $m$ exactly when the knot masses do
not consume the whole weight anywhere. The proof of
Theorem~\ref{thm:limit2} has two ingredients: a two-sided Bloch inequality
for two-periodic coefficients, sharp with a two-periodic null vector, and
a localization by a partition of unity (Appendix~\ref{primer:ims}), a
smooth division of a profile into windows whose overlaps cost an explicit
error, that reduces the lower bound to frozen-coefficient windows, where
the inequality applies; the upper bound is the two-periodic null profile
under a smooth envelope at a near-minimizer. The computed constants
approach the closed forms for every prior tested, including the
sign-changing prior $v - v^2 + v^3$, whose limit at the top knot, $0.1657$,
lies well below the value $0.2400$ a formula without the oscillation would
give (Appendix~\ref{app:arc3}).

\subsection{Four or more bidders}\label{sec:general-4}

For four or more bidders the limit is expected to be the same closed
form; what is proved follows the conjecture.

\begin{conjecture}[knot masses, $n \ge 4$]\label{conj:knot}
For $n \ge 4$ and $F \in C^5$ with $f > 0$ on $[0,1]$, the best constant of the
piecewise-linear certificate at $\bstar_m$ tends to $\min_{[0,1]}
E_F^{\mathrm{osc}}$ as $m \to \infty$, with $E_F^{\mathrm{osc}}$ as displayed in
Section~\ref{sec:general-23}.
\end{conjecture}

\begin{proposition}[what the conjecture needs]\label{prop:reduction}
Let $n \ge 3$, $F \in C^5[0,1]$ with $f > 0$ on $[0,1]$, and let $\bstar_m$ be
Galerkin solutions with slopes $\sigma_j$ such that (i) on every $[v_0, 1]$,
uniformly in the pieces (the top piece included), $\sigma_j = \bstar{}'(\mathrm{mid}_j) +
(-1)^{m-j}(\bstar{}''(1)h/6)\hat Q(v_j) + o(h)$, the slope law of
Theorem~\ref{thm:oscillation2}; and (ii) the slopes are bounded above and
below uniformly in $m$, and $\varepsilon(v_0) := \limsup_m \max_{v_k \le
v_0}v_k^{n-1}|\sigma_{k+1} - \sigma_k|/h \to 0$ as $v_0 \to 0$. Then $c_m \to
\min_{[0,1]}E_F^{\mathrm{osc}}$, and $c_m^\vartheta \to \min E_F^{\mathrm{osc}}
/\vartheta$ for every continuous $\vartheta > 0$ on $[0,1]$.
\end{proposition}

What separates the conjecture from a theorem is the slope law on compact
subsets of $(0,1]$, whose proof for $n \le 3$ passes through the bottom. At
the bottom the law's own amplitude against the bid is $\asymp h^{(4-n)/3}
k^{-(n+2)/3}$, bounded exactly for $n \le 4$; its value at the first knot,
$A_F := |\bstar{}''(1)|h^2\hat Q(v_1)/(12\bstar(v_1))$, is $m$-independent at
$n = 4$, while for $n \ge 5$ a layer forms at the bottom, about
$h^{-(n-4)/(n+2)}$ knots wide, in which the law fails. The layer is
universal: near $v = 0$ every prior's Galerkin rows are the uniform
prior's, driven by the checkerboard arriving from the top, so the layer is
the uniform-prior $n$-bidder system forced at its top row, a function of
$n$ and the arriving amplitude alone, with two branches (the first knot
pushed down or up) that saturate. In what follows the \emph{two-scale
form} of a knot error is the decomposition of
Theorem~\ref{thm:oscillation2}, a checkerboard with the law's amplitude
plus a smooth part plus a remainder, and the \emph{universal layer at
amplitude $\pm A$} is the solution of that bottom system driven by a
checkerboard of amplitude $A$ and the given sign. For four bidders the law
is proved under a smallness condition on the top curvature, and the
condition is removed, for amplitudes up to a verified bound, by proving
the layer itself.

\begin{theorem}[four bidders, small top curvature]\label{thm:four}
Let $n = 4$ and $F \in C^5[0,1]$ with $f > 0$ on $[0,1]$. There are
$\varepsilon_0, h_0 > 0$, depending on $\min f$ and $\norm{F}_{C^5}$ only,
such that for $|\bstar{}''(1)| \le \varepsilon_0$ and $h = 1/m \le h_0$ the
Galerkin system has a solution $\bstar_m$, unique in an $h$-independent
ball, in a weighted norm, around the first Newton iterate from the
interpolant, whose knot error has the two-scale form with $|\theta_k| \le
C/k$ and the remainder bounds of Appendix~\ref{app:arc3}. The slope
law~(i) holds on every $[v_0, 1]$ with error $O(h^2\log m)$ and at the
bottom knots within a relative $O(1/k)$, condition~(ii) holds, and $c_m
\to \min_{[0,1]}E_F^{\mathrm{osc}}$, $c_m^\vartheta \to \min
E_F^{\mathrm{osc}}/\vartheta$, with $E_F^{\mathrm{osc}}$ as in
Theorem~\ref{thm:limit2} at $n = 4$.
\end{theorem}

\begin{theorem}[four bidders, bounded layer amplitude; computer-assisted]\label{thm:layer}
Let $n = 4$ and $F \in C^5[0,1]$ with $f > 0$ on $[0,1]$. If $A_F \le A_0 :=
8.2419$, there are $K_0(F)$ and $h_0(F)$ such that for $h \le h_0$ the
Galerkin system has a solution $\bstar_m$, unique in an $h$-independent ball
around a Newton iterate from the composite of the universal layer at
amplitude $\pm A_F$ and the first iterate from the interpolant; its
first $K_0/2$ knots are the layer's within a relative $C(F)/(K_0k^2)$, and
beyond $K_0$ its knot error has the two-scale form with $|\theta_k| \le
C(F)/k$ and $|\rho_k| \le C(F)(1/k + h\log m)h^2\hat Q(v_k)$. The slope
law~(i) holds on every $[v_0, 1]$, condition~(ii) holds, and $c_m \to
\min_{[0,1]}E_F^{\mathrm{osc}}$, $c_m^\vartheta \to \min
E_F^{\mathrm{osc}}/\vartheta$.
\end{theorem}

The computer-assisted inputs are three, each verified in interval
arithmetic (Appendix~\ref{primer:interval}): the constants of the tail
lemmas' invariant regions, the Krawczyk sweep of the layer's first knots
with its box-boundary tests, and the matching's mode constants; the
one-step identities, the inductions they close and the matching argument
are analytic. What each verification establishes, its code and its run
records are in Appendix~\ref{app:proofs-arc3}. The first part of the
proof is the layer: in scaled tail variables the uniform prior's
four-bidder rows are an exact one-step map whose smooth part contracts, an
invariant region closes under three inequalities that interval arithmetic
verifies from a hand-off index on, the knot index beyond which the
one-step map is a contraction, and below it a parametric Krawczyk sweep,
an interval Newton test that verifies a root in each box, over $1.3$
million boxes verifies existence, nondegeneracy and the slope bounds, so
that the verified first-knot values form an interval reaching every
amplitude up to $A_0$. The second is the matching: Newton--Kantorovich
from the composite, in norms adapted to the matching index, with the
layer's variational solution as the bottom-pinned mode and the inland rows
supplying the top-pinned one. At five bidders the layer's existence is
verified by the same route for $|A| \le 6.8423$, with its weighted value
verified negative beyond $|A| = 0.61$ (Proposition~\ref{prop:layer5},
Appendix~\ref{app:arc3}); the matching is not redone there, and what
remains for $n \ge 5$ is the layer for every amplitude.

\begin{proposition}[the universal layer at five bidders; computer-assisted]\label{prop:layer5}
Let $n = 5$. For every far-field amplitude $|A| \le 6.8423$, the amplitude
of the checkerboard arriving from the interior, there is a trajectory of
the uniform prior's five-bidder rows, existing for all $k$, with slopes in
$[0.2691, 2.2128]\,\bstar{}'$, far field $(-1)^kP_\infty k^{-4/3} +
O(k^{-5/3})$, and a bounded variational solution whose alternating content
at the hand-off index is bounded below with one sign; the two signs of $A$
are the two branches. The full statement and the verification are in
Appendices~\ref{app:arc3} and \ref{app:proofs-arc3}.
\end{proposition}

\subsection{Positivity at every $m$: the exact local bound and the rate}\label{sec:general-finite}

Whether $\min E_F^{\mathrm{osc}} > 0$ implies positivity of the certificate
for every $m$, not only for large $m$, is open as a theorem; for all large
$m$ it follows from the limit theorems, and where the rates below are
theorems the certificate exceeds its limit by half the leading term, so
what is open is a finite range of $m$, closed for one prior. What is exact
at finite $m$ is a local bound: split each interior knot mass of the
decomposition identity (Section~\ref{sec:general-pl}) in halves between
its two pieces and put the boundary term on the top piece, so that $Q_m =
\sum_jP_j$ with $P_j$ a quadratic form in $(\eta_{j-1}, \eta_j)$, and let
$M_j = \int_{\text{piece }j}f\eta^2$.

\begin{proposition}[the exact local bound]\label{prop:local}
Let $n = 2$ and let $\bstar_m$ be a Galerkin solution with positive slopes,
$Q_m$ the certificate's quadratic form at it. Then $c_m \ge L_m := \min_j
\lambda_{\min}(P_j, M_j)$ for every $m$; $c_m$ itself is the supremum of
$\min_j\lambda_{\min}$ over all divisions of the knot diagonal entries
between adjacent pieces, and the supremum is attained; and under the
hypotheses of Theorem~\ref{thm:oscillation2}, $L_m \to \min E_F^{\mathrm{osc}}$.
\end{proposition}

The last statement gives the lower bound of Theorem~\ref{thm:limit2}
without localization (Appendix~\ref{app:proofs-arc3}). The margin can be
nonlocal: when the minimizer of $E_F^{\mathrm{osc}}$ is interior, the local
bound lies below the limit while $c_m$ lies above it. Its mechanism is a
wave packet, a profile alternating in the local two-periodic pattern under
a slowly varying envelope. The frozen problem at $v$ is two-periodic in
the knot index; its lower Bloch band is $E_F^{\mathrm{osc}}(v) +
\kappa(v)\theta^2 + O(\theta^4)$ with $\kappa = \tfrac32g_*t_*/(1 + t_* +
t_*^2) \ge 0$, $g_*$ the Bloch constant and $t_*$ the null-profile ratio,
and a packet with a slowly varying envelope sees $\theta$ as $-ih\,d/dv$.
This is the semiclassical reading (Appendix~\ref{primer:semiclassical}):
$c_m$ is, to leading order, the ground-state energy of
$-\kappa(v)h^2d^2/dv^2 + E_F^{\mathrm{osc}}(v)$ with a Dirichlet wall at
$v = 1$, and the approach to the limit is the ground-state energy of the
well at the minimizer.

\begin{theorem}[the rate at an interior minimizer]\label{thm:interior}
Let $n \in \{2, 3, 4\}$ and $F \in C^6[0,1]$ with $f > 0$ on $[0,1]$, for
$n = 4$ with the layer amplitude $A_F \le A_0$ of Theorem~\ref{thm:layer}
or $|\bstar{}''(1)| \le \varepsilon_0$ of Theorem~\ref{thm:four}, and
let $E = E_F^{\mathrm{osc}}$ attain its minimum $E_{\min}$ at
a unique point $\hat v \in (0,1)$ with $E''(\hat v) > 0$ and $\kappa(\hat v)
> 0$. Then
\[
  c_m = E_{\min} + h\sqrt{\kappa(\hat v)E''(\hat v)/2} + O\bigl(h^{3/2}\log^{3/2}(1/h)\bigr).
\]
\end{theorem}

When the minimizer sits at the top knot the well is one-sided, cut by the
Dirichlet wall, and the rate is the Airy one.

\begin{theorem}[the rate at a top-knot minimizer]\label{thm:airy}
Let $n \in \{2, 3, 4\}$ and $F \in C^6[0,1]$ with $f > 0$ on $[0,1]$ ($C^5$
suffices for $n = 2$), for $n = 4$ with the layer amplitude $A_F \le A_0$
of Theorem~\ref{thm:layer}, and let $E = E_F^{\mathrm{osc}}$ attain
its minimum over $[0,1]$ at $v = 1$ only, with $E'(1) < 0$ and
$\bstar{}''(1) \ne 0$. Then
\[
  c_m = E(1) + a_1\,\kappa(1)^{1/3}\,|E'(1)|^{2/3}\,h^{2/3} + O(h).
\]
For $n = 4$ the same holds under the hypotheses of Theorem~\ref{thm:four}
in place of the layer's, with $O(h\log m)$ in place of $O(h)$.
\end{theorem}

The proofs rest on a second-order amplitude law (Theorem~\ref{app:H2}): on
every compact subset of $(0,1]$ the alternating amplitude of $\bstar_m -
I_h\bstar$ follows its law with the exact constant to relative $O(h^2)$ and
the smooth part is $h^2$ times a $C^2$ function. Given it, the knot masses
have no first-order correction, the piece integrand's first-order
correction is alternating and cancels across cells, and the two rates are
the ground energies of the harmonic and the Airy well. Both leading terms
are positive, so the finite-$m$ constant approaches its limit from above:

\begin{corollary}[eventual domination of the limit]\label{cor:eventual}
Let $E = E_F^{\mathrm{osc}}$ in its $n$-bidder form and $E_{\min} = \min_{[0,1]}E$.
In each case there is $m_0(F)$ such that for every $m \ge m_0$:
(a) if $E_{\min} > 0$ then $c_m > 0$, under the hypotheses of
Theorem~\ref{thm:limit2} ($n = 2, 3$), Theorem~\ref{thm:four} or
Theorem~\ref{thm:layer} ($n = 4$);
(b) under the hypotheses of Theorem~\ref{thm:interior}, $c_m \ge E_{\min} +
\tfrac12h\sqrt{\kappa(\hat v)E''(\hat v)/2} > E_{\min}$;
(c) under the hypotheses of Theorem~\ref{thm:airy}, $c_m \ge E_{\min} +
\tfrac12a_1\kappa(1)^{1/3}|E'(1)|^{2/3}h^{2/3} > E_{\min}$.
In (b) and (c) the factor $\tfrac12$ may be replaced by any fraction below
$1$, with $m_0$ depending on it.
\end{corollary}

For a prior in one of these classes the statement ``$c_m > 0$ for every
$m \ge 2$'' is thus equivalent to the same assertion on the finite range
$2 \le m < m_0(F)$. For one prior that range is closed, by a verified
computation of the local bound at each $m$ up to a threshold and a
verified tracking of the existence theorem's constants beyond it:

\begin{corollary}[the pilot prior at every $m$]\label{cor:pilot}
For $n = 2$ and $F(v) = (v + v^2)/2$, $c_m > 0$ for every $m \ge 2$.
\end{corollary}

The verified bounds, the threshold and the run records are
Appendix~\ref{app:arc3}; the three-bidder analogue in the geometry
weighted by $w_F$, and the rate at a zero of the limit function on the
bottom edge, are stated there too.

\subsection*{What the arc proves, and how}
For a general prior the Minty identity survives with the prior in the
integrand once the equilibrium ODE is imposed, so the certificate at the
function-space equilibrium is a weight in closed form. The discretization
the learners run on does not sit at that equilibrium: its own equilibrium
carries a checkerboard seeded at the top knot, whose amplitude the
equilibrium ODE integrates in closed form, and the certificate at it is
the weight less knot masses that alternate with the checkerboard. Proving
the checkerboard is the deep step, a Newton argument in weighted norms
with the Jacobian's modes made explicit; a two-periodic inequality and a
localization then turn the alternating masses into the limit of the
certificate, a closed form in $F$; at four bidders the bottom of the
support needs a universal layer verified by interval arithmetic. The
limit is approached from above at a rate that a semiclassical reading
gives exactly, so a positive limit is a positive certificate from some $m$
on, and for one prior at every $m$. The certificate for a general prior is
therefore the explicit weight corrected by the discretization's own
oscillation, and both are exact. The chain of results, with what each
hands to the next, is Appendix~\ref{app:arc3}.

\section{Conclusion and open problems}
\label{sec:conclusion}

For the first-price auction with uniform values the obstruction to the VI
convergence theory is precise, and so is what replaces it: the operator is
not monotone, but for two bidders its Gateaux derivative at the equilibrium
is $\Ltwo$-coercive with best constant exactly one in every piecewise-linear
space, the Minty inequality holds on a ball whose radius shrinks like
$m^{-5/2}$ with an explicit extremal profile, and independent learners see
the same constant once the cap on the win probability is kept. For a
general prior the certificate at the true equilibrium is an explicit
weight, and the certificate the learners see differs from it by the
discretization's oscillation, whose closed-form limit decides positivity
for two, three and, under a verified layer, four bidders.

Open, in roughly the order we would attack them:
\begin{enumerate}[leftmargin=*,itemsep=2pt]
\item \emph{The knot masses for five or more bidders, and for four with
  large top curvature} (Conjecture~\ref{conj:knot}). What separates the
  conjecture from a theorem is the slope law on compact subsets of
  $(0,1]$ (Proposition~\ref{prop:reduction}); the bottom knots play no
  part. At $n = 4$ the smallness condition of Theorem~\ref{thm:four} is
  removed for layer amplitudes up to $A_0$ by the verified layer
  (Theorem~\ref{thm:layer}); beyond $A_0$ the same proof applies once the
  verified range is extended, which is a matter of the tail lemmas'
  closing amplitude at a larger hand-off index, not of the sweep. For
  $n \ge 5$ the law fails in a bottom layer of about $h^{-(n-4)/(n+2)}$
  knots; the layer is universal and verified for bounded amplitudes
  (Proposition~\ref{prop:layer5}), and the missing step is its existence
  up to the saturation and the matching. In the geometry weighted by $w_F$
  the layer sends the certificate to $-\infty$ for $n \ge 5$, numerically,
  so the weighted corollary stops at four bidders.
\item \emph{The constants of the Minty radius.} The limit, the extremal
  direction, and the rate and depth to their leading terms are proved
  (Theorems~\ref{thm:radius}, \ref{thm:direction}); the bounded $O(1)$
  terms are not identified. They are sawtooth functions of $\log m$ from
  the integer depth, identified numerically by an explicit finite pinned
  problem; the matching lower bound would need the sharp stability
  constant of the pinned optimum and the end knot's boundary problem in
  closed form.
\item \emph{Positivity at every $m$ as a theorem.} Where a rate is proved
  the certificate exceeds its limit for all large $m$
  (Corollary~\ref{cor:eventual}) and the all-$m$ statement is a finite
  computation, done for one prior with every constant verified
  (Corollary~\ref{cor:pilot}); the thresholds $m_0(F)$ in general, and the
  certificate's sign at the bottom edge for four bidders, where the
  constant is prior-dependent, are open.
\item \emph{More bidders in the weighted geometry.} For $n \ge 3$ the
  $\Ltwo$ constant tends to $\min E_F^{\mathrm{osc}} \le 0$; in the geometry
  weighted by $w_F$ it is exactly one for the power priors and tends to an
  explicit limit otherwise, approached at the semiclassical rate
  (Corollary~\ref{cor:weighted}, Theorem~\ref{app:interiorw}). Whether
  the learners' form at the $m$-space equilibrium is coercive in that
  geometry, as Section~\ref{sec:twoplayer} shows it is in $\Ltwo$ at the
  function-space one, is not treated here.
\end{enumerate}

Monotonicity is replaced, for this auction, by a certificate with exact
constants; what remains is to prove those constants in those places where
they are so far only computed.

\bibliographystyle{plainnat}
\bibliography{references}

\appendix
\section{The arcs: maps and chains}
\label{app:arcs}

One subsection per arc. Each begins with the arc's map, the chain of
results as a figure and in prose, and then takes the chain result by
result.

\subsection{Arc I: the uniform prior}
\label{app:arc1}

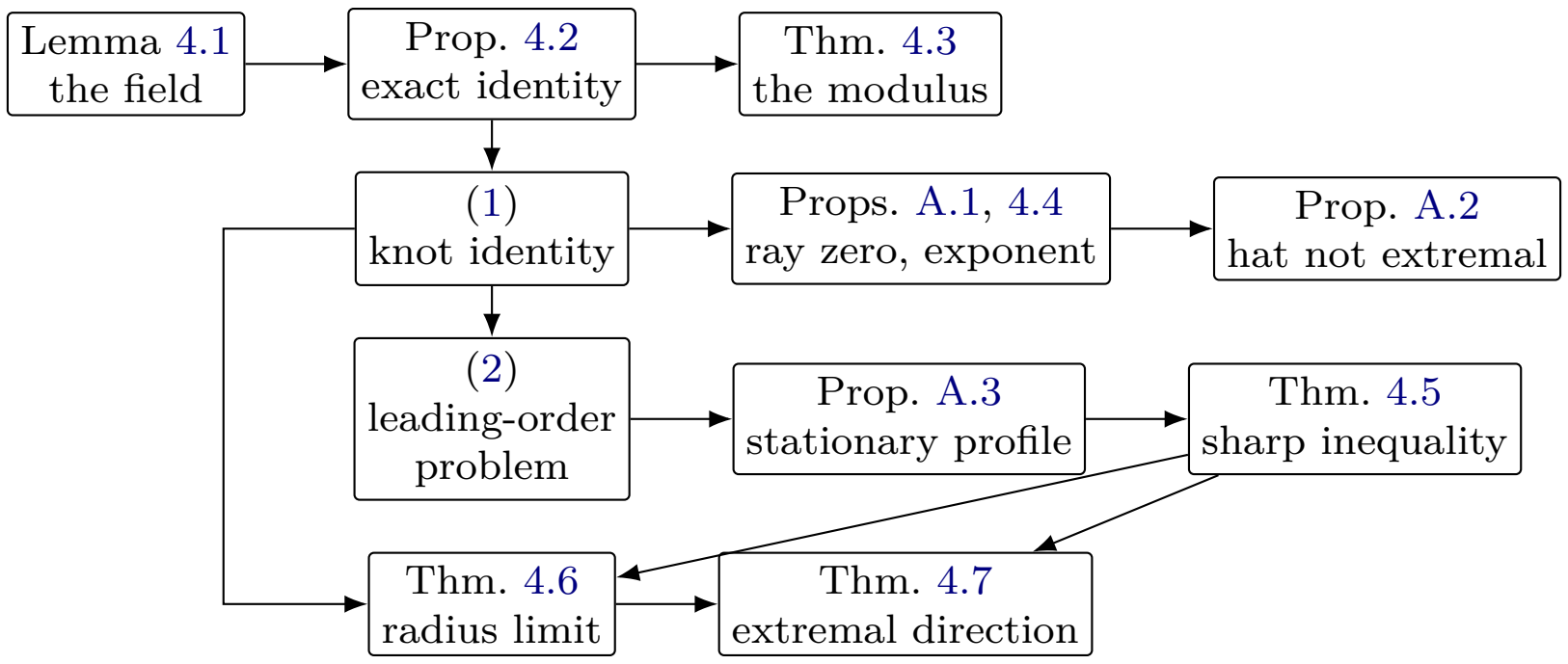
\begin{figure}[t]
\centering
\begin{tikzpicture}[
  node distance=3.5mm and 7mm,
  box/.style={draw, rounded corners=1pt, font=\scriptsize, inner sep=2.5pt, align=center},
  >={Latex[length=1.6mm]}, every path/.style={->, thin}]
\node[box] (field) {Lemma~\ref{lem:field}\\the field};
\node[box, right=of field] (minty) {Prop.~\ref{prop:minty}\\exact identity};
\node[box, right=of minty] (mod) {Thm.~\ref{thm:modulus}\\the modulus};
\node[box, below=of minty] (knots) {\eqref{eq:knots}\\knot identity};
\node[box, right=of knots] (ray) {Props.~\ref{prop:ray}, \ref{prop:radius}\\ray zero, exponent};
\node[box, right=of ray] (notex) {Prop.~\ref{prop:notextremal}\\hat not extremal};
\node[box, below=of knots] (lead) {\eqref{eq:leading}\\leading-order\\problem};
\node[box, right=of lead] (stat) {Prop.~\ref{prop:stationary}\\stationary profile};
\node[box, right=of stat] (sharp) {Thm.~\ref{thm:sharp}\\sharp inequality};
\node[box, below=of lead] (radius) {Thm.~\ref{thm:radius}\\radius limit};
\node[box, right=of radius] (dir) {Thm.~\ref{thm:direction}\\extremal direction};
\draw (field) -- (minty);
\draw (minty) -- (mod);
\draw (minty) -- (knots);
\draw (knots) -- (ray);
\draw (ray) -- (notex);
\draw (knots) -- (lead);
\draw (lead) -- (stat);
\draw (stat) -- (sharp);
\draw (sharp) -- (radius);
\draw (sharp) -- (dir);
\draw (radius) -- (dir);
\draw (knots.west) -- ++(-9mm,0) |- (radius.west);
\end{tikzpicture}
\caption{The chain of Arc I. Rows: the identity and the modulus; the
knot identity and the exponent of the radius; the leading-order shape
problem and its sharp inequality; the limit and the direction. The
long edge is the exact knot identity, expanded with explicit
remainders, entering the limit theorem.}
\label{fig:arc1-map}
\end{figure}

\entry{The map.} Figure~\ref{fig:arc1-map} is the chain. Two things
organize it: everything is read off one exact
formula, the Minty pairing at a perturbed bid as a single integral, and on
the ball that matters the slopes are $O(1/m)$, so the problem of the
constant is grid-free. The field is a closed form (Lemma~\ref{lem:field}),
so the Minty pairing is one exact integral (Proposition~\ref{prop:minty}).
Its second-order coefficient is the certificate, and one integration by
parts gives the weight and the boundary term that keep the constant at $1$
(Theorem~\ref{thm:modulus}). Piece by piece, the same integration by parts
turns the full pairing into the knot identity \eqref{eq:knots}, whose only
negative terms sit at slope drops; the hat at the top knot gives an exact
zero and the exponent $5/2$ (Propositions~\ref{prop:ray},
\ref{prop:radius}), but it is not extremal (Proposition~\ref{prop:notextremal}).
On the relevant ball the slopes are $O(1/m)$, so the third-order expansion
\eqref{eq:leading} is relatively exact and the constant is a
grid-independent shape problem, whose stationary point is the alternating
geometric profile (Proposition~\ref{prop:stationary}) and whose sharp
inequality is proved by a potential (Theorem~\ref{thm:sharp}). The exact
knot identity, expanded with explicit remainders, Abel summation to remove
the position weight, and the sharp inequality applied to each tail with
free ends give the limit and its logarithmic rate
(Theorem~\ref{thm:radius}); a stability version of the inequality, chain
stability, gives the convergence of the minimizers and the depth of the
peak (Theorem~\ref{thm:direction}).

\entry{The entries.} The rest of this subsection takes the map box by
box. The order is the body's; where the chain branches, the branch the
body takes first comes first, and a result the body does not state
follows the result it extends. Each entry says what the result
establishes, why the chain needs it, and how it is proved, and ends
with its pointers: the primer entry for anything unusual, the proof in
Appendix~\ref{app:proofs}, and what Appendix~\ref{app:lean} checks by
machine. An entry that opens with a statement is a result the body does
not state, or states only in compact form; the statement comes first and
the entry glosses it.

\entry{Lemma~\ref{lem:field}, the pointwise field.}
\emph{What.} The field is rational in the bid and its slope: minus a
power of the value, plus a multiple of the gap between value and bid over
the slope. Along the diagonal $z = b(v)$ the win-probability cap is inert,
and at the linear equilibrium the field vanishes.
\emph{Why.} Proposition~\ref{prop:minty} substitutes a perturbation
into this formula; nothing in the arc uses the operator except through
it.
\emph{How.} Differentiate the payoff density $(v - z)W(z)^{n-1}$ in the
bid; $W$ is Lipschitz, so differentiable almost everywhere, and the
exceptional set pulls back to a null set along the diagonal.
Primer: Appendix~\ref{primer:vi}. Proof: Appendix~\ref{app:proofs-field}.
Machine-checked: the derivative and the diagonal substitution
(Appendix~\ref{app:lean}).

\entry{Proposition~\ref{prop:minty}, the exact identity.}
\emph{What.} The Minty pairing $\ip{\Fop(\bstar + \eta)}{\eta}$ is one
explicit integral, exact for every admissible perturbation, with the
perturbed slope in the denominator.
\emph{Why.} Its expansion in the size of the perturbation gives the
certificate (second order) and, integrated by parts piece by piece, the
knot identity; the limit theorem expands it with explicit remainders.
\emph{How.} Write the gap and the slope at $\bstar + \eta$ as the
equilibrium values shifted by $\eta$ and $\eta'$, and pair with $\eta$.
Primer: Appendices~\ref{primer:vi}, \ref{primer:gateaux}. Proof:
Appendix~\ref{app:proofs-field}. Machine-checked in full
(Appendix~\ref{app:lean}).

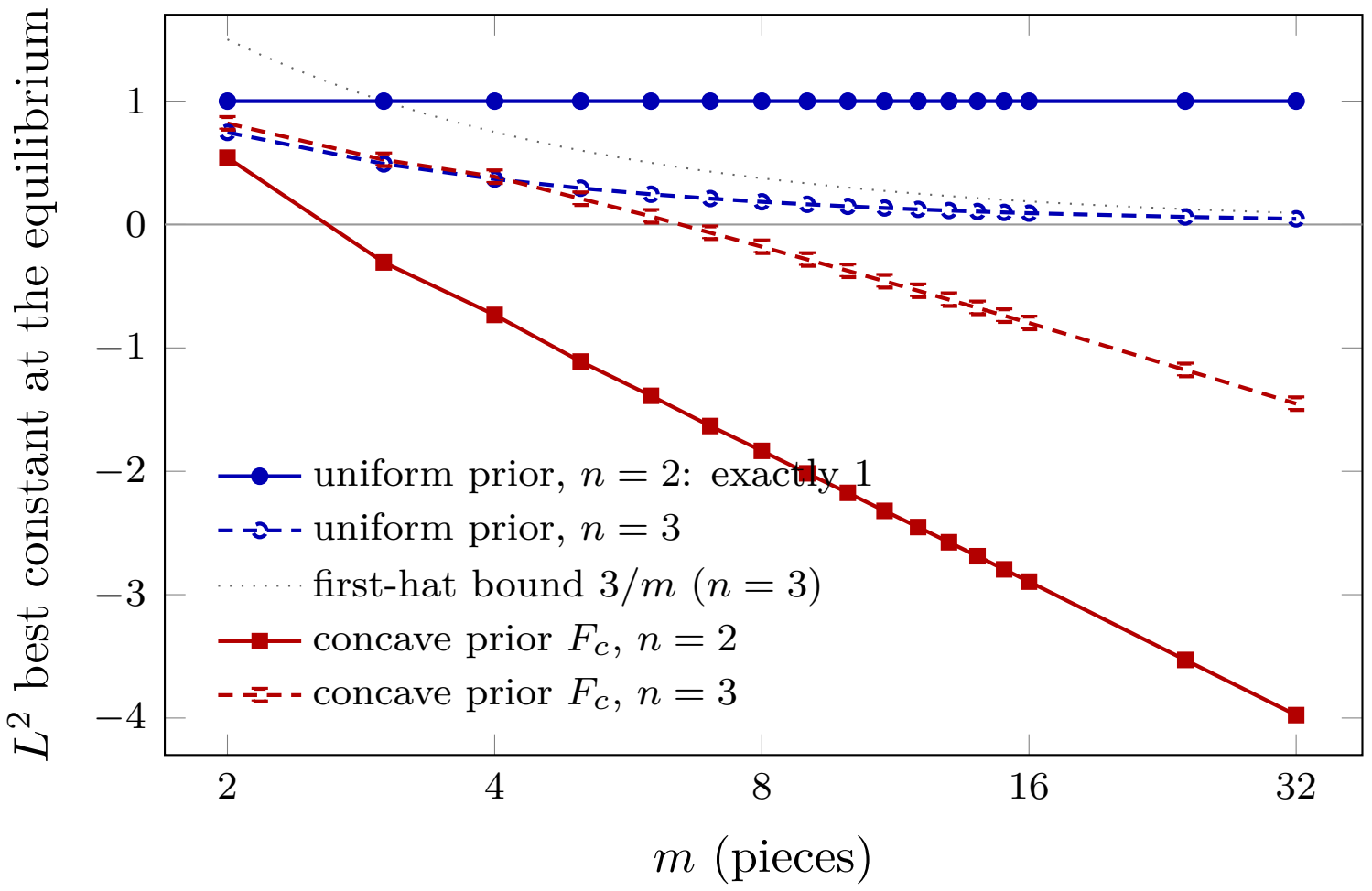
\begin{figure}[t]
\centering
\begin{tikzpicture}
  \begin{axis}[
    width=0.62\textwidth, height=0.42\textwidth,
    xmode=log, log basis x=2,
    xmin=1.7, xmax=38, ymin=-4.3, ymax=1.7,
    xtick={2,4,8,16,32}, xticklabels={2,4,8,16,32},
    minor xtick={},
    ytick={-4,-3,-2,-1,0,1},
    xlabel={$m$ (pieces)}, ylabel={$L^2$ best constant at the equilibrium},
    tick label style={font=\scriptsize}, label style={font=\small},
    legend style={font=\scriptsize, at={(0.03,0.03)}, anchor=south west, cells={anchor=west}, draw=none, fill=none},
    legend cell align=left,
  ]
    \addplot[color=black!40, thin, solid, forget plot] coordinates { (1.7, 0) (38, 0) };
    \addplot[color=blue!70!black, thick, mark=*, mark size=1.4pt] coordinates {
      (2, 1)
      (3, 1)
      (4, 1)
      (5, 1)
      (6, 1)
      (7, 1)
      (8, 1)
      (9, 1)
      (10, 1)
      (11, 1)
      (12, 1)
      (13, 1)
      (14, 1)
      (15, 1)
      (16, 1)
      (24, 1)
      (32, 1)
    };
    \addlegendentry{uniform prior, $n = 2$: exactly $1$}
    \addplot[color=blue!70!black, thick, densely dashed, mark=o, mark size=1.4pt] coordinates {
      (2, 0.746464)
      (3, 0.491639)
      (4, 0.368354)
      (5, 0.29466)
      (6, 0.245548)
      (7, 0.21047)
      (8, 0.184161)
      (9, 0.163699)
      (10, 0.147329)
      (11, 0.133935)
      (12, 0.122774)
      (13, 0.11333)
      (14, 0.105235)
      (15, 0.0982193)
      (16, 0.0920806)
      (24, 0.0613871)
      (32, 0.0460403)
    };
    \addlegendentry{uniform prior, $n = 3$}
    \addplot[color=black!60, thin, dotted] coordinates {
      (2, 1.5)
      (2.18102, 1.37551)
      (2.37841, 1.26134)
      (2.59368, 1.15666)
      (2.82843, 1.06066)
      (3.08442, 0.97263)
      (3.36359, 0.891905)
      (3.66802, 0.817881)
      (4, 0.75)
      (4.36203, 0.687753)
      (4.75683, 0.630672)
      (5.18736, 0.578329)
      (5.65685, 0.53033)
      (6.16884, 0.486315)
      (6.72717, 0.445953)
      (7.33603, 0.40894)
      (8, 0.375)
      (8.72406, 0.343877)
      (9.51366, 0.315336)
      (10.3747, 0.289165)
      (11.3137, 0.265165)
      (12.3377, 0.243157)
      (13.4543, 0.222976)
      (14.6721, 0.20447)
      (16, 0.1875)
      (17.4481, 0.171938)
      (19.0273, 0.157668)
      (20.7494, 0.144582)
      (22.6274, 0.132583)
      (24.6754, 0.121579)
      (26.9087, 0.111488)
      (29.3441, 0.102235)
      (32, 0.09375)
    };
    \addlegendentry{first-hat bound $3/m$ ($n = 3$)}
    \addplot[color=red!70!black, thick, mark=square*, mark size=1.3pt] coordinates {
      (2, 0.541896)
      (3, -0.30732)
      (4, -0.732748)
      (5, -1.11054)
      (6, -1.3882)
      (7, -1.63326)
      (8, -1.83459)
      (9, -2.01672)
      (10, -2.17494)
      (11, -2.32109)
      (12, -2.45239)
      (13, -2.57545)
      (14, -2.68836)
      (15, -2.79519)
      (16, -2.89456)
      (24, -3.52926)
      (32, -3.9764)
    };
    \addlegendentry{concave prior $F_c$, $n = 2$}
    \addplot[color=red!70!black, thick, densely dashed, mark=square, mark size=1.3pt] coordinates {
      (2, 0.819729)
      (3, 0.524591)
      (4, 0.387681)
      (5, 0.209207)
      (6, 0.0663474)
      (7, -0.0665296)
      (8, -0.180029)
      (9, -0.28336)
      (10, -0.375102)
      (11, -0.459698)
      (12, -0.53681)
      (13, -0.608838)
      (14, -0.675661)
      (15, -0.738683)
      (16, -0.797837)
      (24, -1.17887)
      (32, -1.45199)
    };
    \addlegendentry{concave prior $F_c$, $n = 3$}
  \end{axis}
\end{tikzpicture}
\caption{The $\Ltwo$ best constant of the certificate at the equilibrium
on the $m$-piece space. For the uniform prior it is exactly $1$ at
$n = 2$ (Theorem~\ref{thm:modulus}) and at $n = 3$ lies below the
first-hat bound $3/m$ of the proof; for the concave prior $F_c$ of
Section~\ref{sec:general-pl} it is negative beyond $m = 2$ at $n = 2$ and
beyond $m = 6$ at $n = 3$. Data from the computation recorded in
Appendix~\ref{app:numerics}.}
\label{fig:modulus}
\end{figure}

\entry{Theorem~\ref{thm:modulus}, the exact quadratic form.}
\emph{What.} The second-order coefficient of the pairing is the quadratic
form $Q_n$, a weighted $\Ltwo$ mass plus a boundary term at the top value;
for $n = 2$ the weight is $1$ and the form dominates the squared norm,
with equality exactly on directions with zero top value, which exist on
every grid with $m \ge 2$ pieces. The constant is therefore exactly $1$,
uniformly in $m$. For $n \ge 3$ the weight $\frac n2 v^{n-2}$ vanishes at
the bottom and the hat at the first knot gives the upper bound $\frac
n2(2/m)^{n-2}$, so the constant decays with the grid; at $n = 3$ it is
about $1.47/m$ (Appendix~\ref{app:numerics}). Figure~\ref{fig:modulus}
shows the constant against $m$ for the uniform prior and, as a preview of
Arc~III, for the concave prior $F_c$ of Section~\ref{sec:general-pl}, where
it turns negative beyond $m = 2$ at $n = 2$ and beyond $m = 6$ at $n = 3$.
\emph{Why.} This is the certificate at the equilibrium; the rest of
the arc asks on what neighbourhood it holds.
\emph{How.} Along a ray the denominator of the identity contributes from
third order on, so the second-order term is the numerator's integral over
the equilibrium slope; one integration by parts moves the derivative off
$\eta$ and leaves the boundary term. The form extends to $H^1$ by density,
which claims nothing about differentiability in directions of unbounded
slope. Primer: Appendix~\ref{primer:gateaux}. Proof:
Appendix~\ref{app:proofs-field}; the regularity note is
Appendix~\ref{app:regularity}. Machine-checked: the split, the
integration by parts, $Q_2 \ge \norm\eta^2$ with its equality case, the
first-hat bound and the $m = 1$ ratio (Appendix~\ref{app:lean}).

\entry{The knot identity \eqref{eq:knots}.}
\emph{What.} For a piecewise-linear $\eta$ the full pairing is a positive
integral plus a top term minus a sum of knot terms, and a knot term is
negative only where the slope drops.
\emph{Why.} It is the lens for everything about the ball: the hat's
exact zero, the lower bound and the limit are all read off it.
\emph{How.} Integrate the identity of Proposition~\ref{prop:minty} by parts
on each piece and sum by parts across the knots. Primer:
Appendix~\ref{primer:knots}. Machine-checked on a model of the
piecewise-linear space (Appendix~\ref{app:lean}).

\entry{Propositions~\ref{prop:ray} and \ref{prop:radius}, the hat and
the exponent.}
\begin{proposition}[the top-hat ray, exactly]
\label{prop:ray}
For $n = 2$, let $\eta = a\,h_{m-1}$, the hat at $v_{m-1}$ (so $\eta(1) = 0$,
slopes $+am$ and $-am$ on the last two pieces), with $0 < a < 1/(2m)$. Then
$\ip{\Fop(\bstar+\eta)}{\eta} = 0$ exactly at
\[
  a^* = \frac{1}{6m(m-1)}, \qquad \norm{\eta}_{\Ltwo} = \frac{\sqrt{2/(3m)}}{6m(m-1)},
\]
positive for $a < a^*$ and negative for $a > a^*$.
\end{proposition}

\noindent\emph{What.} Along the top hat the pairing vanishes exactly at $a^* =
1/(6m(m-1))$, which gives $r_m \le C_+m^{-5/2}$; crude bounds on the knot
terms give $r_m \ge m^{-5/2}/288$. The exponent $5/2$ is settled.
\emph{Why.} The exponent fixes the scale on which the constant is a
shape problem; the ray's zero is the first probe of the boundary and the
upper curve of Figure~\ref{fig:radius}.
\emph{How.} The hat has two nonzero slopes, so the knot identity is one
bracket, linear in the hat's height after a positive factor is removed.
The lower bound uses inverse inequalities for piecewise-linear functions
and the admissibility of slopes below $1/4$. Proof:
Appendix~\ref{app:proofs-field} and \ref{app:lower}. Machine-checked in
full, line by line, both constants (Appendix~\ref{app:lean}).
\entry{Proposition~\ref{prop:notextremal}, the hat is not extremal.}
\begin{proposition}[the top hat is not extremal]
\label{prop:notextremal}
At $m = 2$ the knot vector with $\eta_1 = 77/1000$ and $\eta_2 = -29/2000$
has
\[
  \norm\eta^2 = \frac{43807}{24000000} < \frac{1}{432}, \qquad
  \ip{\Fop(\bstar+\eta)}{\eta} = -\frac{9653}{4975632000} < 0,
\]
where $1/432$ is the square of the ray value $\sqrt{2/(3\cdot 2)}/(6\cdot 2
\cdot 1)$ of Proposition~\ref{prop:ray} and the pairing is evaluated by
\eqref{eq:knots}. Hence the first violation along the top-hat ray is not the
Minty radius.
\end{proposition}

\noindent\emph{What.} Already at $m = 2$ a rational knot vector has smaller norm
than the hat's zero and a negative pairing, so the hat's zero is not the
radius.
\emph{Why.} It closes the wrong path: the boundary of the ball has
to be found by a shape problem over all knot vectors, and the
counterexample's alternating tail says what shape to expect.
\emph{How.} Evaluate the knot identity at the vector. Machine-checked
(Appendix~\ref{app:lean}).
\entry{The leading-order problem \eqref{eq:leading}.}
\emph{What.} On the relevant ball the slopes are $O(1/m)$, so the knot
identity expands as the mass $\nu$ of the knot vector minus a cubic
functional $J$, with a relative error $O(1/m)$: the constant of the
radius is the maximum of $J/\nu^{3/2}$ over all finitely supported
sequences, a problem with no grid in it.
\emph{Why.} It turns the constant into a shape problem whose answer
is a profile; the limit theorem then shows the reduction is exact in the
limit by carrying the remainders.
\emph{How.} Expand the inverse-slope factors of the knot identity to
third order and collect. Primer: Appendix~\ref{primer:knots}.
Machine-checked: the expansion's cubic coefficient at two bidders
(Appendix~\ref{app:lean}).

\entry{Proposition~\ref{prop:stationary}, the stationary profile.}
\begin{proposition}[a stationary profile in closed form]
\label{prop:stationary}
Let $q = 2 - \sqrt3$, the root in $(0,1)$ of $q^2 - 4q + 1 = 0$. The
two-sided profile $x_j = (-q)^{|j|}$, $j \in \mathbb Z$, satisfies
$\partial\nu/\partial x_j = \tfrac13(x_{j-1} + 4x_j + x_{j+1}) = 0$ and
$\partial J/\partial x_j = 0$ for every $j \ne 0$, hence is a stationary point
of $R$, with
\[
  \nu = \frac{1}{\sqrt3}, \qquad J = \frac{2}{\sqrt3}, \qquad
  R = \frac{J}{\nu^{3/2}} = 2\cdot 3^{1/4}.
\]
With the end knot pinned at zero ($x_{-1} = 0$) the one-sided profile
$x_j = (-q)^j$, $j \ge 0$, is stationary with $\nu = (2+\sqrt3)/6$,
$J = \tfrac12 + 1/\sqrt3$ and $\nu^{3/2}/(4J) = (1+\sqrt3)/24$.
\end{proposition}

\noindent\emph{What.} The two-sided alternating geometric profile $(-q)^{|j|}$,
$q = 2 - \sqrt3$, is a stationary point of $J/\nu^{3/2}$, with both sums in
closed form and value $2\cdot 3^{1/4}$; the one-sided profile pinned at
zero is stationary too, with the constant $(1 + \sqrt3)/24$ that the ray
with the peak one knot below the top realizes.
\emph{Why.} It is the candidate maximizer that Theorem~\ref{thm:sharp}
proves is the maximizer, and the pinned constants $\tilde c_k$ of
Theorem~\ref{thm:radius} are its truncations.
\emph{How.} The mass stencil and the gradient of $J$ vanish off the peak
by $q^2 = 4q - 1$; at the peak Euler's identity for the homogeneous $J$ and
$\nu$ aligns the two gradients. Proof: Appendix~\ref{app:proofs-field}.
Machine-checked: the stencil, the gradient, the sums
(Appendix~\ref{app:lean}).
\entry{Theorem~\ref{thm:sharp}, the sharp inequality.}
\emph{What.} $J \le 2\cdot 3^{1/4}\nu^{3/2}$ for every finitely supported
sequence, with equality exactly on the alternating geometric profile; in
the stronger sup-norm form, $J \le 3\norm x_\infty\nu - \norm x_\infty^3/\sqrt3$.
\emph{Why.} The constant $2\cdot 3^{1/4}$ is the radius limit's
constant; the sup-norm form, applied to each tail with free ends, is what
the limit theorem uses, and its slacks are what the direction theorem
keeps.
\emph{How.} A potential argument. Normalize by the sup norm, write
$3\nu - J$ as a sum of piece costs $\varphi(x_j, x_{j+1})$, and guess the
potential from the answer: the minimum of the cost of a geometric tail
and the cost of jumping to the peak and continuing geometrically. The
Bellman inequality, one step's cost plus the potential at its end is at
least the potential at its start, reduces to four polynomial inequalities
on rectangles with explicit factorizations in $\mathbb Q(\sqrt3)$;
telescoping along the two chains from the peak gives the bound, and zero
slack at every step forces both chains geometric. Primer:
Appendix~\ref{primer:bellman}. Proof: Appendix~\ref{app:sharp}.
Machine-checked in full except the equality case
(Appendix~\ref{app:lean}).

\entry{Theorem~\ref{thm:radius}, the radius limit.}
\begin{theorem}[the radius limit, quantitative form]
\label{thm:radius-full}
For $n = 2$, $m^{5/2} r_m \to 1/(8\cdot 3^{1/4}) = 0.09498$, which is
$3^{5/4}/(4\sqrt2) = 0.698$ times the top-hat constant $\sqrt{2/3}/6$.
Quantitatively,
\[
  r_m \;\ge\; \Bigl(1 - \frac{18}{m}\Bigr)\frac{m^{-5/2}}{8\cdot 3^{1/4}}
  \quad (m \ge 3),
\]
and for every $k \ge 1$ there are directions $P^{(m,k)}$, the geometric
profile with the peak $k$ knots below the top and the end knot pinned, with
$r_m \le \rho_m^{(k)}$ and $m^{5/2}\rho_m^{(k)} \to \tilde c_k$, where
$\tilde c_k = \nu^{3/2}/(4J)$ of the profile $(-q)^{|j|}$, $j \le k-1$,
pinned to zero at $j = k$ ($\tilde c_1 = (1+\sqrt3)/24$, $\tilde c_2 =
0.09565$, $\tilde c_3 = 0.09504$, $\tilde c_4 = 0.094983$, with $\tilde c_k -
1/(8\cdot 3^{1/4}) = O(q^{2k})$). Quantitatively, $m^{5/2}\rho_m^{(k)} \le
\tilde c_k(1 + (2.4k + 18)/m)$ for $m \ge 100$ and $k \le m/8$, and hence
\[
  -\frac{18}{m} \;\le\; \varepsilon_m := 8\cdot 3^{1/4}\,m^{5/2}r_m - 1 \;\le\;
  \frac{\log m + 23}{m} \qquad (m \ge 100).
\]
The logarithm is real and its coefficient is sharp: with $L := 1/(2\log(1/q))
= 0.3797$,
\[
  \frac{L\log m - 18}{m} \;\le\; \varepsilon_m \;\le\; \frac{L\log m + 13}{m}
  \qquad \text{for all large } m
\]
(the upper bound from $m = 400$), so $m^{5/2}r_m = c\,(1 + (L\log m + O(1))/m)$.
\end{theorem}

\noindent\emph{What.} $m^{5/2}r_m \to 1/(8\cdot 3^{1/4})$, which is $0.698$ times
the hat's constant, with the quantitative form above: a lower bound
$(1 - 18/m)$ times the limit for $m \ge 3$, upper bounds along the pinned
geometric rays, and the logarithmic rate with its sharp coefficient $L =
1/(2\log(1/q))$.
\emph{Why.} It is the size of the ball, and its rate feeds the
direction theorem's slack budget.
\emph{How.} Expand the exact knot identity, not the reduction, with
remainders that are relatively $O(1/m)$ on the ball; write the knot sum
as a position-weighted $J$; remove the position weight by Abel summation
into an average over tails (Appendix~\ref{primer:abel}); apply the
free-end form of the sharp inequality, $J(y) \le 2\cdot 3^{1/4}\nu(y)^{3/2}
+ (1 + \sqrt3/6)\norm y_\infty(y_0^2 + y_N^2)$, to each tail. The upper bound
is the first zero along the geometric ray with the peak $k$ knots below
the top and the end knot pinned; the logarithm comes from balancing the
pinning loss $q^{2k}$ of that ray against its position cost $k/m$, and
keeping both costs in the slack budget gives the matching lower bound.
Proof: Appendix~\ref{app:radius}, the rate in Appendix~\ref{app:rates}.
Machine-checked: the free-end inequality and two sum identities of the
expansion (Appendix~\ref{app:lean}).
The $O(1)$ is a bounded sawtooth in $\log m$ from the integer depth of the
peak, measured $1.05 \pm 0.16$ for $m = 10^3$ to $10^6$
(Appendix~\ref{app:numerics}) and not identified; it is the second open
problem of Section~\ref{sec:conclusion}.

\entry{Theorem~\ref{thm:direction}, the extremal direction.}
\begin{theorem}[the extremal direction, quantitative form]
\label{thm:direction-full}
For $n = 2$, let $\eta_m$ attain the radius ($\ip{\Fop(\bstar+\eta_m)}{\eta_m} = 0$,
$\norm{\eta_m} = r_m$), let $y$ be its knot vector divided by its sup norm,
$j_0$ the peak and $d_m = m - j_0$ its depth, and set $\varepsilon_m =
m^{5/2}r_m\cdot 8\cdot 3^{1/4} - 1 \to 0$ and $\bar\varepsilon_m =
5.3(\varepsilon_m + 14/m)/(1 + \varepsilon_m)$. Then for all large $m$:
$y_{j_0} = +1$; the squared defects $\sum(y_{i+1} + qy_i)^2$ of the geometric
recursion along the right chain from the peak, and along any fixed number
of steps of the left chain, are at most $36\,\bar\varepsilon_m(1 + o(1))$, so
$y_{j_0+i} \to (-q)^{|i|}$ for every fixed $i$; the end knot satisfies $y_m^2
\le 2\nu(y)(\varepsilon_m + 14/m)/m \to 0$; and $d_m \le 1.3\,\bar\varepsilon_m
m = o(m)$ while $d_m \ge \log(1/\bar\varepsilon_m)/(2\log(1/q)) - O(1) \to
\infty$. With the rate of Theorem~\ref{thm:radius-full}, $\bar\varepsilon_m \le
5.3(\log m + 37)/m$, so the depth of the peak is $\Theta(\log m)$:
\[
  0.379\log m - 0.38\log(\log m + 37) - O(1) \;\le\; d_m \;\le\; 7\log m + 255
  \qquad \text{for all large } m,
\]
and in fact, with $L = 1/(2\log(1/q))$ as in Theorem~\ref{thm:radius-full},
\[
  L\log m - 4 \;\le\; d_m \;\le\; L\log m + 29 \qquad \text{for all large } m.
\]
\end{theorem}

\noindent\emph{What.} The minimizers that attain the radius converge to the
alternating geometric profile, with the peak $L\log m + O(1)$ knots below
the top and the end knot tending to zero; the quantitative form above
bounds the defects from the geometric recursion and the depth of the
peak.
\emph{Why.} It closes the arc: the first violation is a known
profile, and the same balance that gives the rate gives the depth.
\emph{How.} Chain stability, a stability version of the potential
argument. Keep the Bellman slacks that the telescoping discards; a step
is bad when its slack is small relative to its squared defect from the
geometric recursion, a bad step's predecessor is never bad, and pairing
each bad step with its predecessor shows the total slack along a chain
from a peak dominates a fixed fraction, $0.056$, of the total squared
defect. A minimizer's slack budget, read off the limit theorem's
expansion, is $O(\bar\varepsilon_m)$, so the defects near the peak are
small; the position weight turns the budget into $d_m = o(m)$, and the
truncation of the chain at the end knot, which must tend to zero, forces
$d_m \to \infty$. Primer: Appendix~\ref{primer:bellman}. Proof:
Appendix~\ref{app:direction}, the quantitative clauses in
Appendix~\ref{app:rates}. Machine-checked: the chain-stability lemma with
its constant (Appendix~\ref{app:lean}).
The constants are far from the truth. The computed minimizers
(Appendix~\ref{app:numerics}) are the profile $(-q)^{|j|}$ to three digits
on the side away from the boundary, with the peak one knot below the top
for $m \le 8$, two for $12 \le m \le 128$ and three for $m = 192, 256$,
exactly where $\tilde c_k(1 + k/m)$ is smallest, and $d_m = \lceil L\log m
+ 0.15\rceil$ for $m = 64$ to $10^6$; $m\varepsilon_m$ lies between $2.0$
and $3.2$ for $3 \le m \le 256$, which is $L\log m + 1.0 \pm 0.2$, so the
logarithm of Theorem~\ref{thm:radius} is invisible at $m \le 256$ and
unmistakable by $m = 10^6$, where $m\varepsilon_m = 6.22$.

\subsection{Arc II: the learners}
\label{app:arc2}

\begin{figure}[t]
\centering
\begin{tikzpicture}[
  box/.style={draw, rounded corners=1pt, font=\scriptsize, inner sep=2.5pt, align=center},
  >={Latex[length=1.6mm]}, every path/.style={->, thin}]
\node[box] (mod) at (0,0) {Thm.~\ref{thm:modulus}\\the weight (Arc~I)};
\node[box] (two) at (3.4,0) {Prop.~\ref{prop:twoplayer}\\two-bidder form};
\node[box] (cor) at (6.8,0) {Cor.~\ref{cor:twoplayer}\\constant $1$};
\node[box] (npl) at (3.4,-1.3) {Prop.~\ref{prop:nplayer}\\$n$-bidder form};
\node[box] (gen) at (6.8,-1.3) {Prop.~\ref{prop:generalform}\\general prior};
\node[box] (arc3) at (10.2,-1.3) {Prop.~\ref{prop:general}\\the weight (Arc~III)};
\draw (mod) -- (two);
\draw (two) -- (cor);
\draw (two) -- (npl);
\draw (npl) -- (gen);
\draw (gen) -- (arc3);
\end{tikzpicture}
\caption{The chain of Arc II. The two-bidder form is computed first and
its constant read off; the $n$-bidder and general-prior forms repeat the
computation, and the last hands its weight to Arc~III.}
\label{fig:arc2-map}
\end{figure}
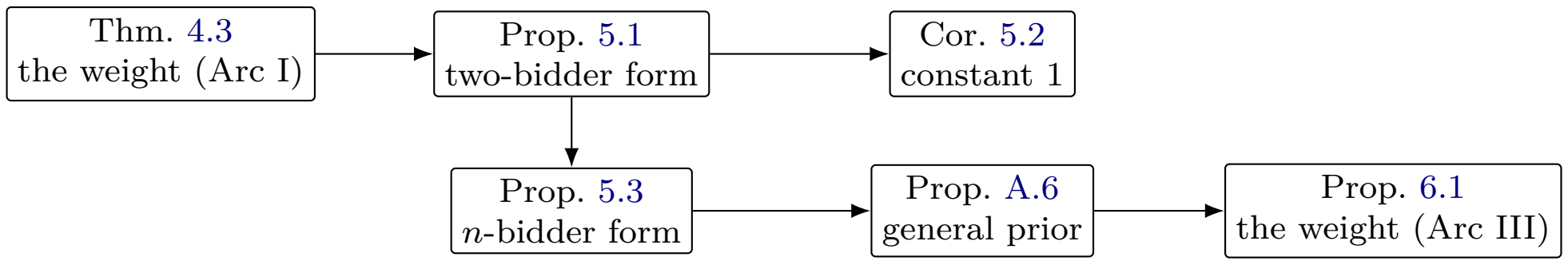

\entry{The map.} Figure~\ref{fig:arc2-map} is the chain. One thing
organizes it: the learners' field is only
one-sidedly differentiable at the symmetric equilibrium, because of the
cap on the win probability, and the object to compute is therefore a
one-sided directional form, an interior part plus a cap part, each exact.
The two-bidder form is interior part plus the square of the larger top
value (Proposition~\ref{prop:twoplayer}); its interior matrix has
eigenvalues $1$ and $7$, the smaller on symmetric directions, so the
constant is $1$ (Corollary~\ref{cor:twoplayer}). With $n$ bidders the
capped field is linear in the number of capped opponents, so overlapping
cap regions add and the form has one cap term per pair, with the
single-population weight on symmetric directions
(Proposition~\ref{prop:nplayer}). With a general prior the same expansion
under the equilibrium ODE gives a form whose symmetric eigenvalue is the
weight $w_F$ of Arc~III and whose anti-symmetric eigenvalue exceeds it
(Proposition~\ref{prop:generalform}).

\entry{The entries.} The rest of this subsection takes the map box by
box. The order is the body's; where the chain branches, the branch the
body takes first comes first, and a result the body does not state
follows the result it extends. Each entry says what the result
establishes, why the chain needs it, and how it is proved, and ends
with its pointers: the primer entry for anything unusual, the proof in
Appendix~\ref{app:proofs}, and what Appendix~\ref{app:lean} checks by
machine. An entry that opens with a statement is a result the body does
not state, or states only in compact form; the statement comes first and
the entry glosses it.

\entry{Proposition~\ref{prop:twoplayer}, the two-bidder form.}
\emph{What.} The one-sided second-order form of the two-bidder operator at
the symmetric equilibrium: $4\int(\eta_1^2 + \eta_2^2) - 6\int\eta_1\eta_2$
plus twice the square of the larger top value.
\emph{Why.} Its matrix gives the constant; its structure, interior
part plus cap term, is what the $n$-bidder and general-prior
computations repeat.
\emph{How.} A bidder whose top bid exceeds the other's wins surely on a
region of length of order the perturbation where the field is $1$
instead of the interior formula's value, an order-one jump, so the region
contributes at second order. The interior part is an implicit-function
computation of the inverse bid, $u = v + 2\varepsilon(\eta_1 - \eta_2)$ to
first order, plus one integration by parts; the cap part is the region's
length times the jump, and adding it to the interior boundary term
$2\eta_1(1)\eta_2(1)$ gives $2\max(\eta_1(1),\eta_2(1))^2$. Primer:
Appendix~\ref{primer:gateaux}. Proof: Appendix~\ref{app:proofs-arc2}.
Machine-checked: the interior part; the cap term, a one-sided limit
over the cap region, is not (Appendix~\ref{app:lean}).

\entry{Corollary~\ref{cor:twoplayer}, the constant.}
\emph{What.} $B \ge \norm{(\eta_1,\eta_2)}^2$, with equality iff the two
perturbations coincide and have zero top value; the infimum over the unit
sphere is attained on every common grid and equals $1$.
\emph{Why.} It is the single-population constant: the certificate
of Arc~I transfers to independent learners, and the $\Ltwo$ norm is a
Lyapunov function for the first-order dynamics at the equilibrium.
\emph{How.} The matrix $\bigl(\begin{smallmatrix}4&-3\\-3&4\end{smallmatrix}\bigr)$
has eigenvalues $1$ and $7$, the cap term is nonnegative, and a nonzero
symmetric direction with zero top value exists once $m \ge 2$. Proof:
Appendix~\ref{app:proofs-arc2}. Machine-checked in full given the
proposition (Appendix~\ref{app:lean}).

\entry{Proposition~\ref{prop:nplayer}, the $n$-bidder form.}
\emph{What.} The same structure with $n$ bidders: a weighted integral
with the matrix $\frac{n^2}{n-1}$ on the diagonal and
$-\frac{n(n+1)}{(n-1)^2}$ off it, plus one cap term per pair; on
symmetric directions it is $nQ_n$, and in general it dominates
$\frac n2\int v^{n-2}\sum\eta_i^2$.
\emph{Why.} The single-population weight $\frac n2v^{n-2}$ of
Theorem~\ref{thm:modulus} is what $n$ learners see, so the certificates of
Arc~III, computed for the single-population operator, are theirs.
\emph{How.} Several cap regions coexist near the top, one per ordered
pair, and overlap; the capped field is linear in the number of capped
opponents, so the regions add, and each pair contributes
$\frac{n}{(n-1)^2}\max(a_i,a_j)^2$ once its interior boundary term is
included. The eigenvalues follow from writing $\sum_{i<j}\eta_i\eta_j$
through $(\sum\eta_i)^2$. Proof: Appendix~\ref{app:proofs-arc2}.
Machine-checked: the interior part and the eigenvalues; the cap terms
as for two bidders (Appendix~\ref{app:lean}).

\entry{Proposition~\ref{prop:generalform}, the form for a general prior.}
\begin{proposition}[the learners' form for a general prior]
\label{prop:generalform}
Let $F$ have density $f$ with $f(1) > 0$ and symmetric equilibrium
$\bstar$, let $\eta_i$ be piecewise linear on a common grid with
$\eta_i(0) = 0$ and $a_i := \eta_i(1)$, and let $B_n^F$ be the one-sided
directional form of $\sum_i\ip{\Fop_i}{\eta_i}$ in $\Ltwo(f\,dv)$ at
$(\bstar,\dots,\bstar)$. With $K = -F\,(f/\bstar{}')'/f$ and $A =
fF^{n-1}/\bstar{}'$,
\[
  B_n^F(\eta) = \int_0^1 \Bigl[\sum_i d\,\eta_i^2 + \sum_{i<j} o\,\eta_i\eta_j\Bigr]dv
  + \frac{2c_F}{n-1}\sum_{i<j}\max(a_i,a_j)^2,
\]
\[
  d = fF^{n-2}\Bigl(K + \frac{nf}{\bstar{}'}\Bigr), \qquad
  o = -\frac{2fF^{n-2}}{n-1}\Bigl(K + \frac{f}{\bstar{}'}\Bigr) - \frac{A'}{n-1}, \qquad
  c_F = \frac{f(1)}{2\bstar{}'(1)} = \frac{A(1)}{2}.
\]
The symmetric eigenvalue $d + (n-1)o/2$ of the integrand's matrix is the
weight $w_F$ of Proposition~\ref{prop:general}, so $B_n^F = nQ_F$ on
symmetric directions; the anti-symmetric eigenvalue $\lambda_\perp = d -
o/2$ satisfies, with $s = v - \bstar$,
\[
  w_F = \frac{F^{n-1}(F - fs)}{2(n-1)s^2}, \qquad
  \lambda_\perp - w_F = \frac{nF^{n-1}(F + fs)}{2(n-1)^2 s^2} > 0, \qquad
  \frac{\lambda_\perp}{w_F} = \frac{(2n-1)(n-1) + \bstar{}'}{(n-1)(n-1-\bstar{}')},
\]
so wherever $w_F > 0$ the ratio exceeds $(2n-1)/(n-1)$.
\end{proposition}

\noindent\emph{What.} With the equilibrium ODE imposed the learners' form for a
prior $F$ has symmetric eigenvalue exactly the weight $w_F$ of
Proposition~\ref{prop:general}, so it is $nQ_F$ on symmetric directions,
and anti-symmetric eigenvalue $\lambda_\perp > w_F$ with the ratio in
closed form; the cap constant is $c_F = f(1)/(2\bstar{}'(1))$.
\emph{Why.} It hands Arc~III its weight from the learners' side:
whatever Arc~III proves about $w_F$ is about what the learners see, and
$w_F < 0$ exactly where $\bstar{}' > n-1$.
\emph{How.} The expansion of Proposition~\ref{prop:nplayer} with the
prior in the field and the ODE used to eliminate the gap; the cap term by
the same argument with $f(1) > 0$. Proof:
Appendix~\ref{app:proofs-arc2}. Not machine-checked.
\subsection{Arc III: general priors and bidder counts}
\label{app:arc3}

\begin{figure}[t]
\centering
\begin{tikzpicture}[x=2.75cm, y=1.25cm,
  box/.style={draw, rounded corners=1pt, font=\scriptsize, inner sep=2pt, align=center, text width=2.2cm},
  >={Latex[length=1.6mm]}, every path/.style={->, thin}]
\node[box] (general) at (0,0) {Prop.~\ref{prop:general}\\identity, weight $w_F$};
\node[box] (power) at (1,0) {Thm.~\ref{thm:power}\\power priors};
\node[box] (knots) at (0,-1) {knot masses at $\bstar_m$\\(Section~\ref{sec:general-pl})};
\node[box] (osc) at (1,-1) {Prop.~\ref{prop:oscillation}\\oscillation law};
\node[box] (osc2) at (0,-2) {Thm.~\ref{thm:oscillation2}\\two-scale form, $n \le 3$};
\node[box] (eosc) at (1,-2) {Bloch values\\$E_F^{\mathrm{osc}}$};
\node[box] (limit2) at (2,-2) {Thm.~\ref{thm:limit2}\\the limit};
\node[box] (weighted) at (3,-2) {Cor.~\ref{cor:weighted}\\weighted geometry};
\node[box] (bottom3) at (4,-2) {Prop.~\ref{app:bottom3}\\bottom-edge rate};
\node[box] (four) at (0,-3) {Thm.~\ref{thm:four}\\$n = 4$, small curvature};
\node[box] (layer) at (1,-3) {Thm.~\ref{thm:layer}\\$n = 4$, the layer};
\node[box] (layer5) at (2,-3) {Prop.~\ref{prop:layer5}\\the layer at $n = 5$};
\node[box] (reduction) at (3,-3) {Prop.~\ref{prop:reduction}\\bottom knots inert};
\node[box] (conj) at (4,-3) {Conj.~\ref{conj:knot}\\$n \ge 4$};
\node[box] (H2) at (0,-4) {Thm.~\ref{app:H2}\\second-order law};
\node[box] (interior) at (1,-4) {Thm.~\ref{thm:interior}\\interior rate};
\node[box] (airy) at (2,-4) {Thm.~\ref{thm:airy}\\top-knot rate};
\node[box] (local) at (3,-4) {Prop.~\ref{prop:local}\\exact local bound};
\node[box] (interiorw) at (1,-5) {Thm.~\ref{app:interiorw}\\weighted rate};
\node[box] (eventual) at (2,-5) {Cor.~\ref{cor:eventual}\\positive from $m_0$ on};
\node[box] (pilot) at (3,-5) {Cor.~\ref{cor:pilot}\\one prior, every $m$};
\node[box, dashed] (coda) at (4,-5) {$f(1) = 0$, outside the chain\\(leading order)};
\draw (general) -- (power);
\draw (general) -- (knots);
\draw (knots) -- (osc);
\draw (osc) -- (osc2);
\draw (osc2) -- (eosc);
\draw (eosc) -- (limit2);
\draw (limit2) -- (weighted);
\draw (weighted) -- (bottom3);
\draw (limit2) -- (reduction);
\draw (reduction) -- (conj);
\draw (osc2) -- (four);
\draw (four) -- (layer);
\draw (layer) -- (layer5);
\draw (four) -- (H2);
\draw (H2) -- (interior);
\draw (H2.south east) to[out=-25,in=-155] (airy.south west);
\draw (interior) -- (interiorw);
\draw (interior) -- (eventual);
\draw (airy) -- (eventual);
\draw (eventual) -- (pilot);
\draw (local) -- (pilot);
\end{tikzpicture}
\caption{The chain of Arc III, one row per subsection of Section~6, the
rates split over two rows. The second-order amplitude law rests on the
two-scale form of the row above it (Theorem~\ref{thm:oscillation2} for two
and three bidders, Theorem~\ref{thm:layer} for four); the exact local
bound rests on the knot-mass decomposition of the second row. The dashed
box is the case outside the chain, a density vanishing at the top.}
\label{fig:arc3-map}
\end{figure}
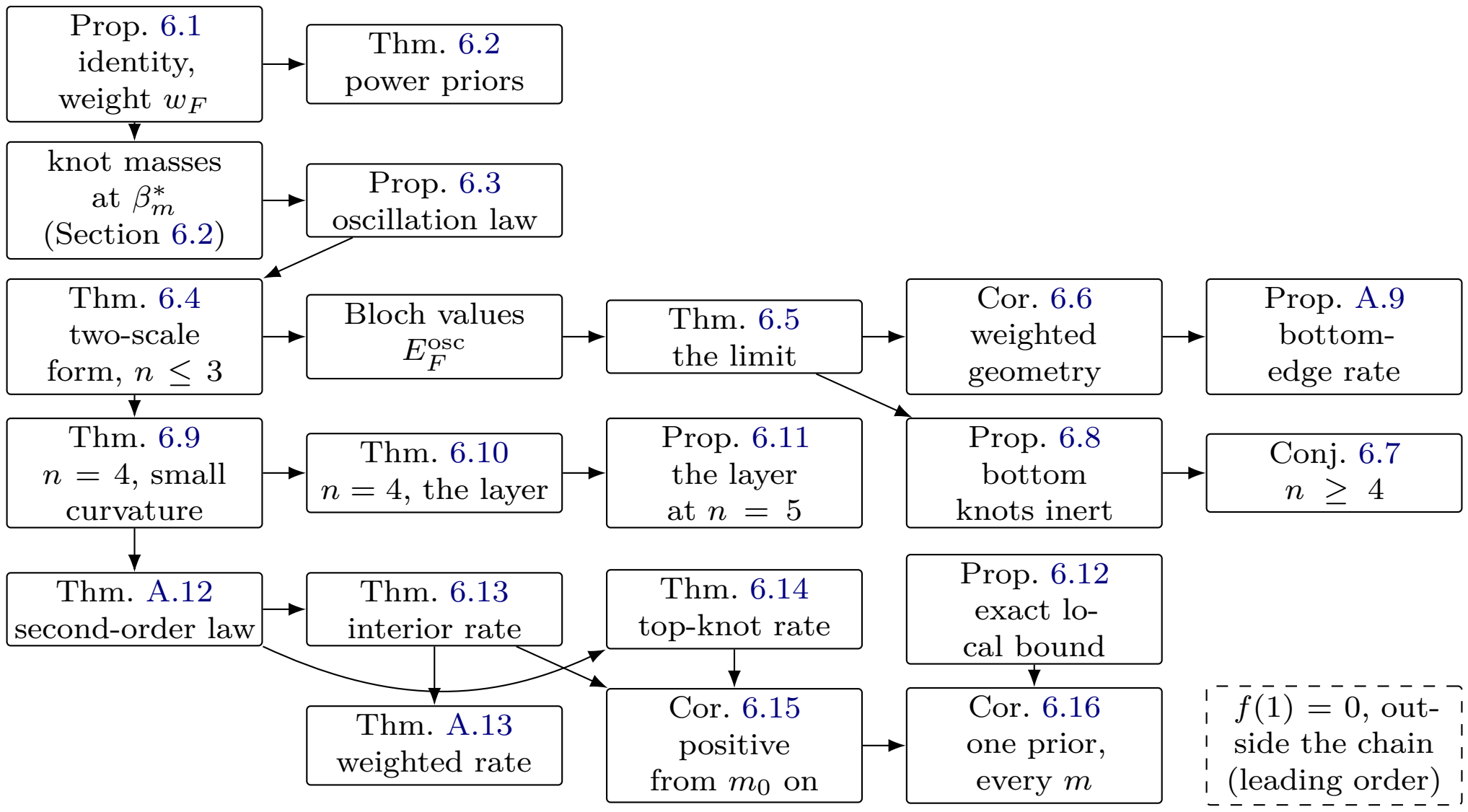

\entry{The map.} Figure~\ref{fig:arc3-map} is the chain. Two things
organize it. First, there are two
certificates: the one at the function-space equilibrium, whose weight
$w_F$ is explicit, and the one at the $m$-space equilibrium $\bstar_m$,
which differs by knot masses. Second, $\bstar_m$ is not the interpolant of
$\bstar$: its error is a checkerboard with a law, and because the knot
masses alternate with it the limit of the $m$-space certificate is a
closed form in $F$ that no smooth-slope formula gives. The Minty identity
generalizes with the prior's distribution and density in the integrand
once the equilibrium ODE is imposed, and its second-order coefficient is
a weight (Proposition~\ref{prop:general}); the equilibrium is linear
exactly for the power priors, for which the discretized certificate is
this one (Theorem~\ref{thm:power}). Otherwise the half hat at the top
seeds a checkerboard that propagates down with an explicit amplitude,
and the knot masses alternate with it (Proposition~\ref{prop:oscillation}).
Proving that two-scale form is the deep step: Newton--Kantorovich in
weighted norms, with the Jacobian's two modes and its Green's function
made explicit and two Newton steps taken with the structure of each
correction tracked (Theorem~\ref{thm:oscillation2}). A two-sided Bloch
inequality and a localization then give the limit of the certificate in
closed form, the characterization theorem (Theorem~\ref{thm:limit2}), in
the $\Ltwo$ geometry and in the geometry weighted by $w_F$
(Corollary~\ref{cor:weighted}), with a universal rate where the limit's
only zero is the bottom edge (Proposition~\ref{app:bottom3}). At four
bidders the argument needs class norms and, to remove its smallness
condition, the universal bottom layer verified by interval arithmetic
and matched to the interior (Theorems~\ref{thm:four}, \ref{thm:layer};
the layer alone at five bidders, Proposition~\ref{prop:layer5}); the
bottom knots play no part in the limit (Proposition~\ref{prop:reduction}),
so what remains for more bidders is the slope law inland
(Conjecture~\ref{conj:knot}). Finally, the amplitude law to second order
(Theorem~\ref{app:H2}) and the semiclassical reading give the rates with
positive leading terms (Theorems~\ref{thm:interior}, \ref{thm:airy},
\ref{app:interiorw}), splitting the knot masses between their pieces
gives an exact lower bound at every $m$ (Proposition~\ref{prop:local}),
and a positive limit therefore means a positive certificate from some $m$
on, and for one prior at every $m$ (Corollaries~\ref{cor:eventual},
\ref{cor:pilot}).

\entry{The entries.} The rest of this subsection takes the map box by
box. The order is the body's; where the chain branches, the branch the
body takes first comes first, and a result the body does not state
follows the result it extends. Each entry says what the result
establishes, why the chain needs it, and how it is proved, and ends
with its pointers: the primer entry for anything unusual, the proof in
Appendix~\ref{app:proofs}, and what Appendix~\ref{app:lean} checks by
machine. An entry that opens with a statement is a result the body does
not state, or states only in compact form; the statement comes first and
the entry glosses it.

\entry{Proposition~\ref{prop:general}, the identity and the weight.}
\emph{What.} The Minty pairing for a prior $F$ as one exact integral,
its second-order coefficient $Q_F$ as a weight $w_F$ times $\eta^2$ plus a
boundary term, and the criterion: the derivative is coercive uniformly
over the piecewise-linear spaces iff $\operatorname{ess\,inf}w_F/f > 0$.
\emph{Why.} Everything in the arc is a statement about $w_F$ and
about what the discretization adds to it.
\emph{How.} The field with a prior is $-F^{n-1} + (n-1)F^{n-2}f(v-b)/b'$;
the equilibrium ODE $\bstar{}' = (n-1)(f/F)(v-\bstar)$ eliminates the gap,
and the two terms combine as for the uniform prior; one integration by
parts with the weight $fF^{n-1}/\bstar{}'$ gives $Q_F$, and the ODE
rewrites $w_F$ as $F^{n-1}(F - f(v-\bstar))/(2(n-1)(v-\bstar)^2)$, so $w_F
< 0$ exactly where $\bstar{}' > n-1$, which a log-concave $F$ never
achieves. Hats with zero top value placed near the infimum of $w_F/f$
show the criterion is sharp. Primer: Appendices~\ref{primer:vi},
\ref{primer:gateaux}. Proof: Appendix~\ref{app:proofs-general}.
Machine-checked: the identity under the ODE, $Q_F$, both forms of $w_F$,
the sign criterion, the log-concavity bound, and the knot-mass
decomposition at a piecewise-linear base point (Appendix~\ref{app:lean}).

\entry{Theorem~\ref{thm:power}, the power priors.}
\emph{What.} The equilibrium is linear iff $F = v^a$; then $\bstar$ lies
in every $m$-piece space, there are no knot masses, and the certificate
is uniform in $m$ iff $\kappa = a(n-1) \le 1$, with explicit constants and
the decay rate $m^{-(\kappa-1)}$ otherwise.
\emph{Why.} It is the one family where the two certificates
coincide, the uniform prior at $a = 1$ among them; for every other prior
the $m$-space equilibrium is a different point.
\emph{How.} A linear equilibrium means $\int_0^vF^{n-1} = c\,vF^{n-1}$;
differentiating gives a first-order ODE whose solutions are the powers.
The constants are a direct computation, and for $\kappa > 1$ the first hat
gives the upper bound and the monotone weight the matching lower one.
Proof: Appendix~\ref{app:proofs-general}. Numerically $\bstar$ is the
unique equilibrium of the $m$-piece space for these priors
(Appendix~\ref{app:numerics}). Machine-checked in its uniform case only
(Appendix~\ref{app:lean}).

\entry{The knot masses at $\bstar_m$ (Section~\ref{sec:general-pl}).}
\emph{What.} At the $m$-space equilibrium the certificate's form
decomposes exactly into a smooth part, a boundary term and point masses
at the interior knots, $-\tfrac12(v_k - \bstar_m(v_k))(n-1)F^{n-2}f(v_k)^2
(1/\sigma_{k+1}^2 - 1/\sigma_k^2)\eta(v_k)^2$, negative where the slopes
decrease and of the same order in $m$ as a hat's $\Ltwo$ mass.
\emph{Why.} The masses are what the discretization adds to $w_F$;
their size depends on the slopes of $\bstar_m$, which the next entry
shows oscillate.
\emph{How.} Integrate the second-order form by parts piece by piece with
the weight $G_1(v-b)/\sigma_j^2$ and sum by parts. Primer:
Appendix~\ref{primer:knots}. Machine-checked (Appendix~\ref{app:lean}).

\entry{Proposition~\ref{prop:oscillation}, the oscillation law.}
\emph{What.} If the knot error of $\bstar_m$ is a slowly varying
checkerboard plus a slowly varying smooth part, the checkerboard's
amplitude is $\bstar{}''(1)h^2/12$ times the profile $\tilde Q(v)/\tilde
Q(1)$, $\tilde Q = \bstar{}'^{4/3}/(F^{(n-1)/3}f^{2/3})$, so the slopes
oscillate by $\pm\bstar{}''(1)(h/6)\hat Q$ and the knot masses alternate as
$1 \pm 2R(v)$ with $R(1) = \tfrac16$.
\emph{Why.} It gives the alternating coefficients that the limit of
the certificate is computed from; the next two theorems prove its
hypothesis.
\emph{How.} The interpolant's Galerkin residual cancels against every full
hat at leading order but not against the half hat at the top, which only
the top piece's slope can balance; the Jacobian at the interpolant is
tridiagonal, each row has a root near $-1$ whose modulus is $1$ plus $h$
times the logarithmic derivative of $\tilde Q$, a computer-algebra
identity, and the product of these roots is the alternating mode
(Appendix~\ref{primer:recurrences}); under the hypothesis the nonlinear
remainder is beyond the law's order, and the top row fixes the amplitude.
The hypothesis is observed with $\gamma = 1$ (Appendix~\ref{app:numerics}).
Primer: Appendices~\ref{primer:galerkin}, \ref{primer:recurrences}. Proof:
Appendix~\ref{app:oscillation}. Machine-checked: the Liouville--Green
step's residual identity (Appendix~\ref{app:lean}).

\entry{Theorem~\ref{thm:oscillation2}, the two-scale form for two and
three bidders.}
\begin{theorem}[the oscillation law for two and three bidders, full form]
\label{thm:oscillation2-full}
Let $n \in \{2, 3\}$ and $F \in C^5[0,1]$ with $f > 0$ on $[0,1]$. There are $h_0 > 0$
and $C$ such that for $h = 1/m \le h_0$ the Galerkin system has a solution
$\bstar_m$, unique in a ball of $h$-independent radius around the second
Newton iterate from the interpolant in the norm $\max_k|\delta_k|/(h^2\hat
Q(v_k))$, $\hat Q = \tilde Q/\tilde Q(1)$, and its knot error is
\[
  \delta_k = (-1)^{m-k}\,\frac{\bstar{}''(1)h^2}{12}\,\hat Q(v_k)\,(1 + \theta_k)
  + u_k + \rho_k, \qquad |\theta_k| \le \frac{C}{k} + C h^{(4-n)/3},
\]
with $|u_k| \le Ch^2\hat Q(v_k)$, $|u_k - u_{k-1}| \le C(1/k + hv_k^{-(n-1)/3})
h^2\hat Q(v_k)$ and $|\rho_k| \le Ch^{(4-n)/3}h^2\hat Q(v_k)$, uniformly in
$k$. In particular the hypothesis of Proposition~\ref{prop:oscillation}
holds, in the weakened form with the remainder $\rho_k$, with $\gamma = 1$
on every $[v_0, 1]$; the oscillation law holds with error $O(h^{(4-n)/3})$
there, $O(h^{1/3})$ for three bidders, and for two the remainder is in fact
$O(h)$, the $h^{2/3}$ of the uniform bound being $hv_k^{-1/3}$ at the bottom
knot; and the amplitude follows $\hat Q$ down to the bottom knot with
relative error $O(1/k)$.
\end{theorem}

\noindent\emph{What.} For $n \le 3$ the Galerkin system has a solution on every fine
grid, unique in a ball of $h$-independent radius, whose knot error is the
law's checkerboard with relative error $O(1/k) + O(h^{(4-n)/3})$, plus a
smooth part, plus a remainder; the full form is above.
\emph{Why.} It is the hypothesis of the oscillation law, so the
limit theorem's coefficients are known; it is also where the argument's
limits show, since its bookkeeping stops at four bidders.
\emph{How.} Newton--Kantorovich (Appendix~\ref{primer:newton}) in norms
in which the law has unit size: the error against $h^2\hat Q$, the
forcing against $h^3$ times a power of position. The inverse of the
Jacobian at the interpolant is bounded through its two pinned modes and
its Green's function, by a discrete Liouville--Green analysis
(Lemmas~\ref{app:L1}, \ref{app:L2}, Theorem~\ref{app:T6}); the first
Newton step's residual is slowly varying because the quadratic
remainder's cross term is odd about each piece midpoint
(Lemmas~\ref{app:L6}, \ref{app:L7}); the Jacobian at the first iterate
keeps a bounded inverse (Lemma~\ref{app:L5}); after the second step the
Kantorovich number is $O(h)$ and the standard theorem closes. The two
obstacles are the bottom of the support, where the Jacobian's entries
degenerate, and the $O(1)$ Lipschitz constant along alternating
directions, which is why no contraction is available at the interpolant.
Proof: Appendix~\ref{app:global}, Steps A to D. Not machine-checked.
\entry{The Bloch values and the limit function $E_F^{\mathrm{osc}}$.}
\emph{What.} With two-periodic knot masses the knot part of the
certificate per unit mass ranges, over all profiles, over an explicit
interval attained by two-periodic profiles, and its infimum $2\tau_F -
\sqrt{\tau_F^2 + 3\zeta_F^2}$ added to the smooth value gives the limit
function $E_F^{\mathrm{osc}} = w_F/f + \tau_F - \sqrt{\tau_F^2 + 3\zeta_F^2}$,
explicit in $F$, $f$, $f'$ and $v - \bstar$.
\emph{Why.} It is the candidate limit; the theorem proves it is the
limit.
\emph{How.} The two-sided Bloch inequality (Lemma~\ref{app:bloch}): each
piece's $2\times2$ form is positive semidefinite exactly at the lower
value $g_-$, by the pencil's characteristic polynomial, and the
piece-split sum equals the knot sum with end weights $\tfrac12$. Primer:
Appendix~\ref{primer:bloch}. Proof: Appendix~\ref{app:limit}.
Machine-checked in full (Appendix~\ref{app:lean}).

\entry{Theorem~\ref{thm:limit2}, the characterization theorem.}
\emph{What.} For two and three bidders the best constant of the
piecewise-linear certificate at $\bstar_m$ converges to
$\min_{[0,1]}E_F^{\mathrm{osc}}$; so the certificate fails uniformly in $m$
whenever that minimum is negative, which happens whenever the smooth-slope
value $\operatorname{ess\,inf}(w_F - \kappa_F)/f$ is.
\emph{Why.} It is the arc's main result for $n \le 3$; everything
after it is about more bidders, finite $m$, and the rate.
\emph{How.} A localization by a partition of unity whose squares sum to
one (Appendix~\ref{primer:ims}): the knot and boundary terms distribute
over windows exactly and a piece term pays $(h/\text{width})^2$ times the
mass. On a window away from the bottom the coefficients are frozen at
the centre at the cost of the width, the Bloch inequality bounds the
windowed quotient below by the limit function at the centre, and the top
window's boundary term dominates its half knot; a bottom window is
handled by continuity at zero. The upper bound is the two-periodic null
profile at a near-minimizer under a smooth bump. Proof:
Appendix~\ref{app:limit}. Machine-checked: the Bloch inequality; the
localization is analytic (Appendix~\ref{app:lean}).

\entry{Corollary~\ref{cor:weighted}, the weighted geometry.}
\begin{corollary}[the certificate in the weighted geometry, full form]
\label{cor:weighted-full}
Let $n \in \{2, 3\}$ and $F$ as in Theorem~\ref{thm:limit2}. For a continuous
weight $\vartheta > 0$ on $[0,1]$, or one vanishing at $v = 0$ like $w_F/f$ at
$n = 3$, which the proof covers, the best constant $c_m^\vartheta$ of the
piecewise-linear certificate at $\bstar_m$ relative to $\int\vartheta\eta^2f$
tends to $\min_{[0,1]}E_F^{\mathrm{osc}}/\vartheta$. Assume $w_F > 0$ on
$(0,1]$; with $\vartheta = w_F/f$, of order $v^{n-2}$ at the bottom and
positive there when $f(0) > 0$,
\[
  \lim_{m\to\infty} c_m^{w_F/f} = \min_{v\in[0,1]}\Bigl[1 + \bigl(\tau_F - \sqrt{\tau_F^2 + 3\zeta_F^2}\bigr)\frac{f}{w_F}\Bigr],
\]
the bracket being read with its continuous extension, $1$, at $v = 0$. For
the power priors $F = v^a$, $a > 0$, outside the hypotheses above unless
$a = 1$ but elementary, the $m$-space equilibrium is $\bstar$ itself for
every $n$ and $m$, and at it $c_m^{w_F/f} = 1$ exactly, attained precisely
on $\{\eta(1) = 0\}$.
\end{corollary}

\noindent\emph{What.} Relative to a continuous positive weight $\vartheta$ the
constant tends to $\min E_F^{\mathrm{osc}}/\vartheta$; with $\vartheta = w_F/f$
the limit is $\min[1 + (\tau_F - \sqrt{\tau_F^2 + 3\zeta_F^2})f/w_F]$, at
most $1$, and exactly $1$ for the power priors at every $m$ and $n$.
\emph{Why.} For three or more bidders the $\Ltwo$ limit is at most
zero, so the weighted geometry is where a positive certificate lives; the
rate theorems for $n = 3$ are stated there.
\emph{How.} The same proof with a weighted mass: on a window the
weighted mass is the weight at the centre times the mass. For a weight
vanishing at the bottom like $w_F/f$ at $n = 3$ every error has to be
small relative to the weight, which the integrand's structure gives.
Proof: Appendix~\ref{app:limit}.
\entry{Proposition~\ref{app:bottom3}, the bottom-edge rate at three
bidders.}
\begin{proposition}[the bottom-edge rate at three bidders]\label{app:bottom3}
Let $n = 3$, $F \in C^5[0,1]$ with $f > 0$ on $[0,1]$, and $E_F^{\mathrm{osc}} >
0$ on $(0,1]$ (so $w_F > 0$ there), and let $c_m$ be the certificate at the
Galerkin solution of Theorem~\ref{thm:oscillation2}. Let $\lambda^*$ be the
ground value of the semi-infinite pencil
\[
  \lambda^* := \inf_{x \ne 0\ \text{finitely supported}}\frac{\ip{\mathbb Tx}{x}}{\ip{\mathbb Mx}{x}},
  \qquad \mathbb T := \operatorname{tridiag}\Bigl(\frac{2k+1}8,\ k,\ \frac{2k+1}8\Bigr)_{k \ge 1},
  \qquad \mathbb M := \operatorname{tridiag}\Bigl(\frac16, \frac23, \frac16\Bigr),
\]
the uniform prior's three-bidder Galerkin rows at the bottom in the units
$h^2$ and $h$ (their symmetrized Jacobian and Gram matrix on the knots $1,
2, \dots$, exactly, for every $m$). Then $\lambda^* = 1.4732899625$, the
infimum approached by the truncations of the $\ell^2$ ground state, which is
localized at the first knot (successive ratios $-0.138, -0.202, -0.224,
\dots \to -(2 - \sqrt3)$), and with $\kappa_3 := \lambda^*/(3/2) =
0.9821933083$,
\[
  c_m = \kappa_3\cdot\tfrac32f(0)^2\cdot h\,(1 + O(h^{1/3})) \qquad (h \to 0),
\]
where $\tfrac32f(0)^2 = \lim_{v \to 0}w_F/(fv)$. The constant $\kappa_3$ is the
same for every such prior; the prior enters through $f(0)^2$ only.
\end{proposition}

\noindent\emph{What.} When $E_F^{\mathrm{osc}} > 0$ on $(0,1]$ at $n = 3$ the only
zero of the limit function is the bottom edge, the certificate is positive
at every $m$ with no positive limit, and it decays like $h$ with a
universal constant: $c_m = \kappa_3\cdot\tfrac32f(0)^2\,h\,(1 + O(h^{1/3}))$,
$\kappa_3 = 0.9821933083$, the prior entering through $f(0)^2$ only.
\emph{Why.} It closes the case neither rate theorem covers, and it
shows the bottom of the support has its own universal spectral problem,
the one the layer of Section~\ref{sec:general-4} lives in.
\emph{How.} Near the bottom every prior's rows are the uniform prior's;
for the uniform prior at $n = 3$ the certificate is exactly a pencil of
two tridiagonal matrices in the units $h^2$ and $h$, whose ground value
$\lambda^*$ is approached by truncations of a geometrically decaying
ground state; a spectral-gap split separates the bottom block from the
rest, where the limit function is positive. Primer:
Appendix~\ref{primer:pencils}. Proof: Appendix~\ref{app:proof-bottom3}.
Machine-checked: the uniform prior's rows and the pencil's entries
(Appendix~\ref{app:lean}).
\entry{Theorem~\ref{thm:four}, four bidders with small top curvature.}
\begin{theorem}[four bidders, small top curvature, full form]\label{thm:four-full}
Let $n = 4$ and $F \in C^5[0,1]$ with $f > 0$ on $[0,1]$. There are
$\varepsilon_0, h_0 > 0$, depending on $\min f$ and $\norm{F}_{C^5}$ only,
such that for $|\bstar{}''(1)| \le \varepsilon_0$ and $h = 1/m \le h_0$ the
Galerkin system has a solution $\bstar_m$, unique in an $h$-independent
ball around the first Newton iterate from the interpolant in the class
norm of Appendix~\ref{app:global}, with knot error
\[
  \delta_k = (-1)^{m-k}\,\frac{\bstar{}''(1)h^2}{12}\,\hat Q(v_k)\,(1 + \theta_k)
  + u_k + \rho_k, \qquad |\theta_k| \le \frac{C}{k},
\]
$|u_k| \le Ch^2\hat Q(v_k)$, $|u_k - u_{k-1}| \le (C/k)h^2\hat Q(v_k)$ and
$|\rho_k| \le C((\bstar{}''(1)^2 + h\log^2m)/k + h\log m)\,h^2\hat Q(v_k)$,
uniformly in $k$. The slope law~(i) of Proposition~\ref{prop:reduction}
holds on every $[v_0, 1]$ with error $O(h^2\log m)$ and at the bottom
knots within a relative $O(1/k)$, condition~(ii) holds, and $c_m \to
\min_{[0,1]}E_F^{\mathrm{osc}}$, $c_m^\vartheta \to \min
E_F^{\mathrm{osc}}/\vartheta$, with $E_F^{\mathrm{osc}}$ as in
Theorem~\ref{thm:limit2} at $n = 4$.
\end{theorem}

\noindent\emph{What.} For $n = 4$ and $|\bstar{}''(1)| \le \varepsilon_0$ the Galerkin
solution exists with the two-scale form, unique in a ball around the
first Newton iterate, the slope law holds on compacts with error $O(h^2
\log m)$ and at the bottom knots within a relative $O(1/k)$, and the
limit theorem follows through Proposition~\ref{prop:reduction}; the full
form with the remainder bounds is above.
\emph{Why.} It extends the characterization theorem to four bidders
under a condition that the next theorem removes.
\emph{How.} At four bidders the weighted inverse bound of
Theorem~\ref{thm:oscillation2} becomes logarithmic and the first
iterate's alternating content is order one. Class norms repair both: on
forcings, a quotient norm over decompositions into a smooth part, an
alternating part and a top entry, with the grid-over-position weight in
the difference slot; on corrections, the matching structure. On these
classes the inverse is bounded uniformly at the interpolant and at the
first iterate, the quadratic remainder needs no structure beyond a
pointwise bound, and the residual at the first iterate has class norm
$\bstar{}''(1)^2$ plus a logarithmic correction, which is where the
smallness enters, twice. The condition is the argument's, not the
problem's: the measured Kantorovich numbers are below $\tfrac12$ for
$|\bstar{}''(1)| \le 0.53$ from the first iterate and for every prior tested
from the second (Appendix~\ref{app:numerics}). Proof:
Appendix~\ref{app:four}. Not machine-checked.
\entry{Theorem~\ref{thm:layer}, the universal layer and the matching.}
\emph{What.} For $n = 4$ and a layer amplitude $A_F \le A_0 = 8.2419$ at
the first knot, the Galerkin solution exists with the universal layer on
its first $K_0/2$ knots and the two-scale form beyond, so the slope law
and the limit hold without the smallness condition.
\emph{Why.} It is the four-bidder characterization theorem in
practice: the measured curvatures of real priors exceed $\varepsilon_0$,
and their amplitudes lie under $A_0$.
\emph{How.} Two parts. The layer: near the bottom every prior's rows are
the uniform prior's, whose four-bidder rows are an exact forward
recursion from the first knot; in scaled tail variables its one-step map
contracts the smooth part, and an invariant region closes under three
inequalities that interval arithmetic verifies from the hand-off index
$1601$ on; below it a parametric Krawczyk sweep (Appendix~\ref{primer:interval})
over $1\,301\,442$ boxes, each carried from knot $401$ to $1601$ by a
verified iteration, with a point test at every box boundary identifying
adjacent branches, verifies existence, nondegeneracy and the slope
bounds, so that the verified first-knot values form an interval
reaching within $0.1\%$ of the two saturations. The matching:
Newton--Kantorovich from the composite of the layer and the first inland
iterate, in norms adapted to the matching index, with the layer's
variational solution as the bottom-pinned mode and the inland rows
supplying the top-pinned one, their Casoratian bounded below from the
verified alternating content; each step gains $1/K$ inland, so the
argument closes once $K$ exceeds a constant of the prior. The three
verified inputs are the invariant-region constants, the sweep with its
boundary tests, and the matching's mode constants; the one-step
identities, the inductions and the matching argument are analytic.
Proof: Appendix~\ref{app:layer}, with its verification table; the run
records, Appendix~\ref{app:numerics}. Machine-checked: the
inductions, as statements conditional on the verified constants; the
one-step identities are computer algebra (Appendix~\ref{app:lean}).

\entry{Proposition~\ref{prop:layer5}, the layer at five bidders.}
\begin{proposition}[the universal layer at five bidders, full form; computer-assisted]\label{prop:layer5-full}
Let $n = 5$ and $A_0^{(5)} := 6.8423$. For every far-field amplitude $|A| \le
A_0^{(5)}$ there is a trajectory of the uniform prior's five-bidder rows, a
forward recursion from the first knot, existing for all $k$, with (i) slopes
in $[0.2691, 2.2128]\,\bstar{}'$ on every piece; (ii) the far field $e_k =
(-1)^kP_\infty k^{-4/3} + k^{-5/3}U_k + \varrho_k$ with $|P_\infty| \le 6.085$,
$|U_k| \le 89.8$ and $|\varrho_k| \le 72k^{-7/3}$ for $k \ge 401$, the amplitude
$A = -P_\infty/(\tfrac45z_\infty)$, $z_\infty \in [1.08199, 1.08224]$ the
bottom-pinned linear mode's far-field constant; (iii) the amplitude
continuous in the trajectory's parameter, zero at the unforced trajectory,
and reaching $\pm A_0^{(5)}$ with one sign on each branch (the first knot
pushed down or up); (iv) a bounded variational solution, $|dS_k/dS_1| \le
284.9$ for $k \le 401$, whose alternating content at the hand-off index
$401$ is at least $0.1464$ with one sign over the whole range. The verified
first-knot values form the interval $[0.3544, 1.2707]$ in units of the bid.
The same leaves verify the layer's weighted value on its first $400$
knots: positive for $|A| < 0.428$, negative for $0.607 < |A| \le 7.0$, and
within the verified bands' width of $-1.33|A| - 0.05$ for $|A| \ge 4$.
\end{proposition}

\noindent\emph{What.} At $n = 5$ the layer exists for every far-field amplitude
$|A| \le 6.8423$, with explicit slope bounds, far field, continuity in its
parameter and a bounded variational solution; its weighted value is
verified negative beyond $|A| = 0.61$.
\emph{Why.} It is the first part of Theorem~\ref{thm:layer}'s
route at five bidders, where the layer is $h^{-(n-4)/(n+2)}$ knots wide and
the law fails inside it; the matching is not redone, and what remains for
$n \ge 5$ is the layer for every amplitude, which is Conjecture~\ref{conj:knot}'s
missing step there.
\emph{How.} The same route with the five-bidder scaled variables, in
which the smooth part is a contracting fixed point: the invariant regions
close in one stage at the hand-off index $400$, and the Krawczyk sweep
passes on $53\,604$ leaves with no failure and no gap. Proof:
Appendix~\ref{app:layer5}, with its verification table; the run records,
Appendix~\ref{app:numerics}.
\entry{Proposition~\ref{prop:reduction}, the bottom knots play no part.}
\emph{What.} For $n \ge 3$ the limit $c_m \to \min E_F^{\mathrm{osc}}$ follows
from the slope law on compact subsets of $(0,1]$ plus a crude bound on
slope differences near the bottom.
\emph{Why.} It locates what separates Conjecture~\ref{conj:knot}
from a theorem: the law inland, not the bottom layer, which may fail
across a number of knots growing slower than a power of $m$.
\emph{How.} Abel's bound for alternating sums (Appendix~\ref{primer:abel})
turns the slope law into uniform convergence of $\bstar_m$; inland the
coefficients are the limit's; the localization is unchanged; on bottom
windows the piece integrand is bounded by a power of position and a knot
mass by the slope difference times $v_k^{n-1}$, so the windowed quotient
is at least a small negative amount, which suffices because the limit
function vanishes at the bottom for $n \ge 3$. Proof:
Appendix~\ref{app:limit}. Machine-checked: Abel's bound
(Appendix~\ref{app:lean}).

\entry{Conjecture~\ref{conj:knot}.}
\emph{What.} For $n \ge 4$ the limit of the certificate is the same closed
form. \emph{Where it stands.} Proved at $n = 4$ for $A_F \le A_0$
(Theorem~\ref{thm:layer}) and for small top curvature
(Theorem~\ref{thm:four}); at $n \ge 5$ the layer is verified for bounded
amplitudes (Proposition~\ref{prop:layer5}) and the law's own amplitude
at the bottom grows like $h^{(4-n)/3}$, so the missing step is the layer
for every amplitude and the matching. It is the first open problem of
Section~\ref{sec:conclusion}.

\entry{Theorem~\ref{app:H2}, the second-order amplitude law.}
\begin{theorem}[the second-order amplitude law]\label{app:H2}
Let $F \in C^6[0,1]$ with $f > 0$ on $[0,1]$, and $n \in \{2, 3\}$, or $n = 4$
with the layer amplitude $A_F \le A_0$ of Theorem~\ref{thm:layer} (or
$|\bstar{}''(1)| \le \varepsilon_0$ of Theorem~\ref{thm:four}). For every
compact $I = [v_0, 1] \subset (0,1]$ there are $C$ and $h_0$ such that the
knot error admits the two-scale decomposition \eqref{eq:law}, $\delta_k =
\varepsilon_k\,\bstar{}''(1)h^2\hat Q(v_k)(1 + \rho_k)/12 + u_k$ with
$\varepsilon_k$ the checkerboard's sign, with, for $v_k \in I$ and $h \le
h_0$,
\[
  |\rho_k| \le Ch^2, \qquad |\Delta u_k| \le Ch^3, \qquad |\Delta^2u_k| \le Ch^4:
\]
the alternating amplitude follows its law to relative $O(h^2)$, with the
exact constant, and the smooth part is $h^2$ times a $C^2$ function on $I$.
\end{theorem}

\noindent\emph{What.} On every compact subset of $(0,1]$ the alternating amplitude
of $\bstar_m - I_h\bstar$ follows its law with the exact constant to
relative $O(h^2)$, and the smooth part is $h^2$ times a $C^2$ function.
\emph{Why.} The rates need the decomposition's coefficients to
second order near the minimizer, since a first-order error in the
amplitude would be a first-order term in the rate; the one coefficient
that could carry it is the amplitude, which the two-scale form controls
only to relative $O(1/k)$.
\emph{How.} An explicit second-order approximation and
Newton--Kantorovich from it: the exact pinned solutions of the Jacobian at
the interpolant for the alternating part, the regular solution of a
first-order ODE for the smooth part, and a residual of fifth order on
compacts by three finite identities, that the law annihilates the
three-term recurrence to third order, that a two-scale vector has no
fourth-order term in a Galerkin row by the hat's symmetry, and that the
top row's third-order balance forces the amplitude's first-order
correction to zero (Lemmas~\ref{app:I1} to \ref{app:I7}). At three
bidders the bottom knots need one change; at four the law is bootstrapped
from the solution of Theorem~\ref{thm:layer}. Proof:
Appendix~\ref{app:proof-H2}. Machine-checked: the symbolic identities of
Lemmas~\ref{app:I1} and \ref{app:I4}, in computer algebra
(Appendix~\ref{app:numerics}).
\entry{Proposition~\ref{prop:local}, the exact local bound.}
\emph{What.} Splitting each knot mass in halves between its two pieces
gives $c_m \ge L_m = \min_j\lambda_{\min}(P_j, M_j)$ at every $m$, a
piece-by-piece bound computable in interval arithmetic; $c_m$ is the
supremum of such bounds over all divisions, attained; and $L_m \to \min
E_F^{\mathrm{osc}}$.
\emph{Why.} It is the finite-$m$ tool: the pilot corollary is its
sign, verified at every $m$. It also gives the limit theorem's lower
bound without localization, and its frozen pencils carry the band
curvature $\kappa$ that the rates need.
\emph{How.} A quotient of sums is at least the least quotient; for the
supremum, the pivots of the tridiagonal matrix of the form minus a value
times the mass split it exactly into positive semidefinite $2\times2$
blocks, which are the piece pencils for some division. The frozen pencil
is the local smooth value plus three times the Bloch pencil's least
eigenvalue, and the lower band's curvature is read off the same pencil at
a small phase. Primer: Appendix~\ref{primer:pencils}. Proof:
Appendix~\ref{app:limit}. Machine-checked: the quotient bound, the pivot
step of the supremum, the pencil's characteristic polynomial, the band's
second-order expansion, $\kappa \ge 0$ (Appendix~\ref{app:lean}).

\entry{Theorem~\ref{thm:interior}, the rate at an interior minimizer.}
\emph{What.} When $E_F^{\mathrm{osc}}$ has a unique interior minimizer with
positive curvature and $\kappa(\hat v) > 0$, $c_m = E_{\min} +
h\sqrt{\kappa E''/2} + O(h^{3/2}\log^{3/2}(1/h))$.
\emph{Why.} The leading term is positive, so the certificate
approaches its limit from above; this is what turns a positive limit into
a positive certificate with a margin.
\emph{How.} The semiclassical reading (Appendix~\ref{primer:semiclassical}):
the frozen problem's lower Bloch band is $E_F^{\mathrm{osc}}(v) +
\kappa\theta^2 + O(\theta^4)$, a wave packet sees $\theta$ as $-ih\,d/dv$,
and near a nondegenerate minimizer the ground state is a Gaussian of
width $\sqrt h$ with the harmonic energy. The upper bound is the Rayleigh
quotient of that packet on the local null profile; the lower bound
follows the Airy argument with the Gaussian in place of the Airy
function, on a window with both ends free; both need the coefficients to
second order, which Theorem~\ref{app:H2} supplies. Proof:
Appendix~\ref{app:proof-interior}. Machine-checked: the pencil identities
(Appendix~\ref{app:lean}).

\entry{Theorem~\ref{app:interiorw}, the rate in a weighted geometry.}
\begin{theorem}[the interior rate in a weighted geometry]\label{app:interiorw}
Let $n \in \{2, 3\}$, $F \in C^6[0,1]$ with $f > 0$ on $[0,1]$, and $\vartheta$
either a weight in $C^2[0,1]$ positive on $[0,1]$ or $\vartheta = w_F/f$ with
$w_F > 0$ on $(0,1]$. Let $E_\vartheta := E_F^{\mathrm{osc}}/\vartheta$ attain its
minimum $\mu_0$ over $[0,1]$ at a single point $\hat v \in (0,1)$ with
$E_\vartheta''(\hat v) > 0$ and $\kappa(\hat v) > 0$. Then
\[
  c_m^\vartheta = \mu_0 + h\sqrt{\frac{\kappa(\hat v)E_\vartheta''(\hat v)}{2\vartheta(\hat v)}} + O\bigl(h^{3/2}\log^{3/2}(1/h)\bigr).
\]
In particular, for $n = 3$ and $\vartheta = w_F/f$ with $\mu_0 > 0$, $c_m^\vartheta >
\mu_0 > 0$ for every $m \ge m_0$.
\end{theorem}

\noindent\emph{What.} The same rate relative to a weight $\vartheta$, including
$w_F/f$ at three bidders: $c_m^\vartheta = \mu_0 + h\sqrt{\kappa
E_\vartheta''/(2\vartheta(\hat v))} + O(h^{3/2}\log^{3/2}(1/h))$, so a positive
weighted limit gives $c_m^\vartheta > \mu_0$ for every $m \ge m_0$.
\emph{Why.} It is where the positive case lives at three bidders,
and the form of Corollary~\ref{cor:eventual}'s case (b) there.
\emph{How.} Nothing in the mechanism changes: the weight enters the
mass, the potential becomes $E - \mu_0\vartheta$, and the packet's weighted
mass is $\vartheta(\hat v)$ times its mass. Proof:
Appendix~\ref{app:proof-interiorw}.
\entry{Theorem~\ref{thm:airy}, the rate at a top-knot minimizer.}
\emph{What.} When the minimum sits at $v = 1$ only, with $E'(1) < 0$ and
$\bstar{}''(1) \ne 0$, $c_m = E(1) + a_1\kappa(1)^{1/3}|E'(1)|^{2/3}h^{2/3} +
O(h)$, $a_1$ the first Airy zero.
\emph{Why.} The other half of the rate picture, with the same
positive sign; the Dirichlet wall at the top makes the well one-sided and
the exponent $2/3$.
\emph{How.} Upper bound: the Airy packet, the Airy function shifted to
vanish at the top knot under a cutoff, on the two-periodic null profile at
the top, whose Rayleigh quotient is the eigenvalue plus one grid by
Riemann sums. Lower bound: every knot vector is an envelope times the
null profile, and by the null relations each piece form at envelope times
profile is exactly a coefficient times the envelope's increment squared,
with no smallness; paired into cells with the harmonic-mean inequality
this is a discrete Dirichlet form with coefficient $\tfrac12\kappa(1 + t +
t^2)$; a dichotomy on the envelope's roughness handles the mass's cross
terms, and Picone's identity with an Airy supersolution gives the first
zero on a window whose outer end is free. Proof: Appendix~\ref{app:limit}.
Machine-checked: the envelope identity, the harmonic-mean inequality,
the cell weights and the kinetic constant (Appendix~\ref{app:lean}).

\entry{Corollary~\ref{cor:eventual}, positive from some $m$ on.}
\emph{What.} Where the limit theorems apply, $E_{\min} > 0$ gives $c_m > 0$
for $m \ge m_0(F)$; where a rate is proved, $c_m$ exceeds $E_{\min}$ by
half the leading term from some $m$ on.
\emph{Why.} It reduces "positive at every $m$" to the finite range
$2 \le m < m_0(F)$, a computation at each $m$ once the Galerkin solution
there is verified.
\emph{How.} Convergence, and a remainder that is eventually at most half
the leading term. Proof: in the statement. Not a formalization target;
$m_0(F)$ is not explicit in general (Appendix~\ref{app:lean}).

\entry{Corollary~\ref{cor:pilot}, one prior at every $m$.}
\emph{What.} For $F = (v + v^2)/2$ at $n = 2$, $c_m > 0$ for every $m \ge 2$.
\emph{Why.} It is the one place where the finite range is closed,
and the end of the positivity story.
\emph{How.} Only the sign of the local bound is needed. The constants of
the existence theorem's Newton--Kantorovich step, the inverse's norm, the
Lipschitz constant and the second iterate's residual, were written out
lemma by lemma (Lemma~\ref{app:L6} with constants is machine-checked) and
verified in interval arithmetic for every $m \ge 40000$, where the
Kantorovich number is below $\tfrac12$ from $m_0 = 2989$ on and each piece
pencil is enclosed from the verified class of the solution, giving $c_m
\ge 0.470$; below $40000$ a driver covers every $m$, giving $c_m \ge
0.4969$, with no floating-point step trusted. The same computation covers
two further priors at $n = 2$ on the finite range. The verified threshold
is far above the true one: in every case computed $c_m > E_{\min}$ at
every $m$ on every tested prior with a positive limit, and the slack is in
the bounds of the existence theorem's solution class. Proof:
Appendix~\ref{app:pilot}, with its verification table; the run records,
Appendix~\ref{app:numerics}.

\entry{A density vanishing at the top, outside the chain.}
\label{app:general-top}
\emph{What.} For a density vanishing at the top the certificate does not
merely lose its uniformity in $m$, it diverges: for $F(v) = 2v - v^2$ it
is negative and of order $m^2$ at every $m$ computed, and at leading order
$c_m/m^2$ tends to $-2E_0/(|f'(1)|(n-1)(1-\bstar(1)))$ with the universal
$E_0 = 1.022199$, that is, $-1.533298$ and $-0.958311$ at $n = 2$ and $3$.
Nothing here is a theorem; the mechanism is derived at leading order and
matched to computation.
\emph{Why.} Nothing in the chain uses it. Every theorem of the arc
assumes $f > 0$ on the closed interval, so that the equilibrium slope is
positive at the top and the top piece obeys the oscillation law; this
case shows what that assumption protects, which is why it sits outside
the map.
\emph{How.} With $f(1) = 0$ the equilibrium slope is of order $1 - v$ near
the top, so the top slopes of $\bstar_m$ are of order $h$, their
oscillation is of relative order one and does not vanish with $m$, and
the knot masses there, of order $m^2$ against a mass $\int f\eta^2$ that
the vanishing density makes small, dominate the certificate; the
minimizer is a fixed profile on the top five knots, with the same
structure for $n = 2$ and $n = 3$. The order and the sign come from this
structure alone. For the coefficient, in the scaled coordinate $u =
(1-v)/h$ the equilibrium equations at the top knots become a two-term
recursion for the scaled slopes, $c_1 = \lambda/2$, $c_{k+1} = \lambda
B_{k,k+1}/(A_k - \lambda B_{k,k}/c_k)$ with $\lambda = 2(n-1)(1 -
\bstar(1))$, and the coefficient of $m^2$ is the bottom eigenvalue of the
pencil of knot masses and smooth parts in that scaling
(Appendix~\ref{primer:pencils}). The derivation takes the top knot value
of $\bstar_m$ to converge to $\bstar(1)$, which fixes $\lambda$; that
convergence is the unproved step, a global property of the equilibrium,
expected, and consistent with the computation, in which the constants
pass the derived values between $m = 512$ and $1024$ and return, as a top
value converging only like $m^{-2/3}$ produces. What the step decides is
the value of $\lim c_m/m^2$, not the divergence, and nothing else in the
arc rests on it. The computed constants are Appendix~\ref{app:numerics}.

\section{Proofs}
\label{app:proofs}

The proofs, one subsection per arc, in the order of the chains of
Appendix~\ref{app:arcs}. Traditional proofs are written out. A
computer-assisted proof ends with a verification table whose rows are the
inputs its analytic argument takes as given: what is verified, the
method, the inputs, the outputs with their enclosures, the code and run
records, and the cost.

\subsection{Arc I: the uniform prior}
\label{app:proofs-arc1}

\subsubsection{The field, the identity and the modulus}
\label{app:proofs-field}

\begin{proof}[Proof of Lemma~\ref{lem:field}]
$W$ is Lipschitz, hence differentiable at almost every $z$, and
$\varphi_z(v,z) = -W^{n-1} + (v-z)(n-1)W^{n-2}W'(z)$ there. Along $z = b(v)$
one has $W = v$ (the cap never binds) and $W'(b(v)) = 1/b'(v)$ at every $v$
with $b'(v) > 0$ whose image $b(v)$ is a point of differentiability of $W$;
the exceptional set in $z$ is null, and its preimage under the Lipschitz map
$b^{-1}$ is null in $v$.
\end{proof}

\begin{proof}[Proof of Proposition~\ref{prop:minty}]
Substitute $v - b = v/n - \eta$ and $b' = (n-1)/n + \eta'$ into
Lemma~\ref{lem:field}: $g = v^{n-2}[-v((n-1)/n + \eta') + (n-1)(v/n - \eta)] /
((n-1)/n + \eta') = -v^{n-2}(v\eta' + (n-1)\eta)/((n-1)/n + \eta')$, and
$\ip{\Fop,\eta} = -\int g\eta\,dv$.
\end{proof}

\begin{proof}[Proof of Theorem~\ref{thm:modulus}]
In Proposition~\ref{prop:minty} the denominator contributes from the third
order in $t$ on, so the $t^2$ coefficient is
\[
  \frac{n}{n-1}\int v^{n-2}(v\eta' + (n-1)\eta)\eta\,dv = \frac{n}{n-1}
  \Bigl[(n-1)\int v^{n-2}\eta^2 + \tfrac12\int v^{n-1}(\eta^2)'\Bigr].
\]
Integration by parts is valid for absolutely continuous $\eta$ with
$\eta(0) = 0$: $\tfrac12\int v^{n-1}(\eta^2)' = \tfrac12\eta(1)^2 -
\tfrac{n-1}{2}\int v^{n-2}\eta^2$.
Collecting terms gives $Q_n$. The inequality and its equality case for $n = 2$
are immediate; for $n \ge 3$ the hat at the first knot has weight $v^{n-2}
\le (2/m)^{n-2}$ on its support, which gives the upper bound on the best
constant. The extension to $H^1$ is the regularity note at the end of this
subsection.
\end{proof}

\begin{proof}[Proof of Proposition~\ref{prop:ray}]
With $w_{m-1} = 1/(1/2 + am)$, $w_m = 1/(1/2 - am)$ and $w \equiv 2$
elsewhere, the bracket in \eqref{eq:knots} equals $(w_{m-1} + w_m)a^2/(3m) -
v_{m-1}(w_m - w_{m-1})a^2$ with $v_{m-1} = 1 - 1/m$, $w_{m-1} + w_m =
1/(1/4 - a^2m^2)$ and $w_m - w_{m-1} = 2am/(1/4 - a^2m^2)$. Multiplying by
$(1/4 - a^2m^2)/a^2 > 0$ leaves $1/(3m) - 2a(m-1)$, which is decreasing in $a$
and vanishes at $a^* = 1/(6m(m-1))$.
\end{proof}

\begin{proof}[Proof of Proposition~\ref{prop:radius}]
The upper bound is Proposition~\ref{prop:ray}, whose radius is at most
$(\sqrt{2/3}/3)\,m^{-5/2}$ for $m \ge 2$ since $(1 - 1/m)^{-1} \le 2$, so
$C_+ = \sqrt{2/3}/3$. The lower bound with the explicit constant $1/288$ is
the next subsubsection.
\end{proof}

\begin{proof}[Proof of Proposition~\ref{prop:stationary}]
A direct computation with $q^2 = 4q - 1$: for $j \ge 1$, $x_{j-1} + 4x_j +
x_{j+1} = (-q)^{j-1}(1 - 4q + q^2)$ and $2\,\partial J/\partial x_j =
-x_j^2(1+q)^2(q^2 - 4q + 1)/q^2$, both zero, and symmetrically for $j \le
-1$. At $j = 0$ both gradients are nonzero, but Euler's identity for the
homogeneous $J$ and $\nu$ gives $x_0\,\partial J/\partial x_0 = 3J$ and
$x_0\,\partial\nu/\partial x_0 = 2\nu$, so $\nabla J = (3J/2\nu)\nabla\nu$ and
the profile is stationary. The sums are geometric series with ratios $q^2$
and $-q^3$, evaluated with $1 - q + q^2 = 3q$, $1 - q^2 = 2\sqrt3\,q$ and
$1 + q^3 = 3q(1+q)$. The identities for $j \ne 0$ and the sums are
machine-checked (Appendix~\ref{app:lean}).
\end{proof}

\subsubsection{The sharp inequality of the leading-order problem}
\label{app:sharp}

We prove Theorem~\ref{thm:sharp}. Write $q = 2 - \sqrt3$, $a_0 = 1 - \sqrt3$,
$M = \norm{x}_\infty$ and $\varphi(a,b) = a^2 + ab + b^2 - \tfrac12(b-a)^2(a+b)$,
which is symmetric in $(a,b)$.

\emph{Reduction.} For $\nu \ge 0$ and $t > 0$, $t^3 - 3\nu t + 2\nu^{3/2}
= (t - \sqrt\nu)^2(t + 2\sqrt\nu) \ge 0$, so $2\nu^{3/2} \ge 3\sqrt c\,\nu -
c^{3/2}$ for every $c > 0$; with $c = M^2/\sqrt3$ this is $2\cdot 3^{1/4}
\nu^{3/2} \ge 3M\nu - M^3/\sqrt3$. It therefore suffices to prove the
sup-norm form, and by homogeneity only for $M = 1$: $3\nu(x) - J(x) \ge
1/\sqrt3$ whenever $\max_j|x_j| = 1$. From the definitions, $3\nu - J =
\sum_j \varphi(x_j, x_{j+1})$.

\emph{The potential.} Let $V_{\mathrm g}(a) = \tfrac{\sqrt3}{2}a^2 -
a^3/\sqrt3$, the cost $\sum_{k\ge0}\varphi(x_k,x_{k+1})$ of the geometric
tail $x_k = a(-q)^k$; $V_{\mathrm j}(a) = \varphi(a,1) + \sqrt3/6$, the cost
of jumping to the peak value and continuing geometrically; and $W =
\min(V_{\mathrm g}, V_{\mathrm j})$. Since $V_{\mathrm g} - V_{\mathrm j} =
-(2\sqrt3 - 3)(a - a_0)(a + 2 + \sqrt3)^2/6$, $W = V_{\mathrm g}$ on $[a_0, 1]$
and $W = V_{\mathrm j}$ on $[-1, a_0]$; $W(0) = 0$, $W(1) = \sqrt3/6$,
$W(-1) = 1 + \sqrt3/6$.

\emph{The Bellman inequality} $\varphi(a,b) + W(b) \ge W(a)$ on $[-1,1]^2$.
As $W(a)$ is at most either branch and $W(b)$ is the branch of $b$, it
reduces to four polynomial inequalities. On $[a_0,1]^2$,
$\varphi(a,b) + V_{\mathrm g}(b) - V_{\mathrm g}(a) = (b + qa)^2 L(a,b)$ with
$L = (\tfrac12 + \tfrac{\sqrt3}{3})(a - b) + 1 + \tfrac{\sqrt3}{2}$, increasing
in $a$ and decreasing in $b$, so $L \ge L(a_0,1) = 0$. On $[-1,a_0]\times[a_0,1]$,
$\varphi(a,b) + V_{\mathrm g}(b) - V_{\mathrm j}(a) = (b - 1)\,Q(a,b)$ with
$Q = a^2/2 + a(b+3)/2 - (\tfrac{\sqrt3}{3} + \tfrac12)b^2 + (\tfrac{\sqrt3}{6} +
\tfrac12)b + \tfrac{\sqrt3}{6} + \tfrac12$; $Q$ is convex in $a$ and concave in
$b$, and on $a \in \{-1, a_0\}$ its maxima over $b$ are $-\tfrac58 +
\tfrac{\sqrt3}{4} < 0$ and $0$ (the latter only at $b = -qa_0$), so $Q \le 0$
and the product is nonnegative. On $[a_0,1]\times[-1,a_0]$ and on
$[-1,a_0]^2$ the remaining two polynomials have no critical point inside
the rectangle and their edge restrictions are minimized at corners, giving
the minima $6 - 3\sqrt3$ and $12 - 6\sqrt3$, both positive (exact arithmetic in
$\mathbb{Q}(\sqrt3)$; the numerical record is in Appendix~\ref{app:numerics}).

\emph{Telescoping.} Let the maximum $|x_{j_0}| = 1$ be attained at $j_0$.
Along the right chain, $\sum_{j \ge j_0}\varphi(x_j,x_{j+1}) \ge
\sum_{j\ge j_0}[W(x_j) - W(x_{j+1})] = W(x_{j_0}) - W(0) = W(x_{j_0})$ by the
Bellman inequality and the finite support, and the left chain gives the
same by the symmetry of $\varphi$. Hence $3\nu - J \ge 2W(x_{j_0}) \ge 2W(1) =
1/\sqrt3$, since $W(-1) > W(1)$.

\emph{Equality.} Equality forces $\nu = M^2/\sqrt3$, $x_{j_0} = 1$, and
equality in every Bellman step. From a current value $a \in (a_0, 1]$ the
successor cannot lie below $a_0$ (that case has slack $\ge 6 - 3\sqrt3$), so
the first case applies, and since $L$ vanishes only at $(a_0, 1)$ and $a \ne
a_0$, zero slack forces $b = -qa \in (a_0, 1]$; inductively from $a = 1$ both
chains are geometric, $x = c(-q)^{|j-j_0|}$, which attains equality by
Proposition~\ref{prop:stationary}. \qed

\subsubsection{The radius lower bound with constants}
\label{app:lower}

We prove the lower bound of Proposition~\ref{prop:radius} with explicit
constants: for $n = 2$ and every $m \ge 2$,
\[
  r_m \;\ge\; \frac{1}{288}\,m^{-5/2}.
\]

Let $\eta$ be piecewise linear on the grid $v_j = j/m$ with $\eta(0) = 0$,
knot values $\eta_j = \eta(v_j)$ and slopes $\eta_j' = m(\eta_j - \eta_{j-1})$
on piece $j$, and write $\rho = \norm{\eta}_{\Ltwo}$.

\emph{Inverse inequalities.} On a piece where $\eta$ runs linearly from $p$
to $q$, $\int\eta^2 = (p^2 + pq + q^2)/(3m) \ge \max(p^2,q^2)/(4m)$, since
$p^2 + pq + q^2 \ge \tfrac34\max(p^2,q^2)$. Hence $\norm{\eta}_\infty^2 \le 4m\rho^2$
and, summing over pieces with $p^2 + pq + q^2 \ge (p^2+q^2)/2$,
$\sum_{j=1}^{m-1}\eta_j^2 \le 3m\rho^2$. Also $|\eta_j'| \le 2m\norm{\eta}_\infty
\le 4m^{3/2}\rho$.

\emph{Admissibility.} If $\rho \le 1/(16m^{3/2})$ then $\norm{\eta'}_\infty
\le 1/4$, so $\eta' > -1/2$ a.e., and $|\eta(1)| \le 2\sqrt m\rho \le 1/(8m)$ keeps
$1/2 + \eta(1) \le \bar b$; Proposition~\ref{prop:minty} applies and
$w_j = 1/(1/2 + \eta_j') \in [4/3, 4]$ on every piece.

\emph{The knot terms.} In \eqref{eq:knots}, $|w_{j+1} - w_j| = |\eta_{j+1}' -
\eta_j'|\,w_jw_{j+1} \le 16\cdot 2\norm{\eta'}_\infty \le 128\,m^{3/2}\rho$, and
$v_j \le 1$, so
\[
  \Bigl|\sum_{j=1}^{m-1} v_j(w_{j+1} - w_j)\eta_j^2\Bigr| \le 128\,m^{3/2}\rho
  \sum_{j=1}^{m-1}\eta_j^2 \le 384\,m^{5/2}\rho^3.
\]

\emph{The positive part.} $\int w\eta^2 \ge \tfrac43\rho^2$ and $w_m\eta(1)^2
\ge 0$.

\emph{Conclusion.} The bracket in \eqref{eq:knots} is at least $\tfrac43\rho^2 -
384\,m^{5/2}\rho^3$, which is positive whenever $\rho < 1/(288\,m^{5/2})$. Since
$1/(288m^{5/2}) \le 1/(16m^{3/2})$ for every $m \ge 1$, admissibility holds on
that ball as well, and $r_m \ge m^{-5/2}/288$. \hfill$\square$

The upper bound $r_m \le (\sqrt{2/3}/6)\,m^{-5/2}\,(1 - 1/m)^{-1}$ is
Proposition~\ref{prop:ray}.

\subsubsection{The radius limit}
\label{app:radius}

We prove Theorem~\ref{thm:radius}. Let $\eta$ be piecewise linear on the
$m$-grid with knot values $x_0 = 0, x_1, \dots, x_m$ ($x_m = \eta(1)$ free),
$\rho = \norm{\eta}$, and write $\varepsilon_j = 2m(x_j - x_{j-1})$, $w_j =
2/(1 + \varepsilon_j)$, $I_j = (x_{j-1}^2 + x_{j-1}x_j + x_j^2)/(3m)$ (so
$\nu = \nu(x) = m\sum_j I_j = m\rho^2$), $M = \max_j|x_j|$, $\delta =
\max_j|\varepsilon_j|$, $v_j = j/m$, $c = 1/(8\cdot 3^{1/4})$, $R = 2\cdot
3^{1/4}$ (so $4Rc = 1$) and $K = 1 + \sqrt3/6$.

\emph{The expansion.} The bracket of \eqref{eq:knots}, which equals
$2\ip{\Fop(\bstar+\eta)}{\eta}$, is $B = \sum_j w_jI_j - \sum_{j<m}
v_j(w_{j+1} - w_j)x_j^2 + w_mx_m^2$. With $2/(1 +
\varepsilon) = 2 - 2\varepsilon + 2\varepsilon^2/(1 + \varepsilon)$ and
$w_{j+1} - w_j = 2(\varepsilon_j - \varepsilon_{j+1})/((1 + \varepsilon_j)(1 +
\varepsilon_{j+1}))$, and since $\sum_j \varepsilon_jI_j = \tfrac23\sum_j(x_j -
x_{j-1})(x_{j-1}^2 + x_{j-1}x_j + x_j^2) = \tfrac23 x_m^3$ telescopes,
\[
  B = \frac{2\nu}{m} + \frac{2x_m^2}{1 + \varepsilon_m} - \frac43 x_m^3
    + 2\sum_j \frac{\varepsilon_j^2 I_j}{1 + \varepsilon_j}
    - 4m\sum_{j=1}^{m-1} v_j\,c_j\,g_j,
  \quad c_j = (2x_j - x_{j-1} - x_{j+1})x_j^2,\ \
  g_j = \frac{1}{(1 + \varepsilon_j)(1 + \varepsilon_{j+1})}.
\]
Dropping the nonnegative sum, $|g_j - 1| \le 2\delta/(1 - \delta)^2$ and
$|c_j| \le 4Mx_j^2$ give
\begin{equation}
\label{eq:Blower}
  B \ge \frac{2\nu}{m} + \frac{2x_m^2}{1 + \delta} - \frac43 Mx_m^2 - 4mS
    - \frac{32\,m\,\delta M}{(1 - \delta)^2}\sum_{j<m}x_j^2,
  \qquad S := \sum_{j=1}^{m-1} v_jc_j.
\end{equation}

\emph{The knot sum as a weighted $J$.} Splitting $c_j = (x_j - x_{j-1})x_j^2
+ (x_j - x_{j+1})x_j^2$ and shifting the second sum,
\[
  S = 2J_v(x) + \frac1m\sum_{i=1}^m (x_i - x_{i-1})x_{i-1}^2 - x_m^2(x_m - x_{m-1}),
  \qquad J_v(x) = \frac12\sum_{i=1}^m v_i(x_i - x_{i-1})^2(x_i + x_{i-1}),
\]
and since $v_i = \sum_{k \le i}1/m$, Abel summation gives $J_v(x) =
\frac1m\sum_{k=1}^m J(x_{k-1}, \dots, x_m)$, an average of the $J$ of the
tails, each starting and ending at a free value.

\emph{Free-end form of Theorem~\ref{thm:sharp}.} For every finite sequence
$y_0, \dots, y_N$ with $\mu = \max|y_i|$,
\begin{equation}
\label{eq:freeend}
  J(y) \le R\,\nu(y)^{3/2} + K\mu\,(y_0^2 + y_N^2).
\end{equation}
Indeed, for $z = y/\mu$ with $|z_{j_0}| = 1$, the Bellman inequality
telescopes along the two chains from $j_0$ to $3\nu(z) - J(z) \ge
2W(z_{j_0}) - W(z_0) - W(z_N) \ge \sqrt3/3 - K(z_0^2 + z_N^2)$, where $K =
\sup_{0 < |a| \le 1}W(a)/a^2 = 1 + \sqrt3/6$ (on $[a_0, 1]$, $W/a^2 =
\sqrt3/2 - a/\sqrt3$ is decreasing with value $K$ at $a_0$; on $[-1, a_0]$,
$Ka^2 - V_{\mathrm j}(a) = \tfrac12(a + 1)(a - a_0)(a - a_2)$ with $a_2 = 2.155$,
nonnegative there); multiplying by $\mu^3$ and using the tangent bound
$3\mu\nu - \mu^3/\sqrt3 \le R\nu^{3/2}$ gives \eqref{eq:freeend}.

\emph{The weighted inequality.} Applying \eqref{eq:freeend} to each tail
($\mu \le M$, $\nu(\text{tail}) \le \nu$) and $\sum_{j=0}^{m-1}x_j^2 \le
3\nu$,
\begin{equation}
\label{eq:Sbound}
  S \le 2R\nu^{3/2} + 6(K + 1)\frac{M\nu}{m} + 2(K + 1)Mx_m^2.
\end{equation}

\emph{The lower bound.} Let $\rho \le (1 - \theta)cm^{-5/2}$, $\theta \in
[0, 1]$. Then $m^2\sqrt\nu \le (1 - \theta)c$, $M \le 2\sqrt\nu \le 2c/m^2$
and $\delta \le 4mM \le 8c/m = 0.76/m$, so the identity \eqref{eq:knots}
applies ($\eta' > -1/2$, and $|\eta(1)| \le 2c/m^2 < 0.022$ keeps $1/2 +
\eta(1) \le \bar b$). Inserting
\eqref{eq:Sbound} into \eqref{eq:Blower},
\[
  B \ge \frac{2\nu}{m}\bigl[1 - 4Rm^2\sqrt\nu\bigr] - 24(K + 1)M\nu
    - \frac{96\,m\delta M\nu}{(1 - \delta)^2}
    + x_m^2\Bigl[\frac{2}{1 + \delta} - \frac43 M - 8(K + 1)mM\Bigr].
\]
The first bracket is at least $\theta$ since $4Rc = 1$; the two error terms
are at most $[48(K + 1)c + 1536c^2/(1 - \delta)^2]\,\nu/m^2 = [10.43 +
13.86/(1 - \delta)^2]\,\nu/m^2$; the end-knot bracket is at least $0.4$ for
$m \ge 3$. Hence $B > 0$ once $\theta > [5.22 + 6.93/(1 - 0.76/m)^2]/m$, in
particular for $\theta = 18/m$, and $r_m \ge (1 - 18/m)cm^{-5/2}$.

\emph{The upper bound.} Fix $k \ge 1$ and let $P = P^{(m,k)}$ have $x_j =
(-q)^{|j - (m-k)|}$ for $1 \le j \le m - 1$, $x_0 = x_m = 0$. Along the ray
$tP$, $t = \tau/m^2$, the expansion gives $m^5B(tP)/(2\tau^2) = \nu(P) -
2\tau S(P) + O(1/m)$ as $m \to \infty$, uniformly for $\tau$ in compact sets
($\delta \le 2\tau(1+q)/m$, no end terms), and $S(P) \to 2J_k$, $\nu(P) \to
\nu_k$, the values of the profile $(-q)^{|j|}$, $j \le k - 1$, pinned to
zero at $j = k$, with the infinite left tail (the terms $a_i$ of $J$ decay
like $q^{3|i - (m-k)|}$, so the position correction $\frac1m\sum(m - i)a_i$,
the middle term of the identity for $S$, and the truncation at $x_0 = 0$
vanish). By the intermediate value theorem the bracket has a zero at some
$\tau_m \to \nu_k/(4J_k)$, at norm $\rho_m$ with $m^{5/2}\rho_m \to
\nu_k^{3/2}/(4J_k) = \tilde c_k$; a zero at norm $\rho_m$ gives $r_m \le
\rho_m$, hence $\limsup m^{5/2}r_m \le \tilde c_k$. As $k \to \infty$, $\nu_k \to 1/\sqrt3$
and $J_k \to 2/\sqrt3$ (Proposition~\ref{prop:stationary}), so $\tilde c_k \to
c$. \qed

\subsubsection{The extremal direction}
\label{app:direction}

We prove Theorem~\ref{thm:direction}. For a step $(a, b) \in [-1,1]^2$ let
$s(a,b) = \varphi(a,b) + W(b) - W(a) \ge 0$ and $d(a,b) = (b + qa)^2$. The four
cases of Appendix~\ref{app:sharp} give: on $(a_0, 1]^2$, $s = d\,L(a,b)$, and
$L < 0.3$ forces $a < -0.4535$, $b > 0.7215$; on $(a_0,1]\times[-1,a_0]$,
$s \ge 0.8038$ and $d \le 1.431$; on $[-1,a_0]^2$, $s \ge 1.6077 \ge d$; on
$[-1,a_0]\times(a_0,1]$, $s = (1-b)(-Q)$ with $-Q \ge 0.538\,d$ (from $Q(a,b)
= Q(a_0,b) + (a - a_0)(a + b + 3 + a_0)/2$, $Q(a_0,b) = -(\sqrt3/3 +
\tfrac12)(b + qa_0)^2$ and $0 \le a_0 - a \le 0.268$), so $s/d < 0.3$ forces
$b > 0.44$. Call a step \emph{bad} if $s < 0.3\,d$; a bad step has $a <
-0.45$ and $d \le 0.774$.

\emph{Chain stability.} For $z_0 = 1$ and $z_1, \dots, z_N \in [-1,1]$,
$\sum_i s(z_i, z_{i+1}) \ge \kappa\sum_i d(z_i,z_{i+1})$ with $\kappa = 0.056$.
Indeed, the predecessor $(a', a)$ of a bad step is never bad and has slack at
least $0.0539$: before a bad step of the first kind, $a \in (a_0, -0.4535)$,
and either $a' > a_0$, where $a + qa' \le -0.1856$ and $L(a',a) \ge 1.566$,
or $a' \le a_0$, where $s' \ge 1.4535\cdot 1.077\,(a + qa_0)^2 \ge 0.66$; before
a bad step of the fourth kind, $a \le a_0$ and the predecessor is of the
second or third kind. Pairing each bad step with its predecessor, a pair has
ratio $\ge s_g/(d_g + 0.774) \ge 0.0565$ since $s_g \ge \max(0.3\,d_g,
0.0539)$, and unpaired steps have ratio $\ge 0.3$. A bad step is not the
first step, since its first entry is below $-0.45$.

\emph{The slack identity.} For a finite sequence $y$ with $\mu = \max|y_i|
> 0$, $z = y/\mu$ and peak $|z_{j_0}| = 1$, telescoping with the slacks kept
gives $3\mu\nu(y) - J(y) \ge \mu^3[\sqrt3/3 + \Gamma] - K\mu(y_0^2 + y_N^2)$
with $\Gamma = 2W(z_{j_0}) - \sqrt3/3 + \sum_i s_i \ge 0$, where $\Gamma \ge 2$
if $z_{j_0} = -1$ and $\Gamma \ge \kappa(D_{\mathrm r} + D_{\mathrm l})$ if
$z_{j_0} = +1$, $D_{\mathrm r}, D_{\mathrm l}$ the squared defects along the
two chains from the peak (the left one reversed, $\varphi$ being symmetric).
With the tangent bound, $J(y) \le R\nu(y)^{3/2} - \mu^3\Gamma + K\mu(y_0^2 +
y_N^2)$.

\emph{The slack budget.} A minimizer exists (the space is finite-dimensional
and the bracket is continuous on the closed ball of radius $cm^{-5/2}$, where
the knot identity holds) and has bracket zero. With $x$ its knots and
$m^2\sqrt\nu = c(1 + \varepsilon_m)$, so that $M \le 2c(1 + \varepsilon_m)/m^2$
and $\delta \le 0.76(1 + \varepsilon_m)/m$, \eqref{eq:Blower} with $B = 0$
gives $4mS \ge 2\nu/m + 1.5x_m^2 - 13.86(1 + \varepsilon_m)^2\nu/(m^2(1-\delta)^2)$, and
the slack identity on each tail refines \eqref{eq:Sbound} to $S \le
2R\nu^{3/2} - \Delta + 6(K+1)M\nu/m + 2(K+1)Mx_m^2$ with
\[
  \Delta = \frac2m\sum_{k=1}^m\bigl[\mu_k^3\Gamma_k + R(\nu^{3/2} - \nu_k^{3/2})\bigr] \ge 0,
\]
$\mu_k, \nu_k, \Gamma_k$ the sup, the $\nu$ and the deficit of the tail
$x_{k-1}, \dots, x_m$. Since $8mR\nu^{3/2} - 2\nu/m = (2\nu/m)\varepsilon_m$,
this yields, for all large $m$ (the error constants carry factors $1 +
\varepsilon_m$ and are below their stated values once $\varepsilon_m \le 0.05$
and $m \ge 100$),
\[
  \Delta \le \nu^{3/2}\bar\varepsilon_m, \qquad x_m^2 \le 1.74\,(\nu/m)(\varepsilon_m + 14/m),
  \qquad \bar\varepsilon_m = 5.3(\varepsilon_m + 14/m)/(1 + \varepsilon_m) \ge 2R(\varepsilon_m + 14/m)/(1 + \varepsilon_m),
\]
and $\bar\varepsilon_m \to 0$ by Theorem~\ref{thm:radius}. Let $\hat\nu =
\nu/M^2$ and $y = x/M$; comparing $S \ge 2R\nu^{3/2}/(1+\varepsilon_m) -
O(\nu/m^3)$ with $J_k \le 3M\nu + 2KM^3$ gives $2R\hat\nu^{3/2} \le (6\hat\nu +
4K)(1 + o(1))$, so $\hat\nu \le 2.5$ for large $m$.

\emph{The peak.} Tails with $k \le j_0$ contain the peak, so $\mu_k = M$ and
$\Gamma_k \ge 2W(y_{j_0}) - \sqrt3/3$; tails with $k > j_0$ miss the piece
$(x_{j_0-1}, x_{j_0})$, whose share of $\nu$ is at least $M^2/4 \ge
\nu/(4\hat\nu)$, so $\nu^{3/2} - \nu_k^{3/2} \ge 0.146\nu^{3/2}$. Hence
$\Delta \ge 0.292R\,(d_m/m)\nu^{3/2}$, i.e. $d_m/m \le 1.3\,\bar\varepsilon_m$;
and if $y_{j_0} = -1$ also $\Delta \ge 4(j_0/m)M^3$, i.e. $j_0/m \le
\hat\nu^{3/2}\bar\varepsilon_m/4 \le \bar\varepsilon_m$, impossible for
$\bar\varepsilon_m < 0.43$. So $y_{j_0} = +1$ and $d_m = o(m)$.

\emph{The shape.} For $k \le j_0$, chain stability on the right chain and on
the reversed left chain of the tail gives $\Gamma_k \ge \kappa[D_{\mathrm r}
+ D_{\mathrm l}^{(\ge k)}]$, so $\Delta \ge (2\kappa M^3/m)[j_0 D_{\mathrm r} +
\sum_{k \le j_0}D_{\mathrm l}^{(\ge k)}]$ and, with $\Delta \le 4M^3
\bar\varepsilon_m$, $D_{\mathrm r} \le 36\,\bar\varepsilon_m(1 + o(1))$ and
likewise for any fixed number of left steps. The recursion $y_{j_0+i+1} =
-qy_{j_0+i} + e_i$ then gives $|y_{j_0+i} - (-q)^i| \le \sqrt{D_{\mathrm r}/(1 -
q^2)}$, and the same on the left.

\emph{The depth.} On the right chain $y_m = (-q)^{d_m} + \sum_{l<d_m}
(-q)^{d_m-1-l}e_l$, so $q^{d_m} \le |y_m| + \sqrt{D_{\mathrm r}/(1 - q^2)} \to
0$ by the two bounds above, which gives $d_m \to \infty$ at the stated rate.
\qed

\subsubsection{The rates}
\label{app:rates}

We prove the quantitative clauses of Theorems~\ref{thm:radius} and
\ref{thm:direction}, stated in full as Theorems~\ref{thm:radius-full} and
\ref{thm:direction-full}. The pinned profile $y_j = (-q)^{|j|}$ ($j \le k-1$),
$y_k = 0$, with the infinite left tail, has by the geometric sums of
Proposition~\ref{prop:stationary}
\[
  \nu_k = \frac{1}{\sqrt3} + \frac{q^{2k-1}}{6}, \qquad
  J_k = \frac{2}{\sqrt3} - (-1)^{k-1}\Bigl(\frac{1}{\sqrt3} - \frac12\Bigr)q^{3k-3},
\]
so $\tilde c_k/c = (1 + (\sqrt3/6)q^{2k-1})^{3/2}/(1 - (-1)^{k-1}0.067\,q^{3k-3})
\le 1 + 1.9\,q^{2k}$ for $k \ge 2$ (expand to second order; $q^{3k-3} \le
q^{2k-1}$). For the ray $P = P^{(m,k)}$ with $1 \le k \le m/8$ and $m \ge
100$: $\nu(P) = \nu_k + (q/6)q^{2(m-k-1)} \le \nu_k + 0.001/m$;
$\sum_{j<m}|c_j| \le 2(1+q) + 12q^3/(1-q^3) \le 3$ (the peak and the
geometric decay $|c_j| = 6q^{3l}$ at distance $l$); and $S(P) \ge 2J_k - (2.5k
+ 2)/m$, from the identity for $S$ with $x_m = 0$: $J_v = J(P) - \frac1{2m}
\sum(m - i)a_i$ with $|a_i| \le 1.177\,q^{3l}$ and $m - i = k + l$ left of the
peak, $\le k - 1 - l$ right of it, so $\sum(m-i)|a_i| \le 2.4k$; the middle
term is $\ge -1.4/m$; and $J(P) - J_k =
O(q^{3(m-k)})$. Along $tP$ with $t = \tau/m^2$, $\tau \le 1$, the expansion
of Appendix~\ref{app:radius} with $\delta = 2\tau(1+q)/m \le 0.0254$ gives
\[
  \frac{m^5 B(tP)}{2\tau^2} \le (\nu_k + 0.001/m)(1 + 6.6/m^2)
  - 2\tau\Bigl[2J_k - \frac{2.5k + 19}{m}\Bigr],
\]
which vanishes at $\tau^* \le 0.2$; hence $r_m \le \norm{t^*P}$ and
\[
  m^{5/2}r_m \le \tau^*\sqrt{\nu(P)} \le \tilde c_k\,
  \frac{(1 + 0.0026/m)(1 + 6.6/m^2)}{1 - (2.5k + 19)/(2J_km)}
  \le \tilde c_k\Bigl(1 + \frac{2.4k + 18}{m}\Bigr),
\]
using $2J_k \ge 1 + 2/\sqrt3$, $(2.5k + 19)/m \le 0.51$ and $1/(1-x) \le 1 + 2x$
for $x \le 1/2$. The
choice $k = \lceil\log m/(2\log(1/q))\rceil$ ($q^{2k} \le 1/m$, $k \le
0.38\log m + 1 \le m/8$) gives $m^{5/2}r_m \le c(1 + 1.9/m)(1 + (0.92\log m +
20.4)/m) \le c(1 + (\log m + 23)/m)$. In Theorem~\ref{thm:direction} this
makes $\bar\varepsilon_m \le 5.3(\log m + 37)/m$, and the two depth bounds
$d_m \le 1.3\,\bar\varepsilon_m m$ and $d_m \ge \log(1/\bar\varepsilon_m)/
(2\log(1/q)) - O(1)$ become the stated ones ($1/(2\log(1/q)) = 0.3797$); the
depth clauses hold for all large $m$ because the step $\hat\nu \le 2.5$ of
Appendix~\ref{app:direction} is asymptotic (quantified, it closes from
about $m = 1300$), while the error constants of the slack budget are below
their stated values from $m = 1000$ on, where $\varepsilon_m \le 0.03$. The
numerical record is in Appendix~\ref{app:numerics}. \qed

\paragraph{The sharp rate and depth.} We prove the two-sided bounds with
$L = 1/(2\log(1/q))$. \emph{Upper bound.} The position-weighted sum is exactly
$J_v = J(P) - \frac1{2m}\sum_i(m-i)a_i$ with $\sum_i(m-i)a_i = 2kJ(P) +
\sum_i(p-i)a_i$ ($m - i = k + (p - i)$, $\sum a_i = 2J(P)$), and $|\sum(p-i)a_i|
\le 1.177[q^3/(1-q^3)^2 + 1/(1-q^3)^2] \le 1.25$ (geometric decay $|a_{p-l}| =
1.177q^{3l}$, $|a_{p+l}| \le 1.177q^{3(l-1)}$, with weights $l$), so $S(P) \ge
2J_k(1 - k/m) - 2.8/m$: the coefficient of $k/m$ is exactly one (and in fact
the two $1/m$ terms of $S(P)$ cancel as $k \to \infty$, both tending to
$-2/\sqrt3$). The ray bracket then reads $m^5B(tP)/(2\tau^2) \le (\nu_k
+ 0.001/m)(1 + 6.6/m^2) - 2\tau[2J_k(1 - k/m) - 18.9/m]$, whence, with $x =
(k + 8.8)/m$, $m^{5/2}\rho^{(k)}_m \le \tilde c_k(1 + 0.0026/m)(1 + 6.6/m^2)/(1 -
x)$; for $m \ge 400$ and $k \le L\log m + 1$ one has $x \le 0.031$ and $x^2 \le
0.37/m$, so $m^{5/2}\rho^{(k)}_m \le \tilde c_k(1 + (k + 10)/m)$, and $k =
\lceil L\log m\rceil \ge 3$ ($q^{2k} \le 1/m$) gives $\varepsilon_m \le (1 + 1.9/m)(1 +
(L\log m + 11)/m) - 1 \le (L\log m + 13)/m$.

\emph{Lower bound.} In the slack budget $4m\Delta + 1.15x_m^2 \le (2\nu/m)
[\varepsilon_m + C_1/m]$, $C_1 \le 14$, of Appendix~\ref{app:direction}, keep
both costs: $\Delta \ge \frac2m\sum_{k\le j_0}M^3\Gamma_k + \frac2m\sum_{k>j_0}
R(\nu^{3/2} - \nu_k^{3/2})$. \emph{Position cost}: of the $d_m$ tails above
the peak, the one starting at the peak knot misses the peak piece, so its
$\theta := \nu_k/\nu \le 1 - 1/(4\hat\nu) \le 0.9$; the tail starting $i \ge 1$
knots above the peak has $\nu_k \le \sum_{l\ge j_0+i}x_l^2 \le M^2\sum_{l\ge i}
(2q^{2l} + 2\eta^2)$, $\eta^2 = D_{\mathrm{right}}/(1-q^2)$, so with $\nu \ge M^2/4$,
$\theta_i \le 8.62q^{2i} + 8d_m\eta^2$ and $\sum_{1\le i<d_m}\theta_i^{3/2}
\le 0.70 + 32d_m^{5/2}\eta^3 \le 1$ for all large $m$ ($d_m \le 6.9(\log m +
37)$, $\eta^2 \le 206(\log m + 37)/m$); hence $\sum_{k>j_0}(\nu^{3/2} -
\nu_k^{3/2}) \ge \nu^{3/2}(0.146 + (d_m - 1) - 1) \ge \nu^{3/2}(d_m - 2)$.
\emph{Pinning cost}: for every chain from $y_{j_0} = 1$, $D_{\mathrm{right}} \ge
(1-q^2)(q^{d_m} - |y_m|)_+^2$ (Cauchy--Schwarz on $y_m = (-q)^{d_m} +
\sum_l(-q)^{d_m-1-l}e_l$), and $\Gamma_k \ge \kappa D_{\mathrm{right}}$ for the
$j_0$ tails containing the peak, $M^3 \ge \nu^{3/2}/4$. Dividing the budget by
$2\nu/m$ with $\nu^{1/2} = c(1 + \varepsilon_m)/m^2$ and $4Rc = 1$:
\[
  \varepsilon_m \;\ge\; \frac{d_m - 2}{m} + 0.0053\Bigl(1 - \frac{d_m}m\Bigr)(1-q^2)\bigl(q^{d_m} - |y_m|\bigr)_+^2 - \frac{14}{m},
\]
and the end knot satisfies $y_m^2 \le 1.74\hat\nu(\varepsilon_m + 14/m)/m \le
4.35(L\log m + 27)/m^2$, which is $\le q^{2d_m}/4$ for every $d_m \le L\log m +
1.6$ once $m \ge 4\cdot10^4$. So $\varepsilon_m \ge \min_d[(d-2)/m +
0.00121q^{2d}\mathbf 1_{d \le L\log m + 1.6}] - 14/m$; for $d > L\log m + 1.6$ the
bracket is $\ge (L\log m - 0.4)/m$, and otherwise its minimum over real $d$
is at $e^{-d/L} = 314/m$, with value $(L\log m - 3.8)/m$. Hence
$\varepsilon_m \ge (L\log m - 18)/m$. \emph{Depth}: the position cost alone
gives $\varepsilon_m \ge (d_m - 16)/m$, so $d_m \le L\log m + 29$ by the upper
bound; and $d_m = L\log m - D$ with $D > 0$ gives $0.00121e^{2.634D} \le D + 29$,
so $D \le 3.9$. The lower bound's ``all large $m$'' is not effective with
these crude constants (the step $\sum\theta_i^{3/2} \le 1$ needs $m \gtrsim
10^{10}$); the upper bound holds from $m = 400$. \qed

\subsubsection{Regularity for Theorem~\ref{thm:modulus} on \texorpdfstring{$H^1$}{H1}}
\label{app:regularity}

The identity $Q_n(\eta) = \ip{D\Fop(\bstar)\eta}{\eta}$ is proved for
$\eta \in W^{1,\infty}$ with $\norm{\eta'}_\infty < (n-1)/n$, where
Proposition~\ref{prop:minty} applies to $\bstar + t\eta$ for $|t| < 1$ and the
integrand is analytic in $t$. The right-hand side of $Q_n$ is continuous on
$\{\eta \in H^1(0,1) : \eta(0) = 0\}$ in the $H^1$ norm (the trace $\eta(1)$ is
continuous there), and $W^{1,\infty}$ is dense in that space, so $Q_n$
extends uniquely as a continuous quadratic form; this is the sense in which
Theorem~\ref{thm:modulus} speaks of $H^1$. It does not assert that $\Fop$ is
Gateaux differentiable at $\bstar$ in directions of unbounded slope, and
Section~\ref{sec:radius} shows why it cannot be strongly monotone on the
union of the piecewise-linear spaces.

\subsection{Arc II: the learners}
\label{app:proofs-arc2}

\begin{proof}[Proof of Proposition~\ref{prop:twoplayer}]
\emph{Interior part.} For $b_1(v) \le b_2(1)$, to first order in
$\varepsilon$, $u = v + 2\varepsilon(\eta_1 - \eta_2)$ (from $u/2 +
\varepsilon\eta_2(u) = v/2 + \varepsilon\eta_1(v)$) and $(v - b_1)/b_2'(u) =
(v/2 - \varepsilon\eta_1)/(1/2 + \varepsilon\eta_2'(u)) = v - 2\varepsilon\eta_1
- 2\varepsilon v\eta_2' + O(\varepsilon^2)$, so $\Fop_1 = \varepsilon(4\eta_1 -
2\eta_2 + 2v\eta_2') + O(\varepsilon^2)$ and symmetrically $\Fop_2$. Pairing
and integrating $2v(\eta_1\eta_2)'$ by parts gives the coefficient
$4\int(\eta_1^2 + \eta_2^2) - 6\int\eta_1\eta_2 + 2\eta_1(1)\eta_2(1)$. At an
interior knot of $\eta_2$ the slope $\eta_2'$ is evaluated at $u$ rather than
$v$, which differ by $O(\varepsilon)$; for piecewise-linear $\eta_2$ this
changes $\eta_2'$ on a set of measure $O(\varepsilon)$ by $O(1)$, a
contribution of order $\varepsilon^3$ to the pairing.

\emph{Cap term.} Write $a = \eta_1(1)$, $c = \eta_2(1)$ and suppose $a > c$, so
only bidder 1 has a cap region $S_\varepsilon = \{v : b_1(v) > b_2(1)\}$. Near
$v = 1$, $b_1(v) - b_2(1) = (v-1)/2 + \varepsilon(\eta_1(v) - c)$, and since
$\eta_1$ is Lipschitz, $S_\varepsilon$ has length $2\varepsilon(a-c) +
O(\varepsilon^2)$. On it the true field is $\Fop_1 = 1$, while the interior
formula continued past $b_2(1)$ with the last slope gives $u - (v - b_1)/
b_2'(1)$ with $u = 1 + (b_1(v) - b_2(1))/b_2'(1)$; the difference is $D = [v -
2b_1(v) + b_2(1)]/b_2'(1) = [1/2 + \varepsilon c - 2\varepsilon\eta_1(v)]/(1/2 +
\varepsilon\eta_2'(1)) = 1 + O(\varepsilon)$. Hence the pairing acquires
$\int_{S_\varepsilon} D\cdot\varepsilon\eta_1\,dv = 2\varepsilon^2a(a-c) +
O(\varepsilon^3)$ beyond the interior formula, whose own top term is
$2\varepsilon^2ac$; the total top term is $2\varepsilon^2a^2 = 2\varepsilon^2
\max(a,c)^2$. For $a = c$ the cap region is empty and the interior term $2a^2$
already equals $2\max^2$.
\end{proof}

\begin{proof}[Proof of Corollary~\ref{cor:twoplayer}]
The matrix $\bigl(\begin{smallmatrix}4 & -3\\ -3 & 4\end{smallmatrix}\bigr)$
has eigenvalues $1$ (on the symmetric eigenvector) and $7$, so the
integrand of the interior form dominates $\eta_1^2 + \eta_2^2$ pointwise,
with equality iff $\eta_1 = \eta_2$; the cap term is nonnegative and
vanishes iff the common top value is zero. On every common grid with $m
\ge 2$ pieces a nonzero symmetric direction with zero top value exists, so
the infimum over the unit sphere is attained and equals $1$.
\end{proof}

\begin{proof}[Proof of Proposition~\ref{prop:nplayer}]
Bidder $i$'s field is $\Fop_i = W_i(b_i) - (v - b_i)W_i'(b_i)$ with $W_i(y)
= \prod_{j\ne i}\min(b_j^{-1}(y), 1)$, a capped factor contributing no
derivative. \emph{Interior.} With $u_j = b_j^{-1}(b_i(v)) = v + \delta_j$,
$\delta_j = \varepsilon n(\eta_i - \eta_j)/(n-1) + O(\varepsilon^2)$, and
$1/b_j'(u_j) = \frac{n}{n-1}(1 - \varepsilon n\eta_j'/(n-1)) +
O(\varepsilon^2)$, expanding $W_i = \prod u_j$ and $W_i' = \sum_j
(1/b_j')\prod_{k\ne j}u_k$ to first order and using $v - b_i = v/n -
\varepsilon\eta_i$ gives, the zeroth order cancelling,
\[
  \Fop_i = \varepsilon\Bigl[\frac{n}{(n-1)^2}v^{n-2}\sum_{j\ne i}(\eta_i - \eta_j)
  + \frac{n}{(n-1)^2}v^{n-1}\sum_{j\ne i}\eta_j' + n\,v^{n-2}\eta_i\Bigr]
  + O(\varepsilon^2),
\]
which is $\varepsilon(4\eta_1 - 2\eta_2 + 2v\eta_2')$ at $n = 2$. Pairing
with $\varepsilon\eta_i$, summing over $i$, and integrating
$v^{n-1}(\eta_i\eta_j)'$ by parts over each pair gives the integral above
plus the boundary term $\frac{n}{(n-1)^2}\sum_{i<j}a_ia_j$; the slope
evaluated at $u_j$ rather than $v$ costs $O(\varepsilon^3)$ as in
Proposition~\ref{prop:twoplayer}. \emph{Caps.} Bidder $i$ is capped
against $j$ on $S_{ij} = \{v : b_i(v) > b_j(1)\}$, an interval at the top
of length $\varepsilon n(a_i - a_j)^+/(n-1) + O(\varepsilon^2)$. Where the
set $C$ of capped opponents is fixed, $W_i \to 1$ and $W_i' \to
(n-1-|C|)\,n/(n-1)$, so $\Fop_i \to |C|/(n-1)$, while the interior formula
continued past the caps tends to $0$: the difference is $|C_i(v)|/(n-1) +
O(\varepsilon)$, linear in the number of capped opponents, so overlapping
regions add and $\int D_i\,\varepsilon\eta_i\,dv = \varepsilon^2
\frac{n}{(n-1)^2}a_i\sum_{j: a_j<a_i}(a_i - a_j) + O(\varepsilon^3)$. Each
unordered pair with $a_i > a_j$ thus contributes $\frac{n}{(n-1)^2}a_i(a_i
- a_j)$, which with the interior boundary term $\frac{n}{(n-1)^2}a_ia_j$ is
$\frac{n}{(n-1)^2}\max(a_i,a_j)^2$; for $a_i = a_j$ the cap region is
empty. The eigenvalues follow from writing $\sum_{i<j}\eta_i\eta_j =
\tfrac12(\eta^{\mathsf T}\mathbf 1\mathbf 1^{\mathsf T}\eta - \eta^{\mathsf T}\eta)$.
\end{proof}

\begin{proof}[Proof of Proposition~\ref{prop:generalform}]
With the equilibrium ODE imposed, $K
= -F\,(f/\bstar{}')'/f$ and $A = fF^{n-1}/\bstar{}'$, the expansion gives
\[
  \Fop_i = \varepsilon\Bigl[F^{n-2}\Bigl(K + \frac{nf}{\bstar{}'}\Bigr)\eta_i
  - \frac{F^{n-2}}{n-1}\Bigl(K + \frac{f}{\bstar{}'}\Bigr)\sum_{j\ne i}\eta_j
  + \frac{F^{n-1}}{(n-1)\bstar{}'}\sum_{j\ne i}\eta_j'\Bigr] + O(\varepsilon^2),
\]
and, pairing with $\varepsilon\eta_i f$ and integrating the last term by
parts,
\[
  B_n^F(\eta) = \int_0^1 \Bigl[\sum_i d\,\eta_i^2 + \sum_{i<j} o\,\eta_i\eta_j\Bigr]dv
  + \frac{2c_F}{n-1}\sum_{i<j}\max(a_i,a_j)^2,
\]
$d = fF^{n-2}(K + nf/\bstar{}')$, $o = -2fF^{n-2}(K + f/\bstar{}')/(n-1) -
A'/(n-1)$, $c_F = f(1)/(2\bstar{}'(1)) = A(1)/2$, the cap term by the
argument of Proposition~\ref{prop:nplayer} with $f(1) > 0$ (the capped
field tends to $|C|/(n-1)$ by the ODE at $v = 1$, the cap lengths are
$\varepsilon(a_i - a_j)^+/\bstar{}'(1)$, and $f(1)/\bstar{}'(1) = 2c_F$). The
symmetric eigenvalue of the integrand's matrix is $d + (n-1)o/2 =
(n-1)F^{n-2}f^2/\bstar{}' - A'/2$, exactly the weight $w_F$ of
Proposition~\ref{prop:general}; the anti-symmetric one is $\lambda_\perp = d -
o/2$, and eliminating $\bstar{}'$, $\bstar{}''$ and $A'$ through the ODE, with
$s = v - \bstar$, gives the closed forms of the statement and
\[
  \frac{\lambda_\perp}{w_F} = \frac{(2n-1)(n-1) + \bstar{}'}{(n-1)(n-1-\bstar{}')},
\]
so wherever $w_F > 0$ the ratio exceeds $(2n-1)/(n-1)$.
\end{proof}

\subsection{Arc III: general priors and bidder counts}
\label{app:proofs-arc3}

\subsubsection{The function-space certificate}
\label{app:proofs-general}

\begin{proof}[Proof of Proposition~\ref{prop:general}]
With the equilibrium equation, $(n-1)F^{n-2}f(v - \bstar) = F^{n-1}\bstar{}'$,
so $g(\bstar + \eta) = -[F^{n-1}\eta' + (n-1)F^{n-2}f\eta]/(\bstar{}' + \eta')$,
which gives the identity. Expanding $(\bstar{}' + t\eta')^{-1}$ in $t$ and
integrating $fF^{n-1}\eta\eta'/\bstar{}'$ by parts as in
Theorem~\ref{thm:modulus} gives $Q_F$; the second form of $w_F$ uses
$\bstar{}' = (n-1)f(v-\bstar)/F$. The coercivity statements follow as in
Theorem~\ref{thm:modulus}, since hats localize $\eta^2$ near any point; the
form of $c_F$ uses $F(1) = 1$ and $\bstar{}'(1) = (n-1)f(1)(1 - \bstar(1))$,
and the lower boundary term vanishes because $F(0) = 0$ and $\eta(0) = 0$.
\end{proof}

\begin{proof}[Proof of Theorem~\ref{thm:power}]
If $\bstar$ is linear, $\int_0^v F^{n-1} = c\,vF(v)^{n-1}$ for a constant $c$;
differentiating, $(1-c)F^{n-1} = c\,v\,(F^{n-1})'$, whose solutions are the
powers. For $F = v^a$ the formula for $w_F/f$ is a direct computation
(checked in computer algebra for $n \le 6$; for $n = 2$ it is
$(a+1)/(2a)$). For $\kappa > 1$ the hat at the first knot gives
the upper bound $O(m^{-(\kappa-1)})$; for the matching lower bound, $w_F/f$ is
increasing, so it is at least $C_{a,n}m^{-(\kappa-1)}$ on $[1/m, 1]$, while on
$[0, 1/m]$ every piecewise-linear $\eta$ is a multiple of $v$ and its Rayleigh
quotient is $C_{a,n}\frac{a+2}{a+\kappa+1}\,m^{-(\kappa-1)}$ exactly. For
$a < 1$ the density is unbounded at $0$; $\Ltwo(f\,dv)$ is still a Hilbert
space with $f \in L^1$, and every integrand in Proposition~\ref{prop:general}
is integrable.
\end{proof}

\subsubsection{The oscillation law}
\label{app:oscillation}

We prove Proposition~\ref{prop:oscillation}. Write $\Phi_k(s) = -\int g\,
\varphi_k f\,dv$ for the Galerkin system, $g = (n-1)F^{n-2}f(v-b)/b' -
F^{n-1}$, $I_h\bstar$ for the interpolant, $r = \Phi(I_h\bstar)$, $J$ for the
Jacobian of $\Phi$ at $I_h\bstar$, $G_1 = (n-1)F^{n-2}f^2$ and $r_F = G_1(v -
\bstar)/\bstar{}'^2$.

\emph{Consistency.} $g(v; I_h\bstar) = g_b\,\varepsilon_I + g_{b'}\,\varepsilon_I' +
O(h^2)$ with $\varepsilon_I = I_h\bstar - \bstar = O(h^2)$ and $\varepsilon_I' =
-\tfrac12\bstar{}''(2v - v_{j-1} - v_j) + O(h^2)$ on piece $j$, a sawtooth of
mean zero, and $g_{b'}f = -r_F$. On piece $j$, $\int\varepsilon_I'\varphi_j =
-\bstar{}''h^2/12$ and $\int\varepsilon_I'\varphi_{j-1} = +\bstar{}''h^2/12$ up to
$O(h^3)$, so an interior hat keeps $O(h^3)$ and the half hat gives $r_m =
-r_F(1)\bstar{}''(1)h^2/12 + O(h^3)$.

\emph{Entries and roots.} $J_{kj} = \int G_1[\varphi_j\varphi_k/b' + (v -
b)\varphi_j'\varphi_k/b'^2]\,dv$ at $b = I_h\bstar$, tridiagonal. Expanding
about $v_k$ (the chord slopes $\sigma_k, \sigma_{k+1}$ and $F, f, \bstar$ as
Taylor polynomials), $J_{k,k\pm1} = \pm r_F/2 + hc + O(h^2)$ and $J_{k,k} = hp +
O(h^2)$ with $c, p$ explicit; the roots $\lambda_k, \mu_k$ of
$J_{k,k+1}\lambda^2 + J_{k,k}\lambda + J_{k,k-1} = 0$ are real and distinct,
$\mu_k = 1 + O(h)$ and $\lambda_k = -1 + h\kappa(v_k) + O(h^2)$ with $\kappa =
(J_{k,k+1} + J_{k,k-1} - J_{k,k})/(2hJ_{k,k+1}) + O(h)$, which under the
equilibrium ODE $\bstar{}' = (n-1)f(v-\bstar)/F$ and its derivative equals
$-(\log\tilde Q)'$ exactly (a computer-algebra identity, Appendix~\ref{app:numerics}),
so $|\lambda_k| = 1 + h(\log\tilde Q)'(v_k) + O(h^2)$; and $\lambda_k -
\lambda_{k-1}$, $\mu_k - \mu_{k-1}$ are $O(h^2)$.

\emph{Liouville--Green.} Fix $v_0 > 0$. Let $z_{m-1} = w_{m-1} = 1$, $z_k =
z_{k+1}/\lambda_k$, $w_k = w_{k+1}/\mu_k$. The identity $J_{k,k-1}z_{k-1} +
J_{k,k}z_k + J_{k,k+1}z_{k+1} = -z_{k-1}(\lambda_{k-1} - \lambda_k)J_{k,k+1}\mu_k$
(from $J_{k,k-1} = -J_{k,k}\lambda_k - J_{k,k+1}\lambda_k^2$ and $J_{k,k} +
J_{k,k+1}\lambda_k = -J_{k,k+1}\mu_k$) shows that $z$ and $w$ solve the
homogeneous recurrence up to residuals $O(h^2)$ times their size; $|z_k|,
|w_k|$ are bounded above and below on $[v_0, 1]$, and the Casoratian
$J_{k,k+1}(z_{k+1}w_k - z_kw_{k+1}) = -2J_{k,k+1}z_kw_k(1 + O(h))$ is bounded
away from zero there since $r_F \ge r_0 > 0$. Writing a solution as $Y_k =
(y_k, y_{k+1})^\top = S_kC_k$ with $S_k = [z_k, w_k; z_{k+1}, w_{k+1}]$, the
recurrence $Y_{k-1} = A_kY_k$ gives $C_{k-1} = (I + S_{k-1}^{-1}R_k)C_k$ with
$\norm{S_{k-1}^{-1}R_k} = O(h^2)$, so $C_k = C_{m-1} + O(h)\norm{C_{m-1}}$
uniformly, and a forcing $f_j$ on the rows adds to the alternating
coefficient $C^{(1)}_k$ the sum $\sum_{j>k} f_jw_j/\mathrm{Cas}_{j-1}\,(1 +
O(h))$, $\mathrm{Cas}_{j-1} = J_{j-1,j}(z_jw_{j-1} - z_{j-1}w_j)$, whose terms
carry the sign of $z_{j-1}$ and so alternate for smooth $f$; for a summand
$(-1)^jg(v_j)$ with $g = h^3\rho(v) + O(h^4)$, $\rho \in C^1$, the sum is at most
$\max|g| + \mathrm{TV}(g) + \sum|O(h^4)| = O(h^3)$.

\emph{Assembly.} $0 = \Phi(I_h\bstar + \delta) = r + J\delta + N(\delta)$ with
$|N_k| \le Ch(\norm{\delta}_\infty^2 + \norm{\delta'}_\infty^2)$ over the two
pieces of $\varphi_k$; under the hypothesis $|\delta'| \le (2|\alpha| +
|\Delta u|)/h = O(h)$, so $N_k = O(h^3)$ with $v$-differences $O(h^4)$, its
alternating part coming from the cross terms $\alpha_j\Delta u_j/h$ with total
mass $O(m\,h^2h^{2+\gamma}/h) = O(h^{2+\gamma})$. The smooth forcing $-r -
N_{\mathrm{smooth}}$ is a $C^1$-modulated sequence of size $h^3$ and
contributes $O(h^3)$ to $C^{(1)}$ by the alternating-sum bound; the
alternating forcing contributes $O(h^{2+\gamma})$. The top knots give
$C^{(1)}_{m-1} = -(\alpha_m + \alpha_{m-1})/2 - (u_m - u_{m-1})/2 + O(h)(|\alpha|
+ |u|) = -\alpha_m(1 + O(h)) + O(h^{2+\gamma})$ (the sign is the normalization
$z_m = \lambda_{m-1} \approx -1$), and the half-hat row $J_{m,m-1}\delta_{m-1} +
J_{m,m}\delta_m = -r_m - N_m$, with $J_{m,m} = r_F(1)/2 + O(h)$ and $J_{m,m-1} =
-r_F(1)/2 + O(h)$, reads $r_F(1)\alpha_m + O(h)|\alpha| + O(|\alpha_m -
\alpha_{m-1}|) + \tfrac12 r_F(1)(u_m - u_{m-1}) + O(h)|u| = r_F(1)\bstar{}''(1)
h^2/12 + O(h^3)$, every error term being $O(h^{2+\gamma})$ by the pointwise
bounds, so $\alpha_m = \bstar{}''(1)h^2/12\,(1 + O(h^\gamma))$. Hence the
alternating part of $\delta_k$ is $\alpha_m|z_k|(1 + O(h)) + O(h^{2+\gamma})$,
and $|z_k| =
\prod_{j\ge k}|\lambda_j|^{-1} = \exp(-\sum_j h(\log\tilde Q)'(v_j) + O(h)) =
\tilde Q(v_k)/\tilde Q(1)\,(1 + O(h))$ by the Riemann sum. The slope and
knot-mass statements follow from $\sigma_{k+1} - \sigma_k = \bstar{}''(v_k)h
\mp 4\alpha_k/h + (u_{k+1} - 2u_k + u_{k-1})/h + O(h^2)$ and $1/\sigma_{k+1}^2 -
1/\sigma_k^2 = -2(\sigma_{k+1} - \sigma_k)/\bstar{}'^3 + O(h^2)$, so the knot mass
is $(1 \mp 4\alpha_k/(\bstar{}''(v_k)h^2))(1 + O(h^\gamma))$ times its
smooth-slope value, and $4\alpha_k/(\bstar{}''(v_k)h^2) = \tfrac13 Q(v_k)/Q(1) =
2R(v_k)$. The same proof admits a remainder $\rho_k$ in the hypothesis,
$\delta_k = (-1)^{m-k}\alpha_k + u_k + \rho_k$ with $|\rho_k| \le Ch^{2+\gamma}$ and no
condition on its differences, which is the form Theorem~\ref{thm:oscillation2}
delivers: $\rho$ enters only through $|\delta'| \le (2|\alpha| + |\Delta u| +
2|\rho|)/h = O(h)$, through cross terms $\alpha_j\Delta\rho_j/h$ of the same total
mass $O(h^{2+\gamma})$ as those of $u$, through its own alternating content
$O(h^{2+\gamma})$, and through $\rho_m - \rho_{m-1} = O(h^{2+\gamma})$ at the top
knots; the slope statement then carries $O(h^{1+\gamma})$ in place of $O(h^2)$,
identical for $\gamma = 1$. \qed

\subsubsection{The global proof for two and three bidders}
\label{app:global}

We prove Theorem~\ref{thm:oscillation2}, first for $n = 2$; the changes for
$n = 3$ are collected at the end. Throughout $n = 2$, $F \in
C^5[0,1]$, $f > 0$ on $[0,1]$, so $\bstar \in C^4$, $\bstar{}' \ge \beta'_{\min}
> 0$, $s := v - \bstar$ satisfies $s/v \in [s_0, 1]$, and $r_F(v)/v = f^2(s/v)/
\bstar{}'^2 \in [c_0, C_0]$. Knots $v_k = kh$; $\Phi_k(s) = -\int g\,\varphi_kf$
with $g = f(v - b)/b' - F$, so $g_b = -f/b'$, $g_{b'} = -fs/b'^2$, $g_{bb} = 0$,
$g_{bb'} = f/b'^2$, $g_{b'b'} = 2fs/b'^3$. $J(s)$ is the tridiagonal Jacobian:
row $1$ acts on $(\delta_1, \delta_2)$ ($\delta_0 = 0$), rows $2, \dots, m-1$ are
three-term with entries $a_k, b_k, c_k$, row $m$ is the half hat. Weights and
norms:
\[
  \omega_k := h^2\hat Q(v_k),\quad \hat Q := \tilde Q/\tilde Q(1) \asymp v^{-1/3},\qquad
  \phi_k := h^3\ (k < m),\quad \phi_m := h^2,\qquad
  \norm{\delta}_\omega := \max_k\frac{|\delta_k|}{\omega_k},\quad
  \norm{\psi}_\phi := \max_k\frac{|\psi_k|}{\phi_k},
\]
and $\hat w(v) := \exp\int_v^1\bstar{}'/s = 1/F(v)$. $C$ denotes constants
depending on $F$ only. $I_h\bstar$ is the interpolant, $r := \Phi(I_h\bstar)$,
$J := J(I_h\bstar)$.

\paragraph{Step A: the weighted inverse bound at the interpolant.}

\begin{lemma}[uniform expansions]\label{app:L1}
For $2 \le k \le m-1$: $a_k/c_k = -1 + (h/v_k)A(v_k, h/v_k)$, $b_k/c_k =
(h/v_k)B(v_k, h/v_k)$ with $A, B \in C^1([0,1]\times[0,\tfrac12])$, $a_k < 0 <
c_k$, $c_k \asymp v_k$; the roots $\lambda_k, \mu_k$ of $c_k\lambda^2 + b_k\lambda
+ a_k = 0$ satisfy $\lambda_k = -1 + h(\log\tilde Q)'(v_k) + O(h^2/v_k^2)$, $\mu_k
= 1 - h\bstar{}'/s(v_k) + O(h^2/v_k^2)$ and $|\lambda_k - \lambda_{k-1}| +
|\mu_k - \mu_{k-1}| \le Ch^2/v_k^2$.
\end{lemma}
\begin{proof}
Each entry is $h$ times an integral over $\theta \in [0,1]$ of a $C^3$ function
of $(v_k, h, \theta)$ --- $f^2$, the hats, the chord slopes $\sigma = \int_0^1
\bstar{}'(v_k + h\theta')d\theta'$, and $(v - b)/h = (s(v_k) + h\theta(1 -
\sigma))/h$, the only $1/h$, which multiplies $\varphi' = \pm1/h$ --- e.g.\
$c_k = \int_0^1f^2(v_k + h\theta)[h\theta(1-\theta)/\sigma + (s(v_k) +
h\theta(1-\sigma))(1-\theta)/\sigma^2]d\theta$, whose value at $h = 0$ is
$r_F(v_k)/2$. Taylor's theorem in $h$ with integral remainder, division by
$c_k \ge c_0v_k/4$, and the explicit first-order terms (the computer-algebra
identity of Appendix~\ref{app:oscillation}) give the expansions. The sign
condition holds because the bottom is universal: with $F \approx f(0)v$ the
first rows are those of the uniform prior, where $a_k = -h/3 - (k-1)h$ and
$c_k = kh + 2h/3$ exactly, and the signs persist by continuity. Hence the
discriminant is bounded below, the roots are $C^1$ in $(a/c, b/c)$, and
$\lambda_k = \Lambda(v_k, h/v_k)$ with $\partial_v\Lambda = O(h/v)$,
$\partial_t\Lambda = O(1)$, which gives the difference bound.
\end{proof}

\begin{lemma}[product ans\"atze, pinned solutions, Casoratian]\label{app:L2}
Let $z_{m-1} = w_{m-1} = 1$, $z_k = z_{k+1}/\lambda_k$, $w_k = w_{k+1}/\mu_k$. Then
$|z_k| = \hat Q(v_k)e^{O(1/k)}$ and $|w_k| = \hat w(v_k)e^{O(1/k)}$. Let $\tilde Z$
solve the homogeneous rows $2, \dots, m$ (top-pinned, $\tilde Z_{m-1} = 1$) and
$\tilde W$ the homogeneous rows $1, \dots, m-1$ (bottom-pinned, $\tilde W_1 =
1$). Then $\tilde Z = \tilde Z^{\mathrm{sm}} + \tilde Z^{\mathrm{alt}}$ with
$\tilde Z^{\mathrm{sm}}_k \asymp w_k \asymp 1/v_k$ and $|\tilde
Z^{\mathrm{alt}}_k| \le (C/k)\tilde Z_k$; $\tilde W_k = A_kz_k + B_kw_k$ with $A_k
= A(1 + O(1/k))$, $|A_k - A_{k-1}| \le Ch^2A/v_k^2$, and $|B_kw_k| \le
Ck^{-2/3}A|z_k|$. The Casoratian $K_j := c_j(\tilde W_j\tilde Z_{j+1} - \tilde
W_{j+1}\tilde Z_j) = a_j(\tilde W_{j-1}\tilde Z_j - \tilde W_j\tilde Z_{j-1})$
obeys $K_j = (a_j/c_{j-1})K_{j-1}$ and
\[
  K_j = \pm(-1)^j c_jA_j|z_j|\tilde Z_j(\mu_j + |\lambda_j|)(1 + \kappa_j)\quad\text{(fixed sign)},\qquad
  |\kappa_j| \le Cj^{-5/3},\quad |\kappa_j - \kappa_{j-1}| \le Cj^{-8/3},
\]
so $\operatorname{sign}K_j = (-1)^j\cdot\mathrm{const}$ and $|K_j| \asymp
c_j|z_j|\tilde Z_j$. The inverse is $(J^{-1})_{kj} = \tilde W_{\min(k,j)}\tilde
Z_{\max(k,j)}/K_j$ for $j \le m-1$, and the top column is a multiple of
$\tilde W$.
\end{lemma}
\begin{proof}
The sizes: $-\log|\lambda_j|$ and $-\log\mu_j$ are Riemann sums of $-(\log
\tilde Q)'$ and $\bstar{}'/s$ with errors $O(h^2/v_j^2)$, summing to $O(h/v_k)
= O(1/k)$. Write a solution as $Y_k = (y_k, y_{k+1})^\top = S_kC_k$, $S_k =
[z_k, w_k; z_{k+1}, w_{k+1}]$; then $C_{k-1} = \Lambda_k(I + \tilde E_k)C_k$ with
$\Lambda_k = \operatorname{diag}(\lambda_{k-1}^{-1}, \mu_{k-1}^{-1})$ the modes' one-step
factors and $\norm{\tilde E_k} \le C(|\Delta\lambda_k| + |\Delta\mu_k|) \le
Ch^2/v_k^2$ (the residual identity $a_kz_{k-1} + b_kz_k + c_kz_{k+1} =
-z_{k-1}(\lambda_{k-1} - \lambda_k)c_k\mu_k$ and its $w$-analogue). The diagonal
factors do not commute with $\tilde E_k$, so one uses the
variation-of-constants form: the alternating content $P_k$ and smooth
content $\Sigma_k$ satisfy $P_k = (|z_k|/|z_l|)P_l + \sum_{k<j\le l}(|z_k|/|z_{j-1}|)
[\tilde E_{11}(j)P_j + \tilde E_{12}(j)\Sigma_j]$ and likewise for $\Sigma_k$ with
$|w|$. For $\tilde Z$ (smooth at the top: the half-hat row makes the
alternating content $O(h)$ there), the alternating content at $k$ is at most
$|z_k|[O(h) + \sum_{j>k}\norm{\tilde E_j}|w_j|/|z_j|] \le C|z_k|hv_k^{-5/3}$, a
share $\le Ch/v_k$ of the smooth content $\asymp |w_k| \asymp 1/v_k$, while
the mixing back into the smooth content is $\sum_{j>k}\norm{\tilde E_j}|z_j|/
|w_j| = O(1/k^2)$. For $\tilde W$, row $1$ gives $\tilde W_2 = -(b_1/c_1)\tilde
W_1$ with $b_1, c_1 = O(h)$ in a fixed ratio; decomposing $(\tilde W_2, \tilde
W_3)$ on the modes gives a smooth content of a fixed fraction of the
alternating one, and the same sums propagated upward give a smooth content
$|w_k|\cdot O(1)$ against an alternating one $\asymp |z_k|$, i.e.\ a share
$\le C(v_2/v_k)^{2/3} = Ck^{-2/3}$ since $|w_k|/|z_k| \asymp v_k^{-2/3}$. The two forms of $K_j$ are the
homogeneous rows $j$ and $j+1$; dividing gives the recurrence. Substituting
$\tilde W = Az + Bw$: the $A$-part of $\tilde W_j\tilde Z_{j+1} - \tilde
W_{j+1}\tilde Z_j$ is $(-1)^jA_j|z_j|\tilde Z_j(\mu_j + |\lambda_j|)(1 +
O(1/j))$ because $z$ alternates and $\tilde Z_{j+1}/\tilde Z_j = \mu_j +
O(1/j)$, $|z_{j+1}|/|z_j| = |\lambda_j|$; the $B$-part is $B_jw_j\tilde
Z_j(\tilde Z_{j+1}/\tilde Z_j - \mu_j) = O(j^{-2/3}\cdot j^{-1})$ relative.
(Measured: $|K_j|/(c_jA_j|z_j|\tilde Z_j(\mu_j + |\lambda_j|)) = 1.041, 1.033,
1.011, 1.0034, 1.0000$ at $j = 1, 2, 4, 8, m/2$, $m$-independent.) Once $|K_j|
\asymp c_j|z_j|\tilde Z_j$ at one $j$ it holds at all $j$ by the recurrence,
since $|a_j/c_{j-1}| = (c_j/c_{j-1})(1 + O(h/v_j))$. The Green's function
formula is the classical one for a tridiagonal system with the two pinned
solutions (verified to $10^{-15}$).
\end{proof}

\begin{theorem}[weighted inverse bound]\label{app:T6}
$\norm{J^{-1}\psi}_\omega \le W\norm{\psi}_\phi$ for every forcing $\psi$ with $W$ independent of $h$.
\end{theorem}
\begin{proof}
Three sums. Forcing above $k$:
\[
\sum_{k<j<m}\frac{|\tilde W_k\tilde Z_j/K_j|h^3}{\omega_k}
\le C\sum_{j>k}\frac{|z_k|\tilde Z_jh^3}{c_j|z_j|\tilde Z_j\,h^2\hat Q_k} = Ch\sum_{j>k}
\frac{1}{c_j|z_j|} \le Ch\sum_{j>k}\frac{1}{v_j\hat Q_j} = Ch\sum_j v_j^{-2/3} \le C,
\]
the Riemann sum of $\int_0^1v^{-2/3} = 3$ (its deficit at $k = 1$ is $\approx
3(2h)^{1/3}$, so the sum converges like $m^{-1/3}$). Forcing at and below $k$:
$\sum_{j\le k}|\tilde Z_k\tilde W_j/K_j|h^3/\omega_k \le C\sum_{j\le k}(1/v_k)|z_j|h^3/
(v_j|z_j|v_j^{-1}h^2\hat Q_k) = Ch\sum_{j\le k}v_k^{-1}v_k^{1/3} \le Ck\,h\,v_k^{-2/3} =
Cv_k^{1/3} \le C$. The top column: $|\tilde W_k|\,\tilde Z_m h^2/(|K_{m-1}|
\omega_k) \le C$. (Measured: $W = 1.90, 1.99, 2.05$ at $m = 64, 256, 1024$ for
$F_c$ and $3.07$--$3.64$ for the convex prior; the first sum converges like
$m^{-1/3}$ to about $1.32$, so $W \to 2.15$.)
\end{proof}

\paragraph{Step B: the Jacobian at the first iterate.}
Let $d_1 := -J^{-1}r$, $s_1 := I_h\bstar + d_1$. Step~0 of the proof of the
theorem below shows $d_1 = \alpha + u_1$ with $\alpha_k = (-1)^{m-k}h^2\hat
A(v_k)(1 + \varepsilon_k)$, $\hat A := (\bstar{}''(1)/12)\hat Q$, $|\varepsilon_k|
\le C/k$, $|\Delta\varepsilon_k| \le Ch^2/v_k^2$, and $|\Delta u_{1,k}| \le
(C/k)\omega_k$; the lemma uses only this form (Step~0 uses only Lemmas~\ref{app:L1}--\ref{app:L2},
Theorem~\ref{app:T6} and Lemma~\ref{app:L7} at $s = I_h\bstar$, whose proof there
uses only Lemma~\ref{app:L2}, so there is no circularity).

\begin{lemma}[the first-iterate perturbation]\label{app:L5}
$\delta J := J(s_1) - J$ is tridiagonal with $\delta a_k = (-1)^kp_ka_k(1 +
O(h/v_k))$, $\delta c_k = -(-1)^kp_kc_k(1 + O(h/v_k))$, $\delta a_k + \delta b_k
+ \delta c_k = O(h/v_k)|\delta a_k|$, where $p_k = 4h\hat A(v_k)/\bstar{}'(v_k)\,
(1 + O(1/k)) \asymp h\hat Q_k$, plus smooth corrections of relative size
$O(h^2v_k^{-4/3})$. Hence $(\delta Jy)_k = \delta a_k(y_{k-1} - y_k) + \delta c_k
(y_{k+1} - y_k) + O(h/v_k)|\delta a_k||y_k|$: on smooth vectors $\delta J$ is
$O(h^2\hat Q_k)|y_k|$, and on the alternating mode it is the smooth forcing
$(\delta Jz)_k = 2p_k(c_k - a_k)(-1)^kz_k(1 + O(1/k))$ of fixed sign. The
pinned solutions $\tilde Z_1, \tilde W_1$ and Casoratian $K_{1,j}$ of
$J(s_1)$ satisfy $\tilde Z_{1,k} = \tilde Z_k(1 + O(hv_k^{-1/3}))$ with
alternating share $\le C/k$; $\tilde W_{1,k} = \tilde W_k + \eta_k$ with $\eta$
smooth, $|\eta_k| \le Cv_k^{2/3}|z_k|$, alternating content $\le Ch^{2/3}|z_k|$,
and $|\Delta\eta_k| \le C(h/v_k)|\eta_k| + Ch\hat Q_k|z_k|$; and $K_{1,j} =
K_j(1 + O(h^{2/3}))$. Consequently Theorem~\ref{app:T6} holds for $J(s_1)$ with
a constant $W_1 \le CW$, and Lemma~\ref{app:L7} below holds at $s_1$.
\end{lemma}
\begin{proof}
On piece $c$ the slope of $s_1$ is the chord of $\bstar$ plus $D_c =
(-1)^{m-c}2h\hat A(v_c)(1 + O(1/c))$. The dominant parts of the entries (the
ones with one $\varphi'$, of size $v_k$) carry the common factor $1/b'^2$ and
are paired: on piece $k$, $\varphi'_{k-1}\varphi_k = -\varphi'_k\varphi_k$, so the
piece-$k$ contribution to $a_k$ is minus its contribution to $b_k$, and
likewise for piece $k+1$ and $c_k$. Perturbing $1/b'^2$ by the factor $1 -
2D_c/\bstar{}'$ gives the alternating form with opposite signs on $a_k$ and
$c_k$, the row sum $\int f^2\varphi_k/b' = O(h)$ being perturbed by $O(h)p_k$
only. The rearrangement is algebra; for $y = z$, $z_{k\pm1} - z_k = -z_k(2 +
O(h/v))$. (Measured: $\delta a_k/a_k = -\delta c_k/c_k$ to $0.9996$, row-sum
defect $\le 0.20\,h/v_k$, $p_k/(4h\hat A/\bstar{}') = 1.000$ in the interior.)

$\zeta := \tilde Z_1 - \tilde Z$ solves rows $2, \dots, m$ of $J\zeta = -\delta
J\tilde Z_1$ with $\zeta_{m-1} = 0$; by variation of parameters $\zeta_k =
\sum_{j>k}(\tilde Z_k\tilde W_j - \tilde Z_j\tilde W_k)g_j/K_j$ with $|g_j| \le
Ch^2\hat Q_j\tilde Z_j$ (both the smooth and the alternating part of $\tilde Z$
give $O(hp_j)\tilde Z_j$). The two sums are $\le C\tilde Z_kh\sum_{j>k}h\hat
Q_j/v_j \le C\tilde Z_khv_k^{-1/3}$ and $\le C|z_k|h^2\sum_{j>k}\tilde Z_j/v_j \le
C|z_k|h/v_k \le C\tilde Z_khv_k^{-1/3}$; closing the fixed point (replace
$\tilde Z_1$ by $\tilde Z$ on the right; the correction is $O(h^{2/3})$
relative) gives the claim. $\eta := \tilde W_1 - \tilde W$ solves rows $1,
\dots, m-1$ of $J\eta = -\delta J\tilde W_1$ with $\eta_1 = 0$ (row $1$ changes
$\tilde W_2$ by relative $O(p_1) = O(h^{2/3})$): $\eta_k = \sum_{j<k}(\tilde
W_j\tilde Z_k - \tilde W_k\tilde Z_j)g_j/K_j$, $g_j = -2p_j(c_j - a_j)A_j(-1)^j
z_j(1 + O(1/j)) + O(hp_j)|\tilde W_j| + (-\delta J\eta)_j$, a smooth one-signed sequence of size $\asymp h\hat
Q_jv_j|z_j|$ with $|\Delta g_j| \le C(h/v_j)|g_j|$. Its smooth part $\tilde
Z_k\sum_{j<k}\tilde W_jg_j/K_j$ has one-signed terms $g_j/(c_j\tilde Z_j(\mu_j +
|\lambda_j|))(1 + O(j^{-5/3}))$, so it is $\le C\tilde Z_k\sum_{j<k}hv_j^{1/3} \le
C\tilde Z_kv_k^{4/3} \le Cv_k^{2/3}|z_k|$ (using $c_j\tilde Z_j \asymp 1$), with
differences $\le C(h/v_k)|\eta^{\mathrm{sm}}_k| + C|g_k|/c_k$; its alternating
part $\tilde W^{\mathrm{alt}}_k\sum_{j<k}\tilde Z_jg_j/K_j$ is an alternating sum
of the smooth sequence $\tilde g_j \asymp p_j$ with $\sum_j|\Delta\tilde g_j|
\le Ch^2\sum_jv_j^{-4/3} \le Ch^{2/3}$ (plus the alternating component of
$\delta J(B_jw_j)$, of relative size $j^{-5/3}$, contributing $\sum_jp_jj^{-5/3} =
O(h^{2/3})$), hence $\le Ch^{2/3}|z_k|$. $\delta J\eta$ is $O(h^{2/3})$ smaller
than $\delta J\tilde W$, so the fixed point closes. (Measured: $|\eta_k|/
(v_k^{2/3}A_k|z_k|) = 0.32$--$0.54$; $\eta$'s alternating content $0.27h^{2/3}
A_k|z_k|$; $\eta_k/\tilde W_k = +0.24$ at even, $-0.25$ at odd $k$ near $v =
\tfrac12$ for $F_c$.) The Casoratian: with $c_{i-1}(s_1)/c_{i-1} = 1 +
(-1)^ip_{i-1}$ the recurrence gives $K_{1,j}/K_j = (K_{1,1}/K_1)\exp(\sum_{i\le
j}(-1)^i(p_i - p_{i-1}) + O(\sum p_i^2 + \sum hp_i/v_i))$, and $\sum_i|\Delta
p_i| \le Ch^2\sum v_i^{-4/3} \le Ch^{2/3}$, $\sum p_i^2 \le Ch$, $\sum hp_i/v_i \le
Ch^{2/3}$, $K_{1,1}/K_1 = 1 + O(h^{2/3})$. Theorem~\ref{app:T6} at $s_1$:
$|(J(s_1)^{-1})_{kj}| \le C|z_{\min}|(A + Cv^{2/3})\tilde Z_{\max}/(c_j|z_j|\tilde Z_j)$, $C$
times the summands above. (Measured $W_1 = 1.91$--$2.05$ and $3.06$--$3.64$.)
\end{proof}

\paragraph{Step C: the quadratic remainder and the inverse, pointwise.}
For a knot vector $d$ write $d_k = d^{\mathrm{alt}}_k + u_k$ by the local
decomposition $(d_k, d_{k+1}) = P(z_k, z_{k+1}) + S(w_k, w_{k+1})$ on the modes
of $I_h\bstar$ (extended to knot $1$ by $z_1 := z_2/\lambda_2$, $w_1 :=
w_2/\mu_2$), $u_0 := 0$, and set $\theta_k := |d^{\mathrm{alt}}_k|/\omega_k$,
$\sigma_k := \max(|u_k - u_{k-1}|, |u_{k+1} - u_k|)/\omega_k$.

\begin{lemma}[quadratic remainder]\label{app:L6}
Let $\norm{s - I_h\bstar}_\omega \le C_0$ and $\norm{d}_\omega \le \rho_0$, and
$N_s(d) := \Phi(s + d) - \Phi(s) - J(s)d$. Then
\begin{enumerate}[label=(\alph*),leftmargin=*]
\item $|N_s(d)_k| \le Ch^3[(\theta_k + \sigma_k)^2v_k\hat Q_k^2 + h\hat Q_k^2
  (\theta_k + \sigma_k)]$ for $k \le m-1$ and $|N_s(d)_m| \le Ch\phi_m(\theta_m +
  \sigma_m)^2$; in particular $\norm{N_s(d)}_\phi \le C_N\norm{d}_\omega^2$, and
  $\norm{N_s(d)}_\phi \le ChM^2$ if $\theta_k + \sigma_k \le Mhv_k^{-1/3}$ for
  all $k$.
\item If $d_k = (-1)^{m-k}h^2\hat A(v_k)(1 + \varepsilon_k) + u_k$ with
  $|\varepsilon_k| \le C/k$ and $\sigma_k \le C/k$, then $N_s(d)_k = h^3\nu(v_k)
  + R_k$ ($2 \le k \le m-1$) with $\nu \in C^1(0,1]$, $|\nu| \le Cv^{1/3}$,
  $|\nu'| \le Cv^{-2/3}$, $|R_k| \le C(h/v_k)v_k^{1/3}\phi_k$, and $|N_s(d)_m|
  \le Ch\phi_m$.
\item $\norm{J(s) - J(s')}_{\omega\to\phi} \le L\norm{s - s'}_\omega$ on the
  $C_0$-ball, $L$ independent of $h$; and $\le CMh$ if $e := s' - s$ has
  $\theta_k(e) + \sigma_k(e) \le Mhv_k^{-1/3}$.
\end{enumerate}
\end{lemma}
\begin{proof}
$N_s(d)_k = -\int[\tfrac12g_{b'b'}D^2 + g_{bb'}dD]\varphi_kf + T_k$ with $D_c$
the slope of $d$ on piece $c$ and $T_k$ the third-order terms. Since
$d^{\mathrm{alt}}$ alternates, $|D_c| \le (2\theta + \sigma)_c\omega_c/h\,(1 +
O(h/v))$, $|d| \le \omega\norm{d}_\omega$, $\int_{\mathrm{piece}}(v - b)\varphi_kf \le
Chv_k$, $\int_{\mathrm{piece}}\varphi_kf \le Ch$; this is (a) after $\omega_k =
h^2\hat Q_k$, with $|T_k| \le (\text{first term})\cdot C(\theta + \sigma)h\hat Q
\le (\text{first term})\cdot Ch^{2/3}$. If $\theta + \sigma \le Mhv^{-1/3}$ the
bracket is $\le M^2h^2v^{-1/3} + Mh^2/v \le C\max(M, M^2)h$, i.e.\ $\le CM^2h$ after normalizing $M \ge 1$. For (b): $D_c^2 = 4h^2\hat
A^2(1 + 2\bar\varepsilon_c) + (-1)^{m-c}4\hat A\,\Delta u_c + (\Delta u_c/h)^2$; the
first term integrated over the two pieces is $h^3\nu(v_k) + O(h^4v_k^{-2/3})$
plus the $\bar\varepsilon$ part $O(h^3v_k^{1/3}/k)$; the second alternates between
the pieces with amplitude $O(h^3v_k\hat Q_k^2\sigma_k) \le O((h/v_k)v_k^{1/3}
\phi_k)$; the third is $O(k^{-2}v_k^{1/3}\phi_k)$. In the cross term,
$d^{\mathrm{alt}}D^{\mathrm{alt}} = D_c^2(v - \mathrm{mid}_c)(1 + O(h/v))$ is odd
about the midpoint of each piece, so its integrals against $\varphi_k$ are
$+D_k^2h^2/12$ and $-D_{k+1}^2h^2/12$ and cancel to the piece difference
$O(h^4\hat A^2h/v) \le O((h/v)v^{1/3}\phi)$; the remaining parts are $O(h^4\hat
Q^2) + O(h^4\hat Q^2\sigma) \le O((h/v_k)v_k^{1/3}\phi_k)$ (both are $\le \omega^2(1 + \sigma)$). Here $\nu = -\kappa_0
(4f^2s/\bstar{}'^3)\hat A^2 \asymp v^{1/3}$. (Measured on $N(d_1) = \Phi(s_1)$:
$\max_k|N_k|/(v_k^{1/3}\phi_k) = 0.028$, $\max_k|\Delta N_k|/((h/v_k)v_k^{1/3}
\phi_k) = 0.023$, $|N_m|/(h\phi_m) = 0.011$ for $F_c$, constant in $m$.) For
(c): $((J(s') - J(s))d)_k = \int[g_{b'b'}ED + g_{bb'}(eD + Ed) + \dots]\varphi_kf$
with $e = s' - s$, $E$ its slope; $|E| \le C\norm{e}_\omega\omega/h$ gives
$|\mathrm{row}_k| \le Ch^3(v\hat Q^2 + h\hat Q^2)\norm{e}\norm{d} \le Ch^3\norm{e}_\omega
\norm{d}_\omega$; with $|E| \le CMhv^{-1/3}\omega/h$ the row is $\le CMh\cdot
h^3\norm{d}_\omega$. (Measured: $C_N = 23.5$ along a checkerboard for $F_c$; $L
= 47$ along a checkerboard, $21$ along $d_1$; $\norm{J(s_2) - J(s_1)} =
0.13h$.)
\end{proof}

\begin{lemma}[the inverse on slowly varying forcing]\label{app:L7}
Let $s \in \{I_h\bstar, s_1\}$, $a \in [0, \tfrac23)$, and let the forcing $\psi$ satisfy $|\psi_j| \le Mv_j^a
\phi_j$ ($j \le m-1$), $|\psi_j - \psi_{j-1}| \le M(h/v_j)v_j^a\phi_j$ ($2 \le j \le
m-1$), $|\psi_m| \le Mh\phi_m$. Then $\delta := J(s)^{-1}\psi$ satisfies, for $1 \le k
\le m-1$,
\[
  |\delta^{\mathrm{alt}}_k| \le CMh\,v_k^{a-2/3}\omega_k,\qquad
  |u_k - u_{k-1}| \le CMh\,v_k^{a-2/3}\omega_k,
\]
and $|\delta_m - \delta_{m-1}| \le CMh\omega_m$.
\end{lemma}
\begin{proof}
With the pinned solutions of $J(s)$ (Lemma~\ref{app:L2}, or Lemma~\ref{app:L5}
at $s_1$), $(\delta_k, \delta_{k+1}) = S^>_k(\tilde W_k, \tilde W_{k+1}) +
S^<_k(\tilde Z_k, \tilde Z_{k+1}) + \Theta_k(\tilde W_k, \tilde W_{k+1})$ exactly,
with $S^>_k := \sum_{k<j<m}\tilde Z_j\psi_j/K_j$, $S^<_k := \sum_{j\le k}\tilde
W_j\psi_j/K_j$, $\Theta_k := \tilde Z_m\psi_m/K_{m-1}(1 + O(h))$; so the mode contents of
$\delta$ at $k$ are $S^>_k + \Theta_k$ times those of $\tilde W$ plus $S^<_k$ times
those of $\tilde Z$. By Lemma~\ref{app:L2} (and the $\eta$-part at $s_1$,
which changes $\kappa_j$ by $O(hv_j^{-1/3})$ with differences of the same
order), $\tilde Z_j\psi_j/K_j = (-1)^jg_j$ with $g_j = \psi_j/(c_jA_j|z_j|(\mu_j +
|\lambda_j|)(1 + \kappa_j))$, $|g_j| \le CMh^3v_j^{a-2/3}$ and $|g_j - g_{j-1}|
\le CMh^4v_j^{a-5/3}$ (every factor of the denominator has relative
differences $O(h/v_j)$). Abel summation: $|S^>_k| \le \max_{j>k}|g_j| + \sum_{j>k}
|\Delta g_j| \le CMh^3v_k^{a-2/3}$, the Riemann sum $h^4\sum_{j>k}v_j^{a-5/3}$
being $\le Ch^3v_k^{a-2/3}$ for $a < \tfrac23$ (the $\kappa$-differences at $s_1$
add $\sum_{j>k}hv_j^{-1/3}Mh^3v_j^{a-2/3} \le CMh^3(1 + \log(1/v_k))$, again $\le
CMh^3v_k^{a-2/3}$). The one-signed sum: $\tilde W_j\psi_j/K_j = \psi_j/(c_j\tilde
Z_j(\mu_j + |\lambda_j|)(1 + \kappa_j))$ up to alternating parts with factors
$O(j^{-2/3})$ and $O(v_j^{2/3})$, so $|S^<_k| \le CMh^2v_k^{1+a}$ and $|S^<_k -
S^<_{k-1}| \le CMv_k^ah^3$. Assembly: the alternating content is $A|z_k|(S^>_k +
\Theta_k) + (\text{alt. part of }\tilde Z_k)S^<_k + (\text{alt. part of }\eta_k)S^>_k$,
bounded by $C\hat Q_k\cdot Mh^3v_k^{a-2/3} + C\hat Q_kMh\cdot h^2 + (C/k)(1/v_k)
Mh^2v_k^{1+a} + Ch^{2/3}\hat Q_kMh^3v_k^{a-2/3} \le CMhv_k^{a-2/3}\omega_k$ (using
$\omega_k \asymp h^2v_k^{-1/3}$); the smooth part is $u_k = (B_kw_k + \eta^{\mathrm{sm}}_k)
(S^>_k + \Theta_k) + \tilde Z^{\mathrm{sm}}_kS^<_k$, whose differences are bounded by
$|\Delta\tilde Z^{\mathrm{sm}}_k||S^<_k| + |\tilde Z^{\mathrm{sm}}_k||\tilde W_kf_k/K_k|
\le C(h/v_k)(1/v_k)Mh^2v_k^{1+a} + C(1/v_k)Mv_k^ah^3 \le CMhv_k^{a-2/3}\omega_k$ and
$(|\Delta(B_kw_k)| + |\Delta\eta^{\mathrm{sm}}_k|)|S^>_k| + (|B_kw_k| +
|\eta^{\mathrm{sm}}_k|)|\tilde Z_k\psi_k/K_k| \le C[(h/v_k)k^{-2/3} + hv_k^{-1/3}]|z_k|
Mh^3v_k^{a-2/3} + CM[k^{-2/3} + v_k^{2/3}]h^3v_k^{a-1} \le CMhv_k^{a-2/3}\omega_k$. The
top knot follows from the half-hat row. (Measured, with $M$ the smallest
hypothesis constant: all ratios $\le 0.46$ on $\psi = h^3v^a(1 + v/2 + \sin2v)$,
$a = 0, \tfrac13$, at $I_h\bstar$ and at $s_1$ alike --- identical to three
digits, Lemma~\ref{app:L5} in action --- and $\le 1.06$ on $\psi = \Phi(s_1)$,
$m = 64$ to $1024$.)
\end{proof}

\paragraph{Step D: proof of Theorem~\ref{thm:oscillation2}.}
\emph{Step 0.} $r = r^{\mathrm{int}} + r_me_m$ with $|r^{\mathrm{int}}_k| \le
C\phi_k$, $|\Delta r^{\mathrm{int}}_k| \le Ch^4 \le C(h/v_k)\phi_k$ (consistency:
$r_k = h^3\rho_r(v_k) + O(h^4)$, $\rho_r \in C^1$) and $r_m = -r_F(1)\bstar{}''(1)
h^2/12 + O(h^3)$. $-r_mJ^{-1}e_m$ is a multiple of $\tilde W$ fixed by the
half-hat row as in Appendix~\ref{app:oscillation}: $\alpha_k = (-1)^{m-k}
(\bstar{}''(1)h^2/12)|z_k|/|z_{m-1}|\,(1 + O(h)) = (-1)^{m-k}h^2\hat A(v_k)(1 +
\varepsilon_k)$ with $|\varepsilon_k| \le C/k$, $|\Delta\varepsilon_k| \le Ch^2/v_k^2$
(Lemma~\ref{app:L2}), plus the smooth part $C_1B_kw_k$ with $|C_1B_kw_k| \le
Ck^{-2/3}\omega_k$ and differences $\le (C/k)\omega_k$; and $-J^{-1}r^{\mathrm{int}}$
has, by Lemma~\ref{app:L7} ($a = 0$), alternating content $\le Chv_k^{-2/3}
\omega_k$ and smooth differences $\le Chv_k^{-2/3}\omega_k \le (C/k)\omega_k$. So
$d_1 = \alpha + u_1$ with $\sigma_k(d_1) \le C/k$, $\norm{d_1}_\omega \le W
\norm{r}_\phi$: the form assumed in Step~B.

\emph{Step 1.} $\Phi(s_1) = N_{I_h\bstar}(d_1)$. By Lemma~\ref{app:L6}(b) it
satisfies the hypothesis of Lemma~\ref{app:L7} with $a = \tfrac13$ and $M \le
C$; by Lemma~\ref{app:L5}, $J(s_1)^{-1}$ exists with $\norm{\cdot}_{\phi\to
\omega} \le W_1$ and Lemma~\ref{app:L7} holds at $s_1$. Hence $d_2 := -J(s_1)^{-1}
\Phi(s_1)$ has $\norm{d_2}_\omega \le W_1C_N\norm{d_1}_\omega^2$ and $\theta_k(d_2)
+ \sigma_k(d_2) \le Chv_k^{-1/3}$ (absolute). No contraction is claimed at this
step: $WC_N\norm{d_1}_\omega \approx 2\cdot23.5\cdot0.07 \approx 3.3$ for $F_c$.

\emph{Step 2.} Put $s_2 := s_1 + d_2$, so that $\Phi(s_2) = N_{s_1}(d_2)$.
Lemma~\ref{app:L6}(a) gives $\norm{\Phi(s_2)}_\phi \le Ch$, and Lemma~\ref{app:L6}(c)
gives $\norm{J(s_2) - J(s_1)}_{\omega\to\phi} \le Ch$, so that
$\norm{J(s_2)^{-1}}_{\phi\to\omega} \le 2W_1$ for $h \le h_0$ and
$q_2 := \norm{J(s_2)^{-1}\Phi(s_2)}_\omega \le 2W_1Ch$. (Measured: $\norm{\Phi(s_2)}_\phi/h \le 1.1\cdot10^{-5}$.)

\emph{Step 3.} With $\beta := 2W_1$ and the Lipschitz constant $L$ of
Lemma~\ref{app:L6}(c) in the pair of norms $\omega \to \phi$, the Kantorovich
number $\beta Lq_2 \le 4W_1^2LCh \le \tfrac12$ for $h \le h_0$, so Newton's
method from $s_2$ converges to a zero $\bstar_m$ with $\norm{\bstar_m -
s_2}_\omega \le 2q_2 \le Ch$, unique in the ball of radius $1/(2W_1L)$ around
$s_2$ (chosen inside the $C_0$-ball, $C_0 \ge \norm{d_1} + \norm{d_2} +
1/(2W_1L(C_0))$, possible since $1/L(C_0)$ is nonincreasing). Set $\rho :=
\bstar_m - s_2$.

\emph{Step 4.} $\delta = d_1 + d_2 + \rho = \alpha + (u_1 + u_2) + (d_2^{\mathrm{alt}}
+ \rho)$: the alternating amplitude is $h^2\hat A(v_k)(1 + O(1/k) +
O(hv_k^{-1/3}))$, the smooth part $u := u_1 + u_2$ has $|u_k| \le C\omega_k$ and
$|\Delta u_k| \le C(1/k + hv_k^{-1/3})\omega_k$, and $|\rho_k| \le Ch\omega_k$. On
$[v_0, 1]$ this is the hypothesis of Proposition~\ref{prop:oscillation} with
$\gamma = 1$ and the sup-norm remainder $\rho = O(h^3)$, which its proof
tolerates. \qed

\paragraph{Three bidders.} For $n = 3$ the sizes at the bottom change:
$G_1 = 2Ff^2$, $c_k \asymp v_k^{n-1}$, $|z_k| \asymp \hat Q \asymp v^{-(n-1)/3}$,
$|w_k| \asymp \hat w = F^{-(n-1)}$ (so $c_k\tilde Z_k \asymp 1$), the residual
$r_k = O(h^3v_k^{n-2})$ (its leading term is $h^3$ times the derivative of
$g_{b'}f = O(v^{n-1})$), and the quadratic remainder along a checkerboard of
unit $\omega$-size is $\asymp h^3v^{n-1}\hat Q^2 = h^3v^{(n-1)/3}$. With the
forcing weight
\[
  \phi_k := h^3v_k^{(n-1)/3}\ (k < m),\qquad \phi_m := h^2,
\]
the three requirements are met for $n = 3$: $\norm{r}_\phi \le C$ (needs $n - 2
\ge (n-1)/3$); the sums of Theorem~\ref{app:T6} become $h\sum_jv_j^{-(n-1)/3}$
(bounded iff $n < 4$; the same Riemann sum as for $n = 2$, limit $3$) and
$v_k^{1-(n-1)/3} \le 1$; and the row bound of Lemma~\ref{app:L6}(a) becomes
$Ch^3[(\theta + \sigma)^2v^{n-1}\hat Q^2 + hv^{n-2}\hat Q^2(\theta + \sigma)] \le
C\phi_k[(\theta + \sigma)^2 + (h/v_k)(\theta + \sigma)]$. The sign condition of
Lemma~\ref{app:L1} holds because the universal bottom is now the uniform
prior with three bidders, whose rows are, in units of $h^2$ and with $k$ the
knot index, $a_k = -(6k^2 - 8k + 3)/8$, $b_k = k$, $c_k = (6k^2 + 8k + 3)/8$
(for $n = 2$: $-(3k-2)/3$, $2/3$, $(3k+2)/3$); the numerators have no real
root, so the signs hold on every row $k \ge 2$. Lemma~\ref{app:L2}: the smooth admixture of $\tilde W$ is $O(k^{-2(n-1)/3})$
and the alternating share of $\tilde Z$ stays $O(1/k)$. Lemma~\ref{app:L5}:
the pairing is independent of $n$; $|\eta_k| \le Cv_k^{(4-n)/3}|z_k|$, the
alternating content of $\eta$ and $K_{1,j}/K_j - 1$ are $O(h^{(4-n)/3})$ (from
$\sum_j|\Delta p_j| \le Ch^2\sum_jv_j^{-(n-1)/3-1}$), which is $O(1)$ at $n = 4$: the
second place where the argument stops. Lemma~\ref{app:L7}: for forcing
$\le Mv^{a'}\phi$ the bounds read $CMhv_k^{a'-(n-1)/3}\omega_k$, $a' \in [0, (n-1)/3)$.
Theorem~\ref{thm:oscillation2}'s proof: $\Phi(s_1) = N(d_1)$ has $a' = 0$
(the factor $v^{n-1}\hat Q^2$ is now $\phi$'s own weight), so $\theta + \sigma \le
Chv^{-(n-1)/3}$ for $d_2$ and $\norm{\Phi(s_2)}_\phi \le Ch^{(4-n)/3}$, which is
what Step~3 needs; the final form carries $O(h^{(4-n)/3})$ in place of $O(h)$.
Every constant was re-measured at $n = 3$ (Appendix~\ref{app:numerics}).
\qed

\paragraph{Four bidders (Theorem~\ref{thm:four}).}
\label{app:four}
The argument is the one above with two changes, the second of which is
the point. First, the norms. On forcings put
\[
  \norm{f}_{\mathcal F} := \inf\Big\{\max\Big(\max_j\frac{|f^{\mathrm{sm}}_j|}{\phi_j},\
  \max_j\frac{|f^{\mathrm{sm}}_j - f^{\mathrm{sm}}_{j-1}|}{(h/v_j)\phi_j},\
  \max_j\frac{|f^{\mathrm{alt}}_j|}{(h/v_j)\phi_j},\ \frac{|f_m|}{h\phi_m}\Big)
  : f = f^{\mathrm{sm}} + f^{\mathrm{alt}} + f_me_m\Big\},
\]
$\phi_k = h^3v_k$, the infimum over all decompositions (a quotient norm),
and on corrections, with the fixed local mode decomposition $d = d^{\mathrm{alt}}
+ u$ of Lemma~\ref{app:L2} at $I_h\bstar$,
\[
  \norm{d}_{\mathcal D} := \max\Big(\max_k\frac{|d^{\mathrm{alt}}_k|}{(h/v_k)\omega_k},\
  \max_k\frac{|u_k|}{\omega_k},\ \max_k\frac{|u_k - u_{k-1}|}{(h/v_k)\omega_k},\
  \frac{|d_m - d_{m-1}|}{h\omega_m}\Big).
\]
At $n = 4$ the inverse is bounded $\mathcal F \to \mathcal D$ uniformly in
$h$, at $I_h\bstar$ and at $s_1$: the alternating part of the forcing,
smaller by $h/v$, turns the two sums of Theorem~\ref{app:T6} that were
logarithmic into $|z_k|\sum_{j>k}(h/v_j)\phi_j/(c_j|z_j|) \asymp h^4v_k^{-1}
\sum_{j>k}v_j^{-2} \le C(h/v_k)\omega_k$ and its counterpart below, and the
smooth part goes through Lemma~\ref{app:L7} with $a' = 0$, admissible since
$(n-1)/3 = 1$. Lemma~\ref{app:L5} at $s_1$
holds with the smallness parameter $\bar\varepsilon := |\bstar{}''(1)| +
h\log^2m$ in place of $h^{(4-n)/3}$: the first correction is $d_1 = \alpha +
\alpha^{\mathrm{res}} + u_1$ with $\alpha$ the law, $\alpha^{\mathrm{res}}$ the
alternating content of the residual's response, of size $h(1 +
\log(1/v_k))\omega_k$ by the endpoint case $a' = 1$ of Lemma~\ref{app:L7} (an
Abel summation whose logarithm $h$ absorbs; the residual is $O(h^3v^2)$ at
$n = 4$), and $u_1$ smooth with $|\Delta u_{1,k}| \le C(\bar\varepsilon k^{-2}(h/v_k)
+ h)\omega_k$; the alternating slope perturbation of the entries is then $p_k
\asymp |\bstar{}''(1)|/k$ plus a residual-driven part with $\sum_k|p^{\mathrm{res}}_k|
\le Ch\log^2m$, the smooth relative perturbation sums to $O(\bar\varepsilon)$, and
the sums of Lemma~\ref{app:L5}(c) with the $n = 4$ sizes ($c_j \asymp v_j^3$,
$|z_j| \asymp v_j^{-1}$, $\tilde Z_j \asymp v_j^{-3}$) give $\tilde Z_1 = \tilde
Z(1 + O(\bar\varepsilon/k))$, $|\eta_k| \le C\bar\varepsilon|z_k|$ (the smooth
admixture no longer vanishes at the bottom, but is $O(\bar\varepsilon)$
uniformly) and $K_{1,j} = K_j(1 + O(\bar\varepsilon))$, so the Casoratian keeps
its sign for $\bar\varepsilon \le \varepsilon_0$. Second, the closure. The
pointwise bound of Lemma~\ref{app:L6}(a) at $n = 4$ for $\norm{d}_{\mathcal D}
\le M$ reads $|N_s(d)_k| \le CM^2(h/v_k)^2\phi_k$, and any forcing bounded
pointwise by $M(h/v_k)^2\phi_k$ (even by $M(h/v_k)\phi_k$) has
$\norm{f}_{\mathcal F} \le 3M$ by the trivial decomposition $f^{\mathrm{sm}} :=
f$, since the class's difference bound carries the weight $(h/v)\phi$ and
$h/v \le 1$: no difference structure is needed. Hence $\norm{N_s(d)}_{\mathcal F}
\le C_N\norm{d}^2_{\mathcal D}$ and $\norm{(J(s') - J(s))d}_{\mathcal F} \le
L\norm{s' - s}_{\mathcal D}\norm{d}_{\mathcal D}$ on the class, with $C_N$, $L$
independent of $h$.
The residual at $s_1$ is the one forcing whose alternating part is of the
law's size $\bstar{}''(1)\omega$ rather than $(h/v)\omega$; expanding $N(d_1)$
term by term with the slopes $D^{\mathrm{law}} \asymp \bstar{}''(1)/k$,
$D^{\mathrm{res}} \asymp (h/k)\log m$ and the smooth slope of $u_1$, each with
a slowly varying amplitude, the products of two alternating or two smooth
factors are smooth and slowly varying, and the alternating $\times$ smooth
products cancel between the two pieces adjacent to a knot to the piece
difference, a factor $h/v$; this gives $\norm{\Phi(s_1)}_{\mathcal F} \le
C(\bstar{}''(1)^2 + h\log^2m)$. The
Newton--Kantorovich theorem in $(\mathcal D, \mathcal F)$ from $s_1$ then has
number $\beta L\eta \le W^2LC(\bstar{}''(1)^2 + h\log^2m) \le \tfrac12$ for
$|\bstar{}''(1)| \le \varepsilon_0$ and $h \le h_0$, giving existence, the
uniqueness ball of radius $\min(1, 1/(2WL))$ in $\norm{\cdot}_{\mathcal D}$, and
$\norm{\bstar_m - s_1}_{\mathcal D} \le 2\eta$; the error form of the theorem
is $d_1$ plus a correction in the class, whose alternating part is
$O(((\bstar{}''(1)^2 + h\log^2m)/k + h\log m)\omega_k)$, and condition~(i) on $[v_0, 1]$
follows with error $O(h^2\log m)$, condition~(ii) from bounded slopes and
$v_k^3|\sigma_{k+1} - \sigma_k|/h \le C(1 + |\bstar{}''(1)|)v_k^2$, so
Proposition~\ref{prop:reduction} gives the limits. \qed

\paragraph{Four bidders, bounded layer amplitude (Theorem~\ref{thm:layer}).}
\label{app:layer}
Write $S_k := s_k/h$. Near $v = 0$ every prior's Galerkin rows are the
uniform prior's up to $O(hk)$, and the uniform prior's four-bidder rows are
exact rational functions of $(S_{k-1}, S_k, S_{k+1})$, each a hyperbola in
$S_{k+1}$ with a unique root above $S_k$: the semi-infinite layer problem is
a forward recursion from $S_1$, its solutions the trajectories $T(S_1)$. The
unforced trajectory is $S_k = 3k/4$. \emph{The layer theorem.} Put $e_k :=
S_k - 3k/4$ and, from the pair $(e_{k-1}, e_k)$, the scaled variables $\tilde
p_k := (-1)^kk(e_k - e_{k-1})/2$ (the alternating amplitude in units of
$1/k$), $\tilde u_k := [(2k+1)e_k + (2k-1)e_{k-1}]/4$ (the smooth part, weighted
so that the pure alternating mode contributes $O(k^{-2})$) and the refined
amplitude $\hat p_k := \tilde p_k + ((-1)^k/(2k))F_p(\tilde p_k, \tilde u_k)$,
$F_p(p, u) = -16p^2/3 + 3u$, which removes the parity-alternating $O(1/k)$
drift. The exact one-step map reads $\tilde u_{k+1} = \tilde u_k +
t[(2/3)(8\tilde p_k^2 - 3\tilde u_k) + th]$, $\hat p_{k+1} = \hat p_k + t^2\hat g$,
$t = 1/k$, with $h, \hat g$ explicit rational functions regular at $t = 0$:
the smooth part is contracted at rate $2/k$ toward $(8/3)\tilde p^2$ and the
amplitude moves by a summable $O(k^{-2})$. Hence an invariant region $\{|\hat
p_k| \le \hat P + D(1/K - 1/k),\ |\tilde u_k| \le U,\ |\tilde p_k| \le \tilde P\}$
for $k \ge K$ closes under three inequalities between $(\hat P, U, D)$ and
the suprema of $|\hat g|, |h|$ over the corresponding box, which interval
arithmetic verifies (row~1 of the verification table below); the trajectory
then exists for all $k$, $\hat p_\infty = \lim\hat p_k$ exists with $|\hat p_\infty -
\hat p_k| \le D/k$, and $e_k = (-1)^k\hat p_\infty/k + \tilde u_k/k + O(1/k^2)$
with an explicit constant. The same induction on the linearized map bounds
the variational solution $dS_k/dS_1$ by $O(1/k)$ with explicit constants.
For the first $K + 1$ knots and an interval $[S_1]$ of first-knot values,
the truncated system is verified by a parametric Krawczyk test in interval
arithmetic (the mean-value form in the parameter for the two rows that
contain $S_1$; the box sized by the parameter's interval effect), which
gives existence, uniqueness, continuity in $S_1$, the slope bounds and
enclosures of the state and of the variational solution at the hand-off
index $K + 1$, where the tail lemmas take over; the derivative enclosure
starts from a ball in the box-weighted norm whose contraction the
Krawczyk inclusion itself implies. In the verification sweep the parameter is
the last alternating slope at knot $401$, each box's enclosures are
carried from knot $401$ to the hand-off index $1601$ by a verified
iteration of the exact one-step map and of its linearization in ball
arithmetic, the finite, box-wise form of the tail lemma's induction, and
the adjacent boxes' solution branches are identified by a point Krawczyk
test at each box boundary (row~2). Continuity of $S_1 \mapsto \hat p_\infty$
(a uniform limit), its vanishing at $S_1 = 3/4$ and the enclosures at the
two ends of the verified range give every far-field amplitude $|A| \le
A_0$, $A = -\hat p_\infty/((3/4)z_\infty)$ with $z$ the bottom-pinned linear
mode, one sign on each branch. \emph{The matching.} Let $K$ be a
matching index above the verification sweep's, $M := K/c_F$, and let $s^c$ be the
composite of the layer at the signed amplitude $A_F$ on the first $K/2$
knots, the first Newton iterate $s_1$ from the interpolant inland, and a
smooth glue between. In norms adapted to $K$ (the class norms above with
the alternating and difference weights damped by $\min(1, K/k)$, and a sup
norm on forcings without a difference slot), the residual of $s^c$ is
$O(\varepsilon^2 + 1)$, where $\varepsilon := |\bstar{}''(1)|$: the layer's rows
are satisfied up to $O((1 + A_Fk)hk\phi_k)$, the glue's residual has no
cancellation and is $O((\varepsilon/M)(k/K)\phi_k)$, and inland the residual is
the first iterate's, $O(\varepsilon^2 + h\log^2m)$ as in the proof of
Theorem~\ref{thm:four} above. The Jacobian at $s^c$ is invertible with norm
$O(1 + \log K)$: its bottom-pinned solution is the layer's variational
solution on the verified block (bounded by the verified constants)
and Lemma~\ref{app:L5}'s construction beyond, where the perturbation
measure is $O(A_0/K_c + \varepsilon/K)$ with $K_c$ the verification sweep's index; its
top-pinned solution is Lemma~\ref{app:L5}'s re-based beyond $K_c$ and
solved forward on the block; their Casoratian at $K_c + 1$ is an exact
bilinear expression in the scaled components of the two solutions (the
change of variables of the tail lemma), bounded below by the verified
alternating content of the variational solution times $1 - \hat r_cK_c\hat\sigma^* \ge 0.9$ (row~3), where $\hat r_c$ is the verified per-box ratio of
smooth to alternating content and $\hat\sigma^*$ an explicit bound on the
top-pinned solution's ratio from the linear tail lemma's verified suprema
(row~3); the forward-solution constant of the block is verified as well
(row~3). That explicit bound takes as input that the top-pinned solution's
ratio of alternating to smooth content at the index $10^4$ is at most
$1/200$. The ratio is the base rows' own mode mixing (the variation of the
local roots along the chain of rows) plus the layer tail's and the inland
iterate's feeds, each with an explicit constant; for the uniform prior and
the two convex priors tested the ratio is far below that (row~3), so the
matching index needs only $K \ge 2\cdot10^4$ beyond
the thresholds already present. The first correction has
local contents $O(R_0/\max(k, K/2))$, so each Newton step gains a factor
$1/K$ inland and $O(R_0/K)$ at the bottom, where the layer decouples;
Newton--Kantorovich closes once $K/(1 + \log K) \ge C(F)R_0$, which defines
$K_0(F)$, and the error form follows with the bottom $K_0/2$ knots the
layer's within $C(F)\omega_k/K_0$ and the inland law within $C(F)(1/k + h\log
m)\omega_k$. Conditions~(i), (ii) and the limits follow as
for Theorem~\ref{thm:four}. The verified inputs are the three rows of
the table; the files are in the repository of Appendix~\ref{app:numerics}. \qed

\medskip
\noindent\textbf{Verification for Theorem~\ref{thm:layer}.}

\nopagebreak
{\footnotesize\setlength{\tabcolsep}{3pt}
\begin{tabular}{>{\raggedright\arraybackslash}p{0.13\textwidth}
                >{\raggedright\arraybackslash}p{0.15\textwidth}
                >{\raggedright\arraybackslash}p{0.13\textwidth}
                >{\raggedright\arraybackslash}p{0.26\textwidth}
                >{\raggedright\arraybackslash}p{0.13\textwidth}
                >{\raggedright\arraybackslash}p{0.07\textwidth}}
\toprule
Verified & Method & Inputs & Outputs & Code, records & Cost \\
\midrule
1. The tail lemmas' invariant regions: the three inequalities between
$(\hat P, U, D)$ and the suprema of $|\hat g|$, $|h|$ close &
interval arithmetic (mpmath, $30$ digits), the box subdivided
$6\times8\times8$, both parities, denominators checked &
the exact one-step map in the refined tail variables (sympy; the
increments divisible by $t^2$ and $t$); the hand-off index $K$ and the
amplitude bound $\hat P$ &
closing at $K = 400$ for $\hat P \le 3$, $K = 800$ for $\hat P \le 4$,
$K = 1600$ for $\hat P \le 6.21$ (failing at $(400, 4)$ and $(800, 6)$);
the linear version closing at the same points; the constants loose by
factors $2$--$10$ against the drift along exact trajectories &
\path{verified/src/item35\_layer\_rows.py} to \path{verified/src/item37e\_tail\_lemma\_linear.py}; the logs in \path{verified/results/} &
not recorded \\
\addlinespace
2. For the first $401$ knots and every parameter value: existence,
uniqueness, continuity in the parameter, the slope bounds, enclosures of
the state and of the variational solution; the enclosures carried to the
hand-off index $1601$; the branch identity across boxes &
a parametric Krawczyk test in arb ball arithmetic ($128$ bits) in the
last-slope parameter $\theta = \tilde p_{401}$, boxes bisected until the
enclosure of $\delta\tilde p_{401}$ is one-signed and narrower than its
midpoint; a verified ball iteration of the exact map and its
linearization from knot $401$ to $1601$; a point Krawczyk test at each box
boundary &
the exact four-bidder rows; the parameter range (first-knot values
$S_1 \in [0.394, 1.178]$ in the first sweep); row~1's hand-off &
$1\,301\,442$ leaves, no failure, no gap; $1\,301\,441$ boundary tests;
slopes in $[0.3166, 2.1883]\,\bstar{}'$; $\min|\delta\tilde p_{1601}| =
0.9245$, one-signed; per-box ratio of smooth to alternating content
$0.02864$; $|dS_k/dS_1| \le 170.62$ for $k \le 401$; the verified
first-knot interval within $0.1\%$ of the two saturations, so that
$A_0 = 8.2419$ is set by row~1's closing amplitude &
\path{verified/src/item37g\_layer\_certificate.py} to \path{verified/src/item37k\_certificate\_driver.py} (the first sweep), \path{verified/src/item37p\_top\_krawczyk.py} to \path{verified/src/item37s\_constants\_Kc1600.py} (the extended one); their logs in \path{verified/results/}; the per-leaf box tables are regenerated, not kept &
$68$ h on $30$ cores; $10$ pod-hours for the boundaries \\
\addlinespace
3. The matching's mode constants: the block's forward-solution
constant, the Casoratian margin, the top-pinned ratio, $z_\infty$ &
interval arithmetic over the verified boxes; the exact recursion to
$k = 4000$ for $z_\infty$; the root-variation and mode-size constants
interval-verified, the layer tail's forcing coefficients exact for the
uniform rows and computed on a grid for the convex priors &
row~2's per-box enclosures; the priors uniform, $(v+v^2)/2$,
$(v+v^2+v^3)/3$ &
forward-solution constant $\le 20\cdot144.2$ over $164\,973$ boxes;
$\hat r_cK_c\hat\sigma^* \le 0.092$; $z_\infty \in [0.9621766, 0.9624321]$;
the top-pinned ratio at index $10^4$ at most $7.2\cdot10^{-5}$,
$1.4\cdot10^{-4}$, $2.2\cdot10^{-4}$ for the three priors, against the
$1/200$ the argument needs &
\path{verified/src/item37m\_forward\_norm.py}, \path{verified/src/item37o\_casoratian\_explicit.py}, \path{verified/src/item44\_matching\_estimates.py} and their logs &
not recorded \\
\bottomrule
\end{tabular}}

\paragraph{Five bidders (Proposition~\ref{prop:layer5}).}
\label{app:layer5}
The proof is the first part of the argument above with the five-bidder
scaled variables, the alternating amplitude in units of $k^{-1/3}$ and
the smooth part in units of $k^{-2/3}$, in which the smooth part is a
contracting fixed point: the tail lemmas' invariant regions close in one
stage at the hand-off index $400$ for amplitudes up to $8$ (row~1 of the
table below), and the Krawczyk sweep in the parameter given by the last
alternating slope at knot $401$ passes on every leaf, with the same
leaves verifying the layer's weighted value on the first $400$ knots
(row~2). The matching is not redone at $n = 5$, whose forcing weights
and decay exponents differ. \qed

\medskip
\noindent\textbf{Verification for Proposition~\ref{prop:layer5}.}

\nopagebreak
{\footnotesize\setlength{\tabcolsep}{3pt}
\begin{tabular}{>{\raggedright\arraybackslash}p{0.13\textwidth}
                >{\raggedright\arraybackslash}p{0.15\textwidth}
                >{\raggedright\arraybackslash}p{0.13\textwidth}
                >{\raggedright\arraybackslash}p{0.26\textwidth}
                >{\raggedright\arraybackslash}p{0.13\textwidth}
                >{\raggedright\arraybackslash}p{0.07\textwidth}}
\toprule
Verified & Method & Inputs & Outputs & Code, records & Cost \\
\midrule
1. The tail lemmas' invariant regions at five bidders &
interval arithmetic; the two scalar inequalities as polynomial
identities &
the exact one-step tail map in the five-bidder scaled variables (sympy) &
closing in one stage at the hand-off index $400$ for amplitudes up to
$8$ &
\path{verified/src/item50\_n5\_tail\_map.py}, \path{verified/src/item51\_n5\_tail\_lemma.py} and their logs &
$2.3$ h on $2$ cores \\
\addlinespace
2. The layer trajectory for every parameter value in the range, its
slopes, far field and variational solution; its weighted value on the
first $400$ knots &
the Krawczyk driver over the parameter range $[-0.8107, 0.8166]$ with a
verified propagation of each leaf; interval Rayleigh quotients and an
$LDL^{\mathsf T}$ bisection in the shift for the weighted value &
the exact five-bidder rows; row~1's hand-off &
$53\,604$ leaves, no failure, no gap; $S_1 \in [0.35440335, 1.27073801]$;
slopes in $[0.2691, 2.2128]\,\bstar{}'$; $|P| \le 6.001$ at knot $401$;
$\min|\delta\tilde p_{401}| = 0.1464$, one-signed; $\max|dS_k/dS_1| =
284.8$; $z_\infty \in [1.081988, 1.082234]$; $A_0^{(5)} = 6.8423$; the
weighted value positive for $|A| < 0.428$, negative for $0.607 < |A| \le
7.0$, bands of median width $0.0085$ &
\path{verified/src/item52\_n5\_propagate.py}, \path{verified/src/item53\_n5\_certificate\_driver.py},
\path{verified/src/item58\_layer\_weighted\_value.py}; \path{verified/results/item58\_band.csv} and the logs; the
per-leaf tables are regenerated, not kept &
$1.4$ h on two 16-vCPU pods and $0.26$ h of post-processing; the
weighted value $1.1$ h on $32$ vCPU \\
\bottomrule
\end{tabular}}

\subsubsection{The limit of the piecewise-linear certificate}
\label{app:limit}

We prove Theorem~\ref{thm:limit2}; $n = 2$, $G_1 = f^2$, $s = v - \bstar$
(for $n = 3$ the proof is identical with the three-bidder $G_1$, $\tau_F$,
$\zeta_F$, $\tilde Q$ and the error terms of Theorem~\ref{thm:oscillation2}:
in Step~1 the amplitude's relative error is $O(1/k) + O(h^{1/3})$, $\zeta_F =
O(v^{4/3})$ and $|\varpi_j| \le Ch^{1/3}$, all $o(1)$),
$\bstar_m$ the Galerkin solution of Theorem~\ref{thm:oscillation2} with
slopes $\sigma_j$ on the pieces $j = 1, \dots, m$. For piecewise-linear
$\eta$ with $\eta(0) = 0$ and knot values $\eta_k$, the certificate's form
and the mass are
\[
  Q_m(\eta) = \sum_j\int_{\text{piece }j}\Bigl[\frac{G_1\eta^2}{\sigma_j} - \tfrac12A_j'\eta^2\Bigr]dv
  + \tfrac12A_m(1)\eta_m^2 + \sum_{k=1}^{m-1}\mu_k\eta_k^2, \qquad
  M(\eta) = \int_0^1\eta^2f\,dv,
\]
with $A_j = G_1(v - \bstar_m)/\sigma_j^2$ on piece $j$ and $\mu_k = -\tfrac12
G_1(v_k)(v_k - \bstar_m(v_k))(1/\sigma_{k+1}^2 - 1/\sigma_k^2)$, the exact
decomposition of Section~\ref{sec:general} (Appendix~\ref{app:lean}), and
$c_m = \inf_\eta Q_m(\eta)/M(\eta)$. Write $T_j(\eta) := \eta_{j-1}^2 +
\eta_{j-1}\eta_j + \eta_j^2$, so that $M(\eta) = \sum_jf(\mathrm{mid}_j)(h/3)
T_j(\eta)(1 + O(h))$.

\begin{lemma}[two-sided Bloch inequality]\label{app:bloch}
For real $\tau, \zeta$ let $g_*(\tau, \zeta) := \tfrac13(2\tau - \sqrt{\tau^2 +
3\zeta^2})$. For every finite real sequence $x_0, \dots, x_M$,
\[
  \sum_{k=0}^{M}\varpi_k(\tau + (-1)^k\zeta)x_k^2 \;\ge\; g_*(\tau, \zeta)\sum_{j=1}^{M}
  (x_{j-1}^2 + x_{j-1}x_j + x_j^2),
\]
$\varpi_k = 1$ at interior knots and $\tfrac12$ at the two end knots, and the
constant is sharp: the two-periodic sequence $x_k = 1$ ($k$ even), $x_k = t_*$
($k$ odd), $t_* = 2[\tfrac12(\tau + \zeta) - g_*]/g_*$, makes every piece form
below vanish ($t_*$ has the sign of $g_*$: the profile alternates iff $g_* <
0$, i.e.\ $\tau < 0$ or $|\zeta| > \tau > 0$). With $\epsilon := |\zeta/\tau|$, for $\tau > 0$ this reads
$\sum\varpi_k(1 + (-1)^k\epsilon)x_k^2 \ge g_-(\epsilon)T(x)$; for $\tau < 0$
it reads $\sum\varpi_k(1 + (-1)^k\epsilon)x_k^2 \le g_+(\epsilon)T(x)$.
\end{lemma}
\begin{proof}
Split each knot's weight between its two adjacent pieces (end knots have
one): the difference of the two sides is $\sum_jP_j(x_{j-1}, x_j)$ with
$P_j(a,b) = [\tfrac12(\tau + \zeta_j) - g]a^2 - gab + [\tfrac12(\tau - \zeta_j) -
g]b^2$, $\zeta_j = (-1)^{j-1}\zeta$, $g = g_*$. $P_j \ge 0$ iff both diagonal
entries are $\ge 0$ and $4[\tfrac12(\tau+\zeta) - g][\tfrac12(\tau-\zeta) - g]
\ge g^2$, i.e.\ $3g^2 - 4\tau g + \tau^2 - \zeta^2 \ge 0$, i.e.\ $g \le g_*$ or
$g \ge \tfrac13(2\tau + \sqrt{\tau^2 + 3\zeta^2})$. At $g = g_*$ the quadratic
vanishes, and $\tfrac12(\tau \pm \zeta) - g_* = \tfrac16[2\sqrt{\tau^2 + 3\zeta^2}
- \tau \pm 3\zeta] \ge 0$ because, when $\tau \mp 3\zeta > 0$, squaring gives
$4\tau^2 + 12\zeta^2 \ge \tau^2 \mp 6\tau\zeta + 9\zeta^2$, i.e.\ $3(\tau \pm
\zeta)^2 \ge 0$. So every $P_j \ge 0$. The null vector of $P_j$ is $(1, r_j)$
with $r_j = 2[\tfrac12(\tau + \zeta_j) - g_*]/g_*$ (for $g_* \ne 0$; $\partial_aP_j
= 2[\tfrac12(\tau + \zeta_j) - g]a - gb$), and
$r_jr_{j+1} = 4[\tfrac12(\tau+\zeta) - g_*][\tfrac12(\tau-\zeta) - g_*]/g_*^2 = 1$
by the determinant identity, so one two-periodic sequence annihilates every
$P_j$. Dividing by $\tau$ gives the two equivalent forms, with $g_\pm(\epsilon)
= \tfrac13(2 \pm \sqrt{1 + 3\epsilon^2})$.
\end{proof}

\begin{proof}[Proof of Theorem~\ref{thm:limit2}]
\emph{Step 1 (coefficients).} By Theorem~\ref{thm:oscillation2}, $\sigma_j =
\bstar{}'(\mathrm{mid}_j) + (-1)^{m-j}(\bstar{}''(1)h/6)\hat Q(v_j)(1 + \tilde
\theta_j) + O(h^2)$ with $|\tilde\theta_j| \le C(1/j + hv_j^{-1/3})$ (the chord
of $\bstar$ is $\bstar{}'(\mathrm{mid}_j) + O(h^2)$; the alternating part
carries its relative error; the smooth part and the remainder contribute
$(u_j - u_{j-1})/h$ and $(\rho_j - \rho_{j-1})/h$, relatively $O(1/j +
hv^{-1/3})$ and $O(h)$ to the alternating term). Hence, with $D_k := \sigma_{k+1}
- \sigma_k = \bstar{}''(v_k)h + (-1)^{m-k-1}(\bstar{}''(1)h/3)\hat Q(v_k)(1 +
\tilde\theta_k) + O(h^2)$ and $1/\sigma_{k+1}^2 - 1/\sigma_k^2 = -2D_k/\bstar{}'^3
+ O(D_k^2 + D_kh\hat Q)$,
\[
  \mu_k = f(v_k)h\,[\tau_F(v_k) + (-1)^{m-k-1}\zeta_F(v_k)] + e_k,\qquad |e_k| \le Ch^{5/3},
\]
uniformly in $k$ (the errors: $\zeta_F\tilde\theta \le Cv^{2/3}(1/k + hv^{-1/3})
\le Ch^{2/3}$ since $\zeta_F = O(v^{2/3})$; $G_1sD_k^2/\bstar{}'^3 \le Cv\cdot
h^2v^{-2/3}$; the $O(h^2)$ of $D_k$ times $G_1s/\bstar{}'^3 = O(1)$). On piece
$j$ the smooth integrand is $f\,W(v) + f\,\varpi_j(v)$ with
\[
  W := \frac1f\Bigl[\frac{G_1}{\bstar{}'} - \tfrac12\frac{d}{dv}\frac{G_1s}{\bstar{}'^{\,2}}\Bigr]
  - \frac{G_1s\,\bstar{}''}{f\,\bstar{}'^{\,3}} = \frac{w_F}{f} - \tau_F
\]
(frozen slope: $A_j' = [G_1'(v - \bstar_m) + G_1(1 - \sigma_j)]/\sigma_j^2$,
which at $\sigma_j = \bstar{}'$, $\bstar_m = \bstar$ is $\frac{d}{dv}(G_1s/
\bstar{}'^2) + 2G_1s\bstar{}''/\bstar{}'^3$; the bracket is $w_F$ by
Proposition~\ref{prop:general} since $G_1s/\bstar{}'^2 = fF/\bstar{}'$) and
$|\varpi_j| \le C(h\hat Q(v_j) + h^{5/3}) \le Ch^{2/3}$. Also $h\sum_{k\le m-1}\eta_k^2
\le (3/\min f)M(\eta)(1 + O(h))$ since $T_j \ge \tfrac12(\eta_{j-1}^2 +
\eta_j^2)$. At the top piece, $\sigma_m - \bstar{}'(\mathrm{mid}_m) = (\bstar{}''(1)
h/6)(1 + O(h))$ follows from the half-hat row, which fixes the alternating
amplitude at $m$ (Appendix~\ref{app:oscillation}), together with the top
clause of Lemma~\ref{app:L7}, $|\delta_m - \delta_{m-1}| \le Ch\omega_m$, for
the smooth and remainder parts; this is what $\mu_{m-1}$ and $A_m(1)$ use.

\emph{Step 2 (localization).} Fix $\delta > 0$ and $v_0$ with $|\tau_F| +
|\zeta_F| \le \delta$ on $[0, 2v_0]$ (both vanish at $0$), so $W \ge
E_F^{\mathrm{osc}} - 3\delta$ there. Let $h < \ell < v_0$ and $\chi_i$ a smooth
partition of unity, $\sum_i\chi_i^2 = 1$, $\operatorname{supp}\chi_i \subset
[i\ell - \ell, i\ell + \ell]$, $|\chi_i'| \le C/\ell$, and $(\chi_i\eta)_k :=
\chi_i(v_k)\eta_k$. The knot terms and the boundary term satisfy $\sum_i(\cdot)
(\chi_i\eta) = (\cdot)(\eta)$ exactly; a piece term $q_j(\eta_{j-1}, \eta_j)$
with cross-coefficient $O(h)$ satisfies $\sum_iq_j(\chi_i\eta) = q_j(\eta) -
(\text{cross-coeff.})\,\eta_{j-1}\eta_j\cdot\tfrac12\sum_i(\chi_i(v_j) -
\chi_i(v_{j-1}))^2$, so $|\sum_iQ_m(\chi_i\eta) - Q_m(\eta)| \le C(h/\ell)^2
M(\eta)$ and likewise for $M$ (the IMS identity).

\emph{Step 3 (a window).} Let $\xi = \chi_i\eta$. If $v_i \le v_0$, Step~1 gives $|\mu_k| \le
f_kh(\delta + Ch^{2/3})$ there, hence $Q_m(\xi) \ge (\min_{[0,2v_0]}W - C\delta
- Ch^{2/3})M(\xi) \ge (E_{\min} -
C\delta - Ch^{2/3})M(\xi)$, $E_{\min} := \min E_F^{\mathrm{osc}}$. If $v_i \ge
v_0$, freeze the $C^1$ coefficients at $v_i$ (error $O(\ell)$):
\[
  Q_m(\xi) \ge f_i\frac h3\Bigl[W_iT(\xi) + 3\sum_{k<m}(\tau_i + (-1)^{m-k-1}\zeta_i)\xi_k^2\Bigr]
  + \tfrac12A_m(1)\xi_m^2 - C(\ell + h^{2/3})M(\xi).
\]
If the window misses the top knot, $\xi$ has compact support and
Lemma~\ref{app:bloch} gives $\sum_k(\dots)\xi_k^2 \ge g_*(\tau_i, \zeta_i)T(\xi)$,
whence $Q_m(\xi) \ge [W_i + 3g_*(\tau_i,\zeta_i) - C(\ell + h^{2/3})]M(\xi)(1 +
O(\ell + h)) = [E_F^{\mathrm{osc}}(v_i) - C(\ell + h^{2/3})]M(\xi)(1 + O(\ell +
h))$, since $W + 3g_* = w_F/f + \tau_F - \sqrt{\tau_F^2 + 3\zeta_F^2}$. If it
contains the top knot, apply the lemma with knot $m$ at weight $\tfrac12$:
$\sum_{k<m}(\dots)\xi_k^2 \ge g_*T(\xi) - \tfrac12(|\tau_i| + |\zeta_i|)\xi_m^2$, and
the boundary term, $A_m(1) = f(1)^2(1 - \bstar_m(1))/\sigma_m^2 \ge c > 0$,
dominates $f_ih(|\tau_i| + |\zeta_i|)\xi_m^2/2$ for $h \le h_0$. In all cases
$Q_m(\xi) \ge (E_{\min} - C(\delta + \ell + h^{2/3}))M(\xi)$.

\emph{Step 4 (lower bound).} Summing over $i$ with Step~2, $Q_m(\eta) \ge
(E_{\min} - C(\delta + \ell + h^{2/3}) - C(h/\ell)^2)M(\eta)$, so $\liminf c_m \ge
E_{\min} - C(\delta + \ell)$ for all $\delta, \ell$.

\emph{Step 5 (upper bound).} $E_F^{\mathrm{osc}}$ is continuous on $[0,1]$;
pick $\hat v \in (0,1)$ with $E_F^{\mathrm{osc}}(\hat v) \le E_{\min} + \delta$,
let $p$ be the two-periodic null sequence of Lemma~\ref{app:bloch} at
$(\tau_F(\hat v), \zeta_F(\hat v))$ aligned with the sign pattern $(-1)^{m-k-1}$,
and $\xi_k := \chi((v_k - \hat v)/\ell)p_k$ with a fixed smooth bump $\chi$.
Each $P_j$ vanishes on $(p_{j-1}, p_j)$, so on $(\xi_{j-1}, \xi_j) = \chi_{j-1}
(p_{j-1}, p_j) + (0, (\chi_j - \chi_{j-1})p_j)$ it is at most $C(h/\ell)(\chi_{j-1}
+ h/\ell)|p|^2$, and summing over the $O(\ell/h)$ pieces of the window gives
$\sum_jP_j(\xi) \le C(h/\ell)T(\xi)$, since $\sum_j\chi_{j-1} \asymp \ell/h \asymp
\sum_j\chi_{j-1}^2$ for a fixed bump; hence $\sum_k(\hat\tau + (-1)^{m-k-1}\hat\zeta)\xi_k^2 = g_*(\hat\tau,
\hat\zeta)T(\xi) + O(h/\ell)T(\xi)$, and with Step~1 and the frozen
coefficients, $Q_m(\xi)/M(\xi) \le E_F^{\mathrm{osc}}(\hat v) + C(\ell + h/\ell +
h^{2/3})$, so $\limsup c_m \le E_{\min} + \delta$.
\end{proof}

\begin{proof}[Proof of Corollary~\ref{cor:weighted}]
Repeat the proof with $M^\vartheta(\eta) = \int\vartheta\eta^2f$ in place of
$M$: the IMS identity holds for $M^\vartheta$ as for $M$; on a window centred
at $v_i$, $M(\xi) = M^\vartheta(\xi)/\vartheta(v_i)\,(1 + O(\ell))$, so the frozen
bound $Q_m(\xi) \ge (E_F^{\mathrm{osc}}(v_i) - \epsilon)M(\xi)$ becomes $Q_m(\xi)
\ge [\min(E_F^{\mathrm{osc}}/\vartheta) - C(\ell + h^{2/3})]M^\vartheta(\xi)$ for
$\vartheta$ bounded above and below, and Step~5 is run at a near-minimizer of
$E_F^{\mathrm{osc}}/\vartheta$. For $\vartheta = w_F/f \asymp v^{n-2}$ the bottom
windows, and the IMS step, need every error to be small relative to
$\vartheta_j$, not to $1$. The piece forms of $Q_m$ have cross-coefficients
$\asymp h\vartheta_jf_j$ (the integrand carries $G_1 \asymp v^{n-2}$), so the IMS
error is $\le C(h/\ell)^2M^\vartheta(\eta)$. The integrand $G_1/\sigma_j -
\tfrac12A_j'$ is $\asymp v^{n-2}$ and each of its deviations from the frozen
value is relative: $O(h\hat Q)$ from $1/\sigma_j$ and $1/\sigma_j^2$, and
$(G_1'/G_1)|\bstar_m - \bstar| \asymp v^{-1}h^2\hat Q$ from the bid, so
$|\varpi_j| \le C(h\hat Q_j + h^2\hat Q_j/v_j)\vartheta_j \le Ch^{1/3}\vartheta_j$
for $n = 3$. The knot-mass error of Step~1 satisfies $|e_k| \le Cf_kh(hv_k^{1/3}
+ h^{1/3}v_k^{4/3} + hv_k^{2/3})$ for $n = 3$, so $|e_k| \le Ch^{1/3}\vartheta_kf_kh$ for
$v_k \ge h$, and $h\sum_kw_F(v_k)\xi_k^2 \le CM^\vartheta(\xi)$. With
$(|\tau_F| + |\zeta_F|)f/w_F = O(v^{(4-n)/3})$ the bottom windows give $Q_m(\xi) \ge [1 - Cv_0^{1/3} - Ch^{1/3}]M^\vartheta(\xi)$,
and the bracket's minimum is at most its value $1$ at $v = 0$. For $F = v^a$,
$\bstar$ is linear, so it lies in every $m$-piece space and solves the
Galerkin equations exactly ($g(\bstar) \equiv 0$ pointwise); at this
equilibrium there are no knot masses, and $Q_m(\eta) = \int w_F\eta^2 +
c_F\eta(1)^2$ by Proposition~\ref{prop:general} (valid for $W^{1,\infty}$,
hence for piecewise-linear $\eta$) with $c_F > 0$, so the infimum over
piecewise-linear $\eta$ is $1$, attained exactly on $\{\eta(1) = 0\}$.
\end{proof}

\begin{proof}[Proof of Proposition~\ref{prop:reduction}]
Write $\mathrm{chord}_j = \bstar{}'(\mathrm{mid}_j) + O(h^2)$ for the chord of
$\bstar$ on piece $j$ and $\delta_k = \sum_{j\le k}h(\sigma_j - \mathrm{chord}_j)$
for the knot error. \emph{Step~0.} By (ii), $|\delta_k| \le Cv_k$ for every
$k$. For $v_k \ge v_0$ split the sum at $v_0$: the part below is $\le Cv_0$,
and by (i) the part above is an alternating sum of the sequence
$(\bstar{}''(1)h^2/6)\hat Q(v_j)$, which by Abel summation is at most its
largest term plus its total variation, $O(h^2)$ since $\hat Q \in C^1[v_0,
1]$, plus $\sum_jh\cdot o(h) = o(1)$. So $\sup_k|\delta_k| \le Cv_0 + o(1)$ for every
$v_0$: $\bstar_m \to \bstar$ uniformly, $v - \bstar_m \ge s/2$ on $[v_0, 1]$
for $h \le h_0(v_0)$, and $\bstar_m(1) \to \bstar(1) < 1$. \emph{Step~1.}
On $[v_0/2, 1]$, which the interior windows reach, (i) gives $\sigma_{k+1} - \sigma_k = \bstar{}''(v_k)h +
(-1)^{m-k-1}(\bstar{}''(1)h/3)\hat Q(v_k) + o(h)$ and $|\sigma_j - \bstar{}'| =
O(h)$, so, exactly as in Step~1 of the proof of Theorem~\ref{thm:limit2},
$\mu_k = f_kh[\tau_F \pm \zeta_F](v_k) + o(h)$ and the smooth integrand is
$f(W + \varpi_j)$ with $|\varpi_j| \le C(|\sigma_j - \bstar{}'| + \sup|\delta|) =
o(1)$, each deviation from the frozen value being Lipschitz in $(\sigma_j,
\bstar_m)$ for slopes bounded below; and $A_m(1) \to G_1(1)(1 - \bstar(1))/
\bstar{}'(1)^2 > 0$. \emph{Steps~2, 4, 5} are unchanged (bounded slopes give
the $O(h)$ cross-coefficients of the IMS step; the upper bound is taken at
an interior near-minimizer, where (i) holds). \emph{Step~3, the bottom
windows.} On a window in $[0, 2v_0]$ the piece integrand is bounded in
absolute value by $G_1/\sigma_j + \tfrac12|G_1'(v - \bstar_m) + G_1(1 -
\sigma_j)|/\sigma_j^2 \le Cv^{n-2}$, since $|\bstar_m(v)| \le Cv$ by bounded
slopes and $\bstar_m(0) = 0$, and $n \ge 3$; so the piece terms are $\ge -Cv_0M(\xi)$. A knot mass is
$|\mu_k| \le \tfrac12G_1(v_k)|v_k - \bstar_m(v_k)|\,|\sigma_{k+1}^{-2} -
\sigma_k^{-2}| \le Cv_k^{n-1}|\sigma_{k+1} - \sigma_k| \le C(\varepsilon(2v_0) +
o(1))h$ by (ii) (the window lies in $[0, 2v_0]$), so $\sum_k|\mu_k|\xi_k^2 \le
C(\varepsilon(2v_0) + o(1))M(\xi)$. Hence $Q_m(\xi) \ge -C(v_0 +
\varepsilon(2v_0) + o(1))M(\xi) \ge [E_{\min} - C(v_0 + \varepsilon(2v_0)) -
o(1)]M(\xi)$, because $E_{\min} \le
E_F^{\mathrm{osc}}(0) = w_F/f(0) = 0$ for $n \ge 3$. For the weighted statement
divide by $\min\vartheta$ and use $\min(E_F^{\mathrm{osc}}/\vartheta) \le 0$.
Letting $v_0 \to 0$ after $h \to 0$ gives $\liminf c_m \ge E_{\min}$, and
Step~5 the reverse.
\end{proof}

\begin{proof}[Proof of Proposition~\ref{prop:local}]
The first claim is the decomposition identity with each interior knot mass
written as two halves and the piece integrals and masses summed piece by
piece ($\eta_0 = 0$, so $P_1$ is a $1\times1$ form and its least eigenvalue
is the ratio $P_1/M_1$); the quotient of sums is at least the least
quotient. For the second, any division gives $c_m \ge \min_j\lambda_{\min}$
by the same argument; conversely, for $E < c_m$ the matrix $A$ of $Q_m -
EM$ in the knot basis is tridiagonal and positive definite, hence a sum of
positive semidefinite $2\times2$ blocks on the consecutive pairs plus a
positive $1\times1$ block at knot $1$: take the $LDL^{\mathsf T}$ pivots
from the top down, $d_m = a_{mm}$, $d_{k-1} = a_{k-1,k-1} - a_{k-1,k}^2/d_k$,
and the blocks $\bigl[\begin{smallmatrix} a_{k-1,k}^2/d_k & a_{k-1,k}\\
a_{k-1,k} & d_k\end{smallmatrix}\bigr]$ on $(k-1, k)$ with $d_1$ at knot $1$;
they sum to $A$ by the pivot recursion. The off-diagonal entries and the
piece integrals are fixed per piece, so these blocks are the $P_j - EM_j$
for some division of the knot diagonal entries, whose
$\min_j\lambda_{\min}$ is then $\ge E$; so the supremum over divisions is
$c_m$ (attained at $E = c_m$, where the trailing pivots of the singular
Jacobi matrix are still positive and $d_1 = 0$). For the third, Step~1 of the proof of
Theorem~\ref{thm:limit2} gives $\mu_k = f_kh[\tau_F \pm \zeta_F](v_k) +
O(h^{5/3})$, the piece integrand $f(W + O(h^{2/3}))$ and $M_j =
f(\mathrm{mid}_j)(h/3)T_j(1 + O(h))$; freezing the coefficients at $v_j$,
$\lambda_{\min}(P_j, M_j) = W(v_j) + 3\lambda_{\min}(\hat P_j, T_j) +
O(h^{2/3})$ with $\hat P_j(a, b) = \tfrac12(\tau \pm \zeta)a^2 + \tfrac12(\tau
\mp \zeta)b^2$ exactly the piece form of the two-sided Bloch inequality with
its half weights, whose least eigenvalue against $T_j$ is $g_*(\tau, \zeta)$
by the proof of that inequality (the form is positive semidefinite iff $g
\le g_*$, singular at $g_*$); so $\lambda_{\min}(P_j, M_j) = E_F^{\mathrm{osc}}(v_j)
+ O(h^{2/3})$ for $j \le m - 1$; the top piece's boundary term dominates
the omitted $O(h)\eta_m^2$ and gives $\lambda_{\min}(P_m, M_m) \ge
E_F^{\mathrm{osc}}(1) - Ch^{2/3}$ (no more: the penalty is of order $1/h$ per
unit mass, so this pencil tends to the $1\times1$ value with $\eta_m = 0$,
above $E_F^{\mathrm{osc}}(1)$), and the bottom piece gives $W(v_1) +
O(1)(|\tau_F| + |\zeta_F|)(v_1) + O(h^{2/3}) \to E_F^{\mathrm{osc}}(0)$; since
$\min_{j \le m-1}E_F^{\mathrm{osc}}(v_j) \to \min E_F^{\mathrm{osc}}$, $L_m$
converges to the minimum. The band expansion
behind Theorem~\ref{thm:interior}: a Bloch wave $\eta_k = e^{ik\theta}u_k$
with $u = (a, b)$ on the two knots of a cell has per-cell knot energy $(\tau
+ \zeta)a^2 + (\tau - \zeta)b^2$ and mass $2a^2 + 2b^2 + 2ab\cos\theta$, so
the lower band is $W + 3\lambda_-(\theta)$ for the pencil $(\mathrm{diag}(\tau
+ \zeta, \tau - \zeta), [\begin{smallmatrix}2 & \cos\theta\\ \cos\theta &
2\end{smallmatrix}])$, whose determinant at $\theta = 0$ is the quadratic
$3\lambda^2 - 4\tau\lambda + \tau^2 - \zeta^2$ of the Bloch inequality; the
first-order perturbation of the eigenvalue in $\cos\theta = 1 - \theta^2/2$
along the null profile $(1, t_*)$ gives the curvature $\kappa = \tfrac32
g_*t_*/(1 + t_* + t_*^2)$, nonnegative because $g_*t_* = \tau + \zeta - 2g_*
\ge 0$ by the proof of the Bloch inequality.
\end{proof}

\begin{proof}[Proof of Theorem~\ref{thm:airy}, upper bound]
Write $\tau, \zeta, g, t$ for $\tau_F(1)$, $\zeta_F(1) = \tau/3$, $g_*(\tau, \zeta)$
and $t_*(\tau, \zeta)$, and let $p$ be the two-periodic profile with $p_k = 1$
at the knots whose mass carries $\tau + \zeta$ (the knot below the top) and
$p_k = t$ at the others: the null vector of the two-sided Bloch inequality
at $(\tau, \zeta)$, aligned with the masses' sign pattern. Each piece form
$P_j(a, b) = A_ja^2 + B_jb^2 - gab$ of that inequality is positive
semidefinite with $P_j(p_{j-1}, p_j) = 0$, hence $2A_jp_{j-1} = gp_j$ and
$2B_jp_j = gp_{j-1}$. \emph{Step~1.} For an envelope $\chi$ and $\xi_k =
\chi_kp_k$, $P_j(\xi_{j-1}, \xi_j) = (\Delta\chi_j)^2A_jp_{j-1}^2$ exactly
(the $\chi_j^2$ term is $P_j(p) = 0$ and the cross term is $\Delta\chi_j\chi_j
(2A_jp_{j-1}^2 - gp_{j-1}p_j) = 0$), and over a cell $A_+ + A_-t^2 = gt$ by
the two null relations, so $3\sum_jP_j(\xi)/T(\xi) = \tfrac32gt/(1 + t + t^2)
\cdot\sum(\Delta\chi)^2/\sum\chi^2\,(1 + O(\Delta\chi/\chi)) = \kappa(1)
\sum(\Delta\chi)^2/\sum\chi^2$ to leading order. \emph{Step~2.} On a window
$[1 - \ell, 1]$ with $\xi_m = 0$, Step~1 of the proof of
Theorem~\ref{thm:limit2} gives the knot masses $f_kh(\tau_F \pm
\zeta_F)(v_k) + O(h^2)$, the piece integrand $f(W + O(h))$ and $M_j =
f(h/3)T_j(1 + O(h))$; splitting the knot sum at the local values and
writing $P_j^{(v_j)}$ for the local piece form (null vector $p(v_j)$, with
$p(1) = p(v_j) + O(1 - v_j)$), positive semidefiniteness kills the cross
term $P_j^{(v_j)}(\chi p(v_j), \chi\,\delta p_j)$, so $Q_m(\xi)/M(\xi)$ equals
the discrete Rayleigh quotient of $-\kappa(1)h^2d^2/dv^2 + E(1) +
E'(1)(v - 1)$ with weight $f$ up to $O(h + \ell^2 + (h^2/\ell^2)(h/\ell +
\ell))$. \emph{Step~3.} With $u = 1 - v$, $L = (\kappa(1)h^2/|E'(1)|)^{1/3}$ and
$\lambda = a_1|E'(1)|L$, the envelope $\chi(u) = \mathrm{Ai}((u -
\lambda/|E'(1)|)/L)$, cut off at $u = CL(\log(1/h))^{2/3}$, vanishes at $u =
0$ and solves $-\kappa(1)h^2\chi'' + |E'(1)|u\chi = \lambda\chi$, so its
continuous quotient is $\lambda$ up to $O(h^2)$ from the cutoff; the Riemann
sums over knots reproduce it to relative $O(h/L) = O(h^{1/3})$, i.e.
$O(h)$ absolutely, and the Step~2 remainders with $\ell \asymp
L\log^{2/3}(1/h)$ are $O(h)$. Hence $c_m \le Q_m(\xi)/M(\xi) \le E(1) +
\lambda + O(h)$. The argument uses the theorem only through coefficient asymptotics of
relative error $o(h^{2/3})$ on $[v_0, 1]$; Theorem~\ref{thm:four} supplies
$O(h\log m)$ at $n = 4$, while at $n = 3$ the $O(h^{1/3})$ remainder of
Theorem~\ref{thm:oscillation2} does not suffice.
\end{proof}

\begin{proof}[Proof of Theorem~\ref{thm:airy}, lower bound]
Notation as above, with $u := 1 - v$, $A_\pm = \tfrac12(\tau \pm \zeta) - g > 0$
(their product is $g^2/4$), $w_+ = A_+$, $w_- = A_-t^2$. \emph{Two lemmas.}
(i) Every knot vector is $\eta = \chi p$ with $\chi := \eta/p$ ($t \ne 0$ since
$A_+ = gt/2$ and $g \ne 0$), so $P_j(\eta) = (\Delta\chi_j)^2A_jp_{j-1}^2$ for
every $\eta$; and for a cell (two consecutive pieces whose outer knots
carry $p = 1$), with $a := \eta_j/t - \chi_{j-1}$, $b := \chi_{j+1} - \eta_j/t$,
$3[w_+a^2 + w_-b^2] \ge 3\,\tfrac{w_+w_-}{w_+ + w_-}(a + b)^2 = \tfrac12\kappa(1)(1
+ t + t^2)(\chi_{j+1} - \chi_{j-1})^2$, because $w_+ + w_- = gt$ by the null
relations and $w_+w_- = g^2t^2/4$ by the determinant identity (Cauchy--Schwarz);
the same holds with the two piece forms evaluated at any common local
point and the envelope frozen at $v = 1$, since $a + b$ involves the
envelope only at the outer knots, where every profile equals $1$. (ii)
(Airy with a free outer end.) For $\vartheta, b > 0$, $L = (\vartheta/b)^{1/3}$,
$\lambda = a_1bL$ and $\ell \ge (a_1 + x_1)L$, every $\chi \in H^1(0, \ell)$ with
$\chi(0) = 0$ satisfies $\int_0^\ell[\vartheta\chi'^2 + bu\chi^2] \ge \lambda(1 -
\delta)\int_0^\ell\chi^2$ with $\delta = Ce^{-(4/3)x_1^{3/2}}$, $C$ an absolute
constant and $x_1$ large enough: with $x_0 := -a_1(1 -
\delta)$ the solution $\Phi = \mathrm{Bi}(x)\mathrm{Ai}(x_0) - \mathrm{Ai}(x)\mathrm{Bi}(x_0)$
of Airy's equation vanishes at $x_0$, is positive beyond it (Sturm
interlacing with $\mathrm{Ai}$, which has no zero above $x_0$) and, since
$\Phi'' = x\Phi$, has a single interior minimum $x_{\min} =
((3/4)\log(c/\delta))^{2/3}(1 + o(1))$ beyond which it increases; $\varphi(u) :=
\Phi((u - \lambda(1 - \delta)/b)/L)$ then solves $-\vartheta\varphi'' + (bu -
\lambda(1 - \delta))\varphi = 0$ with $\varphi(0) = 0$, $\varphi > 0$ and
$\varphi'(\ell) \ge 0$, and Picone's identity $\int_0^\ell[\vartheta\chi'^2 + (bu -
\lambda(1 - \delta))\chi^2] = \int_0^\ell\vartheta(\chi' - \chi\varphi'/\varphi)^2 +
\vartheta(\varphi'/\varphi)(\ell)\chi(\ell)^2 \ge 0$ gives the claim.
\emph{Step~1.} Fix $v_0 < 1$ with $E - E(1) \ge |E'(1)|(1 - v)/2$ on $[v_0, 1]$
and $c_0 := \min_{[0,v_0]}(E - E(1)) > 0$, $K := C_1\log^{2/3}(1/h)$ (so that
$\delta \le h^2$ in (ii) and $K \ge 4a_1$), $\ell = KL$, and let $\mathcal W$ be the
pieces with $u_j \le \ell$, the top piece included. By the decomposition
identity and Step~1 of the proof of Theorem~\ref{thm:limit2}, $Q_m(\eta) =
\sum_jf_jh[E(v_j)T_j/3 + P_j^{(v_j)}(\eta)] + \tfrac12A_m(1)\eta_m^2 -
\tfrac12f_mh(\tau - \zeta)\eta_m^2 + O(h)M(\eta) + O(h^{2/3})M_{<v_0}(\eta)$ with
$P_j^{(v)}$ the local piece form, positive semidefinite (the half knot mass
at the top knot is the term Lemma~\ref{app:bloch}'s split assigns there
and the identity does not contain); $E(1)$ factors out of the mass up to
$O(h)M$; the pieces outside $\mathcal W$ contribute at least $2\lambda$ times
their mass ($E(v_j) - E(1) \ge |E'(1)|KL/2 = (K/2a_1)\lambda$ for $v_j \ge v_0$,
$\ge c_0$ below); and it remains to bound the window's sum, with the
penalty $\tfrac12A_m(1)\eta_m^2 - Ch\eta_m^2$, below by $(\lambda - Ch)M_{\mathcal
W}$, $M_{\mathcal W} := \sum_{\mathcal W}f_j(h/3)T_j(\eta)$. \emph{Step~2.} Write
$\eta = \chi p$ for the profile at $v = 1$; knot $m$ is a $t$-knot, so the top
cell is the half cell (piece $m$) and the full cells, with both piece forms
evaluated at the cell's midpoint (an $O(h^2)$ change), satisfy (i) with
the local constant $\kappa(v_c)(1 + t(v_c) + t(v_c)^2)/2 \ge (1 - C\ell)\kappa(1)(1
+ t + t^2)/2$ and no mismatch term; $E(v_j) - E(1) \ge |E'(1)|u_j - C\ell^2$
on $\mathcal W$; a leftover piece at the window's edge is covered by its
potential term alone. \emph{Step~3.} With $r := \sum_{\mathcal W}(\Delta\chi_j)^2/
\sum_{\mathcal W}\chi_j^2$ and $T_j(\chi p) = (1 + t + t^2)\chi_j^2 -
\Delta\chi_j\chi_j(2p_{j-1}^2 + p_{j-1}p_j) + (\Delta\chi_j)^2p_{j-1}^2$: if $r \ge
Rh^{2/3}$ ($R$ large, depending on $\tau, \zeta, \kappa(1), |E'(1)|$) the kinetic
sum alone exceeds $2\lambda M_{\mathcal W}$ (the potential terms are $\ge 0$
exactly); otherwise the cross terms are a relative $O(h^{1/3})$ on the
mass, i.e. $O(h)$ absolutely at the $\lambda$ level. \emph{Step~4.} In the
latter case, on the lattice of $1$-knots $u = h, 3h, \dots$ with
$\tilde\chi$ their piecewise-linear interpolant extended by $\tilde\chi(0) := 0$:
the top half cell gives, by the harmonic-mean inequality with weights
$f_mhA_+$ and $\tfrac12A_m(1)t^2 - Ch$ on $(\chi_m - \chi_{m-1}, -\chi_m)$, at least
$f_mhA_+(1 - O(h))\chi_{m-1}^2$, which divided by $f(1)(1 + t + t^2)/3$ is
$\kappa(1)h\chi_{m-1}^2 = \kappa(1)h^2\int_0^h\tilde\chi'^2$ — the Dirichlet
condition at $u = 0$ with the right coefficient (the excess of the
discrete mass near the top, $O(h)(\chi_{m-1}^2 + \chi_m^2)$, is paid by a
spare part of the penalty and by running (ii) with $\vartheta = (1 -
C'h^{2/3})\kappa(1)h^2$, at a cost $O(h^{4/3})$); the full cells give $(1 -
C\ell)\kappa(1)h^2\int_h^\ell\tilde\chi'^2$ exactly ($\sum_c(\chi_{c+} -
\chi_{c-})^2/(2h) = \int\tilde\chi'^2$); the mass is $\int\tilde\chi^2(1 + O(r))$ up to end terms $O(h)(\chi_{m-1}^2 +
\chi_{\mathrm{edge}}^2)$ (the first covered as above, the second by the
slack of the two outermost cells' potential terms, kept out of (ii)) and
the potential sum is $\ge \int|E'(1)|u\tilde\chi^2 - 3h\int\tilde\chi^2$ (on each
cell $\int u\tilde\chi^2 \le (u_c + 2h)h(\chi_c^2 + \chi_{c-2h}^2)$, with no $\sqrt r$
term). So the window's quotient is at least $\int_0^\ell[\kappa(1)h^2\tilde\chi'^2
+ |E'(1)|u\tilde\chi^2]/\int_0^\ell\tilde\chi^2 - Ch \ge \lambda(1 - h^2) - Ch$ by (ii)
with $\vartheta = \kappa(1)h^2$, $b = |E'(1)|$, the outer end free. All
remainders are $O(h)$ ($\ell^2 = O(h^{4/3}\log^{4/3}(1/h))$, $\ell\lambda =
O(h^{4/3}\log^{2/3}(1/h))$), so $c_m \ge E(1) + \lambda - Ch$. At $n = 4$ under
Theorem~\ref{thm:four} the coefficient remainders are $O(h\log m)$ and the
same proof gives $O(h\log m)$.

\emph{Three and four bidders.} Both halves use $\bstar_m$ only through
the coefficients of the decomposition. On the window, and on every compact
$[v_0, 1]$, the second-order amplitude law
(Theorem~\ref{app:H2}, Appendix~\ref{app:interior}: proved for $n = 3$, and for $n = 4$ given the layer) gives the alternating
slope with relative error $O(h^2)$ and the smooth part with increments
$O(h^3)$, $O(h^4)$, so the knot masses and piece integrands have relative
error $O(h)$ there, including the top row, whose half-hat slope is pinned to
$(\bstar{}''(1)h/6)(1 + O(h))$; the null profile, the cell identity and
$\kappa(1)$ are the same expressions in $(\tau, \zeta)$ since $\zeta_F(1) =
\tau_F(1)/3$ at every $n$, and the two lemmas above contain no $n$. Below
$v_0$ only a bound is used: at $n = 3$ the coefficients are within a relative
$O(h^{1/3})$ uniformly (the oscillation law's remainder), and at $n = 4$ the
law holds within $C(F)(1/k + h\log m)$ for $k \ge K_0$ while on the first
$K_0$ knots the slope bounds of Theorem~\ref{thm:layer} alone give knot masses
$\le Ch^3k^3$ and piece integrands $\le Ch^2K_0^2$, so those pieces'
contribution differs from $\sum(E(v_j) - E(1))\cdot(\mathrm{mass})$ by
$O(h^2K_0^3)$ of their mass. Since the local data are $O(v^2)$ at $n \ge 3$, $E$
is continuous at $0$ with $E(0) = 0 > E(1)$, so $c_0 := \min_{[0, v_0]}(E -
E(1)) > 0$ pays the far pieces' errors and the $2\lambda$ of Step~1 for $h \le
h_0(F)$, exactly as the far pieces are paid at $n = 2$; the remainders are
$O(h)$.
\end{proof}

\paragraph{The bottom edge at three bidders.}\label{app:proof-bottom3} For $n \ge 3$ every regular
prior has $E_F^{\mathrm{osc}}(0) = 0$, and for a prior with
$E_F^{\mathrm{osc}} > 0$ on $(0,1]$ the minimum is attained at $v = 0$ only,
with $E \sim C_Fv^{n-2}$: the certificate is positive at every $m$ but has no
positive limit, and neither rate theorem applies. At three bidders the
rate's constant is universal.

Proposition~\ref{app:bottom3}, stated in Appendix~\ref{app:arc3}, gives
its constant.

\begin{proof}[Proof of Proposition~\ref{app:bottom3}]
\emph{The uniform prior.} At $n = 3$ the uniform prior has $\bstar = 2v/3$,
$s = v/3$, $w_F = 3v/2$ and no knot masses, and its quadratic form at the
equilibrium is exactly $Q_m(\eta) = \int(3v/2)\eta^2 + c_F\eta(1)^2$ for every
$m$ (Corollary~\ref{cor:weighted}: $\bstar$ lies in every piece space and
solves the Galerkin equations pointwise). On the knots $k \ge 1$ with the
hats $\varphi_k$, $\int v\varphi_k^2 = (2k/3)h^2$ and $\int v\varphi_k\varphi_{k+1} =
((2k+1)/12)h^2$, which gives $h^2\mathbb T$, and the Gram matrix is
$h\mathbb M$. The rows being scale-free, $c_m/h$ is the ground value of the
pencil restricted to $m$ knots; the ground state decays geometrically, so
the restriction to $K \ge 12$ knots gives $\lambda^*$ to twelve digits
(Appendix~\ref{app:numerics}), and $c_m = \lambda^*h$ exactly for every $m$.

\emph{Localization.} For a general $F$ fix $v_0' > 0$ and a fixed-scale
partition of unity $\chi_0^2 + \chi_1^2 = 1$ with $\chi_0 = 1$ on $[0, v_0'/2]$
and $\chi_0 = 0$ on $[v_0', 1]$; the IMS cost is $C(h/v_0')^2M(\eta)$ (Step~2 of
the proof of Theorem~\ref{thm:limit2}), $O(h)$ relative to the scale $hM$ of
the claim. On the upper part, Steps~2 to 4 of that proof restricted to
$[v_0'/4, 1]$ give $Q_m(\chi_1\eta) \ge (c_0 - o(1))M(\chi_1\eta)$ with $c_0 :=
\min_{[v_0'/4, 1]}E_F^{\mathrm{osc}} > 0$, which is $\gg hM(\chi_1\eta)$.

\emph{The bottom part is the uniform problem to relative $o(1)$.} On the
knots $v_k \le v_0'$ the piece integrand is its smooth-slope value within a
relative $O(h^{1/3})$ of its own size $\asymp v$ (the proof of
Corollary~\ref{cor:weighted}), and the smooth-slope value is $f(0)^3(1 +
O(v))$ times the uniform prior's, since $F = f(0)v(1 + O(v))$, $\bstar =
(2v/3)(1 + O(v))$ and $G_1 = 2Ff^2 = 2f(0)^3v(1 + O(v))$; the knot masses,
zero for the uniform prior, are the slope-law values within
$Ch^{1/3}\vartheta_kf_kh$, and the slope-law values themselves are a relative
$O(v_k^{1/3})$ of $\vartheta_kf_kh$ ($\tau_F = O(v^2)$, $\zeta_F \asymp v^{4/3}$
at $n = 3$); the mass density is $f(0)(1 + O(v))$. So for every $\xi$
supported on those knots, with $P_j(\xi) \ge 0$ the uniform piece forms
($\sum_jP_j = \ip{\mathbb T\xi}{\xi}$), $m_k(\xi) \ge 0$ the mass forms of the two
pieces at knot $k$, and $\epsilon_k := C((kh)^{1/3} + h^{1/3})$,
\[
  \Bigl|\frac{Q_m(\xi)}{f(0)^3h^2} - \ip{\mathbb T\xi}{\xi}\Bigr| \le D(\xi) := \sum_j\epsilon_jP_j(\xi) + \sum_k\epsilon_kk\xi_k^2,
  \qquad
  \Bigl|\frac{M(\xi)}{f(0)h} - \ip{\mathbb M\xi}{\xi}\Bigr| \le E(\xi) := \sum_kCkh\,m_k(\xi),
\]
$D$ and $E$ nonnegative quadratic forms, with $D(\xi) \le 3\epsilon_{\max}
\ip{\mathbb T\xi}{\xi}$ on knots $\le v_0'$ by $\sum_kk\xi_k^2 \le 2\ip{\mathbb T\xi}{\xi}$
($\mathbb T$ is diagonally dominant), $\epsilon_{\max} := C(v_0'^{1/3} + h^{1/3})$.

\emph{The spectral gap.} Let $x^*$ be the pencil's $\ell^2$ ground state,
$\mathbb M$-normalized, and $\tilde x^*$ the ground state of the pencil
restricted to the knots $\le v_0'/h$, which decays geometrically too and
differs from $x^*$ by an exponentially small amount. For $\xi$ supported on
the block write $\xi = a\tilde x^* + y$ with $\ip{\mathbb M\tilde x^*}{y} = 0$, so
that $\ip{\mathbb T\tilde x^*}{y} = 0$ and $\ip{\mathbb Ty}{y} \ge \lambda_2\ip{\mathbb My}{y}$
with $\lambda_2 = 2.99788$ the pencil's second eigenvalue (the block's second
eigenvalue is at least $\lambda_2$ by the min-max principle). Since $D$ and $E$
are nonnegative quadratic forms, $D(\xi) \le 2a^2D(\tilde x^*) + 2D(y)$ and
likewise for $E$, with $D(\tilde x^*) \le Ch^{1/3}$ and $E(\tilde x^*) \le Ch$
($x^*$ decays geometrically, so $\sum_kk^{4/3}(x^*_k)^2 < \infty$), $D(y) \le
3\epsilon_{\max}\ip{\mathbb Ty}{y}$ and $E(y) \le Cv_0'\ip{\mathbb My}{y}$. The mediant
inequality gives
\[
  \frac{Q_m(\xi)}{f(0)^2hM(\xi)} \ge \min\Bigl(\frac{\lambda^* - 2Ch^{1/3}}{1 + 2Ch},\ \frac{\lambda_2(1 - 6\epsilon_{\max})}{1 + 2Cv_0'}\Bigr) = \lambda^* - Ch^{1/3}
\]
once $v_0'$ is a fixed small number and $h \le h_0$, because $\lambda_2/\lambda^*
= 2.035$ leaves room: $\lambda_2(1 - 6\epsilon_{\max}) \ge \lambda^*(1 + 2Cv_0')$
holds for $\epsilon_{\max} \le 0.08$. The test sequence $\tilde x^*$ truncated at
$K = \lceil C\log(1/h)\rceil$ knots gives the reverse, $Q_m(\tilde x^*)/M(\tilde x^*)
\le f(0)^2\lambda^*h(1 + Ch^{1/3})$.

\emph{Assembly.} For every $\eta$ the localization and the block bound
give $Q_m(\eta) \ge f(0)^2\lambda^*h(1 - Ch^{1/3})M(\eta) - Ch^2M(\eta)$, and the
test sequence gives $c_m \le f(0)^2\lambda^*h(1 + Ch^{1/3})$: the factors
$f(0)^3$ of the rows and $f(0)$ of the mass combine to $f(0)^2$, and $f(0)^2\lambda^*
= \kappa_3\cdot\tfrac32f(0)^2$. Without the split along $x^*$ the block bound
gives only $c_m = f(0)^2\lambda^*h(1 + o(1))$; the spectral gap is what turns
the fixed-$v_0'$ error into $h^{1/3}$.
\end{proof}

The remainder $O(h^{1/3})$ is the knot-mass bound's; the rows' own
corrections are $O(v)$ and would give $O(h)$ alone, and the measured
approach of $c_mm/(\tfrac32f(0)^2)$ to $\kappa_3$ (two convex priors,
Appendix~\ref{app:numerics}) sits between the two. $\kappa_3$ is not known in
closed form: the pencil's $\ell^2$ solution is the minimal solution of a
three-term recurrence with coefficients linear in $k$, and $\lambda^*$ is the
value at which it satisfies the first row. At four bidders the bottom
problem is the uniform rows driven by the layer at the amplitude $A_F$,
and the constant is prior-dependent.

\subsubsection{The interior rate}
\label{app:interior}

We prove Theorem~\ref{thm:interior}, its weighted form
(Theorem~\ref{app:interiorw} below) and the second-order amplitude law
the two rest on (Theorem~\ref{app:H2}). Notation as in
Appendices~\ref{app:global} and \ref{app:limit}: $h = 1/m$, $v_k = kh$,
$\varepsilon_k := (-1)^{m-k}$, $\hat Q = \tilde Q/\tilde Q(1)$, $\hat A :=
(\bstar{}''(1)/12)\hat Q$, $\omega_k = h^2\hat Q(v_k)$, $\phi_k$ the forcing
weight of the bidder count in hand, $J = J(I_h\bstar)$ the Jacobian at the
interpolant with its interior rows $R_k(y) = a_ky_{k-1} + b_ky_k + c_ky_{k+1}$,
row $1$ $R_1(y) = b_1y_1 + c_1y_2$ and half-hat row $T(y) = a_my_{m-1} +
b_my_m$, $\tilde W, \tilde Z, K$ its pinned solutions and Casoratian
(Lemma~\ref{app:L2}), $N_s(d) = \Phi(s + d) - \Phi(s) - J(s)d$, $\delta = \bstar_m -
I_h\bstar$ the knot error, $\Delta y_k = y_k - y_{k-1}$, $\Delta^2y_k = y_{k+1} -
2y_k + y_{k-1}$; and the local data of Appendix~\ref{app:limit}: the knot
masses $\mu_k$, the piece forms $P_j^{(v)}$ of Lemma~\ref{app:bloch} with
null profile $p(v)$ (value $1$ at the knots carrying $\tau_F + \zeta_F$,
$t_*(v)$ at the others), $A_\pm$, $g_*$, $\kappa = \tfrac32g_*t_*/(1 + t_* +
t_*^2)$ and $E = E_F^{\mathrm{osc}} = W + 3g_*$ with $W = w_F/f - \tau_F$. By
Theorem~\ref{thm:oscillation2} (and Theorems~\ref{thm:four},
\ref{thm:layer} at $n = 4$) the error has the two-scale form
\begin{equation}
\label{eq:law}
  \delta_k = \varepsilon_k\,\frac{\bstar{}''(1)h^2}{12}\,\hat Q(v_k)\,(1 + \rho_k) + u_k,
\end{equation}
with $|\rho_k| \le C/k$ and $u$ smooth with $|\Delta u_k| \le C(1/k +
hv_k^{-(n-1)/3})\omega_k$; the decomposition is fixed only up to that order.

Theorem~\ref{app:H2}, stated in Appendix~\ref{app:arc3}, asserts that on
every compact $I = [v_0, 1] \subset (0,1]$ the decomposition \eqref{eq:law}
holds with $|\rho_k| \le Ch^2$, $|\Delta u_k| \le Ch^3$ and $|\Delta^2u_k|
\le Ch^4$ for $v_k \in I$ and $h \le h_0$; this subsubsection proves it.

The three bounds are what the local data need: $\rho$ for the first-order
term of the knot masses, the first difference of $u$ for the slopes'
second-order remainder to be bounded, the second for it to vary slowly.
The remainders are attained ($|\rho| \asymp h^2$ and $\Delta^2u \asymp h^4$,
Appendix~\ref{app:numerics}).

\paragraph{The law in the interpolant recurrence.}

\begin{lemma}[the second-order law of the recurrence]\label{app:I1}
Let $F \in C^5$ with $f > 0$, $n \ge 2$, and $J_{k,k-1}, J_{k,k}, J_{k,k+1}$
the entries of row $k$ of $J$. Uniformly on compact subsets of $(0,1]$,
\[
  -J_{k,k+1}\tilde Q(v_k + h) + J_{k,k}\tilde Q(v_k) - J_{k,k-1}\tilde Q(v_k - h)
  = O(h^3)\,\tilde Q(v_k):
\]
$y_k = (-1)^k\tilde Q(v_k)$ solves the homogeneous rows with residual
$O(h^3)|y_k|$ per row. The coefficient of $h^0$ in the expansion of the left
side vanishes identically, those of $h^1$ and $h^2$ under the equilibrium
ODE $\bstar{}' = (n-1)f(v - \bstar)/F$; the coefficient of $h^3$ does not, so
the law is exact to second order and no further.
\end{lemma}

\begin{proof}
A finite computation (Appendix~\ref{app:numerics}). Each entry is $h$ times
an integral over the two pieces adjacent to $v_k$ of $G_1[\varphi_j\varphi_k/\sigma
+ (v - b)\varphi_j'\varphi_k/\sigma^2]$ with the chord slopes $\sigma$ of
$I_h\bstar$; expanding $F$, $f$, $\bstar$ in Taylor data at $v_k$ and
integrating term by term gives the entries to $O(h^3)$ as polynomials in
the data, and $\tilde Q(v_k \pm h)$ in the same data. The $h^0$ coefficient
of the residual is $J_{k,k} - J_{k,k+1} - J_{k,k-1}$ at leading order, the
symbol's zero at $-1$; the $h^1$ coefficient vanishes after substituting
the ODE and its derivatives (this is the root identity $|\lambda_k| = 1 +
h(\log\tilde Q)' + O(h^2)$ of Appendix~\ref{app:oscillation}); the $h^2$
coefficient vanishes after the same substitution up to the fourth
derivative of $\bstar$, as a rational function of the data and $n$, hence
for every $n$. The remainder is $O((h/v_k)^3)\tilde Q(v_k)$ where $h/v_k \le
\tfrac12$, the Taylor data being bounded there and the expansions of
$\tilde Q \propto F^{-(n-1)/3}$ and $G_1 \propto F^{n-2}$ having relative
remainders $O((h/v_k)^3)$ since $F \asymp v$.
\end{proof}

The product of the frozen roots $\prod_{j \ge k}|\lambda_j|^{-1}$ of
Lemma~\ref{app:L2} has a nonzero first-order term; it is the ansatz, not a
solution, and the discrete Liouville--Green amplitude factor cancels that
term exactly. Lemma~\ref{app:I1} is the statement about the mode: an
alternating forcing of size $h^3|z_j|$ per row gives, through the Green's
function, a response $O(h^2)|z_k|$, so the alternating homogeneous solution
of $J$ is $\tilde Q(1 + O(h^2))$ on compacts.

\paragraph{The local data to second order.}

\begin{lemma}[the second-order local data]\label{app:I2}
Assume the bounds of Theorem~\ref{app:H2} on $I = [v_0, v_1] \subset (0,1)$.
Then for every knot $v_k \in I$, piece $j$ with $v_j \in I$ and
piecewise-linear $\eta$ with $\eta_0 = 0$, with $v_{j-\frac12}$ the piece
midpoint, $T_j(\eta) = \eta_{j-1}^2 + \eta_{j-1}\eta_j + \eta_j^2$ and
$c_j := G_1/\sigma_j - \tfrac12A_j'$ the integrand of the decomposition on
piece $j$,
\begin{align*}
  \text{(a)}\quad & \mu_k = f(v_k)\,h\,[\tau_F(v_k) + (-1)^{m-k-1}\zeta_F(v_k)] + O(h^3),\\
  \text{(b)}\quad & \int_{\text{piece }j}c_j\eta^2\,dv = h\bigl[(fW)(v_{j-\frac12}) + \varepsilon_jh\,c_1(v_{j-\frac12})\bigr]\frac{T_j(\eta)}3
    + h^2\,\mathcal G(v_{j-\frac12})\,\frac{\eta_j^2 - \eta_{j-1}^2}{12} + O(h^3)(\eta_{j-1}^2 + \eta_j^2),\\
  \text{(c)}\quad & \int_{\text{piece }j}f\eta^2\,dv = h\,f(v_{j-\frac12})\frac{T_j(\eta)}3
    + h^2f'(v_{j-\frac12})\frac{\eta_j^2 - \eta_{j-1}^2}{12} + O(h^3)(\eta_{j-1}^2 + \eta_j^2),
\end{align*}
with $c_1 := a_1[-\tfrac12G_1/\bstar{}'^2 + (G_1s)'/\bstar{}'^3]$, $a_1 :=
(\bstar{}''(1)/6)\hat Q$, $s = v - \bstar$, and $\mathcal G := G_1'/\bstar{}' -
\tfrac12(G_1s)''/\bstar{}'^2 \in C^1(I)$ the derivative of the frozen-slope
integrand (which is not $(fW)'$). The second-order terms of (b) and (c)
telescope: summed over consecutive pieces they give $O(h^3)\sum_j\eta_j^2 +
O(h^2)$ times the two end values squared.
\end{lemma}

\begin{proof}
By \eqref{eq:law} and the hypothesis the slopes are
\[
  \sigma_j = \bstar{}'(v_{j-\frac12}) + \varepsilon_jh\,a_1(v_{j-\frac12}) + h^2r_j + O(h^3),
  \qquad |r_j| \le C,\ |r_j - r_{j-1}| \le Ch,
\]
the alternating term exact to $O(h^3)$ because $\hat Q(v_k) + \hat Q(v_{k-1})
= 2\hat Q(v_{j-\frac12}) + O(h^2)$ and $\rho = O(h^2)$, and $r_j =
\bstar{}'''(v_{j-\frac12})/24 + (u_j - u_{j-1})/h^3$. (a) The knot mass is
$-\tfrac12G_1(v_k)(v_k - s_k)(1/\sigma_{k+1}^2 - 1/\sigma_k^2)$. Expanding
$1/\sigma^2$ about $\bar\sigma_\pm := \bstar{}'$ at the two midpoints, with
$\varepsilon_{k+1} = -\varepsilon_k$,
\[
  \frac1{\sigma_{k+1}^2} - \frac1{\sigma_k^2}
  = \Bigl[\frac1{\bar\sigma_+^2} - \frac1{\bar\sigma_-^2}\Bigr]
  + 2\varepsilon_kh\Bigl[\frac{a_{1+}}{\bar\sigma_+^3} + \frac{a_{1-}}{\bar\sigma_-^3}\Bigr]
  - 2h^2\Bigl[\frac{r_{k+1}}{\bar\sigma_+^3} - \frac{r_k}{\bar\sigma_-^3}\Bigr]
  + 3h^2\Bigl[\frac{a_{1+}^2}{\bar\sigma_+^4} - \frac{a_{1-}^2}{\bar\sigma_-^4}\Bigr] + O(h^3).
\]
The first bracket is the centred difference $h(1/\bstar{}'^2)'(v_k) +
O(h^3)$, the second is $4\varepsilon_kha_1(v_k)/\bstar{}'(v_k)^3 + O(h^3)$, and
the third and fourth are $h^2$ times differences across one step of
functions with $O(h)$ variation, $O(h^3)$. Multiplying by
$-\tfrac12G_1(v_k)(s(v_k) + O(h^2))$ gives $\mu_k = fh[\tau_F -
\varepsilon_k\zeta_F](v_k) + O(h^3)$, since $\tau_F = sG_1\bstar{}''/(f\bstar{}'^3)$
and $\zeta_F = 2sG_1a_1/(f\bstar{}'^3)$ by the definitions of
Section~\ref{sec:general-23}. The second-order terms cancel by the centring
alone, and the smooth part enters through $r$ only, at $O(h^3)$. (b) On
piece $j$, $c_j = \bar c + \varepsilon_jhc_1 + O(h^2)$ with $\bar c := G_1/\bstar{}'
- \tfrac12(G_1s)'/\bstar{}'^2 = fW$ (the identity $G_1s/\bstar{}'^2 =
fF^{n-1}/\bstar{}'$ from the ODE gives $w_F = G_1/\bstar{}' -
\tfrac12(G_1s/\bstar{}'^2)'$ and $(G_1s/\bstar{}'^2)' = (G_1s)'/\bstar{}'^2 -
2G_1s\bstar{}''/\bstar{}'^3$, so $w_F - f\tau_F = \bar c$ at every $n$) and $c_1
= \partial c_j/\partial\sigma_j$ at $\bstar{}'$ times $a_1$. For a linear $\eta$
on the piece and $g \in C^2$, $\int_{\text{piece}}g\eta^2 = hg(v_{j-\frac12})T_j/3
+ h^2g'(v_{j-\frac12})(\eta_j^2 - \eta_{j-1}^2)/12 + O(h^3)(\eta_{j-1}^2 +
\eta_j^2)$: the linear term of $g$ about the midpoint integrates to zero
against the cross term by symmetry and to $\pm h^2/12$ against the two
squares. Applied to the frozen-slope integrand (midpoint value $fW$,
derivative $\mathcal G$) and, with the factor $\varepsilon_jh$, to $c_1$, this
is (b); applied to $f$ it is (c).
\end{proof}

Given the slopes to second order, the knot masses have no first-order
correction at all, and the piece integrand's first-order correction is
alternating with a bounded smooth amplitude: harmless for slowly varying
envelopes, where it cancels across each cell, and paid by rough ones out of
their kinetic energy. This is the mechanism of the lower bound below.

\paragraph{The harmonic quotient with free ends.}

\begin{lemma}\label{app:I3}
Let $\vartheta, a > 0$, $\lambda := \sqrt{\vartheta a}$, $\varsigma :=
(\vartheta/a)^{1/4}$, and $\ell \ge x_1\varsigma$ with $x_1 \ge 3$. Then for every
$\chi \in H^1(-\ell, \ell)$, with no condition at either end,
\[
  \int_{-\ell}^{\ell}[\vartheta\chi'^2 + ax^2\chi^2]\,dx \;\ge\; \lambda(1 - \delta)\int_{-\ell}^{\ell}\chi^2\,dx,
  \qquad \delta := x_1^2e^{-x_1^2}.
\]
\end{lemma}

\begin{proof}
Scale to $\vartheta = a = 1$. Let $\delta \in (0, \tfrac12)$ and $\varphi$ the
even solution of $-\varphi'' + x^2\varphi = (1 - \delta)\varphi$ with $\varphi(0)
= 1$, $\varphi'(0) = 0$. Then $\varphi > 0$ on $\mathbb R$ by Sturm comparison
with the ground state $e^{-x^2/2}$, whose energy $1$ exceeds $\varphi$'s (a
zero of $\varphi$ would give a Neumann--Dirichlet eigenvalue of $-d^2 + x^2$
below the whole-line ground energy); $\varphi'' = (x^2 - 1 + \delta)\varphi > 0$
for $|x| > 1$ and $\varphi' \to +\infty$ since $\varphi$ is not the ground
state, so $\varphi'$ has exactly one positive zero $x_{\min}$, with
$x_{\min}^2 = \log(1/\delta) + \log x_{\min} + \log(4/\sqrt\pi) + o(1)$ from the
expansion $\varphi = e^{-x^2/2} + \delta\psi_1 + O(\delta^2)$, $\psi_1 \sim
(\sqrt\pi/4)e^{x^2/2}/x$; hence $x_{\min} \le x_1$ for $\delta = x_1^2e^{-x_1^2}$,
$x_1 \ge 3$. Picone's identity on $(-\ell, \ell)$,
\[
  \int_{-\ell}^{\ell}[\chi'^2 + (x^2 - 1 + \delta)\chi^2]\,dx
  = \int_{-\ell}^{\ell}\bigl(\chi' - \chi\varphi'/\varphi\bigr)^2dx + \Bigl[\frac{\varphi'}{\varphi}\chi^2\Bigr]_{-\ell}^{\ell} \;\ge\; 0,
\]
since $\varphi'(\ell) \ge 0$ and $\varphi'(-\ell) = -\varphi'(\ell)$ for $\ell \ge
x_{\min}$.
\end{proof}

\paragraph{Proof of Theorem~\ref{thm:interior}, given Theorem~\ref{app:H2}.}
\label{app:proof-interior}
Throughout, $\lambda := h\sqrt{\kappa E''/2}$ with $\kappa := \kappa(\hat v)$, $E''
:= E''(\hat v)$, $a := E''/2$, $\vartheta := \kappa h^2$, $\varsigma :=
(\vartheta/a)^{1/4} \asymp h^{1/2}$, $x := v - \hat v$, $K := C_1\log^{1/2}(1/h)$
with $C_1$ large, $\ell := K\varsigma$, and $\mathcal W$ the pieces with $|v_j -
\hat v| \le \ell$. $A_\pm$, $t$, $g$, $p$ denote the values at $\hat v$ ($A_+A_-
= g^2/4 > 0$ by $\kappa(\hat v) > 0$); the knots are $1$-knots or $t$-knots
by the profile's parity.

\emph{Upper bound.} Let $\chi(v) := \exp(-x^2/(2\varsigma^2))$, the Gaussian
ground state of $-\vartheta d^2 + ax^2$ with eigenvalue $\lambda$, cut off at
$|x| = \ell$ (where it is $\le h^2$). Define $\xi_k := \chi(v_k)$ at the
$1$-knots and, at each $t$-knot $j$ between the $1$-knots $j \pm 1$, the
harmonic-optimal value $\xi_j := t(v_j)[\chi_{j-1} + (w_-/(w_+ +
w_-))(\chi_{j+1} - \chi_{j-1})]$ with $w_+ = A_+(v_j)$, $w_- = A_-(v_j)t(v_j)^2$,
the equality case of the cell inequality (i) in the proof of
Theorem~\ref{thm:airy}. By the decomposition, Lemma~\ref{app:I2} and that
equality case,
\[
  Q_m(\xi) = \sum_jf_jhE(v_j)\frac{T_j(\xi)}3 + \sum_{\text{cells}}f_jh\frac{\kappa(v_j)}6(1 + t_j + t_j^2)(\chi_{j+1} - \chi_{j-1})^2
  + \sum_j\varepsilon_jf_jh^2w_1(v_j)\frac{T_j(\xi)}3 + O(h^2)M(\xi),
\]
$w_1 := c_1/f$, and $M(\xi) = \sum_jf_jhT_j(\xi)/3 + O(h^2)M(\xi)$ by
Lemma~\ref{app:I2}(c) with its telescoping ($\xi$ has no boundary terms).
On $\mathcal W$ the data vary by $O(|x_j|)$ and $E(v_j) = E_{\min} + ax_j^2 +
O(|x_j|^3)$; for the Gaussian these variations are $O(\varsigma)$ and
$O(\varsigma^3)$ on average, so they cost relative $O(\varsigma)$ on the
$\lambda$-level terms and $O(h^{3/2})M$ absolutely. The mass, the potential
sum and the kinetic sum are Riemann sums on the lattice of $1$-knots of
smooth functions of scale $\varsigma$, each $f(\hat v)(1 + t + t^2)/3$ times the
continuum integral to relative $O(h/\varsigma)$; the alternating term is a
sum over cells of differences $g_j - g_{j+1}$ of a $C^1$-modulated sequence
times $\chi^2$, bounded by $C(h/\varsigma)\sum\chi_j^2$ through the Gaussian's
first absolute moment, i.e.\ $O(h^{3/2})M$. Hence
\[
  \frac{Q_m(\xi)}{M(\xi)} = E_{\min} + \frac{\int[ax^2\chi^2 + \kappa h^2\chi'^2]}{\int\chi^2}\,(1 + O(\varsigma) + O(h/\varsigma)) + O(h^{3/2})
  = E_{\min} + \lambda + O(h^{3/2}),
\]
and $c_m \le Q_m(\xi)/M(\xi)$.

\emph{Lower bound.} Fix $v_0 > 0$ with $[\hat v - v_0, \hat v + v_0] \subset
(0,1)$ and $E - E_{\min} \ge (E''/4)x^2$ there, and $c_0 := \min_{|x| \ge
v_0}(E - E_{\min}) > 0$.

\emph{Step 1 (the exact decomposition and the far pieces).} By the
decomposition, Lemma~\ref{app:I2} on the compact and the coefficient
asymptotics of Theorem~\ref{thm:oscillation2} elsewhere ($O(h)$ for $v \ge
v_0' > 0$, $O(h^{(4-n)/3})$ uniformly; at $n = 4$, $O(h\log m)$ inland and the
layer's first $K_0$ knots as in the proof of Theorem~\ref{thm:airy}), for
every $\eta$ with $\eta_0 = 0$,
\[
  Q_m(\eta) = \sum_jf_jh\Bigl[E(v_j)\frac{T_j}3 + P_j^{(v_j)}(\eta)\Bigr]
  + \sum_{|x_j| \le v_0}\varepsilon_jf_jh^2w_1(v_j)\frac{T_j}3 + \tfrac12A_m(1)\eta_m^2 - \tfrac12f_mh(\tau - \zeta)(1)\eta_m^2 + e(\eta),
\]
with $|e(\eta)| \le Ch^2\sum_{|x_j| \le v_0}f_jhT_j + Ch\sum_{\text{rest}}f_jhT_j$
(the uniform remainder on the far pieces), $P_j^{(v)} \ge 0$ the local
piece forms with the knot masses split in halves at the local values
(Lemma~\ref{app:I2}'s second-order terms telescope to $O(h^2)\sum_jh\eta_j^2$
plus two boundary terms at $|x| = v_0$, which the far slack absorbs), and
likewise $M(\eta) = \sum_jf_j(h/3)T_j + O(h^2)\sum_{|x_j| \le v_0}h\eta_j^2 +
O(h)M_{\mathrm{far}}$. Dropping the nonnegative $P_j$ outside $\mathcal W$,
the far pieces pay: for $|x_j| \ge v_0$, $E(v_j) - E_{\min} \ge c_0 \ge
2Ch^{(4-n)/3} + 2\lambda$; for $\ell < |x_j| < v_0$, $E(v_j) - E_{\min} \ge
(E''/4)\ell^2 = K^2\lambda/2 \ge 2\lambda + 2Ch$ once $K^2 \ge 4 +
4C/\sqrt{\kappa E''/2}$; the alternating term outside $\mathcal W$ is at most
$(\lambda/2)M_{\mathrm{far}}$; the top knot's terms are far. It remains to
show
\begin{equation}
\label{eq:window}
  \sum_{j \in \mathcal W}f_jh\Bigl[(E(v_j) - E_{\min})\frac{T_j}3 + P_j^{(v_j)}(\eta)\Bigr]
  + \sum_{j \in \mathcal W}\varepsilon_jf_jh^2w_1(v_j)\frac{T_j}3
  \;\ge\; \bigl(\lambda - Ch^{3/2}\log^{3/2}(1/h)\bigr)M_{\mathcal W}(\eta).
\end{equation}

\emph{Step 2 (the envelope, the cells, the alternating term).} Write $\eta =
\chi p$ with $\chi := \eta/p$ for the profile at $\hat v$ ($t \ne 0$), and pair
the pieces of $\mathcal W$ into cells with $1$-knots outside and a $t$-knot
in the middle; evaluating both piece forms of a cell at its middle knot (a
change of $O(h^2)$ in the coefficients), the cell inequality gives
$3[P_j^{(v_j)} + P_{j+1}^{(v_j)}](\eta) \ge (\kappa/2)(1 + t + t^2)(1 -
C\ell)(\chi_{j+1} - \chi_{j-1})^2$ with no mismatch term; a leftover piece at
the window's edge has $E(v_j) - E_{\min} \ge \lambda$ and is covered by its
potential term alone. The alternating term: summing the two pieces of a
cell, the $\chi_j^2$ parts differ by $O(h)\chi_j^2 + C|\chi_j -
\chi_{j+1}|(|\chi_j| + |\chi_{j+1}|)$ ($w_1 \in C^1$) and the $\Delta\chi$ parts
are bounded by $C(|\Delta\chi_j||\chi_j| + (\Delta\chi_j)^2)$, so with
$2|\Delta\chi||\chi| \le h^{-1/2}(\Delta\chi)^2 + h^{1/2}\chi^2$,
\[
  \Bigl|\sum_{\mathcal W}\varepsilon_jf_jh^2w_1\frac{T_j}3\Bigr|
  \le Ch^{5/2}\sum_{\mathcal W}\chi_j^2 + Ch^{3/2}\sum_{\mathcal W}(\Delta\chi_j)^2.
\]
The $(\Delta\chi)^2$ term is a relative $O(h^{1/2})$ of the kinetic sum and
is absorbed by a fraction $\theta := C'h^{1/2}$ of it through the single-piece
bound $P_j^{(v_j)}(\eta) \ge \tfrac12w_{\min}(\Delta\chi_j)^2 - C\ell^2(\chi_{j-1}^2 +
\chi_j^2)$, $w_{\min} := \min(A_+, A_-t^2) > 0$, at the cost
$O(K^2h^{3/2})M_{\mathcal W}$; the $\chi^2$ term is $O(h^{3/2})M_{\mathcal W}$.
So \eqref{eq:window} reduces to the same inequality without the
alternating term, with $\lambda - CK^2h^{3/2}$ in place of $\lambda$ and the
kinetic coefficient $(\kappa/2)(1 + t + t^2)(1 - C\ell - Ch^{1/2})$.

\emph{Step 3 (the dichotomy).} Let $r := \sum_{\mathcal W}(\Delta\chi_j)^2/
\sum_{\mathcal W}\chi_j^2$ and $R := R_0K^2$ with $R_0$ large. If $r \ge Rh$,
the kinetic sum alone is at least $[w_{\min}r/(3C(1 + r))]M_{\mathcal W} \ge
2\lambda M_{\mathcal W} + C\ell^3M_{\mathcal W}$ by the single-piece bound,
and \eqref{eq:window} holds. If $r < Rh$, then $\sqrt r \le CKh^{1/2}$ and by
Cauchy--Schwarz on the cross terms of $T_j(\chi p) = (1 + t + t^2)\chi_j^2 -
\Delta\chi_j\chi_j(2p_{j-1}^2 + p_{j-1}p_j) + (\Delta\chi_j)^2p_{j-1}^2$ the mass
is $(1 + t + t^2)\sum f_jh\chi_j^2(1 + O(Kh^{1/2}))/3$, a relative error that
multiplies $\lambda$ and is absolute $O(Kh^{3/2})$; the potential sum's
cross terms are bounded absolutely by $Ca\ell^2\sqrt r\sum f_jh\chi_j^2 =
O(K^3h^{3/2})M_{\mathcal W}$ and its $(\Delta\chi)^2$ terms by $O(K^4h^2)M_{\mathcal W}$.

\emph{Step 4 (the cell lattice and the continuum).} Index the cells by
their $1$-knots $x_c$ (spacing $2h$) and let $\tilde\chi$ be the
piecewise-linear interpolant of $\chi$ on that lattice, on $[-\ell', \ell']$,
$\ell' = \ell - O(h)$. The kinetic sum over the full cells, divided by
$f(\hat v)(1 + t + t^2)/3$, is at least $(1 - C\ell - Ch^{1/2})\kappa h^2\int\tilde\chi'^2$
exactly, since $\sum_c(\chi_c - \chi_{c-2h})^2/(2h) = \int\tilde\chi'^2$; the mass
is $\int\tilde\chi^2(1 + O(r)) + O(h)\chi^2_{\mathrm{edge}}$, the edge terms covered
by the slack of the two outermost cells' potential terms, which are kept out
of Lemma~\ref{app:I3} (applied on the window shortened by $4h$ at each
end); and the potential sum is at least $a\int x^2\tilde\chi^2 -
O(Kh^{3/2})M_{\mathcal W} - C\ell^3M_{\mathcal W}$, since on each cell
$\int x^2\tilde\chi^2 \le (x_c^2 + 5h\ell)h(\chi_c^2 + \chi_{c-2h}^2)$. Collecting,
\eqref{eq:window} follows in the second case from Lemma~\ref{app:I3} with
$\vartheta = \kappa h^2$, $a = E''/2$, $\ell' = (K - O(h/\varsigma))\varsigma$: $\delta
= K^2e^{-K^2} \le h^2$ for $C_1 \ge 2$. The error budget: $e(\eta)$ and the
mass's $f'$ term $O(h^2)M$; the alternating term $O(K^2h^{3/2})M$; the
dichotomy's $\sqrt r$ terms $O(K^3h^{3/2})M$; the Taylor term $\ell^3 =
O(K^3h^{3/2})$; the data's $(1 - C\ell)$ and the Riemann sum $O(Kh^{3/2})$;
Lemma~\ref{app:I3}'s $\delta$ $O(h^2)$. The two $K^3$ terms set the power of
the logarithm: $c_m \ge E_{\min} + \lambda - Ch^{3/2}\log^{3/2}(1/h)$. \qed

Beyond the ingredients of Theorem~\ref{thm:airy}'s proof (the positivity and
null relations of the piece forms, the cell inequality, the far pieces one
by one), the proof uses two things only: Lemma~\ref{app:I2}'s second-order
local data on a window of width $O(h^{1/2}\log^{1/2}(1/h))$ around $\hat v$,
which is where Theorem~\ref{app:H2} enters, and Lemma~\ref{app:I3} in place
of the Airy lemma. No first-order gradient term appears, because the
envelope identity makes the kinetic form exactly $(\Delta\chi)^2A_jp^2$ with
the local null profile and the potential exactly $E(v_j)T_j/3$ plus a
nonnegative local form; the first-order terms are either odd about $\hat v$
or sit on the $\lambda$-level terms as relative $O(\ell)$ factors, or are the
alternating integrand, which cancels across cells. The frozen profile
$p(\hat v)$ in place of the local one would cost a genuine first-order term,
which is why the local profile is used.

\paragraph{Proof of Theorem~\ref{app:H2} for two bidders.}
\label{app:proof-H2}
Throughout $n = 2$, $F \in C^6[0,1]$ (one derivative more than
Theorem~\ref{thm:oscillation2}, for the $h^6$ remainders of
Lemma~\ref{app:I4}), $\hat w := 1/F$, and
\[
  L_sU := \frac Fs(FU)', \qquad
  L_1A := \frac{F}{3fs^2}\bigl[(4Ff + 2Ff's - 9f^2s)A - 3fsFA'\bigr], \qquad
  \nu_3(A) := -\frac{4A^2F^3}{fs^2},
\]
\[
  \rho_3 := -\frac{Ff}{12s^2} + \frac{Ff''}{12} - \frac{Ff'^2}{12f} + \frac{f^2}{6s} - \frac{ff'}{12};
\]
the kernel of $L_1$ is spanned by $\hat Q$ (its logarithmic derivative is
$(\log\tilde Q)'$ by the ODE), that of $L_s$ by $\hat w$. $I = [v_0, 1]$ is a
fixed compact and "on $I$" means for knots $v_k \in I$.

\emph{The idea.} Theorem~\ref{thm:oscillation2} gives \eqref{eq:law} with
the law exact to relative $O(1/k) + O(h)$ and a remainder $O(h\omega)$. One
order more on compacts cannot be had by running its Newton sequence one
order deeper: the checkerboard part of the quadratic remainder contains
$h\hat A\cdot\Delta^2u$, which with the control $\Delta^2u = O(h^3)$ of that
theorem is only $O(h^4)$ and feeds a relative-$O(h)$ alternating response.
Instead, an explicit second-order ansatz $\tilde s$ with residual $O(h^5)$
on compacts is built from the exact pinned solutions of $J$ (so that its
alternating part is a homogeneous solution and its bottom row is
satisfied) and the regular solution $U_2$ of a smooth first-order ODE;
Newton--Kantorovich from $\tilde s$ and a two-pass pointwise bootstrap on
nested compacts, in which the bottom's contribution is a multiple of the
top-pinned solution, then give the three bounds. Three finite identities
make the ansatz second-order exact.

\begin{lemma}[the second-order two-scale identities, $n = 2$]\label{app:I4}
Fix a compact $I' \subset (0,1]$ and $K > 0$. Let $2 \le k \le m-1$ with $v_k
\in I'$, $\varepsilon = \pm1$, and
\[
  s_j = \bstar(v_j) + \varepsilon(-1)^{j-k}h^2A_j + h^2U_j, \qquad j = k-1, k, k+1,
\]
with three-point data $\mathcal A_0 := A_k$, $\mathcal A_1 := (A_{k+1} -
A_{k-1})/(2h)$, $\mathcal A_2 := \Delta^2A_k/h^2$ and $\mathcal U_0, \mathcal U_1,
\mathcal U_2$ likewise, all bounded by $K$. Then $\Phi_k(s) = \sum_{p=3}^5h^p[S_p
+ \varepsilon E_p] + O(h^6)$, with $S_p, E_p$ polynomials in the six data whose
coefficients are $C^1$ functions of $v_k$ on $I'$, the $O(h^6)$ uniform, and:
\begin{enumerate}[label=(\alph*),leftmargin=*]
\item $E_3 = L_1A$ with the data $(\mathcal A_0, \mathcal A_1)$; in particular
  $E_3 = 0$ when they are the values of $C\hat Q$ and $C\hat Q'$ at $v_k$.
\item $S_3 = L_sU + \rho_3 + \nu_3(\mathcal A_0)$ with the data $(\mathcal U_0,
  \mathcal U_1)$; $\rho_3$ is the residual density of the interpolant ($r_k =
  h^3\rho_3(v_k) + O(h^4)$) and $\nu_3$ the $\nu$ of Lemma~\ref{app:L6}(b).
\item $E_4 = 0$ and $S_4 = 0$ identically in all six data: the Galerkin row
  has no $h^4$ term at a two-scale vector.
\item (The top row.) For the half-hat row with $s_j = \bstar(v_j) +
  (-1)^{m-j}h^2A_j + h^2U_j$, $j = m-1, m$, and two-point data, $\Phi_m(s) =
  h^2T_2 + h^3T_3 + O(h^4)$ with
  \[
    T_2 = \frac{F^2}s\mathcal A_0 - \frac{Ff - s(2f^2 - Ff')}{12s}\ \text{ at } v = 1,
    \qquad T_3 = \tfrac12[L_1A + L_sU + \rho_3 + \nu_3(\mathcal A_0)](1):
  \]
  the half-hat row's $h^3$ coefficient is exactly half the interior one, so
  $T_2 = 0$ iff $\mathcal A_0 = \bstar{}''(1)/12 = \hat A(1)$, and with the law's
  data and a smooth part obeying its ODE at $v = 1$, $T_3 = 0$.
\item (Smooth vectors.) For $y_j = Y(v_j)$ with $Y \in C^3$, $(Jy)_k =
  hL_sY(v_k) + O(h^3\norm Y_{C^3})$ uniformly in $k$ (the $h^2$ coefficient
  vanishes), and $T(y) = \tfrac12hL_sY(1) + O(h^2)$; in particular $R_k(\hat w)
  = O(h^3)$ and $T(\hat w) = O(h^2)$.
\end{enumerate}
\end{lemma}

\begin{proof}
A finite computation (Appendix~\ref{app:numerics}). The row $\Phi_k(s) =
\int[fF - f^2(v - b)/b']\varphi_k\,dv$ is a sum over the two adjacent pieces
of $h$ times an integral over $\theta \in [0,1]$ of a function rational in
the piece slope (bounded below for $h \le h_0$); with $F, f, \bstar$ in
Taylor data at $v_k$ to the orders the $h^5$ coefficient needs and the knot
values as above, the expansion through $h^5$ is finite, and its remainder is
$O(h^6)$ uniformly because the row is a $C^6$ function of $(v_k, h)$ and the
data for $F \in C^6$. The $h^0, h^1, h^2$ coefficients vanish; after
substituting the ODE and its derivatives, the rest are as stated: (a), (b)
are read off; (c) is the computed vanishing of $E_4$ and $S_4$ with all six
data free; (d) is the same computation on the single piece $m$; (e) is the
part of (b), (c), (d) linear in the $\mathcal U$-data (the hat is symmetric,
so the piecewise-constant slope's $O(h)$ jump across $v_k$ enters the two
pieces with opposite signs), uniform in $k$ by Lemma~\ref{app:L1}; $L_s\hat w
= 0$.
\end{proof}

(d) is the fact that makes the law's constant exact to relative $O(h^2)$:
the half-hat row at order $h^3$ involves the smooth part only through
$L_sU(1) = -\rho_3(1) - \nu_3(1)$, so no nonlocal quantity enters and what is
left is a local identity. The same computation with the general-$n$ row
and the ODE $\bstar{}' = (n-1)fs/F$ gives the lemma verbatim at $n = 3$ and
$n = 4$, with $L_1A = (F^{n-1}/(3(n-1)fs^2))[(4Ff + 2Ff's - (5n-1)f^2s)A -
3fsFA']$, $L_sU = (F/((n-1)s))(F^{n-1}U)'$ (kernel $\hat w = F^{-(n-1)}$),
$\nu_3(A) = -4F^{n+1}A^2/((n-1)^2fs^2)$ and the corresponding $\rho_3$, $T_2$
(Appendix~\ref{app:numerics}).

\begin{lemma}[the exact modes to second order]\label{app:I5}
Let $\check y_k := \varepsilon_k\hat Q(v_k)$, $Y^{(1)}$ the solution of the
interior rows with $Y^{(1)}_{m-1} = \check y_{m-1}$, $Y^{(1)}_m = \check y_m$, and
$Y^{(2)}$ the solution with $Y^{(2)}_{m-1} = \hat w(v_{m-1})$, $Y^{(2)}_m =
\hat w(1)$. On $I$, uniformly for $h \le h_0$:
\begin{enumerate}[label=(\alph*),leftmargin=*]
\item $Y^{(1)}_k = \varepsilon_k\hat Q(v_k)(1 + h^2\alpha_k) + Y^{(1)\mathrm{rem}}_k$
  with $|\alpha_k| \le C$, $|\Delta\alpha_k| \le Ch$, $|\Delta^2\alpha_k| \le Ch^2$
  and $|Y^{(1)\mathrm{rem}}_k| \le Ch^{8/3}$.
\item $Y^{(2)}_k = \hat w(v_k)(1 + S'_k) + Y^{(2)\mathrm{alt}}_k$ with $|S'_k| \le
  Ch^2$, $|\Delta S'_k| \le Ch^3$, $|\Delta^2S'_k| \le Ch^4$ and
  $|Y^{(2)\mathrm{alt}}_k| \le Ch^3\hat Q_k$; hence $|\Delta Y^{(2)}_k| \le Ch$,
  $|\Delta^2Y^{(2)}_k| \le Ch^2$.
\item $\tilde W = \mathsf aY^{(1)} + B\tilde Z$ and $\tilde Z = \lambda(Y^{(2)} +
  c_hY^{(1)})$ with $|\mathsf a| \asymp 1$, $|B| \le Ch^{2/3}|\mathsf a|$, $\lambda =
  1 + O(h)$, $|c_h| \le Ch^2$.
\end{enumerate}
Consequently, on $I$, $\tilde Z = \tilde Z^{\mathrm{sm}} + \tilde Z^{\mathrm{alt}}$
with $\tilde Z^{\mathrm{sm}} := \lambda Y^{(2)}$ and $|\tilde Z^{\mathrm{alt}}_k| \le
Ch^2\hat Q_k$, and the alternating content of $\tilde W$ is
$\mathsf a\varepsilon_k\hat Q(v_k)(1 + h^2\alpha_k) + O(h^{8/3})\hat Q_k$.
\end{lemma}

\begin{proof}
Variation of constants: for a sequence $\check y$ with residuals
$\mathrm{res}_j := R_j(\check y)$, the solution of the interior rows with
$\check y$'s top two values is
\begin{equation}
\label{eq:VC}
  y_k = \check y_k + \sum_{j=k+1}^{m-1}(\tilde Z_k\tilde W_j - \tilde W_k\tilde Z_j)\frac{\mathrm{res}_j}{K_j}
  =: \check y_k + \tilde Z_k\mathcal S_k - \tilde W_k\mathcal T_k,
\end{equation}
each summand being a homogeneous solution vanishing at $k = j$ with $R_k$
of the sum equal to $-\mathrm{res}_k$ by the second form of $K_k$ in
Lemma~\ref{app:L2}. (a), crude form: by Lemma~\ref{app:I1} $\mathrm{res}_j =
(-1)^{m-j}R(v_j, h)$ with $R = h^3\hat Q\varrho + h^4R_4$, $\varrho \in C^1$; the
summands $\tilde W_j\mathrm{res}_j/K_j$ have the form $(-1)^jg_j$ with $|g_j| \le
Ch^3$ and $\sum_j|\Delta g_j| \le Ch^{8/3}$ (the relative differences of
$\tilde W_j$ are $O(h^{2/3})$ because of its smooth companion), so by Abel
summation $|\mathcal S_k| \le Ch^{8/3}$, and $\tilde Z_j\mathrm{res}_j/K_j$ is
one-signed of size $Ch^3$ with differences $O(h^4)$, so $|\mathcal T_k| \le
Ch^2$, $|\Delta\mathcal T_k| \le Ch^3$, $|\Delta^2\mathcal T_k| \le Ch^4$. (c):
$Y^{(1)}$ and $\tilde Z$ are independent solutions, so $\tilde W = \mathsf aY^{(1)}
+ B\tilde Z$, and the local decomposition on $I$ (Lemma~\ref{app:L2}:
$\tilde W$'s smooth content is at most $C(h/v)^{2/3}$ times its alternating
content) gives $|B| \le Ch^{2/3}|\mathsf a|$, $|\mathsf a| \asymp 1$; for $\tilde Z$,
$T(Y^{(1)}) = T(\check y) = \varepsilon_mr_F(1)\hat Q(1)(1 + O(h)) \ne 0$ (the
half-hat row of Appendix~\ref{app:oscillation}) while $T(Y^{(2)}) = T(\hat w)
= O(h^2)$ by Lemma~\ref{app:I4}(e), so $T(\tilde Z) = 0$ gives $c_h =
O(h^2)$ and $\lambda = 1 + O(h)$. (a), refined: substituting $\tilde W$ in
\eqref{eq:VC}, $Y^{(1)}_k(1 + \mathsf a\mathcal T_k) = \check y_k + \tilde Z_k(\mathcal S_k
- B\mathcal T_k)$, so $h^2\alpha_k := (1 + \mathsf a\mathcal T_k)^{-1} - 1$ has the
stated differences and the second term is $Y^{(1)\mathrm{rem}} =
O(h^{8/3})$. (b): $\mathrm{res}'_j := R_j(\hat w) = h^3\varrho_2(v_j) + O(h^4)$ by
Lemma~\ref{app:I4}(e); the $\mathsf a$-part of $\tilde W_j\mathrm{res}'_j/K_j$ is
one-signed of size $Ch^3$ with differences $O(h^4)$, giving $\mathcal S'_k$ with
$|\mathcal S'_k| \le Ch^2$ and differences $O(h^3)$, $O(h^4)$; the $B$-part is
alternating and Abel-sums to $O(h^{11/3})$; $\mathcal T'_k$ is an alternating sum
of a smooth sequence of size $h^3$. Then \eqref{eq:VC} with $\tilde Z =
\lambda(Y^{(2)} + c_hY^{(1)})$ gives $Y^{(2)}_k = \hat w_k/(1 - \lambda\mathcal S'_k) +
O(h^3\hat Q)$, which is (b).
\end{proof}

\emph{The ansatz.} Let $U_2$ be the solution of $L_sU_2 = -\rho_3 -
\nu_3(\hat A)$ on $(0,1]$ bounded at $0$,
\[
  U_2(v) = -\frac1{F(v)}\int_0^v\frac sF\bigl(\rho_3 + \nu_3(\hat A)\bigr)dv'.
\]
Since $\rho_3 \in C^3[0,1]$ (its two terms singular at $0$ combine to
$(f/(12s^2))(2fs - F)$ with $2fs - F = O(v^2)$) and $\nu_3(\hat A) = v^{1/3}\mathfrak r(v)$
with $\mathfrak r \in C^3[0,1]$, $U_2 = P + v^{1/3}R$ with $P, R \in C^3[0,1]$:
$|U_2^{(p)}(v)| \le Cv^{1/3-p}$ for $p = 1, 2, 3$. Define, with $\mathsf a, B$ from
Lemma~\ref{app:I5}(c),
\[
  \tilde C := \frac{\bstar{}''(1)}{12\mathsf a}, \quad \hat\kappa := \tilde CB, \quad
  \hat\sigma := -\frac{r_1 + (J(h^2U_2))_1}{R_1(\tilde Z)}, \quad
  \hat\delta := h^2\tilde C\tilde W + h^2U_2(v_\cdot) + \hat\sigma\tilde Z, \quad
  \tilde s := I_h\bstar + \hat\delta,
\]
so that $\hat\delta = h^2\tilde C\mathsf aY^{(1)} + h^2U_2 + (h^2\hat\kappa + \hat\sigma)\tilde Z$;
$R_1(\tilde Z) \asymp 1$, $|r_1| \le Ch^3$ and $|(J(h^2U_2))_1| \le Ch^3$ give
$|\hat\sigma| \le Ch^3$, and $|\hat\kappa| \le Ch^{2/3}$.

\begin{lemma}[the residual of the ansatz]\label{app:I6}
(i) $\hat\delta$ has the form Lemma~\ref{app:L5} assumes of the first
correction, and on $I$, $\hat\delta_k = \varepsilon_kh^2\hat A(v_k)(1 + h^2\alpha_k)
+ h^2U_2(v_k) + (h^2\hat\kappa + \hat\sigma)\tilde Z^{\mathrm{sm}}_k + \mathfrak r_k$ with
$|\mathfrak r_k| \le Ch^{14/3}$, $\alpha$ of Lemma~\ref{app:I5}(a).
(ii) On every compact $I' \subset (0,1]$: $|\Phi_k(\tilde s)| \le C(I')h^5$ for
$2 \le k \le m-1$ with $v_k \in I'$. (iii) $|\Phi_m(\tilde s)| \le Ch^4$.
(iv) For $2 \le k \le m-1$: $|\Phi_k(\tilde s)| \le C[h^4 + h^5v_k^{-5/3} +
h^4v_k^{-2/3}]$, so $|\Phi_k(\tilde s)| \le Ch^{10/3}$ for all $k \ge 2$ and
$\sum_{k \ge 2, v_k < v_1}|\Phi_k(\tilde s)| \le C(v_1)h^3$; and $\Phi_1(\tilde s) =
N_{I_h\bstar}(\hat\delta)_1 = O(h^{10/3})$. (v) $\norm{\Phi(\tilde s)}_\phi \le Ch^{1/3}$.
\end{lemma}

\begin{proof}
(i) Globally, $h^2\tilde C\tilde W$ is the law's vector up to normalization
(Step~0 of the proof of Theorem~\ref{thm:oscillation2}), and the rest of
the smooth part, $h^2U_2 + (h^2\hat\kappa + \hat\sigma)\tilde Z$, has increments
$\le (C/k)\omega_k$ ($|\Delta(h^2U_2)| \le h^3|U_2'| \le Chv^{-1/3}\omega$, and
$\tilde Z$'s increments are $O(h/v^2)$); on $I$ the fine form is
Lemma~\ref{app:I5}. (ii) Since $\tilde W$, $\tilde Z$ solve the interior rows
exactly, $\Phi_k(\tilde s) = r_k + (J(h^2U_2))_k + N_{I_h\bstar}(\hat\delta)_k$.
The linear part is $\Phi_k(I_h\bstar + h^2U_2) - N(h^2U_2)_k = h^3[L_sU_2 +
\rho_3](v_k) + O(h^5)$ by Lemma~\ref{app:I4} with no alternating part and the
three-point data of $U_2$ ($S_4 = 0$; $S_3$ has no term quadratic in
$\mathcal U$). The quadratic part is the degree-$\ge 2$ part of
Lemma~\ref{app:I4}'s expansion at any decomposition of $\hat\delta$'s three
knot values with bounded data, i.e.\ $h^3\nu_3(\mathcal A_0) + O(h^5)$ ($E_4 = S_4
= 0$), with $\mathcal A_0 = \hat A(v_k) + O(h^2)$ by (i). Summing, $\Phi_k(\tilde s) =
h^3[L_sU_2 + \rho_3 + \nu_3(\hat A)](v_k) + O(h^5) = O(h^5)$ by the ODE for
$U_2$. (iii) $T(\tilde Z) = 0$; the two-point data of $h^2\tilde C\mathsf aY^{(1)}$
at $m-1, m$ are the law's, and $U_2$ satisfies its ODE at $v = 1$, so
Lemma~\ref{app:I4}(d) gives $r_m + T(\cdot) + N_m(\cdot) = h^2\cdot0 +
h^3\cdot O(h) + O(h^4)$; the extra smooth piece $(h^2\hat\kappa +
\hat\sigma)\tilde Z = O(h^{8/3})$ enters the half-hat row's quadratic terms only
through products with the alternating slope, $O(h^{14/3})$. (iv) Row~1 is
$N_1(\hat\delta)$ by the choice of $\hat\sigma$ and $R_1(\tilde W) = 0$, bounded by
Lemma~\ref{app:L6}(a) at $k = 1$. For $2 \le k \le m-1$ write, using the
ODE, $\Phi_k(\tilde s) = [r_k - h^3\rho_3(v_k)] + [(J(h^2U_2))_k -
h^3L_sU_2(v_k)] + [N_{I_h\bstar}(\hat\delta)_k - h^3\nu_3(\hat A(v_k))]$: the
first bracket is $O(h^4)$ uniformly (the residual's expansion with one more
term, $F \in C^6$); the third is Lemma~\ref{app:L6}(b)'s remainder,
$\le Ch^4v_k^{-2/3}$; the second is the Taylor expansion of $a_kU_2(v_{k-1}) +
b_kU_2(v_k) + c_kU_2(v_{k+1})$ with the entries' expansions of
Lemma~\ref{app:L1}: the $h^3$ terms assemble to $h^3L_sU_2(v_k)$, the $h^4$
terms cancel (Lemma~\ref{app:I4}(e), uniformly in $k$), and the rest is
bounded by $Ch^2[h^3v_k|U_2'''| + h^3|U_2''| + h^3|U_2'| + h^3|U_2|]$ on
$[v_k/2, 3v_k/2]$, which is $\le Ch^5v_k^{-5/3}$ by the bounds on $U_2$ (the
factor $v_k$ from $r_F \asymp v$ in the coefficient of $U_2'''$). (v) follows:
$|\Phi_k(\tilde s)|/\phi_k \le Ch^{1/3}$ for $k \le m-1$ and $\le Ch^2$ at $k = m$.
\end{proof}

The smooth companion $B\tilde Z$ of the bottom-pinned solution is why the
ansatz is built on $\tilde W$ rather than on the law: the solution's smooth
part is $h^2U_2 + \kappa_hh^2/F$ with $\kappa_h \asymp h^{2/3}$, a non-analytic
term in $h$ forced by the bottom row, which a power
series in $h$ could not produce and whose residual at the bottom would not
be summable to $O(h^3)$.

\begin{lemma}[the top-pinned solution at the ansatz; the response on a compact]\label{app:I7}
Lemma~\ref{app:L5} applies at $\tilde s$ (its proof uses only the form of
$d_1$, which $\hat\delta$ has by Lemma~\ref{app:I6}(i)): with $\tilde Z_1$,
$\tilde W_1$, $K_1$ the pinned solutions and Casoratian of $J(\tilde s)$,
Theorem~\ref{app:T6} holds at $\tilde s$ with a constant $W_1$,
Lemma~\ref{app:L7} holds there, and the Green's function $G_1$ of $J(\tilde s)$
satisfies $|G_{1,kj}| \le C\hat Q_kv_j^{-2/3}$ for $j \ge k$, $|G_{1,kj}| \le
C\hat Q_kv_k^{-2/3}$ for $j < k$, $|G_{1,km}| \le C\hat Q_k$.
(a) On $I$, $\tilde Z_1 = \tilde Z_1^{\mathrm{sm}} + \tilde Z_1^{\mathrm{alt}}$ with
$|\tilde Z_1^{\mathrm{alt},k}| \le Ch\hat Q_k$, $|\Delta\tilde Z_1^{\mathrm{sm}}| \le Ch$
and $|\Delta^2\tilde Z_1^{\mathrm{sm}}| \le Ch^2$.
(b) For a forcing $f$, $v_1 \in (0, v_0)$ and $e := J(\tilde s)^{-1}f$: for
$v_k \in I$,
\[
  e_k = \tilde Z_{1,k}\sigma + e'_k, \qquad \sigma := \sum_{v_j < v_1}\frac{\tilde W_{1,j}f_j}{K_{1,j}}, \qquad
  |\sigma| \le C\sum_{v_j < v_1}|f_j|, \qquad |e'_k| \le C(v_1)\Bigl[\sum_{v_j \ge v_1, j < m}|f_j| + |f_m|\Bigr].
\]
\end{lemma}

\begin{proof}
(a) By Lemma~\ref{app:L5}(c)'s construction, $\zeta := \tilde Z_1 - \tilde Z =
\tilde Z\mathcal S - \tilde W\mathcal T$ with $\mathcal S_k = \sum_{j > k}\tilde W_jg_j/K_j$,
$\mathcal T_k = \sum_{j > k}\tilde Z_jg_j/K_j$, $g := -\delta J\tilde Z_1$, $\delta J :=
J(\tilde s) - J$, and the alternating part $\tilde W_k\mathcal T_k$ is $\le
C(h/v_k)|z_k|$. What has to be shown is $|\Delta\mathcal S_k| \le Ch^2$ and
$|\Delta^2\mathcal S_k| \le Ch^{8/3}$ on $I$, i.e.\ bounds on the size and the first
differences of the summands. By Lemma~\ref{app:L5}(b), $\delta J$ applied
to the alternating part of $\tilde Z_1$ (amplitude $\mathfrak a_j$, $|\mathfrak a_j| \le
Ch/v_j$, slowly varying) is one-signed of size $\le Ch^2\hat Q_j$ with
differences $O(h^3\hat Q_j)$, and applied to its smooth part (increments
$O(h)$, second increments $O(h^2)$) is $(-1)^jO(h^3\hat Q_j) + O(h^3)$,
because $a_j + c_j = O(h)$ and the row-sum defect's alternating part comes
from the two pieces' opposite alternating slopes against integrals that
differ by $O(h^2)$. So $g_j = \mathfrak g_j + (-1)^j\mathfrak g'_j$ with $|\mathfrak g_j| \le
Ch^2\hat Q_j$, $|\Delta\mathfrak g_j| \le Ch^3\hat Q_j$, $|\mathfrak g'_j| \le Ch^3\hat Q_j$,
and $\tilde W_j/K_j = [\mathsf aY^{(1)}_j + B\tilde Z_j]/K_j$ is a one-signed
sequence with relative differences $O(h)$ plus a $B$-part $O(h^{2/3})$:
the summands have size $O(h^2)$ and differences $O(h^{8/3})$. The fixed
point ($g$ is defined through $\tilde Z_1$) is Lemma~\ref{app:L5}'s
iteration, each step $O(h^{2/3})$ smaller, so the bounds pass to the
limit. (b) is the Green's function formula at $\tilde s$: for $v_j < v_1 <
v_0 \le v_k$ the entry is $\tilde W_{1,j}\tilde Z_{1,k}/K_{1,j}$ with $|\tilde W_{1,j}/K_{1,j}|
\le C$; the rest is the bound on $G_1$ with $v_j \ge v_1$.
\end{proof}

\emph{Step 0 (the base point).} By Lemma~\ref{app:I6}(i) and
Lemma~\ref{app:L5}, $\norm{J(\tilde s)^{-1}}_{\phi\to\omega} \le W_1$; by
Lemma~\ref{app:L6}(c), $J$ is Lipschitz with constant $L$ on the $C_0$-ball,
which contains $\tilde s$ and the Kantorovich ball below; by
Lemma~\ref{app:I6}(v), $\mathsf a_0 := \norm{J(\tilde s)^{-1}\Phi(\tilde s)}_\omega \le
W_1Ch^{1/3}$.

\emph{Step 1 (Newton--Kantorovich from $\tilde s$).} The Kantorovich number
$W_1L\mathsf a_0 \le CW_1^2Lh^{1/3}$ is $\le \tfrac12$ for $h \le h_0$, so Newton
from $\tilde s$ converges to a zero $b'$ with $\norm{b' - \tilde s}_\omega \le
Ch^{1/3}$. To identify $b'$ with $\bstar_m$, the uniqueness of
Theorem~\ref{thm:oscillation2} is used: $\bstar_m$ is the only zero in the
$\omega$-ball of radius $1/(2W_1L)$ around $s_2$, so it suffices that
$\norm{\tilde s - s_2}_\omega \le Ch^{1/3}$. The top column's response
$-r_mJ^{-1}e_m$ is exactly a multiple of $\tilde W$ whose top amplitude is
the law's to relative $O(h)$, so it differs from $h^2\tilde C\tilde W$ by $O(h)$
in $\omega$-norm; and $\mathfrak d := -J^{-1}r^{\mathrm{int}} + d_2 - h^2U_2 -
\hat\sigma\tilde Z$ satisfies
\[
  J\mathfrak d = -\bigl[r^{\mathrm{int}} + h^3\nu_3(\hat A)\mathbf 1_{2..m-1} + J(h^2U_2) + \hat\sigma R_1(\tilde Z)e_1\bigr]
  - \bigl[N_{I_h\bstar}(d_1) - h^3\nu_3(\hat A)\mathbf 1_{2..m-1}\bigr] - (J(s_1) - J)d_2,
\]
where the first bracket has row $1$ zero (the choice of $\hat\sigma$), rows
$2 \le k \le m-1$ equal to the first two brackets of
Lemma~\ref{app:I6}(iv), $O(h^4 + h^5v_k^{-5/3})$, and row $m$ equal to
$T(h^2U_2) = O(h\phi_m)$, of $\phi$-norm $\le Ch^{1/3}$; the second is
Lemma~\ref{app:L6}(b)'s remainder for $d_1$, of $\phi$-norm $\le Ch^{1/3}$;
the third is $O(h)$ in $\phi$-norm by Lemma~\ref{app:L6}(c)'s second form,
since $d_2$ has local contents $\le Chv^{-1/3}$ (Step~1 of the proof of
Theorem~\ref{thm:oscillation2}). By Theorem~\ref{app:T6}, $\norm{\mathfrak d}_\omega
\le Ch^{1/3}$, so $b' = \bstar_m$ for $h \le h_0$. Set $e := \bstar_m -
\tilde s$: $\norm e_\omega \le Ch^{1/3}$ and
\begin{equation}
\label{eq:E}
  J(\tilde s)e = -\Phi(\tilde s) - N_{\tilde s}(e).
\end{equation}

\emph{Step 2 (the quadratic term).} Lemma~\ref{app:L6}(a) at the base
$\tilde s$ gives, with $\theta_k, \sigma_k$ the local alternating content and
slow variation of $e$, $|N_{\tilde s}(e)_k| \le Ch^3[(\theta_k + \sigma_k)^2v_k\hat Q_k^2
+ h\hat Q_k^2(\theta_k + \sigma_k)\norm e_\omega]$ for $k \le m-1$ and
$|N_{\tilde s}(e)_m| \le Ch\phi_m(\theta_m + \sigma_m)^2$, and $\theta_k + \sigma_k \le
C\max_{|j-k| \le 2}|e_j|/\omega_j$. Hence if $|e_j| \le Mh^p\omega_j$ for all
$j$, then $|N_{\tilde s}(e)_k| \le CM^2h^{3+2p}v_k^{1/3}$ for $k \le m-1$ and
$|N_{\tilde s}(e)_m| \le CM^2h^{3+2p}$.

\emph{Step 3 (the global pointwise bound).} By \eqref{eq:E}, the bound on
$G_1$ and Lemma~\ref{app:I6}(ii)--(iv) with $I' = [v_0/16, 1]$,
\[
  |e_k| \le C\hat Q_k\Bigl[\sum_{v_j < v_0/16}v_j^{-2/3}\bigl(h^4 + h^5v_j^{-5/3} + h^4v_j^{-2/3}\bigr) + \sum_{v_j \ge v_0/16}v_j^{-2/3}h^5 + h^4\Bigr]
  + C\hat Q_k\sum_jv_j^{-2/3}|N_{\tilde s}(e)_j| + C\hat Q_k|N_{\tilde s}(e)_m|.
\]
The first bracket is $\le Ch^{8/3}$ (the sums $\sum_j(jh)^{-2/3}, (jh)^{-7/3},
(jh)^{-4/3}$ are $\le Ch^{-1}, Ch^{-7/3}, Ch^{-4/3}$); with $p = 1/3$ from
Step~1, Step~2 gives $\sum_jv_j^{-2/3}|N_j| \le Ch^{11/3}\sum_jv_j^{-1/3} \le
Ch^{8/3}$. So $|e_k| \le Ch^{2/3}\omega_k$ for all $k$: $p = 2/3$.

\emph{Step 4 (two passes on nested compacts).} Pass~1, on $I_1 := [v_0/5, 1]$
with $v_1 := v_0/8$ in Lemma~\ref{app:I7}(b): $e = \tilde Z_1\sigma_1 + e'_1$ with
$|\sigma_1| \le C[\sum_{v_j < v_0/8}|\Phi_j(\tilde s)| + \sum_{v_j < v_0/8}|N_j|] \le
C[h^3 + h^{13/3}h^{-1}] \le Ch^3$ and $|e'_{1,k}| \le C[\sum_{v_j \ge v_0/8}(h^5 +
h^{13/3}) + h^4 + h^{13/3}] \le Ch^{10/3}$. On $[v_0/4, 1]$ therefore $\theta_k +
\sigma_k \le C[|\tilde Z_1|h^3 + h^{10/3}]/h^2 \le Ch$, and by Step~2 (with $\norm
e_\omega \le Ch^{2/3}$) $|N_{\tilde s}(e)_k| \le Ch^5$ there, $k = m$ included.
Pass~2, on $I_2 := [v_0/2, 1]$ with $v_1 := v_0/4$: $e = \tilde Z_1\sigma + e'$
with $|\sigma| \le C[h^3 + h^{10/3}] \le Ch^3$ and, since on $[v_0/4, 1]$ now
$|N_j| \le Ch^5$, $|e'_k| \le C[\sum_{v_j \ge v_0/4}(h^5 + h^5) + h^4 + h^5] \le
Ch^4$. So on $I \subset I_2$: $e_k = \tilde Z_{1,k}\sigma + e'_k$, $|\sigma| \le
Ch^3$, $|e'_k| \le Ch^4$.

\emph{Step 5 (assembly).} On $I$, by Lemma~\ref{app:I6}(i),
Lemma~\ref{app:I5} and Step~4,
\[
  \delta_k = \varepsilon_kh^2\hat A(v_k)(1 + h^2\alpha_k) + h^2U_2(v_k) + (h^2\hat\kappa + \hat\sigma)\tilde Z^{\mathrm{sm}}_k + \mathfrak r_k
  + \tilde Z_{1,k}^{\mathrm{sm}}\sigma + \tilde Z_{1,k}^{\mathrm{alt}}\sigma + e'_k.
\]
Define $\rho_k$ by $\varepsilon_kh^2\hat A(v_k)\rho_k := \varepsilon_kh^2\hat A(v_k)h^2\alpha_k
+ \tilde Z_{1,k}^{\mathrm{alt}}\sigma$ and $u_k :=$ the rest. Then $|\rho_k| \le
C[h^2 + h\hat Q_k\cdot h^3/(h^2\hat Q_k)] \le Ch^2$ by Lemma~\ref{app:I7}(a),
and $u_k = h^2U_2(v_k) + (h^2\hat\kappa + \hat\sigma)\tilde Z^{\mathrm{sm}}_k +
\tilde Z_{1,k}^{\mathrm{sm}}\sigma + O(h^4)$ has $|\Delta u_k| \le h^3|U_2'| + C(h^{8/3}
+ h^3)h + Ch^3 + Ch^4 \le Ch^3$ and $|\Delta^2u_k| \le h^4|U_2''| + C(h^{8/3} +
h^3)h^2 + Ch^2h^3 + Ch^4 \le Ch^4$, by Lemmas~\ref{app:I5}(b) and
\ref{app:I7}(a) and $U_2 \in C^3$ on $I$. This is Theorem~\ref{app:H2} at $n
= 2$. \qed

No information about the bottom rows beyond Theorem~\ref{thm:oscillation2}'s
is needed: the ansatz's error at the bottom is $O(\omega/k)$, as the
solution's own structure there, and only its sum enters, through the
top-pinned solution of $J(\tilde s)$, whose alternating content is $O(h)$
and whose smooth part has $O(h)$ first and $O(h^2)$ second differences
(Lemma~\ref{app:I7}(a)).

\paragraph{Three bidders.} The argument is the one above with the
three-bidder sizes of Appendix~\ref{app:global} ($\phi_k = h^3v_k^{2/3}$,
$\hat Q \asymp v^{-2/3}$, $\hat w = F^{-2}$, $c_k \asymp v_k^2$), the constants of
Lemma~\ref{app:I4} at $n = 3$, and two genuine changes at the bottom knots.
First, the smooth limit $U_2 = -(1/F^2)\int_0^v(2s/F)(\rho_3 + \nu_3(\hat A))$
is unbounded there, $U_2 = P + v^{-1/3}R$ with $P, R \in C^3[0,1]$ (in
general $U_2 - P \asymp v^{(5-2n)/3}$), still inside the weight: $h^2|U_2(v_k)|
\le Cv_k^{1/3}\omega_k$. Second, the ansatz's residual on the first rows is of
the size of the forcing weight $\phi$ itself ($|\Phi_k(\tilde s)| \le
C(h/v_k)\phi_k$ for $k \ge 2$ and $|\Phi_1(\tilde s)| \le C\phi_1$: the smooth
part's Taylor error, the quadratic remainder's slow-variation error and the
self-interaction at row~1 are all $\phi$-sized because $\phi$'s exponent
$(n-1)/3$ was chosen so that the checkerboard remainder is $\phi$-sized), so
$\norm{\Phi(\tilde s)}_\phi = O(1)$ and the Kantorovich step has to be run on
the response to the residual rather than on its $\phi$-norm. The Green's
function gives that response directly: for a forcing with $|f_j| \le
Mv_j^{a'}\phi_j$, $a' \in (-5/3, -1/3)$, and $|f_m| \le Mh\phi_m$, at any base
with Lemma~\ref{app:L2}'s sizes, the response satisfies $|\delta_k| \le
CMv_k^{a'+1/3}\omega_k$ with local contents $\le CMv_k^{a'+1/3}$ (the sums
$\hat Q_kMh^3\sum_{j \ge k}v_j^{a'-2/3}$ and $v_k^{-2}Mh^3\sum_{j < k}v_j^{a'+2/3}$
are each $\asymp Mv_k^{a'+1/3}\omega_k$); for $a' = -1$, $M = Ch$ this is
$Chv_k^{-2/3}\omega_k$, $O(h^{1/3})$ at the bottom knot and $O(h)$ on compacts.
With it, $\mathsf a_0 \le Ch^{1/3}$ and Step~1 closes, with the top entry's
response $O(h\omega)$ treated separately; the identification $b' = \bstar_m$ is
as at $n = 2$ with the second Newton iterate's local contents $\le
Chv^{-2/3}$. The one further change is in Step~3: splitting $e = -J(\tilde s)^{-1}\Phi(\tilde s)
- J(\tilde s)^{-1}N_{\tilde s}(e)$, the first part has local contents $\le
Chv_k^{-2/3}$ by the response bound and the second $\le Ch^{2/3}$ by the
inverse bound, so the local contents of $e$ decay like $hv_k^{-2/3} + h^{2/3}$,
which makes the bottom sum $\sum_{v_j < v_1}|N_{\tilde s}(e)_j| \le
C(v_1)h^{10/3}$ (the crude uniform bound would give only $h^{8/3}$, and the
alternating error on compacts only relative $h^{5/3}$). Steps~4 and 5 are
then as at $n = 2$ with $|\mathfrak r| = O(h^5)$, $|B| \le Ch^{4/3}$ (the smooth
companion of the bottom-pinned solution is $k^{-4/3}$ at $n = 3$) and
$|\hat\sigma| \le Ch^{11/3}$.

\paragraph{Four bidders.} At $n = 4$ the Green's function's bottom exponent
is $2(n-1)/3 = 2$, so the bottom knots' own response to a $\phi$-sized bottom
residual is of the size of the weight $\omega_1$ there, and the Kantorovich
step from the ansatz does not close. What replaces it is
Theorem~\ref{thm:layer} (or Theorem~\ref{thm:four}), which proves existence
from a different base and gives, more than existence, the local contents
of $\bstar_m - s^c$: $\le C/\max(k, K/2)$ for every $k$. Take as base
$\tilde s := \chi s_L + (1 - \chi)a$, the composite of the layer on the first
$K/2$ knots and the inland ansatz $a := I_h\bstar + h^2\tilde C\mathsf aY^{(1)} +
h^2U_2$ (no bottom-pinned companion and no $\hat\sigma\tilde Z$: their roles are
the layer's; $U_2 = P + v^{-1}R$ is now of the size of the weight at the
bottom, the far field of the layer's smooth part), and $e := \bstar_m -
\tilde s$. No Newton step is run: (i) $e$ has local contents $\le C/\max(k,
K/2)$ and $\norm e_\omega \le C$, from Theorem~\ref{thm:layer}'s Step~4 and
the structure of $a - s_1$ on the inland rows (the law to relative $O(h)$,
smooth parts with increments $\le (C/k)\omega_k$); (ii) the residual of the
base satisfies $\sum_{v_j < v_1, j < m}|\Phi_j(\tilde s)| \le C(v_1)h^3$ (the
layer's rows up to $C(1 + A_Fk)hk\phi_k$ on the verified block, the glue
rows crudely $\le Chv_k^3$, and $|\Phi_k(a)| \le C[h^4v_k + h^4 + h^5/v_k]$ for
$k > K$ by Lemma~\ref{app:I6}(iv)'s expansion with the $n = 4$ entries, in
which the $h^3$ remainder of the row sum carries no factor $v_k$),
$\sum_{v_j \ge v_1}|\Phi_j(\tilde s)| \le C(v_1)h^4$ and $|\Phi_m(\tilde s)| \le Ch^4$;
(iii) the quadratic term satisfies $|N_{\tilde s}(e)_k| \le Ch^4/\max(k, K/2)$,
so its bottom sum is $\le Ch^4\log(1/h)$; (iv) the response on a compact,
Lemma~\ref{app:I7}(b), holds at this base: the inland rows are of the type
of the matching's inverse bound, the bottom-pinned solution on the
verified block is the layer's variational solution, and the Casoratian's
anchor at the hand-off index is re-established for the pair (variational
solution, top-pinned solution of $J(\tilde s)$) exactly as for the composite,
the finitely many bottom ratios $\tilde W_{1,j}/K_{1,j}$ being bounded by the
verified constants. One pass of (iv) with $v_1 = v_0/2$ then gives $e_k =
\tilde Z_{1,k}\sigma + e'_k$ on $I$ with $|\sigma| \le Ch^3$ and $|e'_k| \le
C(v_0)h^4$, and Step~5 reads off the three bounds with
Lemma~\ref{app:I7}(a) at $\tilde s$ (the top-pinned solution is computed from
the top downward and sees the inland rows only). Under
Theorem~\ref{thm:four}'s hypothesis instead of the layer's, the same runs
with the second Newton iterate from the interpolant in place of $s_L$ on
the rows $k < K/2$, whose residual is $\le Ch^4/k$ row by row. \qed

\emph{Proof of Theorem~\ref{thm:interior} at $n = 3, 4$.} The proof above
is run with Theorem~\ref{app:H2} at $n = 3$ or $4$ in place of $n = 2$. Its
inputs are the decomposition with its knot masses and the piece forms of
Lemma~\ref{app:bloch}, which are defined by the local data at every $n$;
Lemma~\ref{app:I3} and the cell inequality, which are $n$-free;
Lemma~\ref{app:I2}, whose proof is general ($G_1 = (n-1)F^{n-2}f^2$ and
$\bar c = fW$ at every $n$ by the identity in its proof); and the
coefficient asymptotics outside the window: on every compact they are
second-order exact by Theorem~\ref{app:H2}, and below, at $n = 3$ the
uniform $O(h^{1/3})$ of Theorem~\ref{thm:oscillation2}, at $n = 4$ the law
within a relative $C(F)(1/k + h\log m)$ down to $K_0h$ and, on the first
$K_0$ knots, the layer's slope bounds alone (knot masses $\le Ch^3k^3$, piece
integrands $\le Ch^2K_0^2$), as in the proof of Theorem~\ref{thm:airy}; the
far slack $c_0$ pays them for $h \le h_0$, the near-window condition on $K^2$
being unchanged. Since $E(0) = 0$ at $n \ge 3$, a unique interior minimizer has
$E_{\min} < 0$: at three and four bidders the interior rate describes the
approach to a negative limit. \qed

\paragraph{The weighted geometry.}\label{app:proof-interiorw} The certificates with a positive limit
at three bidders are the weighted ones of Corollary~\ref{cor:weighted},
$c_m^\vartheta := \inf_\eta Q_m(\eta)/M^\vartheta(\eta)$ with $M^\vartheta(\eta) :=
\int\vartheta\eta^2f\,dv$.

Theorem~\ref{app:interiorw}, stated in Appendix~\ref{app:arc3}, gives the
rate there.

\begin{proof}[Proof of Theorem~\ref{app:interiorw}]
Nothing in the mechanism changes: the weight enters the mass, the
potential becomes $V := E - \mu_0\vartheta \ge 0$, whose double zero at $\hat v$
has curvature $V''(\hat v) = \vartheta E_\vartheta''$, and the packet's weighted mass
is $\vartheta(\hat v)$ times its mass, which turns $\lambda = h\sqrt{\kappa V''/2}$
into $\lambda_\vartheta := \lambda/\vartheta(\hat v)$. Upper bound: Theorem~\ref{thm:interior}'s
test vector with Lemma~\ref{app:I2}(c) applied to $\vartheta f$ in place of
$f$ gives $Q_m(\xi) - \mu_0M^\vartheta(\xi) = [\lambda + O(h^{3/2})]M(\xi)$ and
$M^\vartheta(\xi) = \vartheta(\hat v)M(\xi)(1 + O(\varsigma))$. Lower bound: a
fixed-scale IMS cutoff $\chi_0^2 + \chi_1^2 = 1$ at $v_0' < \hat v - v_0$
separates the bottom at a cost $O(h^2)M^\vartheta$ (the cross coefficients
of the piece forms are $\le Ch\vartheta_jf_j$ under either hypothesis, by
the proof of Corollary~\ref{cor:weighted} for $w_F/f$); on the bottom part
the bottom windows of Theorem~\ref{thm:limit2}'s proof (or
Corollary~\ref{cor:weighted}'s, relative to the weight, for $w_F/f$ at $n =
3$, where $E_\vartheta(0) = 1 > \mu_0$) give $Q_m(\chi_0\eta) - \mu_0M^\vartheta(\chi_0\eta)
\ge (c_0/2)M^\vartheta(\chi_0\eta)$, $c_0 := \min_{|v - \hat v| \ge v_0}(E_\vartheta -
\mu_0) > 0$; on the upper part $\vartheta$ is bounded below and $\vartheta f \in
C^2$, and Theorem~\ref{thm:interior}'s lower bound runs with $V$ in place
of $E - E_{\min}$ (the far pieces paying $V(v_j) \ge c_0\vartheta(v_j) \ge
\vartheta_j(2\lambda_\vartheta + 2Ch)$, the near-window arithmetic with $a = \tfrac12V''(\hat v)$)
and Steps~2 to 4 verbatim, since they use of $E$ only $V(v_j) = ax_j^2 +
O(|x_j|^3)$ on the window; $M_{\mathcal W} = M^\vartheta_{\mathcal W}/(\vartheta(\hat v)(1
+ O(\ell)))$ converts the window inequality. The positivity clause: $\lambda_\vartheta
\asymp h > 0$, and $\mu_0 = 1 - \Delta_F$ by Corollary~\ref{cor:weighted}.
\end{proof}

The hypothesis $\kappa(\hat v) > 0$ is the unweighted band curvature: the
weight does not enter the kinetic term, whose origin is the slope
oscillation of the pieces, only the mass. At $n = 4$ the weight $w_F/f
\asymp v^2$ vanishes at the bottom, where the universal layer's own weighted
value decides, and no theorem is claimed.

\subsubsection{Positivity at every $m$ for the pilot prior}
\label{app:pilot}

\begin{proof}[Proof of Corollary~\ref{cor:pilot}]
Let $F = (v + v^2)/2$ and $n = 2$. For $h \le 1/40000$ the constants of
the Newton--Kantorovich step of Theorem~\ref{thm:oscillation2} are
verified (row~1 of the table below): the Kantorovich number is below
$\tfrac12$ from $m_0 = 2989$ on, so the Galerkin solution exists in the
verified class for every $m \ge 40000$, and on that class every piece
pencil of Proposition~\ref{prop:local} is enclosed with least eigenvalue
at least $0.4702$ (row~2), so $c_m \ge L_m \ge 0.470 > 0$. For $2 \le m
\le 40000$ the Galerkin solution is verified at each $m$ by
Newton--Kantorovich in balls and the local bound is evaluated from the
piece pencils on the ball (row~2), giving $c_m \ge 0.4969 > 0$. The two
ranges overlap, and no floating-point step is trusted in either.
\end{proof}

\medskip
\noindent\textbf{Verification for Corollary~\ref{cor:pilot}.}

\nopagebreak
{\footnotesize\setlength{\tabcolsep}{3pt}
\begin{tabular}{>{\raggedright\arraybackslash}p{0.13\textwidth}
                >{\raggedright\arraybackslash}p{0.15\textwidth}
                >{\raggedright\arraybackslash}p{0.13\textwidth}
                >{\raggedright\arraybackslash}p{0.26\textwidth}
                >{\raggedright\arraybackslash}p{0.13\textwidth}
                >{\raggedright\arraybackslash}p{0.07\textwidth}}
\toprule
Verified & Method & Inputs & Outputs & Code, records & Cost \\
\midrule
1. The constants of the existence theorem for every $h \le 1/40000$: the
inverse's norm, the Lipschitz constant, the residual class, the Abel
constants, the Kantorovich number &
interval arithmetic; the rows exact in $h$ below the index $K_0 = 100$,
the regime above by Taylor coefficients in $h$ with a Cauchy remainder
and enclosures of the alternating and smooth modes &
$F = (v+v^2)/2$, $n = 2$; Lemmas~\ref{app:L5}--\ref{app:L7} with their
constants (the remainder identities of Lemma~\ref{app:L6}
machine-checked, Appendix~\ref{app:lean}) &
$W \le 22.85$, $L \le 31.09$, $W_1 \le 30.87$ ($\bar\eta \le 0.268$,
$\bar\zeta \le 0.011$, $\bar\kappa \le 0.052$); $\sup|r|/h^3 \le 0.0139$;
the class of $\Phi(s_1)$: $M_1, M_\Delta \le 2.25\cdot10^{-4}$; the Abel
constants $C_7^{\mathrm{alt}} \le 54.74$, $C_7^{\mathrm{sm}} \le 81.47$;
$C_2 \le 0.01261$; $m_0 = 2989$; against the measured $W = 3.77$ and
$C_7 \le 1.1$ &
\path{verified/src/item103\_lemma3\_regimeII\_certified\_n2.py} to \path{verified/src/item111\_C2\_explicit\_n2.py} and their logs &
not recorded \\
\addlinespace
2. $c_m > 0$: the piece pencils enclosed on the verified class for
$h \le 1/40000$; the finite range $2 \le m \le 40000$ &
interval arithmetic on the exact tridiagonal pencils (the pieces
$j \le 100$ exactly in $h$, a scaled bottom box, $373$ $v$-boxes, the two
end pieces, both parities of the alternating knot masses); below $40000$
Newton--Kantorovich in balls at each $m$ with the local bound from the
pencils on the ball &
row~1's verified class; the Galerkin solution at each $m \le 40000$ &
$\lambda_j \ge 0.4702$ on every piece against $E_{\min} = 0.4985$, so
$L_m \ge 0.47$ for $m \ge 40000$ (the enclosures validated against the
float solution at $m = 256, 1024$); $c_m \ge 0.4969$ for $2 \le m \le
40000$, also for $v - v^2 + v^3$ and $F_{0.1}$ &
\path{verified/src/item112\_prop13c\_certified\_n2.py} (\path{verified/results/item112\_log.txt});
\path{verified/src/item101\_certified\_range\_n2.py} and its logs to $40000$ &
not recorded \\
\bottomrule
\end{tabular}}

\section{Machine-checked results}
\label{app:lean}

The statements of Sections~\ref{sec:exact}--\ref{sec:general} that are
finite algebra or standard real analysis have been checked with
the Lean~4 proof assistant \citep{moura2021lean} and its library Mathlib
\citep{mathlib2020}, version \MathlibVersion{}. The development consists of
\LeanFileCount{} files and \LeanLineCount{} lines, contains
\LeanTheoremCount{} machine-checked theorems and lemmas, compiles with no
unproved steps, and is the \path{lean/} directory of the repository of
Appendix~\ref{app:numerics}. What is
checked, and what each check takes as given, follows the chain of
Appendix~\ref{app:arcs}, arc by arc and entry by entry; the Lean names
are those of the \path{lean/FirstPriceCertificates/} modules.

{\footnotesize
\begin{longtable}{>{\raggedright\arraybackslash}p{0.26\textwidth}
                  >{\raggedright\arraybackslash}p{0.35\textwidth}
                  >{\raggedright\arraybackslash}p{0.30\textwidth}}
\toprule
Statement & Checked & Taken as given \\
\midrule
\endfirsthead
\toprule
Statement & Checked & Taken as given \\
\midrule
\endhead
\multicolumn{3}{l}{\emph{Arc I: the uniform prior}} \\
\addlinespace[2pt]
Lemma~\ref{lem:field}, the pointwise field & the derivative of $z \mapsto
  (v-z)W(z)^{n-1}$ wherever $W$ is differentiable, and the diagonal
  substitution $W(b(v)) = v$, $W'(b(v)) = 1/b'(v)$ (\path{lemma_field}); the
  field vanishes at $\bstar$ & that $W$ is Lipschitz and the exceptional
  null set pulls back to a null set in $v$ \\
Proposition~\ref{prop:minty}, the exact identity & the pointwise identity
  and its integral form (\path{minty_integrand}, \path{minty_pairing}) & the
  $\Ltwo$ pairing as the definition of $\Fop$ \\
Theorem~\ref{thm:modulus}, the exact quadratic form & the exact $t^2$/$t^3$
  split of the integrand, the integration by parts for $C^1$ directions
  with $\eta(0) = 0$, $Q_n$ by collecting terms, $Q_2 \ge \norm\eta^2$ with
  its equality case (\path{secondOrder_eq_Q}, \path{Q_zero_eq_iff}); the
  same identity for piecewise-linear directions through the model, by the
  integration by parts on each piece and telescoping
  (\path{secondOrder_eq_Q_pl}); for $n \ge 3$, $Q_n(h_1) \le \frac n2
  (2/m)^{n-2}\norm{h_1}^2$ with $\norm{h_1}^2 = 2/(3m) > 0$ for the hat at
  the first knot (\path{Q_firstHat_le}); the remark's $m = 1$ ratio, $Q_2(cv) =
  4\int(cv)^2$ (\path{Q_linear_ratio}) & the identification of the $t^2$
  coefficient with $\ip{D\Fop(\bstar)\eta}{\eta}$; the extension to $H^1$ \\
The knot identity \eqref{eq:knots} & on a model of piecewise-linear
  functions on the $m$-grid (knot values, slopes, the function and its
  slope as functions of $v$, the split of $[0,1]$ into pieces), the
  identity exactly as displayed, from Proposition~\ref{prop:minty} by the
  integration by parts on each piece and summation by parts
  (\path{knots_identity}); $\norm\eta^2$ as the sum of the piece integrals
  & the $\Ltwo$ pairing as the definition of $\Fop$ \\
Proposition~\ref{prop:ray}, the hat & in full on the model: the pairing of
  the hat equals $\tfrac12$ of the paper's bracket, with its exact zero,
  sign and norm (\path{hat_pairing_eq_zero_iff}, \path{hat_pairing_pos_iff},
  \path{hat_pairing_neg_iff}, \path{hat_normSq}) & nothing beyond the
  pairing's definition \\
Proposition~\ref{prop:radius}, the radius lower bound
  (Appendix~\ref{app:lower}) & in full on the model, line by line:
  $\norm\eta_\infty^2 \le 4m\rho^2$, $\sum_j \eta_j^2 \le 3m\rho^2$,
  $|\eta'_j| \le 4m^{3/2}\rho$, $w \in [\tfrac43, 4]$, $|w_{j+1} - w_j| \le
  128 m^{3/2}\rho$, the knot-term and positive-part bounds, and
  $\ip{\Fop(\bstar+\eta)}{\eta} > 0$ for every piecewise-linear $\eta$ with
  $0 < \norm\eta < m^{-5/2}/288$ (\path{field_pairing_pos_of_small}); the upper
  constant: the ray's radius is $(\sqrt{2/3}/6)\,m^{-5/2}(1 - 1/m)^{-1} \le
  (\sqrt{2/3}/3)\,m^{-5/2}$ for $m \ge 2$ (\path{ray_radius_eq},
  \path{ray_radius_le}) & nothing beyond the pairing's definition \\
Proposition~\ref{prop:notextremal}, the hat is not extremal & the rational
  counterexample at $m = 2$ through \eqref{eq:knots}
  (\path{hat_not_extremal_two}) & nothing beyond the knot identity \\
The leading-order problem \eqref{eq:leading} & at two bidders the Minty
  integrand along $t\eta$ is exactly $t^2\cdot2(\eta^2 + v\eta\eta') -
  t^3\cdot4(\eta^2 + v\eta\eta')\eta' + t^4R$ with $R = 4(\eta^2 +
  v\eta\eta')\eta'^2/(\tfrac12 + t\eta')$ (\path{leading_expansion_two}), and
  the $t^3$ term integrates to $-4[\int v\eta\eta'^2 + \tfrac13\eta(1)^3]$ by
  $\int\eta^2\eta' = \eta(1)^3/3$ (\path{integral_sq_mul_deriv},
  \path{leading_cubic_coefficient}) & that the $t^4$ term is $O(t^4)$
  uniformly on the ball; the grid form of the problem (the expansion of
  the knot identity's inverse-slope factors) and the reduction to $J/\nu^{3/2}$ \\
Proposition~\ref{prop:stationary}, the stationary profile & $q^2 = 4q - 1$;
  the mass stencil and $\partial J/\partial x_j$ vanish off the peak
  (\path{stencil_zero}, \path{Jgrad_zero}); at the peak Euler's identity,
  $x_0\partial J/\partial x_0 = 3J$ and $x_0\partial\nu/\partial x_0 = 2\nu$
  (\path{euler_J}, \path{euler_nu}); the one-sided geometric sums $\sqrt3/6$
  and $1/\sqrt3$ (\path{nu_side_hasSum}, \path{J_side_hasSum}); $J^2 =
  4\sqrt3\,\nu^3$ and the constant's square $1/(64\sqrt3)$; the pinned
  profile's $\nu = (2 + \sqrt3)/6$, $J = \tfrac12 + 1/\sqrt3$ and
  $\nu^{3/2}/(4J) = (1 + \sqrt3)/24$ (\path{nuPinned_hasSum},
  \path{JPinned_hasSum}, \path{pinned_constant}) & the leading-order
  reduction \eqref{eq:leading} \\
Theorem~\ref{thm:sharp}, the sharp inequality & in full: the piece cost
  $\varphi$ and the identity $3\nu - J = \sum\varphi$, the potential $W =
  \min(V_{\mathrm g}, V_{\mathrm j})$ with its branches, the four polynomial
  cases of the Bellman inequality as exact identities in $\mathbb{Q}(\sqrt3)$
  followed by sign bounds (\path{case_gg}, \path{case_jg}, \path{case_gj},
  \path{case_jj}, \path{bellman}), the telescoping along the two chains from
  a peak (\path{chain_telescope}, \path{chains_from_peak}), the sup-norm form
  by scaling and the tangent bound (\path{sup_norm_form},
  \path{tangent_bound}), the inequality itself for every finitely supported
  sequence (\path{sharp_inequality}), and the equality case in the sup-norm
  form: a zero-slack step from $(a_0, 1]$ goes to $-qa$, so at equality the
  peak is $+1$ and both chains are geometric (\path{zero_slack_step},
  \path{equality_case}) & the reduction of equality in $J \le 2\cdot3^{1/4}
  \nu^{3/2}$ to the sup-norm form and the scaling \\
Theorem~\ref{thm:radius}, the radius limit & the free-end inequality
  \eqref{eq:freeend}: $W(a) \le (1 + \sqrt3/6)a^2$ on $[-1,1]$ as two exact
  factorizations, the chains from a peak with free ends, the scaled
  sup-norm form with end terms, and \path{free_end_inequality} for any bound
  on the values; two sum identities of the expansion: $\sum_j
  \varepsilon_jI_j = \tfrac23\sum_j(x_j - x_{j-1})(x_{j-1}^2 + x_{j-1}x_j + x_j^2)$
  telescopes to $\tfrac23x_m^3$ and the knot sum $S = \sum v_jc_j$ splits as
  $2J_v + \tfrac1m\sum(x_i - x_{i-1})x_{i-1}^2 - x_m^2(x_m - x_{m-1})$
  (\path{cubic_telescope}, \path{knot_sum_split}) & the rest of the
  expansion of the knot identity, the Abel step and the constants \\
Theorem~\ref{thm:direction}, the extremal direction & the chain-stability
  lemma with the theorem's constant $\kappa = 0.056$ (\path{chain_stability}): the four cases
  quantitatively, the bad-step characterization, the predecessor bounds
  and the pairing, by strong induction; the slack identity $\sum\varphi =
  2W(z_{j_0}) + \sum_{\mathrm r}s + \sum_{\mathrm l}s - W(z_0) - W(z_N)$ with
  $\Gamma \ge 2$ at a $-1$ peak and $\Gamma \ge 0.056\,(D_{\mathrm r} +
  D_{\mathrm l})$ at a $+1$ peak (\path{slack_identity},
  \path{Gamma_ge_defects_of_pos_peak}) & the scaling to $\mu$, the slack
  budget and the tails \\
\addlinespace[4pt]
\multicolumn{3}{l}{\emph{Arc II: the learners}} \\
\addlinespace[2pt]
Proposition~\ref{prop:twoplayer}, the two-bidder form, interior part & the
  implicit derivative $u'(0) = 2(\eta_1 - \eta_2)$ of the inverse bid from
  $u/2 + \varepsilon\eta_2(u) = v/2 + \varepsilon\eta_1(v)$
  (\path{inverse_deriv_of_identity}); the interior field vanishes at
  $\varepsilon = 0$ and its $\varepsilon$-derivative there is $4\eta_1 - 2\eta_2 +
  2v\eta_2'$ (\path{interior_field_deriv}); pairing and integrating
  $2v(\eta_1\eta_2)'$ by parts gives $4\int(\eta_1^2 + \eta_2^2) -
  6\int\eta_1\eta_2 + 2\eta_1(1)\eta_2(1)$ (\path{interior_form}) & the
  inverse function theorem for $u$; that the $O(\varepsilon^2)$ remainders
  integrate to $O(\varepsilon^2)$; the cap term \\
Corollary~\ref{cor:twoplayer}, the constant & the eigenvectors of
  $\bigl(\begin{smallmatrix} 4 & -3\\ -3 & 4\end{smallmatrix}\bigr)$,
  $4(x^2+y^2) - 6xy \ge x^2 + y^2$ with equality iff $x = y$, the integrated
  form and both inequalities for $B$ (\path{B_ge}); the attainment: on the
  grid with $m \ge 2$ pieces the symmetric pair of first hats has zero top
  value and $B = \norm{(h_1, h_1)}^2 > 0$ (\path{B_attained}); the remark:
  on $(\eta, -\eta)$ the interior form is $14\int\eta^2 - 2\eta(1)^2$,
  negative for the top hat once $m \ge 3$ (\path{antisym_interior},
  \path{antisym_top_hat_neg}) & Proposition~\ref{prop:twoplayer} \\
Proposition~\ref{prop:nplayer}, the $n$-bidder form, interior part and
  eigenvalues & the interior field of bidder $i$ through the opponents'
  inverse bids $u_j$ and reciprocal slopes $1/b_j'(u_j)$ vanishes at
  $\varepsilon = 0$ and its $\varepsilon$-derivative there is
  $\frac{n}{(n-1)^2}v^{n-2}\sum_{j\ne i}(\eta_i - \eta_j) +
  \frac{n}{(n-1)^2}v^{n-1}\sum_{j\ne i}\eta_j' + nv^{n-2}\eta_i$
  (\path{field_zero}, \path{field_deriv}); pairing with $\eta_i$, summing
  over the bidders and integrating $v^{n-1}(\eta_i\eta_j)'$ by parts over
  the pairs gives the integral of the proposition plus the boundary term
  $\frac{n}{(n-1)^2}\sum_{i<j}a_ia_j$ (\path{interior_form_n}); the pointwise
  form dominates $\frac n2\sum_i\eta_i^2$ and equals $\frac{n^2}2\eta^2$ on
  symmetric directions, with the cap $n\cdot\frac{n}{2(n-1)}a^2$: $B_n =
  nQ_n$ there (\path{form_ge_half}, \path{form_symmetric},
  \path{cap_symmetric}) & the inverse function theorem for the $u_j$ and
  the two first-order expansions, as hypotheses; that the $O(\varepsilon^2)$
  remainders integrate to $O(\varepsilon^2)$; $C^1$ directions in place of
  piecewise-linear ones; the cap terms \\
Proposition~\ref{prop:generalform}, the form for a general prior & the
  symmetric eigenvalue of the integrand's matrix, $d + (n-1)o/2$, is
  exactly $w_F$ (\path{symmetric_eigenvalue}); under the equilibrium ODE the
  anti-symmetric one satisfies $\lambda_\perp - w_F = nF^{n-1}(F +
  fs)/(2(n-1)^2s^2)$ and $\lambda_\perp/w_F = ((2n-1)(n-1) +
  \bstar{}')/((n-1)(n-1-\bstar{}'))$ (\path{antisym_minus_weight},
  \path{antisym_ratio}); $c_F = f(1)/(2\bstar{}'(1)) = 1/(2(n-1)(1 -
  \bstar(1)))$ (\path{cF_closed_form}) & the expansion that produces $d$,
  $o$ and the cap term; the derivatives $K$, $A'$ as the ODE gives them \\
\addlinespace[4pt]
\multicolumn{3}{l}{\emph{Arc III: general priors and bidder counts}} \\
\addlinespace[2pt]
Proposition~\ref{prop:general}, the identity and the weight & the
  general-prior field in value coordinates and the exact identity
  $-g(\bstar+\eta)\,\eta f = (F^{n-1}\eta' + (n-1)F^{n-2}f\eta)\,\eta
  f/(\bstar{}' + \eta')$ under the equilibrium ODE
  (\path{general_minty_integrand}); the exact $t^2$/$t^3$ split; the
  integration by parts with the weight $fF^{n-1}/\bstar{}'$ and $Q_F$ by
  collecting terms (\path{secondOrder_eq_QF}); the second form of $w_F$
  through the ODE (\path{weight_second_form},
  \path{weight_coeff_second_form}); the closed form $w_F = F^{n-1}(F -
  fs)/(2(n-1)s^2)$ and the criterion $w_F < 0 \iff fs > F$
  (\path{weight_closed_form}, \path{weight_neg_iff}); the log-concavity
  bound, $\int_0^v F^{n-1} \le (F/f)(v)F(v)^{n-1}/(n-1)$ and $\bstar{}' \le 1$
  when $F/f$ is nondecreasing (\path{integral_pow_le_of_logconcave},
  \path{slope_le_one_of_logconcave}); the uniform prior recovers $Q_n$
  (\path{QF_uniform}) & the identification of the $t^2$ coefficient with
  $\ip{D\Fop(\bstar)\eta}{\eta}$; $\eta$ and $fF^{n-1}/\bstar{}'$
  differentiable with continuous derivatives, $\bstar{}' > 0$ everywhere, in
  place of the regularity hypotheses; the $L^2(dF)$ pairing as the
  definition of $\Fop$ \\
Theorem~\ref{thm:power}, the power priors, the algebra & for $F = v^a$ (as
  $fv = aF$) with $\kappa = a(n-1)$ and $s = v/(\kappa+1)$: the equilibrium
  ODE holds with $\bstar{}' = \kappa/(\kappa+1)$ (\path{power_ode}), $w_F/f =
  C_{a,n}v^{\kappa-1}$ with $C_{a,n}$ as displayed, for every $n$
  (\path{power_weight_over_f}), and $C_{a,2} = (a+1)/(2a)$, $C_{1,n} = n/2$,
  $C = (2n-3)/(n-1)$ at $\kappa = 1$ (\path{C_an_two}, \path{C_one_n},
  \path{C_kappa_one}) & that $\bstar$ linear forces a power prior; the
  Rayleigh quotient on $[0, 1/m]$ and the decay $m^{-(\kappa-1)}$ \\
The knot masses at $\bstar_m$ (Section~\ref{sec:general-pl}) & the exact
  decomposition of the second-order form at a piecewise-linear base point
  into the smooth part, the boundary terms and the knot masses
  (\path{knot_mass_identity}, by a piece-wise integration by parts with the
  weight $G_1(v-b)/\sigma_j^2$ and summation by parts) & the $L^2(dF)$
  pairing as the definition of $\Fop$ \\
Proposition~\ref{prop:oscillation}, the oscillation law: the
  Liouville--Green step & with $\lambda_k, \mu_k$ the roots of row $k$'s
  quadratic (Vieta) and $z_{k+1} = \lambda_kz_k$, the row residual of $z$ is
  exactly $-z_{k-1}(\lambda_{k-1} - \lambda_k)J_{k,k+1}\mu_k$
  (\path{liouville_green_residual}) & the consistency and entry expansions,
  the computer-algebra identity $\kappa = -(\log\tilde Q)'$
  (Appendix~\ref{app:numerics}), the roots' asymptotics and the assembly \\
Theorem~\ref{thm:oscillation2}, the two-scale form for two and three
  bidders: Lemmas~\ref{app:L2} and \ref{app:L6} & the two forms of the
  Casoratian agree across a row both modes solve, so $K_j =
  (a_j/c_{j-1})K_{j-1}$ (\path{casoratian_step}); the variation-of-constants
  sum \eqref{eq:VC} has row residual $-\mathrm{res}_k$ (\path{vc_row}: each
  summand a homogeneous solution vanishing at $k = j$); the integrands of
  the quadratic remainder $N_s(d)$ and of the Jacobian difference
  $(J(s+\varepsilon) - J(s))d$ on a piece are exact rational identities in the
  piece slope and the perturbation's value and slope
  (\path{remainder_identity}, \path{jacobian_difference}), and the
  remainder's two terms are bounded pointwise for $|D| < \sigma$
  (\path{remainder_bound}); the uniform prior's two-bidder rows $J_{k,k-1} =
  -h(3k-2)/3$, $J_{k,k} = 2h/3$, $J_{k,k+1} = h(3k+2)/3$ as hat integrals
  (\path{row_a_two}, \path{row_b_two}, \path{row_c_two}) & the sizes of
  Lemma~\ref{app:L2}, the Green's function as the full inverse, the
  integration against $f^2\varphi_k$ termwise, the slopes' nonvanishing on
  the $C_0$-ball, the Newton--Kantorovich step \\
The Bloch values and the limit function $E_F^{\mathrm{osc}}$ & $3g_- = 2\tau
  - \sqrt{\tau^2 + 3\zeta^2}$, with $\zeta = \tau\epsilon$ the value $\tau
  g_\mp(\epsilon)$ by the sign of $\tau$ (\path{gminus_scale_pos},
  \path{gminus_scale_neg}); the sharpness --- the two-periodic profile $1,
  t_*, 1, t_*, \dots$ annihilates every piece form and attains the constant
  (\path{pieceSum_twoPeriodic}) ---; $E_F^{\mathrm{osc}}$ as the smooth value
  plus the infimum (\path{Eosc_eq}) and its top value $w_F/f + (1 +
  2/\sqrt3)\tau_F$ for $\tau_F \le 0$ (\path{Eosc_top_concave}) & the
  coefficient asymptotics of Theorem~\ref{thm:oscillation2} that define
  $\tau_F$ and $\zeta_F$ \\
Theorem~\ref{thm:limit2}, the characterization theorem & the two-sided
  Bloch inequality (Lemma~\ref{app:bloch}) for every finite sequence and
  all real $\tau, \zeta$ (\path{bloch_lower}, \path{bloch_upper}): each
  piece's $2\times2$ form is positive semidefinite exactly at $g_-$
  (\path{det_eq}, \path{diag_nonneg}, \path{quad_nonneg}), and the
  piece-split sum equals the knot sum with end weights $\tfrac12$
  (\path{pieceSum_eq_knotSum}) & the coefficient asymptotics from
  Theorem~\ref{thm:oscillation2} and the localization argument \\
Corollary~\ref{cor:weighted}, the weighted geometry, the power-prior clause
  & a linear base point has equal piece slopes, so the knot masses of the
  decomposition vanish (\path{linear_slopes}, \path{knot_masses_vanish}),
  and the quotient of $\int w_F\eta^2 + c_F\eta(1)^2$ against $\int
  w_F\eta^2$ is at least $1$, exactly $1$ iff $\eta(1) = 0$
  (\path{weighted_quotient}) & that $\bstar$ solves the Galerkin equations
  exactly at a power prior; the limit statement \\
Proposition~\ref{app:bottom3}, the bottom-edge rate at three bidders: the
  uniform prior's rows & $\bstar = 2v/3$ solves the equilibrium ODE with $s =
  v/3$ and $w_F = 3v/2$ (\path{uniform_three_data}); $\int v\varphi_k^2 =
  (2k/3)h^2$, $\int v\varphi_k\varphi_{k+1} = ((2k+1)/12)h^2$ and the Gram
  integrals $2h/3$, $h/6$ (\path{hat_sq_weighted}, \path{hat_cross_weighted},
  \path{hat_gram_sq}, \path{hat_gram_cross}), so the pencil is $h^2\mathbb T$,
  $h\mathbb M$ exactly (\path{bottom_pencil_entries}) & the ground value
  $\lambda^*$ and its truncations (numerical, Appendix~\ref{app:numerics}), the
  comparison of the bottom part with the uniform problem and the spectral
  gap \\
Theorem~\ref{thm:layer}, the universal layer and the matching: the tail
  lemma's invariant region & the tail lemma's induction with the one-step
  map known through its suprema $G$, $H$ (verification row~1): under $D \ge
  (1 + 1/K)G$ and $2U \ge \frac{16}3\tilde P_\infty^2 + H/K$ the region
  $\{|\hat p| \le \hat P_K + D(1/K - 1/k), |\tilde u| \le U, |\tilde p| \le
  \tilde P_\infty\}$ is invariant for $k \ge K$ (\path{step_pHat},
  \path{step_u}, \path{invariant_region}) and the increments of $\hat p$
  telescope, $|\hat p_j - \hat p_k| \le D/k$ (\path{pHat_cauchy}) & the
  one-step map's rational form from the four-bidder rows (computer
  algebra, Appendix~\ref{app:numerics}) and its suprema (interval
  arithmetic); the root-selection step for $\tilde p_{k+1}$ (the
  contraction and connectedness argument), as a hypothesis; the Krawczyk
  sweep; the matching (Appendix~\ref{app:layer}): its chain and response
  identities, and the interval-verified root-variation and mode-size
  constants they are conditional on \\
Proposition~\ref{prop:reduction}, the bottom knots play no part & Abel's
  bound for alternating sums, $|\sum_{j\le N}(-1)^jx_j| \le |x_N| +
  \sum_{j<N}|x_{j+1} - x_j|$, Step~0 of the proof
  (\path{abel\_alternating\_bound}) & the bottom-window estimate and the
  limit argument \\
Proposition~\ref{prop:local}, the exact local bound & the piecewise
  quotient bound $\sum P_j \ge \lambda\sum M_j$ from $P_j \ge \lambda M_j$
  (\path{quotient\_ge\_min}); the pencil $(\mathrm{diag}(\tau+\zeta,
  \tau-\zeta), [\begin{smallmatrix}2&1\\1&2\end{smallmatrix}])$ has
  characteristic polynomial $3g^2 - 4\tau g + \tau^2 - \zeta^2$ with roots
  $g_\mp$ (\path{pencil\_det}, \path{pencil\_det\_gminus}); part~(b)'s pivot
  step: a pivot block is $(ox + dy)^2/d \ge 0$ for $d > 0$, and the
  tridiagonal form with its pivots taken from the top down is $d_1x_1^2$
  plus the blocks on the consecutive pairs, nonnegative when every pivot
  is positive (\path{pivot\_decomposition}, \path{pivot\_form\_nonneg}); the
  band's second derivative: at $\lambda = g_- + \tfrac13\kappa\theta^2$,
  $\cos\theta = 1 - \theta^2/2$ the pencil's determinant is $\theta^4$ times
  an explicit polynomial, the $\theta^2$ terms cancelling by the
  Hellmann--Feynman identity $g^2/(3g - 2\tau) = -gt/(1 + t + t^2)$ at the
  null profile (\path{eigen\_deriv\_identity}, \path{band\_curvature}) & the
  decomposition into piece pencils (the knot-mass identity with the
  halves); in part~(b) the supremum over divisions and its attainment; in
  the band expansion the implicit function theorem and $\cos\theta = 1 -
  \theta^2/2 + O(\theta^4)$ \\
Theorem~\ref{thm:interior}, the rate at an interior minimizer & $g_-t_- =
  \tau + \zeta - 2g_- \ge 0$, so the band curvature $\kappa \ge 0$
  (\path{gminus\_mul\_tminus}, \path{kappa\_nonneg}) & the asymptotics of
  the piece constants and the semiclassical derivation (the band's
  second-order expansion is in the row of Proposition~\ref{prop:local}); the interior constant (Appendix~\ref{app:interior}): the
  symbolic identities of Lemmas~\ref{app:I1} and \ref{app:I4} (computer
  algebra, Appendix~\ref{app:numerics}), the second-order local data, the
  free-end harmonic quotient and the rate theorem itself \\
Theorem~\ref{thm:airy}, the rate at a top-knot minimizer & the envelope
  identity $P(\chi_1a, \chi_2b) = (\chi_2 - \chi_1)^2Aa^2$ from the null
  relations (\path{envelope\_identity}); the harmonic-mean inequality $w_+x^2
  + w_-y^2 \ge (x+y)^2w_+w_-/(w_+ + w_-)$ (\path{harmonic\_mean\_bound}); the
  cell weights $w_+ + w_- = gt$, $w_+w_-/(w_+ + w_-) = gt/4$ and the cell
  kinetic bound $3[w_+\Delta_1^2 + w_-\Delta_2^2] \ge (3gt/4)(\Delta_1 +
  \Delta_2)^2$ with $3gt/4 = \tfrac12\kappa(1 + t + t^2)$
  (\path{cell\_weights}, \path{cell\_kinetic\_bound},
  \path{kinetic\_constant}) & that every knot vector is $\chi p$, the far
  pieces, the dichotomy, the Airy eigenvalue and the Riemann sums \\
Corollary~\ref{cor:eventual}, positive from some $m$ on & nothing (the
  statement is the convergence and the rates with their signs) & the
  theorems it cites; $m_0(F)$ is not explicit in general \\
Corollary~\ref{cor:pilot}, one prior at every $m$ & Lemma~\ref{app:L6}
  with its constants (the remainder identities and the pointwise bound,
  the row of Theorem~\ref{thm:oscillation2} above) & every verified
  constant (interval arithmetic, the verification table of
  Appendix~\ref{app:pilot}): the finite range $m \le 40000$, $m_0 = 2989$
  and $L_m \ge 0.47$ for $m \ge 40000$ by direct enclosure of the piece
  pencils; the finite range for two further priors \\
\bottomrule
\end{longtable}}

The entries of Appendix~\ref{app:arcs} without a row above, in the
chain's order. Not formalization targets: the leading-order problem's
grid-free reduction (which the stationary-profile and sharp-inequality
rows take as given; its cubic coefficient is checked above), the equality case of Theorem~\ref{thm:sharp} beyond
its sup-norm form, the analytic parts of Theorems~\ref{thm:radius} and
\ref{thm:direction}, the Newton--Kantorovich arguments of
Theorems~\ref{thm:oscillation2} and \ref{thm:four},
Conjecture~\ref{conj:knot}, the analytic part of Theorem~\ref{thm:airy}
(the coefficient asymptotics, which for $n = 3, 4$ rest on the second-order
amplitude law and for $n = 4$ on the verified layer, the dichotomy and
the Airy eigenvalue), and the vanishing-density coda. Analytic statements
not yet checked: the cap terms of Propositions~\ref{prop:twoplayer} and
\ref{prop:nplayer} (one-sided limits over the cap regions),
Proposition~\ref{prop:generalform} beyond its eigenvalues,
Theorem~\ref{thm:power} beyond its algebra,
Proposition~\ref{prop:oscillation} beyond its Liouville--Green step,
Proposition~\ref{prop:local}(b) beyond its pivot step, and
Theorem~\ref{app:interiorw}. The computer-assisted results are checked
only at the identities named above, their constants being interval
arithmetic: Theorem~\ref{thm:layer} (the tail lemma's inductions; the
one-step identities are computer algebra), Theorem~\ref{app:H2} (its
symbolic identities, Lemmas~\ref{app:I1} and \ref{app:I4}, are computer
algebra), Proposition~\ref{prop:layer5} (nothing checked),
Proposition~\ref{app:bottom3} (the uniform prior's rows and the pencil's
entries; the ground value is numerical), and
Corollary~\ref{cor:pilot} (Lemma~\ref{app:L6} with its constants).

\section{Computational records}
\label{app:numerics}

Every number in Sections~\ref{sec:exact}--\ref{sec:general} and in the
appendices was produced by code that is published, with its outputs and
run records, in the repository accompanying this paper,
\url{https://github.com/blubin/firstprice-certificates}. The repository's
\path{RECORDS.md} maps each statement of the paper that rests on, or is
illustrated by, a computation to the computation that produced its
numbers: what was computed, the numbers measured, and where the code, the
outputs and the run records are.

The repository holds the Lean development under \path{lean/}, the
verified computations of Appendix~\ref{app:proofs} under
\path{verified/} (the scripts, their outputs and the run records of
the verified constants), and the other computations under
\path{computations/}, one directory per computation with its code and
outputs and, where the paper draws a figure from it, the figure's
generator and data.

A pointer to this appendix in the body or in
Appendices~\ref{app:arcs}--\ref{app:lean} points at the entry of
\path{RECORDS.md} for the result named there. A number the paper calls
measured is a computation's output reported as evidence, not a proved
quantity; a number it calls verified is an interval-arithmetic
enclosure whose verification Appendix~\ref{app:proofs} tabulates. Every
result of the paper is proved in the paper.

\section{The computer's three roles}
\label{app:computer}

This appendix discloses the use of computer assistance in this
research. The authors' view is that the word ``disclosure'' is the
wrong one for something that should be reported with care rather than
hidden: there are now tools, both for the creation of new theory and
for checking it, that make the strongest and deepest results the joint
work of such tools and human insight. In a period awash in
low-quality machine-generated text there is a danger of judging a
careful use of these tools to reach otherwise unreachable insight with
the brush that rightly condemns their careless use. The authors hope
the reader will judge this paper on the former side of that line, and
this appendix says what was done so that the reader can.

Computation entered the \emph{research} of this paper, as distinct
from the editing of the manuscript, in three distinct ways: a language
model proposed theory, both statements and proofs, under the authors'
direction; interval arithmetic verified the finite computations that
several of the proofs take as inputs; and a proof assistant
machine-checked a large part of the proofs. The second and the third
are two kinds of checking, and the appendix refers to them together as
the checks. We discuss each of the three in turn.

\paragraph{Model-assisted theorizing.}
The theory of this paper was developed in conjunction with a language
model, Fable~5.1 \citep{anthropic2026fable}, by a method that such
models make possible and that may be of independent interest.

The authors posed the questions, fixed the objects (the operator, the
spaces, the certificates) and judged what to pursue; the model proposed
and critiqued formal statements and drafted multi-step proofs, drawing
on a wide array of mathematical areas to find solutions to the
problems posed. These areas lie well outside those that the literature
on learning in auctions normally invokes: the potential and Bellman
methods behind the sharp inequality of Arc~I, the discrete
Liouville--Green and Bloch-wave analysis of the oscillation, the IMS
localization and the Airy well behind the rates, Newton--Kantorovich in
weighted and class norms, and interval verification of a universal
layer. Finding that such tools apply to this problem, and how, is a
strength of the latest models: having both breadth and depth, they
find avenues through difficult problems that a human working alone
would be unlikely to find.

The checks establish that a proof is correct; they do not help with
knowing what should be proved in the first place. So even with the
power of the latest models the method requires care, and the care is
less about validity than about getting the theory to go where it
should. That part falls to the human and the agent together, and it
requires a process that lets each play to its strengths.

Specifically, each section of the work went through iterated phases,
the human and the agent working together: specification; numerical
falsification, that is, a search for a counterexample where one could
be reached; and, for what survived, elaboration into a candidate
proof. Next came a further attempt at falsification by a second agent
with a clean context. This step is not primarily about correctness,
which the checks establish later, but about speed: the second agent
finds problems quickly, before an expensive human read and before the
checks, verification for the finite computations and machine checking
for the identities and the leading-order inequalities.

Because the checks are strong, the real challenge for the human is to
find an arc through the domain of inquiry along which a sequence of
results makes sense, and then to work with the agent on the statements
and the analytic arguments of each proof, which can then be checked.
Appendix~\ref{app:arcs} lays out the arcs of the present work step by
step.

\paragraph{Computer-assisted proofs.} Three results
(Theorem~\ref{thm:layer}, Proposition~\ref{prop:layer5},
Corollary~\ref{cor:pilot}) have proofs in which one step is a finite
computation done in interval or ball arithmetic. Such a step is
\emph{not} a numerical experiment: it is a proof of a finite set of
inequalities, and it is as rigorous as the arithmetic library and the
script that calls it, both of which are public
(Appendix~\ref{app:numerics}). The verification table that ends each of
those proofs says what was verified, by what method, from what inputs,
with what outputs and at what cost; the records repository holds the
code and the logs. These computations are one step of each proof: the
analytic argument around the step is not verified by the computer and
is proved like any other statement. For more information about these
approaches see Appendix~\ref{primer:interval}.

\paragraph{Machine-checked statements.} A proof assistant, here
Lean~4 \citep{moura2021lean} with its library Mathlib
\citep{mathlib2020}, re-derives a statement from the axioms of the
library. It establishes exactly the statement as written: more than a
verified computation, since there is no arithmetic library to trust,
and less, since only what was formalized is checked, and a
formalization makes assumptions. Appendix~\ref{app:lean} lists, beside
each checked statement, the assumptions it takes as given. For more
information about what a Lean theorem establishes see
Appendix~\ref{primer:lean}.

\section{Primer}
\label{app:primer}

One entry for each mathematical concept and technique the paper uses:
the definition, what the paper takes from it, and a standard
reference. The arc appendix (Appendix~\ref{app:arcs}) refers here for
anything unusual; the entries are in the order the arcs meet them, and the
appendix is placed last so that the arcs and the proofs come first.

\subsection{Variational inequalities, the Minty pairing and monotonicity}
\label{primer:vi}

A variational inequality on a convex set $K$ in a Hilbert space with an
operator $\Fop$ asks for $\bstar \in K$ with $\ip{\Fop(\bstar)}{b - \bstar}
\ge 0$ for every $b \in K$ (the Stampacchia form). The Minty form asks
instead for $\ip{\Fop(b)}{b - \bstar} \ge 0$ for every $b \in K$. For a
continuous operator every Minty solution is a Stampacchia solution; the
converse holds when $\Fop$ is monotone, $\ip{\Fop(b) - \Fop(b')}{b - b'}
\ge 0$, and then the solution set is convex. Monotonicity comes in three
strengths: monotone, strictly monotone (the inequality strict for $b \ne
b'$, which gives uniqueness), and strongly monotone, $\ip{\Fop(b) -
\Fop(b')}{b - b'} \ge \mu\norm{b - b'}^2$ with $\mu > 0$, which gives
existence, uniqueness, Lipschitz dependence on the data and the
convergence of the projected gradient dynamics at a rate set by $\mu$. The
paper calls $\ip{\Fop(\bstar + \eta)}{\eta}$ the Minty pairing at the
perturbation $\eta$: the Minty form holds on a ball around $\bstar$ iff
the pairing is nonnegative there, and strong monotonicity on the ball iff
it dominates $\mu\norm\eta^2$. What the paper takes from this is the
first implication, Minty to Stampacchia, which lets a local Minty
inequality stand in for the monotonicity that the auction operator lacks.
\citet{kinderlehrer1980introduction} and \citet{facchinei2003finite}.

\subsection{The G\^ateaux derivative and the second-order coefficient of a pairing}
\label{primer:gateaux}

The G\^ateaux derivative of $\Fop$ at $\bstar$ in the direction $\eta$ is
$D\Fop(\bstar)\eta = \lim_{t \to 0}(\Fop(\bstar + t\eta) - \Fop(\bstar))/t$;
when the limit exists for every $\eta$ and is linear in $\eta$ it is the
Jacobian of Section~\ref{sec:certificates}. Its symmetric part is the
quadratic form $\eta \mapsto \ip{D\Fop(\bstar)\eta}{\eta}$, and because
$\Fop(\bstar) = 0$ in the pairing, this form is exactly the coefficient of
$t^2$ in $\ip{\Fop(\bstar + t\eta)}{t\eta}$. The paper computes that
coefficient directly from an exact formula for the pairing, which is why
it never needs the derivative to exist in every direction: the form is
defined wherever the expansion is, and extends by density to the space on
which it is continuous (Appendix~\ref{app:regularity}). When the limit
exists only as $t \downarrow 0$ the form is one-sided and positively
homogeneous of degree two but not quadratic, which is the learners' case
in Arc~II. \citet[Chapter~3]{ortega2000iterative}.

\subsection{Piecewise-linear spaces, hats, the Galerkin equations and the interpolant}
\label{primer:galerkin}

The $m$-piece space is spanned by the hat functions $h_k$ of the grid
$v_k = k/m$, $h_k$ being $1$ at $v_k$, $0$ at the other knots and linear
between, with the half hat at $v_m = 1$; a bid in the space is determined
by its knot values and has a constant slope $\sigma_j$ on each piece. The
Galerkin equations ask that the field, tested against every hat, vanish:
$\int \Fop(b)\,h_k\,dF = 0$ for $k = 1, \dots, m$, which is the first-order
condition of the discretized game and the equation the learners'
dynamics solve. The interpolant $I_h\bstar$ of the function-space
equilibrium matches it at the knots. What the paper takes from this is
that the interpolant is not a Galerkin solution: its residual is of second
order in $h = 1/m$, cancels against every full hat at leading order
because the interpolation error's slope is a sawtooth of mean zero on each
piece, but not against the half hat at the top, which only the slope of
the top piece can balance (Section~\ref{sec:general-pl}). The inverse
inequalities for piecewise-linear functions, $\norm\eta_\infty^2 \le
4m\norm\eta^2$ and $\norm{\eta'}_\infty \le 2m\norm\eta_\infty$ on the
$m$-grid, are the other tool (Appendix~\ref{app:lower}).
\citet[Chapters~0--4]{brenner2008mathematical}.

\subsection{Knot identities and knot masses}
\label{primer:knots}

Integrating the Minty pairing by parts on each piece, where the bid is
linear and the slope constant, and summing by parts across the knots
turns the pairing into a positive integral plus a boundary term at the
top minus a sum of terms at the interior knots, each proportional to the
knot value squared times the jump of a function of the slope across the
knot: the knot identity \eqref{eq:knots} for the uniform prior, and the
decomposition of Section~\ref{sec:general-pl} at a general base point. A
knot term is negative exactly where the slope drops, so it is called a
knot mass. What the paper takes from this is that the masses are of the
same order in $m$ as the $\Ltwo$ mass of a hat, so they compete with the
weight of the certificate on hat-like directions and decide its sign at
the discretized equilibrium. Both identities are machine-checked
(Appendix~\ref{app:lean}); the technique is the summation by parts of any
calculus text, applied piece by piece.

\subsection{Rayleigh quotients, matrix pencils and the $LDL^{\mathsf T}$ decomposition}
\label{primer:pencils}

For symmetric matrices $A$ and $M$ with $M$ positive definite, the pencil
$(A, M)$ has real eigenvalues, the stationary values of the Rayleigh
quotient $x^{\mathsf T}Ax/x^{\mathsf T}Mx$, and its least eigenvalue is the
quotient's infimum; the certificate's best constant $c_m$ is the least
eigenvalue of the pencil of the form against the mass. The paper takes
two things from this. First, a quotient of sums is at least the least of the
quotients: if $A = \sum_jP_j$ and $M = \sum_jM_j$ with every $M_j$
positive semidefinite, then $\lambda_{\min}(A, M) \ge
\min_j\lambda_{\min}(P_j, M_j)$, which is Proposition~\ref{prop:local}'s
local bound. Second, a positive definite tridiagonal matrix splits exactly
into positive semidefinite $2\times2$ blocks on the consecutive pairs plus
a positive $1\times1$ block, by the $LDL^{\mathsf T}$ pivots taken from the
top down, $d_{k-1} = a_{k-1,k-1} - a_{k-1,k}^2/d_k$; that is why the
supremum of the local bound over all divisions of the diagonal between
adjacent pieces is the best constant itself. \citet{parlett1998symmetric}
and \citet[Chapter~4]{golub2013matrix}.

\subsection{Two-periodic coefficients, Bloch waves and the band curvature}
\label{primer:bloch}

A quadratic form on sequences whose coefficients repeat with period two
is diagonalized by Bloch waves, sequences of the form $e^{ik\theta}$ times
a two-periodic profile, with the quasi-momentum $\theta \in [0, \pi]$; for
each $\theta$ the form restricted to such waves is a $2\times2$ pencil
whose two eigenvalues, as functions of $\theta$, are the two bands. The
infimum of the form over all finitely supported sequences, per unit mass,
is the bottom of the lower band, and it is attained in the limit by
two-periodic profiles, the waves at $\theta = 0$. The paper's Bloch
inequality (Lemma~\ref{app:bloch}) is this statement for the knot part
of the certificate, with the bottom of the band in closed form, $g_- =
(2\tau - \sqrt{\tau^2 + 3\zeta^2})/3$, and proved by exhibiting the form as
a sum of piece forms that are positive semidefinite exactly at $g_-$. The
curvature of the lower band at $\theta = 0$, $\kappa$ in
Section~\ref{sec:general-finite}, is the coefficient of $\theta^2$ in its
expansion; it is the kinetic coefficient the semiclassical rates need.
\citet[Section~XIII.16]{reed1978analysis}.

\subsection{Three-term recurrences: roots, discrete Liouville--Green, the Casoratian and the Green's function}
\label{primer:recurrences}

A three-term recurrence $a_ky_{k-1} + b_ky_k + c_ky_{k+1} = \psi_k$ is a
tridiagonal linear system; its homogeneous solutions form a
two-dimensional space. When the coefficients vary slowly with $k$ the two
roots $\lambda_k, \mu_k$ of the local characteristic quadratic
$c_k\lambda^2 + b_k\lambda + a_k = 0$ give approximate solutions, the
products $\prod_{j \le k}\lambda_j$ and $\prod_{j \le k}\mu_j$, with a
relative error controlled by the coefficients' increments: the discrete
Liouville--Green (or WKB) approximation. For the Galerkin Jacobian one
root is near $-1$ and gives the alternating mode, the other near $+1$ and
gives the smooth mode. Two solutions pinned at the two ends, $\tilde W$
(vanishing below the first row) and $\tilde Z$ (vanishing above the last),
have a Casoratian $K_j = c_j(\tilde W_j\tilde Z_{j+1} - \tilde W_{j+1}\tilde
Z_j)$, the discrete Wronskian, which satisfies $K_j = (a_j/c_{j-1})K_{j-1}$
and is nonzero iff the system is invertible; the inverse is then the
Green's function $(J^{-1})_{kj} = \tilde W_{\min(k,j)}\tilde
Z_{\max(k,j)}/K_j$. What the paper takes from this is that bounding the
inverse in a weighted norm reduces to three sums over the Green's function
weighted by the mode sizes, which converge for two and three bidders and
diverge logarithmically at four (Theorem~\ref{app:T6}).
\citet[Chapter~6]{olver1974asymptotics} for the continuous original and
\citet[Chapter~2]{elaydi2005introduction} for the recurrence.

\subsection{Abel summation and alternating sums of slowly varying sequences}
\label{primer:abel}

Abel summation is summation by parts for sequences: $\sum_{j \le N}a_jb_j
= A_Nb_N - \sum_{j < N}A_j(b_{j+1} - b_j)$ with $A_j = \sum_{i \le j}a_i$.
Two uses recur. With $a_j = (-1)^j$, $A_j \in \{-1, 0\}$, so an
alternating sum of a slowly varying sequence is bounded by the last term
plus the total variation, $|\sum_{j \le N}(-1)^jb_j| \le |b_N| + \sum_{j <
N}|b_{j+1} - b_j|$: the alternating mode's response to a slowly varying
forcing is small (Lemma~\ref{app:L7}), and the slope law sums to a
uniformly small knot error (Proposition~\ref{prop:reduction}, where the
bound is machine-checked). With $b_j = j/m$ a position, a
position-weighted sum is the average over all tails of the unweighted sum
of the tail, which is how the radius limit removes the position weight
from the knot sum and applies the sharp inequality to each tail
(Theorem~\ref{thm:radius}). \citet[Chapter~X, \S43]{knopp1951theory}.

\subsection{Newton--Kantorovich in weighted norms and in class norms}
\label{primer:newton}

The Newton--Kantorovich theorem proves the existence of a root of $\Phi$ near a point
$s$ from three numbers: a bound $W$ on the inverse Jacobian's norm, a
bound $\eta$ on the first Newton correction $J(s)^{-1}\Phi(s)$, and a
Lipschitz constant $L$ of the Jacobian on a ball around $s$. If the
Kantorovich number $WL\eta \le \tfrac12$, Newton's iteration from $s$
converges to a root, unique in a ball whose radius the three numbers give,
with the root within about $2\eta$ of $s$. The norms are the user's
choice, and the paper's choice is the point: on the knot error the weight
$h^2\hat Q(v_k)$ makes the oscillation law unit size, on the forcing the
weight $h^3$ times a power of position makes the residual unit size, so
that the theorem is a statement about $h$-independent constants. When no
contraction holds at the natural starting point the paper takes Newton
steps by hand first, tracking the structure of each correction, and
applies the theorem from the second iterate. Class norms
(Appendix~\ref{app:four}) are quotient norms over decompositions of a
forcing into a smooth part, an alternating part and a top entry, each with
its own weight, under which the inverse stays bounded at four bidders
where the plain weighted inverse does not.
\citet[Chapter~XVIII]{kantorovich1982functional} and
\citet[Chapter~12]{ortega2000iterative}.

\subsection{The IMS localization formula}
\label{primer:ims}

For a quadratic form of the kind $\int|\chi'|^2 + \int V|\chi|^2$ and a
smooth partition of unity $\sum_i\chi_i^2 = 1$, the form of $\eta$ equals
the sum of the forms of the pieces $\chi_i\eta$ minus the localization
error $\sum_i\int|\chi_i'|^2|\eta|^2$, which is small when the windows are
wide: the IMS formula, after Ismagilov, Morgan, Simon and Sigal. It
reduces a lower bound on a
form with varying coefficients to lower bounds on windows where the
coefficients are frozen, at a cost of the window width. The paper's
discrete version (Theorem~\ref{thm:limit2}, Step~2) has the knot and
boundary terms distributing exactly over the windows and the piece terms
paying $(h/\ell)^2$ times the mass for a window of width $\ell$, since
their cross-coefficients are $O(h)$. Proposition~\ref{prop:local}'s local
bound later removes the localization from the lower bound altogether; the
upper bound still uses a single window.
\citet[Section~3.1]{cycon1987schrodinger}.

\subsection{Semiclassical ground energies: the harmonic and the Airy well, Picone and Sturm}
\label{primer:semiclassical}

The ground-state energy of $-\kappa h^2\,d^2/dv^2 + E(v)$ as $h \to 0$ is
$\min E$ plus a correction set by the well's shape at the minimizer: at a
nondegenerate interior minimum, where $E = E_{\min} + \tfrac12E''(\hat
v)(v - \hat v)^2 + \dots$, the correction is $h\sqrt{\kappa E''(\hat v)/2}$,
the harmonic oscillator's ground energy, with a Gaussian ground state of
width $\sqrt h$; at a minimum on a Dirichlet wall with $E'(1) < 0$ the
correction is $a_1\kappa^{1/3}|E'(1)|^{2/3}h^{2/3}$, with $a_1 = 2.338\dots$
the first zero of the Airy function $\mathrm{Ai}(-x)$ and an Airy ground
state of width $h^{2/3}$. Upper bounds come from the Rayleigh quotient of
the model ground state under a cutoff; lower bounds from Picone's
identity, $\int[\vartheta\chi'^2 + V\chi^2] = \int\vartheta(\chi' -
\chi\varphi'/\varphi)^2 + [\vartheta(\varphi'/\varphi)\chi^2]$ for any
positive solution $\varphi$ of $-\vartheta\varphi'' + V\varphi = 0$, which
is nonnegative once $\varphi$ is chosen with the right sign of
$\varphi'$ at the free end; the positivity of $\varphi$ is Sturm's
comparison theorem against a solution of known sign. The paper's rates
(Theorems~\ref{thm:interior}, \ref{thm:airy}) are these two model
problems with $E = E_F^{\mathrm{osc}}$ and $\kappa$ the band curvature, the
grid spacing playing the role of $h$.
\citet{helffer1988semiclassical}, \citet[Chapter~XI, Section~3]{hartman2002ordinary},
and \citet[Section~9.9]{olver2010nist} for $a_1$.

\subsection{Potentials and Bellman inequalities along chains}
\label{primer:bellman}

To bound from below a sum of step costs $\sum_j\varphi(x_j, x_{j+1})$
along a chain, find a potential $W$ with $W(0) = 0$ and the Bellman
inequality $\varphi(a, b) + W(b) \ge W(a)$ for every step; then the sum
telescopes to at least $W(x_{j_0})$ for a chain starting at $x_{j_0}$ and
ending at $0$, and the inequality is sharp exactly along steps of zero
slack. The potential is the value function of the dynamic programming
problem "cheapest chain from $a$ to $0$", and it can be guessed from the
answer: the paper's $W$ is the minimum of the cost of a geometric tail
and the cost of jumping to the peak and continuing geometrically
(Theorem~\ref{thm:sharp}). The stability version keeps the slacks
instead of discarding them: a chain whose total slack is small has steps
close to the zero-slack recursion, and pairing each step whose own slack
is small with its predecessor gives a uniform constant
(Theorem~\ref{thm:direction}). Both are elementary once the potential is
found; the four polynomial inequalities that make the Bellman inequality
hold are machine-checked (Appendix~\ref{app:lean}).
\citet[Chapter~1]{bertsekas2017dynamic}.

\subsection{Interval arithmetic, ball arithmetic, the Krawczyk test and invariant regions}
\label{primer:interval}

Interval arithmetic evaluates an expression on intervals so that the
result encloses every value the expression takes on them, with the
floating-point rounding directed outward; ball arithmetic represents an
interval as a midpoint and a radius and is what the arb library
implements at arbitrary precision. An enclosure is a proof: whatever the
expression's value, it lies in the interval. The Krawczyk test verifies
a root of a system $G(x) = 0$ in a box $X$ around an approximate root
$\tilde x$ with an approximate inverse Jacobian $Y$: if the Krawczyk
operator $\tilde x - YG(\tilde x) + (I - YG'(X))(X - \tilde x)$, evaluated
in interval arithmetic, maps into the interior of $X$, then $X$ contains
exactly one root, by the Brouwer and contraction arguments made rigorous
by the enclosure; the parametric version carries a parameter interval
through the same test and gives the root's continuity in it. An invariant
region for a one-step map is a set that the map provably sends into
itself, established by finitely many inequalities between the region's
bounds and the suprema of the map's remainders over the region, which
interval arithmetic verifies; once closed it gives existence for all
steps by induction. The paper's verified computations (Appendix~\ref{app:layer},
\ref{app:pilot}) are of these two kinds, and the tables there say what
each verifies; how to read such a result as a whole is
Appendix~\ref{app:computer}. \citet{moore2009introduction}, \citet{rump2010verification},
\citet{krawczyk1969newton}, \citet{johansson2017arb}.

\subsection{Machine-checked proofs: what a Lean theorem establishes}
\label{primer:lean}

A theorem in Lean~4 over the Mathlib library is a statement whose proof
the kernel has checked from the library's axioms; what it establishes is
exactly its hypotheses and conclusion as written. Appendix~\ref{app:lean}
therefore lists, beside each checked statement, what the Lean statement
takes as given beyond the paper's words: a regularity hypothesis replaced
by differentiability everywhere, the pairing taken as a definition, a
verified constant entered as a hypothesis, a computer-algebra expansion
not reproduced. "Machine-checked in full" means the paper's statement
with no such substitution. A constant produced by interval arithmetic is
never a Lean fact; Lean checks the identities that the arithmetic hulls,
and the verification tables of Appendix~\ref{app:proofs} record the
arithmetic. Every number Appendix~\ref{app:lean} quotes is measured on
the repository's copy of the development.
\citet{moura2021lean} and \citet{mathlib2020}.

\end{document}